# A Community Plan for Fusion Energy and Discovery Plasma Sciences

Report of the 2019–2020 American Physical Society Division of Plasma Physics Community Planning Process

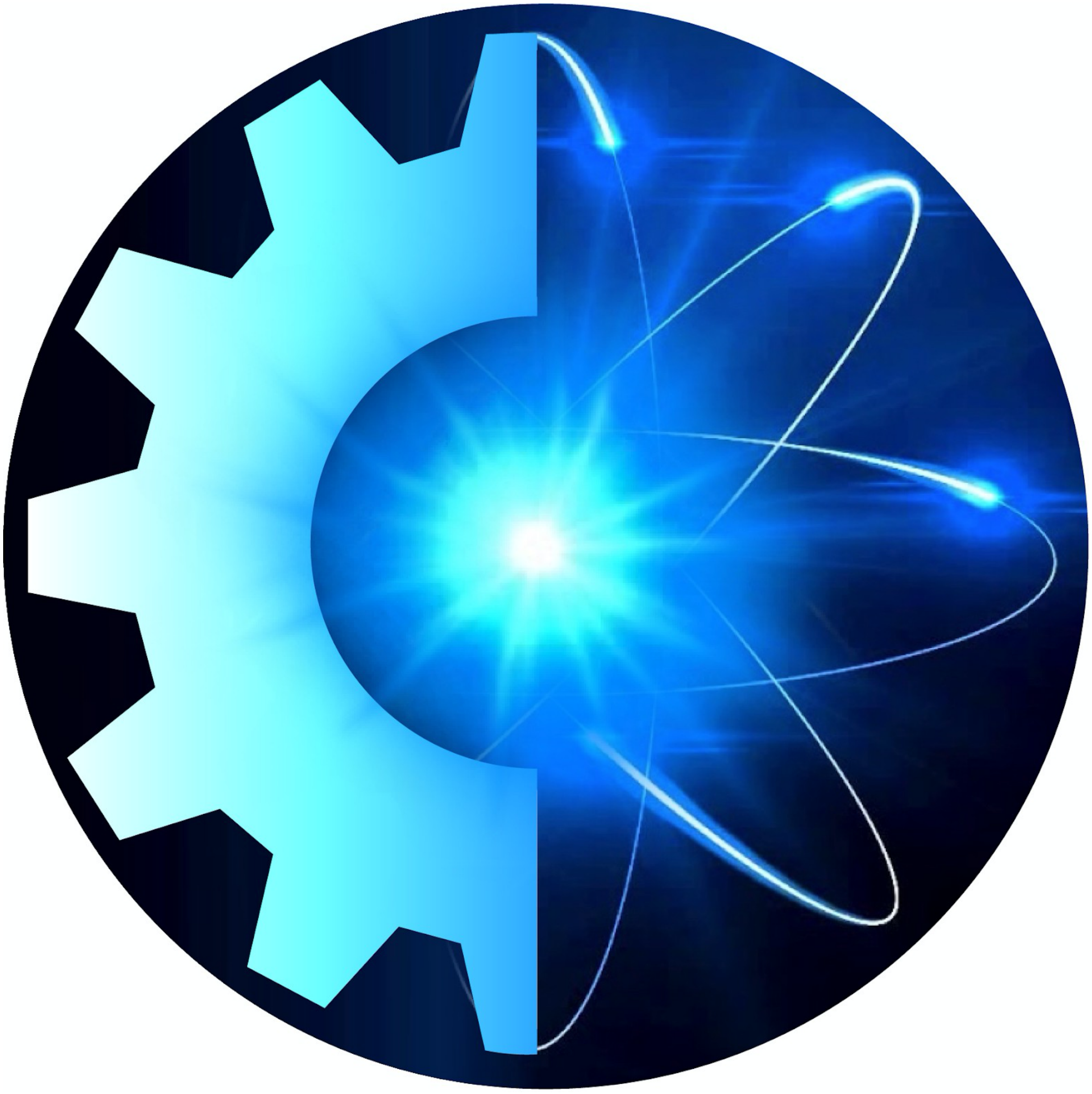

# A Community Plan for Fusion Energy and Discovery Plasma Sciences

## Chairs

| | |
|---|---|
| Scott Baalrud | University of Iowa |
| Nathaniel Ferraro | Princeton Plasma Physics Laboratory |
| Lauren Garrison | Oak Ridge National Laboratory |
| Nathan Howard | Massachusetts Institute of Technology |
| Carolyn Kuranz | University of Michigan |
| John Sarff | University of Wisconsin-Madison |
| Earl Scime (emeritus) | West Virginia University |
| Wayne Solomon | General Atomics |

## Magnetic Fusion Energy

Ted Biewer, ORNL
Dan Brunner, CFS
Cami Collins, General Atomics
Brian Grierson, PPPL
Walter Guttenfelder, PPPL
Chris Hegna, U Wisconsin-Madison
Chris Holland, UCSD
Jerry Hughes, MIT
Aaro Järvinen, LLNL
Richard Magee, TAE
Saskia Mordijck, William & Mary
Craig Petty, General Atomics
Matthew Reinke, ORNL
Uri Shumlak, U Washington

## General Plasma Science

Daniel Den Hartog, U Wisconsin-Madison
Dan Dubin, UCSD
Hantao Ji, Princeton
Yevgeny Raitses, PPPL
Dan Sinars, Sandia
David Schaffner, Bryn Mawr College
Steven Shannon, NC State
Stephen Vincena, UCLA

## Fusion Materials and Technology

John Caughman, ORNL
David Donovan, UT Knoxville
Karl Hammond, U Missouri
Paul Humrickhouse, INL
Robert Kolasinski, Sandia
Ane Lasa, UT Knoxville
Richard Nygren, Sandia
Wahyu Setyawan, PNNL
George Tynan, UCSD
Steven Zinkle, UT Knoxville

## High Energy Density Physics

Alex Arefiev, UCSD
Todd Ditmire, UT Austin
Forrest Doss, LANL
Johan Frenje, MIT
Cliff Thomas, UR/LLE
Arianna Gleason, Stanford/SLAC
Stephanie Hansen, Sandia
Louisa Pickworth, LLNL
Jorge Rocca, Colorado State
Derek Schaeffer, Princeton
Sean Finnegan, LANL

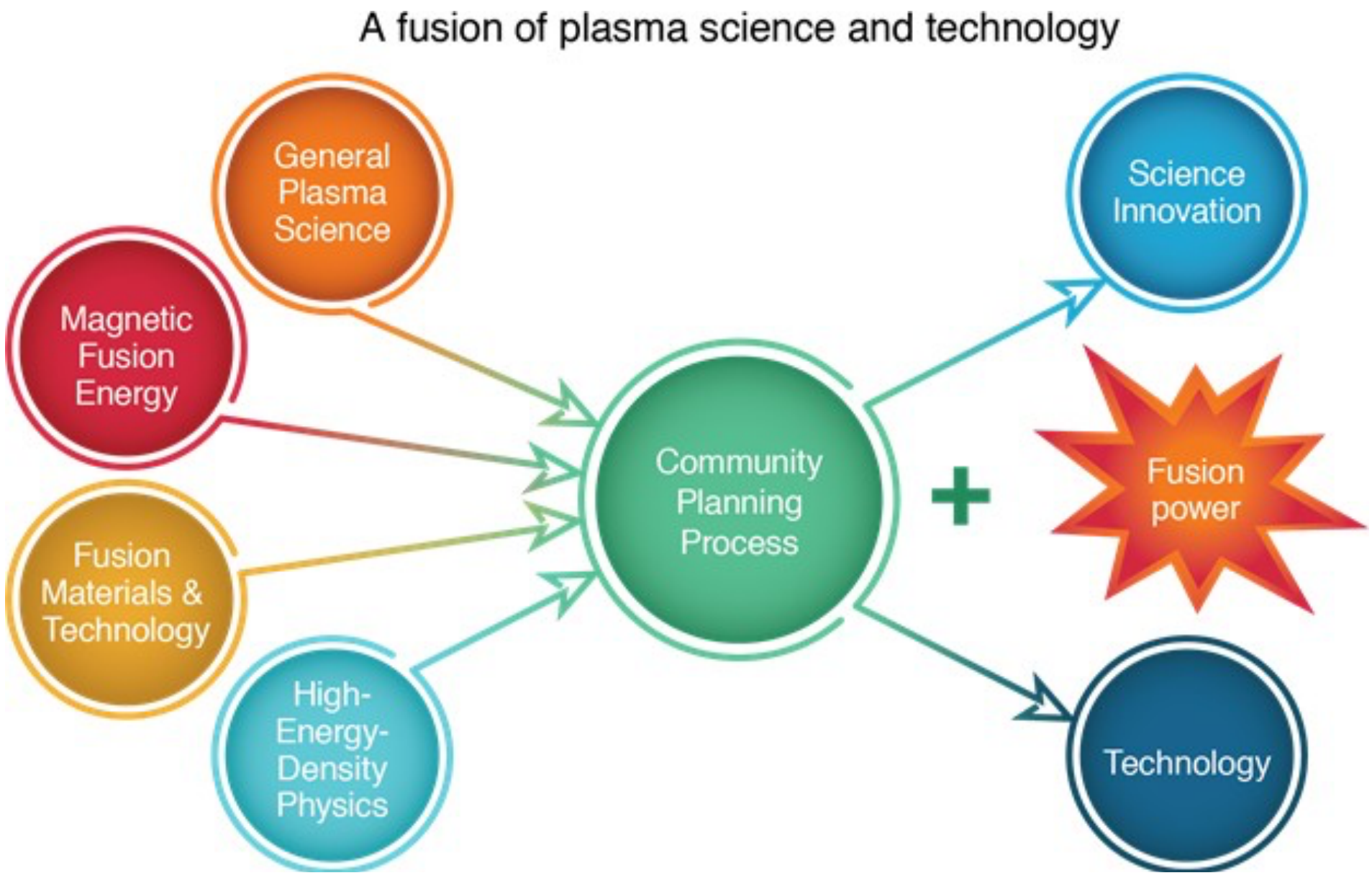

Figure 1. Artwork by Jennifer Hamson LLE/University of Rochester, concept by Dr. David Schaffner, Bryn Mawr College

# Preface

This document is the final report of the Community Planning Process (CPP) that describes a comprehensive plan to deliver fusion energy and to advance plasma science. The CPP was initiated by the executive committee of the American Physical Society Division of Plasma Physics (APS-DPP) to help the Fusion Energy Sciences Advisory Committee (FESAC) fulfill a charge from the U.S. Department of Energy (DOE) to develop a strategic plan for the DOE Office of Fusion Energy Sciences (FES). In this charge, dated Nov. 30, 2018, DOE Deputy Director for Science Dr. Stephen Binkley requested that FESAC "undertake a new long-range strategic planning activity for the Fusion Energy Sciences (FES) program. The strategic planning activity—to encompass the entire FES research portfolio (namely, burning plasma science and discovery plasma science)—should identify and prioritize the research required to advance both the scientific foundation needed to develop a fusion energy source, as well as the broader FES mission to steward plasma science." The CPP represents the first phase in developing a long range strategic plan for FES, and will serve as the basis for the second phase activity conducted by FESAC. It is worth noting that enacting the full scope of the recommendations in the strategic plan in this document will require suitable partnerships with other offices and governmental agencies, as well as with private industry and international partners.

This Community Planning Process has sought to form a consensus plan by the entire U.S. fusion and plasma physics community. The CPP has encouraged and received broad engagement from the entire U.S. fusion and plasma physics community by inviting the involvement of multiple professional societies (including APS, IEEE, ANS, HEDSA, USBPO, UFA, AVS, and others) and hosting frequent town halls, webinars, hundreds of small group discussions among subject matter experts, dedicated workshops, and focus group discussions. Hundreds of whitepapers, initiative proposals, and summary quad charts were submitted by the community throughout the process. This process has been extensively and transparently documented on a dedicated website (https://sites.google.com/pppl.gov/dpp-cpp). We believe that this process has been a success, not only by yielding the plan presented here, but also by bringing a diverse community together to embrace an ambitious vision for the future.

# Executive Summary

Fusion is the fundamental source of energy in the universe. We, and everything around us, are built from elements created by fusion reactions that occurred through the birth and death of stars that lived long ago. Fusion and plasma—the ionized matter that constitutes 99% of the visible universe—are inextricably linked. Fusion in the Sun's burning plasma indirectly powers our daily lives when we capture sunlight, catch the wind, and release ancient solar energy stored in fossil fuels. When harnessed on Earth, burning plasmas will directly provide a source of safe, clean energy capable of powering our society forever. The shared history of fusion and plasma science exemplifies how basic research translates from a deeper understanding of nature to important applications, such as plasma processing in the trillion-dollar microelectronics industry. Foremost of these applications will be the ability to harness fusion energy in a reactor—bringing a star to Earth—as one of the greatest achievements of humankind.

This report details a plan to realize the potential of fusion and plasma science to deepen our understanding of nature and to translate advances into commercialized fusion energy and other technologies that benefit society. It provides a consensus report on behalf of the entire U.S. fusion and plasma science community, which was developed following a community-led process that emphasized inclusivity and transparency at every stage. The following recommendations highlight the key output of this process, which are organized into the two crucial and complementary areas of the U.S. Department of Energy Office of Fusion Energy Sciences (FES): Fusion Science and Technology (FST), and Discovery Plasma Science (DPS), as well as cross-cutting opportunities that span the breadth of fusion and plasma science.

**Fusion Science and Technology (FST)** research holds the promise of providing limitless, clean, sustainable energy to the world. Recent advances, burgeoning private investment, and a renewed urgency to address U.S. energy needs motivate the transition to a mission-driven energy program. This community-driven strategic plan emphasizes exciting new research opportunities in fusion science and technology. It reflects the strong sentiment within the community that research in this area should be driven by the mission to enable construction of a fusion pilot plant (FPP) that produces net electricity and thereby establishes the scientific and technological basis for commercial fusion energy. By developing the innovative science and technology needed to accelerate the construction of a pilot plant at low capital cost, the U.S. will distinguish itself from its international counterparts and lead the way in the commercialization of fusion. To urgently move toward an FPP, cost-effective FPP designs must begin to be developed. The tokamak is presently the leading concept; however, research on other promising concepts, including optimized stellarators, inertial fusion, and alternative concepts, may ultimately lead to an attractive FPP. A prioritized set of strategic objectives needed to achieve this mission is described in this plan. The plan is broadly consistent with the recent National Academies Burning Plasmas report, and collectively establishes three key actions in FST to guide and orient the U.S. fusion program:

- **Accelerate the development of the burning plasma physics basis necessary for a fusion pilot plant.** Understanding burning plasmas, in which heating is dominantly provided by the energy released by fusion reactions, and resolving challenges associated with sustained operation, are critical steps toward achieving fusion energy. The U.S. should sustain full partnership in ITER, as this remains the best option for accessing burning plasmas at the scale of a power plant. To complete the plasma physics basis sufficient for an FPP, we should advance theory and modeling capabilities, utilize existing domestic and international facilities in the near term, and expand opportunities through public/private partnerships to provide access to burning plasma conditions. In addition, the conceptual design of a new U.S. tokamak facility capable of handling power exhaust at conditions typical of an FPP while simultaneously demonstrating the necessary plasma performance should begin immediately, with the goal of beginning research operations on the new facility before the end of the decade.

- **Rapidly expand the fusion materials and technology program.** The community recognizes the need to accelerate research in areas of fusion materials and technology, which apply to nearly any plausible pilot plant design, and likely set the timescale on which any FPP could be successful. The design and construction of a fusion prototypic neutron source (FPNS) should begin immediately to generate world-leading data on the degradation of materials when exposed to neutrons from fusion, in order to evaluate potential solutions for magnets, blankets, and other materials in an FPP. The FPNS should complement an expanded program for the development of structural and functional materials for fusion. Targeted investments should be made in fusion blanket and plasma facing component programs to provide critical new research capabilities and enhance U.S. leadership.

- **Embrace innovation to drive the achievement of economically viable fusion energy.** Research should focus on developing solutions to well-known challenges in fusion energy development by emphasizing exploration and utilization of new, potentially transformative science and technologies. Fully realizing the benefits of innovation requires consideration of the interconnected nature of fusion, which relies on coordinated research in plasma physics, fusion nuclear science, materials science, systems engineering, and many other fields. This should be addressed by establishing a multi-institutional, multidisciplinary program to develop fusion pilot plant concepts to help inform research needs and priorities. Our program must closely partner with private industry to drive innovative technologies that ensure the development of a commercially competitive product.

**Discovery Plasma Science (DPS)** research encompasses the study of the fundamental interactions of particles and light in plasmas, the study of astrophysical plasmas from planetary cores to stars, new theoretical and computational techniques to describe plasmas, and the practical application of plasmas for manufacturing, medicine, and agriculture. Its mission is to develop and verify a fundamental understanding of plasmas and take advantage of their unique properties to engineer technologies that support a growing economy. This work is organized into three Science Drivers: *Explore the Frontiers of Plasma Science*, *Understand the Plasma Universe,* and *Create Transformative Technologies*. In order to establish and maintain U.S. leadership in plasma science, we require world class facilities and reproducible theory, computation, and measurements. Often disciplines are closely identified with the tools used by its practitioners. For this reason we organize DPS into two complementary areas: High Energy Density Plasmas (HEDP), which typically relies on intense lasers or pulsed power, and General Plasma Science (GPS), which uses a broad range of tools. We have three main recommendations in DPS:

- **Build** an intermediate-scale general plasma science facility to study astrophysically-relevant magnetized plasma phenomena, significantly upgrade HED infrastructure, such as, LaserNetUS facilities and the Matter in Extreme Conditions instrument, and co-locate plasma devices at established facilities to leverage community expertise across the plasma science community.

- **Support** world-leading plasma science by ensuring stable funding for a balanced research portfolio, including single and multiple principal investigator scale projects, and those hosted at universities, national laboratories and industry. Leverage expertise outside of the plasma science community to support development of the vital data, methods, and techniques that support plasma science.

- **Collaborate** by developing networks of scientists and facilities to enable a broad range of frontier scientific research, and translate discoveries to advance other areas of science and engineering. Current networks, which include LaserNetUS, and collaborative low-temperature plasma research centers, should be expanded and new collaborative networks, such as ZNetUS and MagNetUSA, should be formed. Support and expand partnerships both within DOE and with other agencies where such collaborations are likely to have high impact. The NSF/DOE Partnership in Basic Plasma Science and Engineering as well as the NNSA/DOE Joint Program in HEDLP are exemplary in this regard, and support for these programs should be continued and expanded. Many other possibilities are emphasized in this report, including connections with the missions of DOE-BES, DOE-NNSA, NASA, NIH, DOD and several other agencies with missions that are advanced by plasma science.

**Cross-cutting opportunities** represent a number of shared challenges and research needs that cut across the wide scope of fusion and plasma science and technology. We highlight four representative recommendations:

- *Harness innovations in advanced scientific computing tools and increase capacity computing to improve fundamental understanding and predictive modeling capabilities.* Plasma and fusion science are drivers of advanced scientific computing and are poised to benefit from and contribute to national priorities for exascale computing resources, machine learning, and quantum information science.

- *Pursue innovations in diagnostic development that advance our understanding of basic plasma science, improve our ability to control fusion plasmas, and enhance survivability in extreme environments***.** Improvements in diagnostic resolution will provide new insights into the fundamental mechanisms governing plasma behavior as well as their interactions with materials. New diagnostics that are resistant to radiation effects are imperative to ensure survivability in a fusion plasma environment.

- *Support public-private partnerships across the full breadth of fusion and plasma science.* The private sector, working alongside government funded research, can create transformational plasma-enabled technologies for improved human health and well-being, including the realization of fusion energy.

- *Embrace diversity, equity, and inclusion, and develop the multidisciplinary workforce required to solve the challenges in fusion and plasma science.* To support the multidisciplinary workforce needed for fusion energy and plasma science, we must increase pathways for undergraduates and technical workers, and increase science literacy by developing community outreach. In so doing, we must commit ourselves to the creation and maintenance of a healthy community climate of diversity, equity, and inclusion, which will benefit the community as a whole and the mission of FES.

We recommend regular community strategic planning every 5–7 years to keep the plan consistent with evolving progress in science, technology, and external factors. This planning activity has provided tremendous benefits by bringing the community together to discuss the common scientific challenges and vision that make FES a coherent program. As a result, the community has embraced the complementary goals of an energy mission to commercialize fusion as soon as possible, and the vigorous pursuit of plasma science to advance our understanding of nature and to develop technology that will benefit our society.

# Statement on Diversity, Equity, and Inclusion

The Discovery Plasma and Fusion Science and Technology community recognizes that having a healthy climate of diversity, equity and inclusion is critical to solve the challenges we face in our field. We acknowledge, as a community, that our current (and historically) unhealthy climate is a serious problem and we commit to taking immediate action to achieve equitable, diverse, and inclusive outcomes. Diversity is expressed in myriad forms, including all ages, socio-economic backgrounds, races, ethnicities, genders, gender identities, gender expressions, national origins, religious affiliations, sexual orientations, family education level, disability status, political perspective—and other visible and nonvisible differences. Equity ensures equal opportunity and the impact of those opportunities in equitable outcomes for all persons; requiring zero tolerance for bias, harassment, and discrimination. Inclusion is the deliberate effort to ensure that our community is a place where differences are welcomed and encouraged, different perspectives are respectfully heard and where every individual feels a sense of belonging.

The limited data available show that our community has serious diversity deficiencies; for example, roughly 9% of the membership of the APS Division of Plasma Physics are women, which is the lowest percentage among all APS divisions. Recent doctoral awardees in plasma physics and nuclear engineering are fewer than 15% women and fewer than 5% underrepresented minorities, below the already poor numbers in all of physics (19% women, 7% underrepresented minorities) and all of engineering (23% women, 9% underrepresented minorities).[1]

If our community is not welcoming and supportive, we will not attract the needed talent and continue to lose people from marginalized groups who, at various stages of their careers, find the barriers and challenges of navigating an unwelcoming community to be an unnecessary burden, resulting in them taking their talents to a different field. It is our shared responsibility to create a healthy climate of diversity, equity, and inclusion (DEI). Such a climate not only creates a fair and safe environment for minoritized groups, but benefits the community as a whole and the mission of FES by leveraging the resources of diversity to advance our collective capabilities. This report outlines several recommendations (CC-WF) that should be acted upon to improve the DEI climate in our field.

[1] NSF Survey of Earned Doctorates (https://ncses.nsf.gov/pubs/nsf20301/data-tables/)

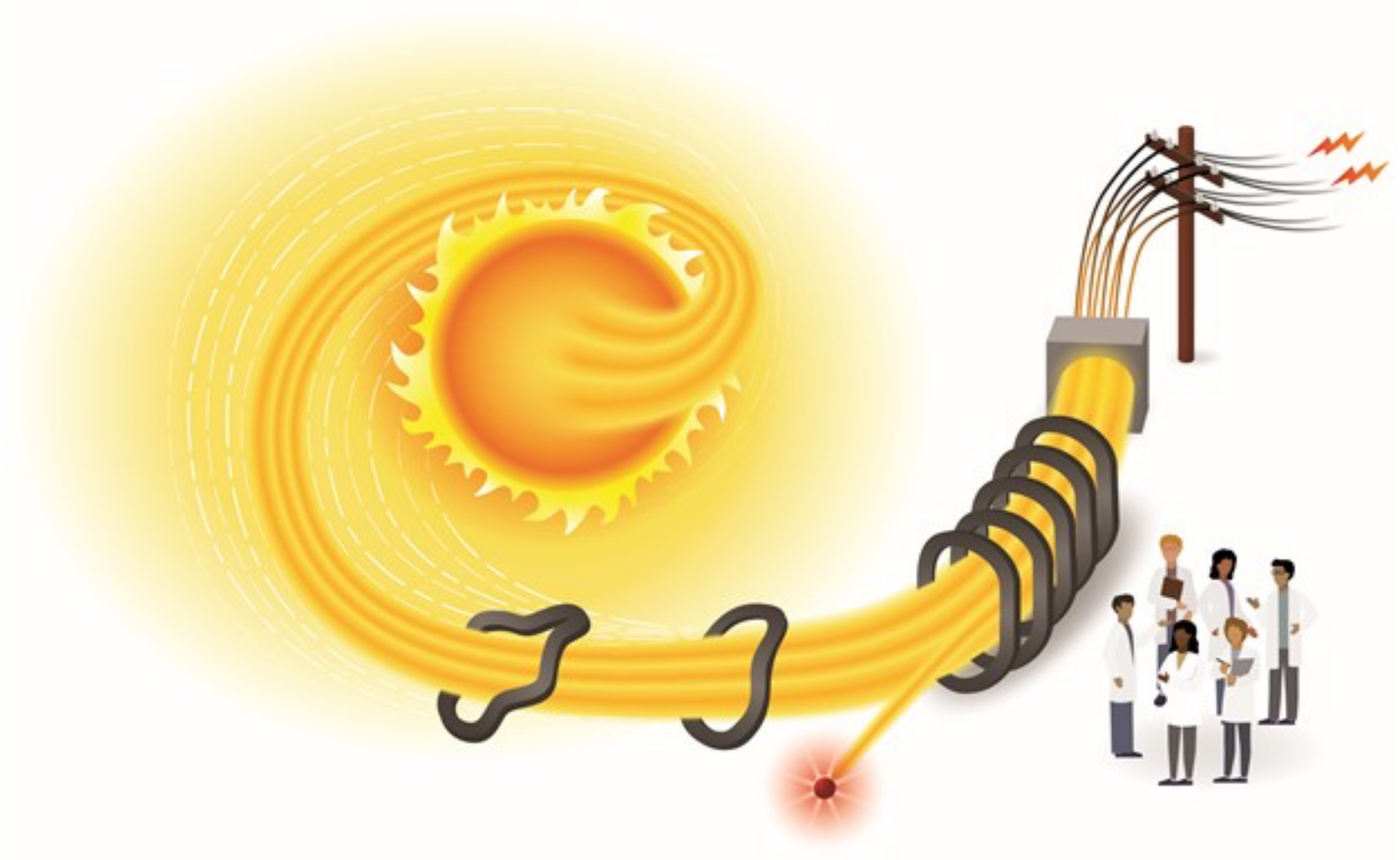

Figure 2. Artwork by Jennifer Hamson LLE/University of Rochester, concept by Dr. Jeffrey Levesque, Columbia University.

# Table of Contents

# Discovery Plasma Science

Discovery Plasma Science (DPS) is an incredibly diverse field of research that advances many areas of science and technology. An indication of its breadth can be gleaned from the enormous range of density and temperature conditions that it encompasses, from near-vacuum trapped ion plasmas at micro-Kelvin temperatures, to high energy density plasmas at several times solid density and tens of millions of degrees. Throughout this range of conditions, advances in DPS research contribute to answering the fundamental questions of science, including understanding symmetries of nature, the evolution of the universe, and how material properties change in extremes of temperature, density, and radiation fields. A characteristic of these discoveries is that they often enable the development of new technologies on short timescales. Plasma science has provided a major impetus to multi-billion-dollar twentieth century technologies such as microelectronics and lighting, and it continues to drive twenty-first century technologies in manufacturing, medicine, agriculture, and national security. As such, we identify the following Vision and Mission for DPS:

## *Vision*

Realize the potential of plasma science to deepen our understanding of nature and to provide the scientific underpinning for plasma-based technologies that benefit society.

## *Mission*

Develop fundamental understanding of the unique dynamical behaviors of plasmas, demonstrate that our understanding is true, and identify opportunities where the unique properties of plasmas can be used to engineer technologies that support a growing and sustainable economy.

Within FES, DPS research is organized into two complementary areas that are primarily associated with the tools that make the science possible:

**High Energy Density Plasma (HEDP)** research is associated with the science enabled by pulsed power devices and high-intensity lasers. These novel technologies produce matter at extreme conditions, reaching energy densities in excess of 100 billion Joules per cubic meter and pressures exceeding a million times atmospheric pressure. This field explores the properties of this extreme state of plasma, recreating conditions of dense astrophysical objects such as the interior of stars, giant planets, and exoplanets. It also explores exotic physical effects associated with the interaction of intense lasers with matter, including the possibility of creating pair plasmas from the vacuum.

**General Plasma Science (GPS)** utilizes a wide range of theoretical tools, and small to mid-scale experimental facilities. Within GPS, basic plasma science seeks to develop accurate theoretical descriptions of the complex emergent behavior of the plasma state, to push it into new regimes that expand our concept of what constitutes a plasma, and to design experiments and diagnostics to explore these states and to validate the models. Plasma astrophysics translates these fundamental discoveries to a better understanding of space and the cosmos, while low temperature plasma science translates discoveries into new technologies that improve our way of life by creating a richer, healthier, and more sustainable future.

Independent of the tools used, the field is also organized by categories of scientific questions that it aims to answer. Although there are many cross-linkages between these, as well as with FST, this report is organized by three *Science Drivers*:

**Explore the Frontiers of Plasma Science:** This driver encompasses research that seeks to advance the foundational frontiers of plasma science and to push its boundaries into new regimes. Basic science is a form of exploration that is motivated primarily by our innate curiosity to understand the world around us. A consistent lesson from the history of science is that the fundamental advances that result from such exploration unlock the potential of engineers to invent new technologies and astronomers to better understand our place in the universe. In this sense, the “explore the frontiers of plasma” science driver underlies and connects with the whole of plasma and fusion science. It includes experiments to measure the basic properties and dynamical behaviors of plasma as well as theory and computational modeling to describe it. This planning process highlights five exciting objectives that will transform the plasma science discipline in the near future:

DPS-A: Understand how intense light couples its energy to matter
DPS-B: Explore how magnetic fields control transport and influence self-organization in plasmas across scales
DPS-C: Advance understanding of plasmas far from equilibrium and at interfaces
DPS-D: Advance understanding of strong coupling and quantum effects in plasmas
DPS-E: Create and explore antimatter plasmas

**Understand the Plasma Universe**: Since more than 99% of the visible universe is in the plasma state of matter, our natural curiosity compels us to understand the role that plasma plays in the origins and dynamic nature of our universe. We are inspired by observations of the cosmos brought to us by a multitude of messengers from beyond our solar system including light, cosmic rays, gravitational waves, and neutrinos. We have also reached out with spacecraft to collect data from the Sun, to all the planets in the solar system, to the shore of interstellar space. In concert with such measurements, there has existed a strong synergy with laboratory experiments that reveal details on the underlying fundamental plasma processes. The early pioneers in fusion research had backgrounds in space plasma science and plasma astrophysics, building on the common challenge to understand processes in hot ionized gas to achieve fusion energy. This is a particularly exciting time to continue this shared tradition, as new observatories become operational, more advanced satellite missions continue, and new plasma regimes are created in laboratories. Together, these ultimately allow us to build better models for the plasma universe. We recommend focusing on three key objectives:

DPS-F: Understand plasma interactions between the Sun, Earth, and other objects in the solar system
DPS-G: Understand the origin and effects of magnetic fields across the universe from star and planet formation to cosmology
DPS-H: Understand the causes and consequences of the most energetic, extreme, and explosive phenomena found in the cosmos

**Create Transformative Technologies**: Fusion energy will represent the ultimate plasma-based technology, but many other transformative technologies have resulted from plasma science, and continue to do so through advances in basic plasma science. Plasma-based technologies have already transformed the microelectronics and materials processing industries, and they are posed to do similar things in energy technologies, healthcare, manufacturing, and agriculture. However, this will only be possible once the underlying science questions needed to engineer these technologies are solved. These are formative questions involving far from equilibrium plasma properties, reaction rates, plasma sources, acceleration and laser-plasma interactions for advanced source development for bright sources of particles and photons, and interactions with materials. However, history has shown that when these hurdles can be overcome, there is often a very rapid translation from new discoveries to technologies. This is why a close connection between the basic science and technology development should be fostered. We recommend that FES focus on four specific objectives:

DPS-I: Develop plasma-based technologies that contribute to a stable national energy infrastructure
DPS-J: Develop plasma-based technologies that enable advanced manufacturing
DPS-K: Develop plasma-based technologies that improve the physical wellbeing of society
DPS-L: Develop plasma-based technologies that provide secondary sources and other new capabilities, to benefit fundamental science, industry, and societal needs.

Completing the science in these drivers and objectives will require the maintenance and growth of a vigorous discovery plasma science enterprise. We recommend a common programmatic structure to support science across DPS: *Build* and upgrade the experimental facilities that provide basic data. *Support* scientists in universities, national laboratories and industry to conduct the research. *Collaborate* with scientists in neighboring disciplines, in order to translate plasma science discoveries to advance other fields, as well as to incorporate technologies from other areas to advance plasma science.

## Criteria

Development and prioritization of the recommendations described in this section were guided by the following rank-ordered criteria:

1. Establish U.S. leadership in plasma science through world class facilities and reproducible theory, computation, and measurements
2. Create transformational applications of plasmas to benefit society
3. Maintain breadth of the research program to benefit from innovation and high risk discovery
4. Engage the entire community of stakeholders, including national laboratories, universities, and industry
5. Capitalize on the potential of interdisciplinary applications of plasma research

We recommend that the prioritization process to be undertaken by FESAC continue to be based on these same criteria. Below are the highest level Programmatic Recommendations in no particular order.

## Programmatic Recommendations

### Build:

#### 1. *Invest in new facilities*

***Invest in an intermediate scale general plasma science facility to investigate the science of solar wind plasmas in the laboratory.***
FES has a timely opportunity to capitalize on the science of current spacecraft missions aimed in part at studying the origin and behavior of the solar wind. Technically feasible, but unbuilt, is a laboratory device capable of performing detailed scientific tests complementary to such missions. To close this clear gap in capabilities, FES should initiate a competitive conceptual design process to build and then support the operation of an intermediate-scale facility capable of producing and diagnosing the properly scaled physics of phenomena inherent to the solar wind.

***Invest in a multi-PW facility that can access intensities beyond the current state of the art with multiple lasers, and in high power lasers with greatly increased repetition rates.***
Novel regimes of light-matter interactions are predicted at ultra-high-laser intensities that are beyond the current state of the art as well in lasers with greatly increased repetition rates. These regimes are likely to benefit multiple applications involving particle and radiation sources and they are critical to advancing our fundamental knowledge of light-matter interactions. Multi-PW, multi-beam laser facilities are required to modernize the domestic research program, making it competitive with the facilities that are currently under construction worldwide. Another frontier that should be supported is to increase repetition rate capability of ultra-high power lasers to exceed greatly that of current lasers, providing testbeds to explore the generation of high-average flux, high-energy particle and photon sources. One possibility includes the Omega EP-OPAL proposal that combines multiple 25 PW 20 fsec beams and multi-kJ (>10 kJ) multi-ns beams.

***Invest in facilities over a broad range of scales.***
A highly functioning plasma program requires a support system with a pyramid-like structure of projects. Each level feeds physical (science), technological (techniques and diagnostics), and experiential (people) resources to the next level. A broad number of single-PI, small scale projects allows for focused physics study, testing ideas and proof of concept as well as free exploration. These smaller experiments are useful for training students as well as for giving junior PI's experience in running labs and managing research groups. These experiments and the PIs/students that are produced support the next level of intermediate scale facilities which include user-facilities and multiple-PI projects. These intermediate scale projects can in turn target science issues and questions that cannot be adequately explored on smaller facilities, as well as pool effort person-power and diagnostic or infrastructure resources. These experiments support large systems, larger numbers of researchers and a broader swath of experiments. The machinery can be higher power/energy and can support a broad diagnostic suite. They are run by professional technicians rather than graduate students or postdoctoral researchers. These are necessary for tackling physics questions in realms inaccessible in smaller devices. The small and intermediate scale in turn support the most expensive large scale projects such as multi-billion dollar space missions (*i.e.* Magnetospheric Multiscale Mission or Parker Solar Probe) for astrophysics, high power laser facilities for extreme plasma conditions, international-level experiments, and technological development in large scale industry for low temperature plasmas. The results from intermediate and small scale experiments can be used to compare to findings from these major endeavours.

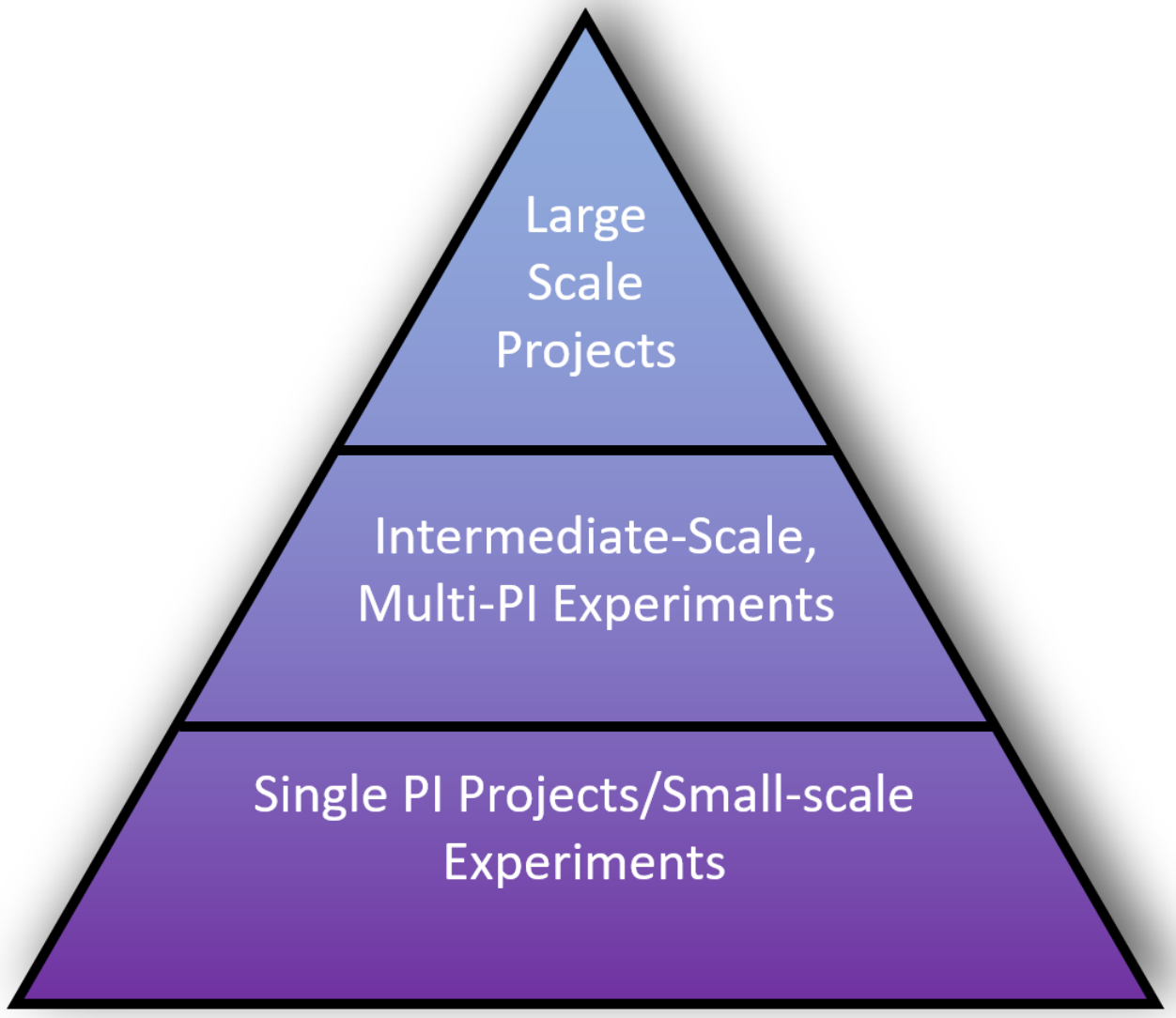

### 2. *Upgrade current facilities*

***Improve and upgrade national HED infrastructure at multiple scales, particularly at LaserNetUS facilities, in power, energy, and repetition rate, with temporally-synchronized multiple beams and precise control of laser parameters.***

These improvements include significantly increased laser and target repetition rate, and increased reproducibility, reliability, and peak power of the existing lasers. Investments that would increase user access to the LaserNetUS machines is desired, which could involve more shot time for users or more shots for each selected user. To address, the LaserNetUS infrastructure technology development should be upgraded in three principal areas:

*Increase repetition rate capability:* This will require technology development for some LaserNetUS facilities to improve laser repetition rate and possibly rep-rated pulsed power devices with advanced thermal management technologies. It will also mandate rep-rated targetry (fabrication, placement), high-data-rate collection and novel diagnostics. This area will likely benefit from applications of machine learning, artificial intelligence, and overall experiment automation.

*Increase precision control of the laser drivers:* This involves increasing the capabilities of existing LaserNetUS lasers in unique ways, such a detailed control of the far field spatial profile, control of temporal, contrast and spectrum, and precision control of laser parameter stability like pulse energy, pulse shape or focal spot.

*Include multi-beam and combined laser-pulsed power capabilities*: Significant scientific advances on multiple frontiers could be explored if LaserNetUS facilities deployed combinations of capabilities to include: multiple, temporally synchronized laser beams into the same interaction region, two or more PW-class beams with flexible interaction geometries for quantum electrodynamics (QED)-plasmas studies, PW-class lasers combined with at least one multi-hundred Joule or kJ nanosecond driver for material compression and heating, and pulsed power devices at the hundreds of kA or few MA current level.

***Couple long pulse multi-kJ and multi-PW lasers with an x-ray free electron laser (XFEL), which can be done at the Matter in Extreme Conditions instrument.***

An XFEL coupled with various drive capabilities including a petawatt (PW) or multi-PW laser with one or more beamlines of synchronized pulses at energy of at least 1 kJ, preferably up to 5 kJ, with precision nanosecond pulse shaping will revolutionize HED plasma science. This will allow creation and interrogation of novel plasma states, and compressed states of matter which can then be probed by the X-ray laser pulses. A key technical requirement here is the application of high repetition rate to build statistics when generating and probing exotic states. Moreover, combining multi-kJ long pulse and ultrafast drivers with an x-ray beam can map the evolution of the key plasma parameters in unprecedented ways. No capability like this currently exists anywhere else in the world.

***Provide upgrades for GPS facilities to leverage current FES investments in frontier-level science.***

Progress on a range of GPS science topics has been possible through the DOE's shrewd investments in world-class facilities such as the Basic Plasma Science Facility. FES clearly recognizes the scientific advantages of capitalizing on existing investments by its recent establishment of LaserNetUS for HEDP research. This network affords access for researchers from around the country who would not normally have the chance to perform experiments at the network's facilities. A similar network is recommended for FES's GPS, magnetized plasma collaborative research facilities (see MagNetUSA.) This new network can also be a vehicle for ensuring long-term viability of these facilities and maximizing returns on FES's investments by pushing cutting-edge operational parameter regimes; advancing the acquisition of broad diagnostic capabilities; and ultimately preserving frontier-level scientific output. In the realm of primarily unmagnetized plasma research, DOE is to be applauded for its recent creation of two new Low Temperature Plasma (LTP) collaborative research facilities. Specific recommendations for aiding user access to these facilities and for maintaining their long-term scientific relevance are to be found below.

### 3. Co-locate facilities

***Co-locate plasma devices at established facilities to leverage community expertise across the plasma science community.***

Significant scientific progress could be made if the plasma science community had a small number of facilities that deployed multiple, temporally synchronized devices in the same interaction region. Facilities with co-located plasma devices will have broad application in multiple science areas and will also be a particularly important tool in understanding dense strongly coupled plasmas and the physics of very intense electromagnetic wave interactions with beams and plasmas. In particular, the co-location of a multi-GeV electron accelerator with a multi-PW optical laser will provide a capability to study strong field QED effects. Co-location of a pulsed power machine at a XFEL or high-energy laser facility could also enable completely novel investigations of magnetized plasmas. This machine could deliver multi-MA current pulses in few hundred ns rise times to produce magnetic fields of up to 100 T in cubic cm volumes. Examples of co-located facilities should include:

-Multi-beam and combined laser-pulsed power capabilities, combining both PW-class, ultrafast lasers and kJ energy nanosecond compression lasers with MA pulse power machines for completely novel magnetized HED plasma studies.

-PW short-pulse and kJ-class long-pulse driver with precision diagnostic capabilities (e.g. an XFEL or future plasma-based advanced sources)

-Multi-PW laser and a dense multi-GeV electron beam for precision investigation of quantum plasmas

-Co-locate high precision diagnostics, including a short pulse beam, and/or beams, at facilities that can compress matter to high densities, temperatures, and pressures.

-Deploy compact pulsed power devices capable of generating large magnetic fields at existing facilities), including pulse shaping capabilities out to 100–500 ns on a 1 MA or more pulsed-power machine.

-Collaborative research facilities to leverage diagnostic capabilities across plasma science

## Support:

### *1. Support steady funding of plasma science*

Discovery Plasma Science is composed of two complementary principal science pillars, High Energy Density Plasmas (HEDP) and General Plasma Science (GPS). Significant funding variations over the past decade present several challenges for these areas. It is challenging to hire students or laboratory staff and to acquire laboratory space or other resources from leaders at these institutions when the “demand signal” from the government for this research varies so much. For the health of the community and long-term research planning, we recommend FES provide consistent stewardship of HEDP and GPS.

As the importance of the science and applications in DPS grow, formalizing funding in these two complementary areas would bring longer-term continuity to the funding profile, and allow the community to plan science priorities on a multi-year time scale. This is particularly important given the need to exploit the proposed mid- to large-scale facilities in aspects of DPS research. Regularizing this research funding for these two areas would enable effective long-term planning, and keep the U.S. at the forefront of HEDLP and GPS.

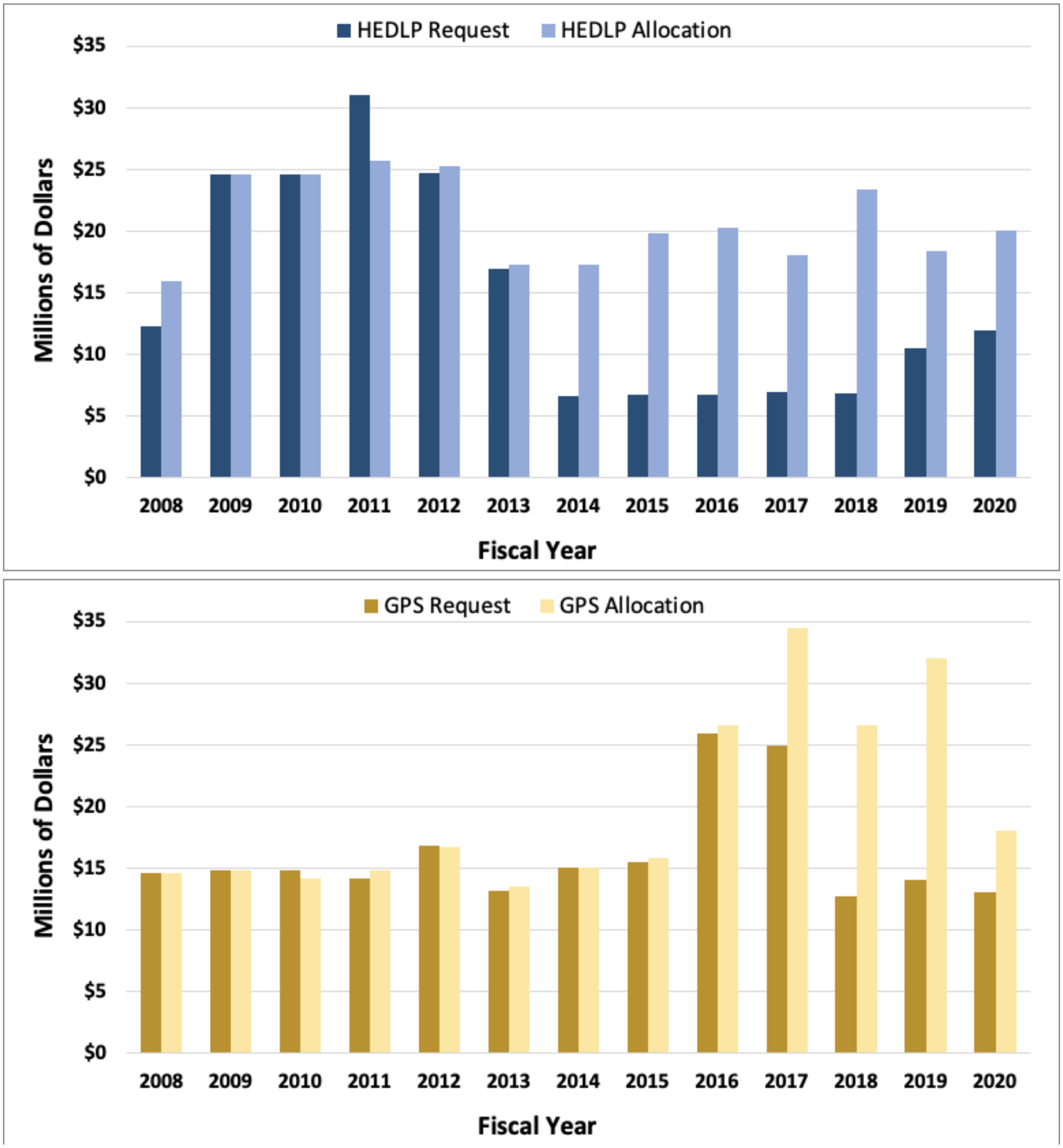

FES budget requests and allocations for HEDLP and GPS. Year-to-year variability and flat or downward trends in funding are inadequate to support healthy DPS research.

### *2. Support fundamental data needs*

> *On two occasions I have been asked, "Pray, Mr. Babbage, if you put into the machine wrong figures, will the right answers come out?" ... I am not able rightly to apprehend the kind of confusion of ideas that could provoke such a question. — Charles Babbage*

Simulations of plasmas, and the analysis of plasma measurements, rely heavily on fundamental data about the constitutive properties of materials (*e.g.* strength, opacity) or information about physical processes, such as cross sections and rate coefficients from atomic, molecular and optical (AMO) properties. As illustrated by the Babbage quote, these are critical needs that enable research throughout plasma science, since if the wrong data or models are used as inputs, we can't expect to get reasonable answers out. For example, spectroscopic diagnostics used in HEDP, low-temperature plasmas, astrophysical observation, basic plasma experiments, and fusion energy research, all rely on accurate emission, opacity, and other radiation rate coefficients. Applications in low-temperature plasmas are also particularly reliant on accurate reaction rate models, such as collision cross sections. The accuracy, completeness, and accessibility of available fundamental data is often a limiting factor determining the resolution of diagnostics or the accuracy of engineering design models. Dedicated support should be provided to expand, improve and increase open access to fundamental AMO and other data in areas where it is most pressing to advance plasma science. This is a need common to many areas of plasma science, but FES is well positioned to take the lead in coordinating the establishment of accessible and verified databases.

### *3. Support science centers*

DOE science centers are a powerful and effective venue for scientists to self-organize themselves to solve time-critical, specific common problems from a wide range of backgrounds. Other successful examples include Physics Frontier Centers sponsored by NSF Physics Division and more recently DRIVE (Diversify, Realize, Integrate, Venture, Educate) Centers sponsored by NASA Heliophysics Division, both of which involve the laboratory plasma physics community. We recommend forming DOE science centers with sufficient flexibility and frequency to address time-critical science problems and allow junior researchers with complementary ideas to join during the center lifetime or to propose new centers. Joint science center ventures (*e.g.* NASA-DOE) should be explored to pool resources and expertise among agencies.

## Collaborate:

### *1. Expand networks*

Networks of scientists and facilities enable a broad range of frontier scientific research, and to translate discoveries to advance other areas of science and engineering. Current networks, which include LaserNetUS, and low-temperature collaborative research centers, should be expanded and new collaborative networks should be formed. This could include expansion of the number of institutions in the existing LaserNetUS network, expanding the capabilities of the existing facilities and possibly broadening the purview of the existing networks to include computational, diagnostic development, target fabrication capabilities, and support for investigators. This could also include fostering collaborative technology development between the laser-based LaserNetUS network and the pulsed power-based ZNetUS to create novel multi-technology drivers. Collaboration with international networks can be beneficial and synergistic.

**Invest further in target fabrication capabilities, and in theory and computation support for LaserNetUS experiments.**

As repetition rate increases across key LaserNetUS platforms, the need for advanced targetry to deliver numerous, precision samples will increase. The burden of this development cost and execution may not be sustainable/possible for the academic or national laboratory communities. Therefore, network infrastructure to share advanced targetry best practices and deployment cost sharing is needed to enable best use of orders of magnitude increased repetition rates. With these high throughput experiments utilizing more novel probes will come the need for careful benchmarking and comparison to state-of-the-art theory and computation. Again, a carefully crafted network to bring together large datasets in concert with model predictions is critical for the success of revealing new physics.

**Establish ZNetUS to coordinate and increase access to DOE supported pulsed power facilities and computational tools used to model these experiments**

We recommend the formation of a ZNetUS consortium of scientists from academia and national laboratories that would support and enable research on pulsed power facilities across the country. Specifically, ZNetUS would help scientists share knowledge about pulsed power technology development, diagnostic advancements, and new modeling and simulation tools using magnetohydrodynamics and particle-in-cell techniques that are relevant to pulsed power. It could also provide guidance for the development of, and access to, a new mid-scale pulsed power facility (3-10 MA), as well as suggesting improvements to the existing Z Fundamental Science Program. Finally, by enabling researchers to share knowledge and/or facility access, it would help expose students at any given institution to a broader range of activities outside their home institution.

**Establish MagNetUSA as a mechanism for strengthening support for a wide range of experimental researchers and for increasing accessibility to DOE supported facilities**

Many GPS level plasma devices exist throughout the country at a wide range of accessible parameter regimes; however, the knowledge regarding the capabilities of each experiment remains fairly insular and access to running experiments on these devices is restricted for many single PIs, particularly early career faculty or researchers at locations outside of plasma physics research hubs. A coordinated network linking these FES-supported devices and their research personnel can expand access to a broader group of people, increasing user pools, improving diversity of experimental ideas, and potentially even fostering the growth of plasma faculty positions in the community. With the foundation of an established network, the ability to coordinate experimental exploration with theory and simulation can be improved. Moreover, training opportunities such as summer/winter schools for graduate students or user-training sessions for prospective PIs can be more easily established. A network can help maintain connections and communication among the researchers on these devices, with an aim toward more clearly seeing gaps in the parameter space so that upgrades to existing facilities can be proposed and implemented judiciously, efficiently, and fairly, in order to keep pace with advancing scientific needs. Similarly, gaps in the diagnostics capabilities can be identified. MagNetUSA can help establish new diagnostic frameworks that can be duplicated at multiple facilities or designed to travel to different sites within the network, promoting accessibility and reproducibility. Essential to the success of collaborative research facilities lies in the support for new and existing users. Creation and support of MagNetUSA would help remove funding barriers for the GPS user community to travel to network facilities, perform research there, and to analyze results at home institutions.

**Support collaborative research networks in low-temperature plasma science**

Recently, FES supported two collaborative low temperature plasma research facilities. These provide scientists and engineers from around the country access to cutting-edge diagnostics, experiments, and simulation codes that are not available in most existing single investigator facilities. Support for this model should be sustained and include pathways for facility upgrades to meet future user needs, as well as support for users to access the facilities including for example, travel grants, funding for construction and adjustment of user experiments to the facility requirements, and to analyze the data obtained. The logistic support should be timely. Support should also include infrastructure that lowers the barrier for researchers outside of the core plasma science community to advance interdisciplinary work critical for the translation of plasma science to disruptive technologies.

In addition to these core user facilities, supporting basic research to advance experimental and computational infrastructure is also recommended. Keeping these facilities at the forefront will require continual evolution, which relies on basic research that does not yet have a technology readiness level that can be supported by a user facility. Support to develop new and novel diagnostic techniques as well as advancing computational infrastructure is encouraged.

**Establish a network program to build new hardware capabilities and support the acquisition of sophisticated diagnostics to be shared between facilities to maximize productivity while minimizing costs.**

Science relies not only on what can be produced, but also what can be measured, therefore continual support advancing diagnostics is needed to support our strategic objectives. These include, but are not limited to, particle velocity distribution function measurements, laser-based diagnostics, and X-ray spectroscopy diagnostics.

**Establish a network program to foster the development of an open source, programming ecosystem for plasma physics and advance computational plasma science.**

Many areas of frontier plasma science require similar computational methods. Sharing and standardizing software allows community wide code development, promotes best programming practices, creates common standards for simulation efforts, and facilitates scientific reproducibility. A common open source codebase and the production of readable, reliable, and maintainable software should be supported. A model in this regard is the “PlasmaPy” project. In addition, modeling efforts in frontier plasma science rely on a wide-range of computing needs, from high-performance to high-throughput. Access to computing facilities specifically aimed at the discovery plasma science community should be increased.

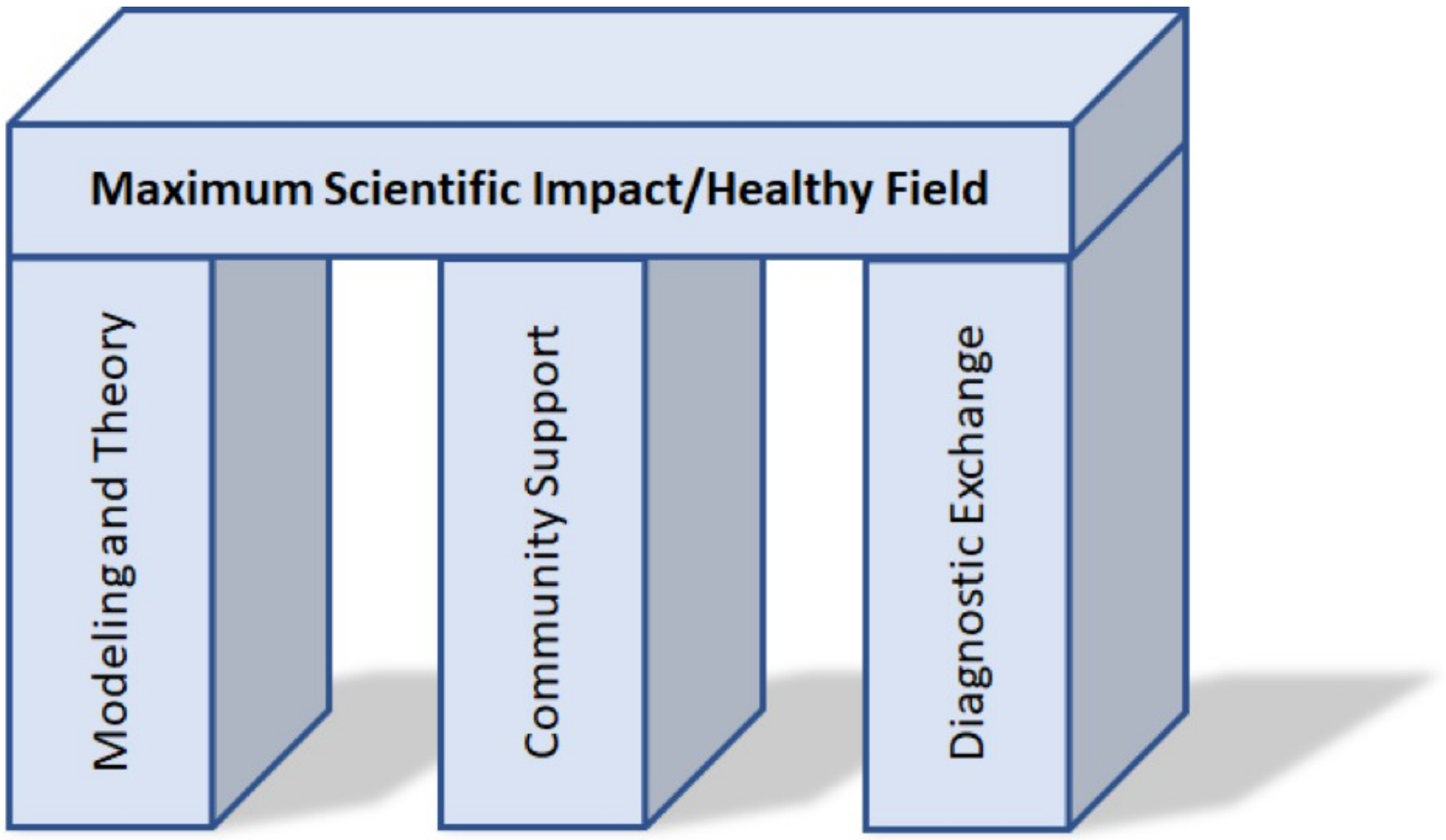

## *2. Expand partnerships*

The interdisciplinary nature of plasma science presents a tremendous opportunity to translate FES-funded research to other areas of science and engineering. Doing so will contribute to the mission of FES by enabling higher impact of the research it supports as a steward of plasma science, while also advancing the mission of other agencies and spawning new industries. A few stakeholders outside of FES already support plasma science either directly, or tangentially through applications. However, much more could be done, and if done carefully, could lead to transformative impacts on science and society. Opportunities that exist throughout DPS are described in more detail throughout this chapter. Each must be considered on a case-by-case basis since the appropriate course of action will differ. However, one common element to any successful partnership is that FES, partnering agencies, and the scientific communities involved must develop a plan together.

The NSF/DOE Partnership in Basic Plasma Science and Engineering is exemplary in demonstrating the impact that a well-formed partnership can have. It has funded many of the most prominent results in fundamental plasma science since its inception in 1997, and continues to drive this field. This partnership should be expanded because doing so will enable the fundamental advancements that extend the boundaries of plasma science to new regimes. The original recommendation made in the NAS/NRC study on Opportunities in Plasma Science and Technology panel report *Plasma Science: From Fundamental Research to Technological Application*, was for $15M ($24.5M in 2019 dollars) per year for university-scale research. This remains an apt target funding level, which is significantly greater than the current level.

The Office of Science-NNSA Joint Program in High-Energy-Density laboratory plasmas operates a partnership with NNSA in HED physics. FESAC provides technical and programmatic advice to FES and NNSA for the joint HEDLP program. The SC-NNSA joint program was the result of a recommendation from an Office of Science and Technology Policy interagency task force and is currently funded at $5M. The specific areas of interest are HED Hydrodynamics, Radiation-Dominated Dynamics and Material Properties, Magnetized HED Plasma Physics, Nonlinear Optics of Plasmas and Laser-Plasma Interactions, Relativistic HED Plasmas and Intense Beam Physics, Warm Dense Matter, High-Z, Multiply Ionized HED Atomic Physics, and Diagnostics for HED Laboratory Plasmas.

Other possible partnerships emphasized in this section are not yet established. It is recommended that FES first support exploratory committees to assess interest and cost-benefit analysis. Some of these opportunities are between DOE and other agencies, including NASA, NIH, USDA, EPA and DOD. Others are between FES and other offices within DOE, including BES, ARPA-E, NNSA, and DOE Advanced Manufacturing. The recommended first step is to support meetings that bring the scientific communities relevant to each potential partnership together in order to identify potential topics of overlapping interest, and to identify the highest-impact target research areas. In cases where important mutually-beneficial opportunities exist, a next step may be to jointly support a committee to make recommendations regarding the appropriate scale and implementation of a partnership program.

| | | DPS-A | DPS-B | DPS-C | DPS-D | DPS-E | DPS-F | DPS-G | DPS-H | DPS-I | DPS-J | DPS-K | DPS-L |
|---|---|---|---|---|---|---|---|---|---|---|---|---|---|
| Build | Invest in an intermediate scale general plasma science facility | | | | | | x | | | | | | |
| | Invest in a multi-PW facility that can access intensities beyond the current state of the art | x | x | | | x | | | x | | | | x |
| | Invest in facilities over a broad range of scales | x | x | x | x | x | x | x | x | x | x | x | x |
| | Improve and upgrade national HED infrastructure at multiple scales particularly at LaserNetUS facilities | x | x | | x | x | x | x | x | | | | x |
| | Couple long pulse multi-kJ and multi-PW lasers with an XFEL, which can be done at the Matter in Extreme Conditions instrument | x | | | x | x | | | x | | | | x |
| | Provide upgrades for GPS facilities to leverage current FES investments in frontier-level science. | | | | | | x | x | | x | x | x | |
| | Co-locate plasma devices at established facilities to leverage community expertise across the plasma science community | x | x | | x | | x | x | x | | | | x |
| Support | Support steady funding of plasma science | x | x | x | x | x | x | x | x | x | x | x | x |
| | Support fundamental data needs | x | x | x | x | x | x | x | x | x | x | x | x |
| | Support science centers | x | | | | | x | x | x | | | | |
| Collaborate | Further investment in target fabrication capabilities, and in theory and computation support for LaserNetUS experiments | x | x | | x | x | x | x | x | | | | x |
| | Establish ZNetUS to coordinate and increase access to pulsed power facilities and necessary computational tools | | x | x | | | x | x | x | | | | |
| | Establish MagNetUSA for a wide range of experimental researchers and for increasing accessibility to DOE supported facilities | | x | | | | x | x | | | | | |
| | Support collaborative research networks in low-temperature plasma science | | | | | | | | | x | x | x | |
| | Establish a network program to build new hardware capabilities and support the acquisition of diagnostics to be shared between facilities | | | | | | x | x | x | x | x | x | |
| | Establish a network program to develop of an open-source, programming ecosystem for plasma physics and advance computational plasma science | x | x | x | x | x | x | x | x | x | x | x | x |
| | Expand partnerships | x | x | x | x | x | x | x | x | x | x | x | x |

Programmatic Recommendations supported by DPS Strategic Objectives

## *DPS-1: Explore the Frontiers of Plasma Science*

Much excitement and opportunity in science comes from exploring new frontiers. History has shown time and again that following our innate curiosity leads to unexpected outcomes that deepen our understanding of nature, and reveal new paths to advanced technology. Modern plasma science is no exception. It is exploring exciting new territory. Some of this is driven by new scientific tools that enable the exploration of new regimes of plasma, such as ultra-high-intensity lasers, advanced light sources, laser cooling techniques, and new diagnostic methods. Others are driven by new theoretical insights, mathematical methods, or computational capabilities. Frontier research topics include the production and trapping of plasmas made from antimatter, plasmas in extreme radiation or magnetic fields, plasma self-organization, as well as strongly coupled and quantum degenerate states of ionized matter that stretch our concept of what constitutes a plasma.

While development of the many applications enabled by plasma dominates the funding landscape, the future of the field relies on a stable funding commitment to foundational exploration. Much of this research is carried out in university physics and engineering departments. Recent trends indicate that the role of plasma research in these departments is declining. It is imperative to the future of plasma science that this trend be reversed, and a stable funding commitment will contribute to that. By its nature, plasma research is not tied to specific objectives, but is more blue-sky exploration. The rich diversity of plasma physics research and its applications also makes it challenging to describe it in an integrated manner. Academic plasma researchers sit in diverse university departments such as physics, applied physics, electrical engineering, mechanical engineering, nuclear engineering, astrophysics, and more. Each subfield of plasma physics is exemplified by different sets of approximations to the complex governing physics equations, different measurement tools, and different research platforms. It is therefore not possible to capture all of the opportunities where advances to the foundational physics can have an impact, but the community planning process has identified a small number of high-priority needs that the Department of Energy and other funding agencies can help nurture.

**Prioritize support for single-investigator research in fundamental discovery plasma science.**

Small-group or single-investigator research in discovery plasma science, *particularly at universities*, is essential to the advancement of the fundamental frontiers of plasma physics, as such research allows us to address foundational curiosity-driven questions that is less subject to programmatic constraints, and to increase the impact of and interest in the work across the broader physics community. While large-scale plasma physics facilities garner significant attention and funding, much of the foundational research in plasma physics does not require huge facilities and is done at universities. There is an opportunity to make more rapid advances in foundational plasma physics at a relatively modest cost by offering more small-scale research grants. Of the 10 top-ranked universities in the U.S. (as defined by U.S. News and World Report), only three (MIT, Princeton, U. Chicago) explicitly list plasma physics research as part of their physics or astrophysics departments. Most schools do not teach plasma physics at the undergraduate level, and many schools do not teach it at the graduate level. This has translated into relatively few students in the American Physical Society's Division of Plasma Physics relative to other Divisions. Because the course offerings at universities are directly related to the skill sets of the professors at those universities, the most direct way to address this is to fund faculty and staff at a range of universities in this field. Foundational plasma physics research is an excellent area for doing so.

## DPS-A: Understand how intense light couples its energy to matter

Simply by shining an intense laser into a plasma, particle energies rivaling those found in cosmic rays can be attained within distances less than a meter. While this novel interaction holds the promise of delivering revolutionary new devices for medicine and industry, it is inherently a highly nonlinear process. Much theoretical and experimental progress must be made, if we are to create and optimize these future compact photon and particle sources based on these unique laser-plasma interactions. At the same time, this research also extends into entirely new directions, like plasma optics, where a plasma is used to combine and synchronize the photons of many laser beams to create a single beam with an intensity that otherwise cannot be achieved with conventional optics. These intense laser beams would, in turn, provide access to new regimes like nonlinear Quantum Electrodynamics (QED), where the very foundations of physics, like creating matter out of pure vacuum using nothing more than intense lasers, can be tested and to provide a much deeper understanding of how our universe works. Just as important, an increased understanding and precise control of laser interactions in under-dense plasma (LPI) can lead to ways of avoiding LPI in situations where it is not desirable, and enhance LPI in useful situations where nonlinear phenomena dominate. Finally, the field is ideally suited to take advantage of the recent explosion of Machine Learning (ML) and "Big Data" with new rep-rated high intensity laser facilities that are currently being planned and built in the U.S. These next-generation laser systems, diagnostics, and analysis tools will enable the study and control of LPI using techniques that were previously impossible, as well as directly impact the creation of compact, bright photon and particle sources.

**Expert Groups**: HEDP

***Recommendations***

**Conduct studies at femtosecond and multi-picosecond time scales, quantify the interaction of relativistic laser beams with matter in circumstances relevant to LPI, and make comparisons to state-of-the-art physics models and simulations.**

More comprehensive evaluations of how high-intensity ($> 10^{18}$ W/cm$^2$) lasers interact with matter needs to be examined at both femtosecond and picosecond time scales. Femtosecond (fs) resolution examinations of how the relativistic interaction of the laser ponderomotive force directly manipulates electrons must be experimentally executed. For instance, X-ray imaging of relativistic HED-plasma interactions at spatial scales of hundreds of nanometers and temporal scales of hundreds of femtoseconds will directly reveal how the solid-density plasma is pushed and expands due to a combination of laser hole-boring and heating. Ultra-fast imaging over a range of X-ray energies and higher laser intensities or different plasma conditions can offer new insights into ways to control and optimize laser interactions.

**Support development of first-principles theoretical and computational models for plasmas with self-consistently included QED effects to stimulate growth of a competitive domestic high-field community.**

QED effects are predicted to have a profound effect on plasmas irradiated by ultra-high-intensity laser pulses (*e.g.* changing the motion of individual particles and creating matter and antimatter). It is critically important to develop a framework that allows one to include these effects into a plasma description. All presently available theoretical and computational frameworks rely on localized rates of standard QED effects obtained in a vacuum rather than in a plasma.

## DPS-B: Explore how magnetic fields control transport and influence self-organization in plasmas across scales

In many cases, particularly in fusion research, magnetic fields are time-varying and consist of both externally imposed fields and fields generated by internal currents. Despite the venerable history of research in this field, there remain a number of outstanding questions. How do magnetic fields affect mass, mixing, momentum, and energy transport in plasma conditions ranging from ultracold non-neutral plasmas to the hot, dense plasmas relevant to inertial fusion energy science? What are the mechanisms behind magnetic field generation and amplification, magnetic reconnection, turbulence, and particle acceleration? What are the processes for generating the strongest magnetic fields on Earth in laboratory plasmas? Transport in highly magnetized, highly collisional strongly coupled plasmas is not well explored; these are regimes in which textbook treatments of transport breakdown. An important objective in HED plasma research is to understand how statistical mechanics couples to quantum mechanics to deliver self-organized materials, and how intense fields can assemble and maintain cohesive states. Relativistic laser–plasma interactions in high-field environments are also not well understood. Plasma transport in strongly magnetized plasmas in which the gyroradius is shorter than the Debye length, such as those produced in non-neutral plasma experiments and magnetized inertial fusion, is an area open for exploration. How intense laser fields propagating in overdense, relativistically transparent plasmas potentially produce MegaTesla fields is a further area of frontier research in this area. These areas are all ripe for theory and computation advances coupled with discovery-driven experiments.

**Expert Groups**: HEDP and GPS

***Recommendations***

**Conduct pulsed-power-based fundamental experiments on magnetic inhibition of transport on university-scale facilities.**

Magnetic fields can strongly decrease cross field transport and diffusion of mass and heat and thereby significantly improve plasmas as fusion energy and radiation sources. This underlies pulsed power-driven fusion schemes such as MagLIF and magnetized laser-driven implosions.

Under HED plasma conditions, challenges arise in modeling magnetic field generation due to several novel field generation processes including the Biermann battery and Weibel instability, and in modeling magnetic field evolution under high heat-flux conditions due to collisional effects such as the Nernst effect in Ohm's law, and finally the role of instabilities and turbulence in cross-field transport of heat and mass. Benchmarking of these processes through concerted effort of simulation with fundamental laboratory experiments will put understanding on firm footing and allow predictions under complex, integrated conditions such as MagLIF or magnetized ICF compressions.

**Support further understanding of the physics of magnetic self-organization by observing and modeling phenomena across a range of energy and spatial scales.**

Examples include creating and analyzing laboratory analogs of astrophysical plasma jets and the magnetic dynamo in stars, galaxies, and the Earth; further analysis of the high-confinement "H" mode in MFE plasmas and the relaxation to Taylor states in such plasmas; examining the connection of vortex crystal states in the turbulent relaxation of non-neutral plasmas (and 2D fluids) to similar plasma phenomena.

**Support further understanding of the physics of how electromagnetic energy couples to plasma across length and time scales, including magnetic turbulence and the transition from nonlinear behavior to turbulence.**

The self-consistent emission and absorption of electromagnetic energy is responsible for a range of plasma phenomena including heating in the solar corona and in *Z*-pinch and magnetized-liner fusion experiments; ionospheric aurora arising from magnetic substorms; and other wave-particle interactions in various contexts. The influence of electromagnetic degrees of freedom on the turbulent state of a driven plasma is fundamental to a range of phenomena, but is incompletely understood. This process underlies transport behavior in both laboratory and astrophysical plasmas, and our lack of fundamental understanding inhibits progress in these fields. These needs transcend essentially all areas of plasma science, including general plasma science and HED plasma.

**Support studies that advance understanding of strongly magnetized plasmas.**

Using magnetic fields to control plasma transport is the basis for much of fusion energy research, and many other applications of plasma physics. Although this is a topic central to plasma physics, current understanding is limited to magnetization strengths that are weak enough that the magnetic field does not alter the motion of particles at the scale at which they interact. Strongly magnetized plasmas, where the gyromotion of particles is at a scale similar to or smaller than the Debye screening length, are fundamentally different. Developing an understanding of this regime is an area ripe for exploration. It is becoming increasingly exciting as extreme magnetic fields are being generated in high energy density plasmas, potentially accessing this regime. It is also encountered in trapped non-neutral plasmas, including antimatter plasmas.

**Support advanced theory that assesses the accuracy of standard magnetized plasma models (*e.g.* MHD fluid dynamics), and develops improved models.**

Magnetohydrodynamic (MHD) theory is often used to describe magnetized plasma but, as a single-fluid theory, has limitations that must be (and are being) addressed in order to study phenomena beyond the model across a range of energy and spatial scales; these include magnetic reconnection in nearly collisionless plasma; collisionless shock formation and stability, field generation by gradients, energetic tails in velocity distributions, and solar wind turbulence. Advancing theory in this area includes supporting the development of coupled models in which a locally kinetic description is matched with a global fluid-like description.

## DPS-C: Advance Understanding of Plasmas Far From Equilibrium and at Interfaces

Laboratory plasmas are often driven by external power and particle sources, and are in contact with surfaces at substantially different temperatures. The result is that at least portions of the plasma can be in states far from equilibrium. For example, low-temperature plasmas are partially-ionized systems in which the ions are close to equilibrium with the neutral gas, but the electrons are hundreds to thousands of times higher in temperature. This can create conditions under which there is both traditional chemistry (between ions and/or neutrals) as well as chemistry associated with high-energy electrons that is unique to the non-equilibrium plasma state. Such plasmas are found in a wide variety of industrial applications, and controlling the electron energy distribution is a critical aspect of many of these devices. Plasmas near surfaces, including both solid and liquid surfaces, are invariably far from equilibrium, as strong electric fields called sheaths form in these regions.

Many open questions remain in our understanding of sheaths, and resolving these questions will be important to advancing applications spanning basic physics to industrial applications. In some extreme cases, plasmas may have velocity flows comparable to or greater than the average velocity of individual particles, or a small number of particles with extremely high velocities relative to the average. Depending on how collisional the particles in the plasma are, these situations can make traditional plasma physics approximations invalid. Moreover, non-equilibrium particle velocity distributions can affect other important plasma parameters such as fusion reactivity (in fusion plasmas, only the highest-velocity particles have enough momentum to overcome the Coulomb repulsion of charged particles and fuse). The frequent presence of far-from-equilibrium states is something that distinguishes plasma physics from many other areas of physics.

**Expert Groups**: GPS
***Recommendations***

**Support studies that advance understanding and control of reaction kinetics, ionization state, and plasma chemistry of non-equilibrium and partially-ionized plasmas.**

While many plasma models assume local thermodynamic equilibrium, where electrons, ions, radiation, and internal atomic and molecular structure are all characterized by the same temperature, most laboratory and astrophysical plasmas are non-equilibrium (*e.g.* non-thermal radiation fields or non-Maxwellian electron velocity distributions). This recommendation necessitates more investment in dedicated experiments, theory, and simulation that address the physics underlying non-equilibrium plasmas, as well as connections to atomic and surface physics and chemistry research. This will aid in understanding the phenomena in integrated experiments.

**Support studies that advance understanding of sheaths and plasma–boundary interactions.**

The interaction of plasmas with material surfaces is complex and only moderately understood. New applications, particularly with regard to liquid surfaces, emitting surfaces, and evaporating surfaces demand new models. Despite its importance in many applications, the plasma boundary transition in magnetized plasmas remains poorly understood. Validation of models is particularly important, and this will require a combination of theory, experiment, and computation.

**Support studies to understand and control the complex, self-consistent effects that locally-trapped particles have on plasma transport and waves (damping and instability) in weakly-collisional plasmas.**

Natural and laboratory plasmas often have several distinct locally trapped particle populations, due to the occurrence of local magnetic and/or electrostatic wells. When subjected to perturbations such as plasma waves or field errors, such configurations can exhibit enhanced “superbanana” transport and wave dissipation: the locally trapped particles respond to the perturbations differently from passing particles, creating discontinuities in the collisionless particle distribution function at the separatrix (or separatrices) between trapped and passing particles; and collisional relaxation of these discontinuities causes enhanced rates of entropy production, wave damping and instability, and transport of particles, momentum, and heat. The term “superbanana” refers to the single-particle drift orbits near the separatrix energy that are perturbed by the waves or field errors. This form of entropy production has an important influence on energy and particle loss in magnetic fusion devices such as stellarators and reversed-field configurations, but can also be studied in smaller, dedicated experiments.

## DPS-D: Advance Understanding of Strong Coupling and Quantum Effects in Plasmas

Plasmas are commonly thought of as hot ionized gases, but they often exist in states with behaviors more akin to ionized liquids, supercritical fluids, and even solids. These strongly-coupled plasmas are dense, cool, or highly-charged systems in which the interaction energy between particles is much larger than their kinetic energy. They behave in fundamentally different ways than weakly-coupled plasmas, and their properties are only beginning to be understood. Strongly-coupled plasmas can be produced in the laboratory using a variety of platforms, including non-neutral plasmas, ultracold neutral plasmas, and dusty plasmas. Such experiments are well-diagnosed and provide precision measurements used to explore the physics of strong coupling. These measurements provide a foundational contribution by testing theoretical models. Some of these experiments may also lead to a platform for quantum computation or quantum simulation.

High-intensity optical and X-ray lasers and the high-energy U.S. flagship facilities NIF, Z, and Omega produce extremely dense plasmas, enabling studies of the response of the entire periodic table to extreme pressures, fields, and temperatures. Under extreme conditions where strong coupling, electron degeneracy, and thermal effects (such as electron ionization) all modify atomic properties, complete, internally consistent computational models are difficult to create and benchmark-quality data are scarce.

For accurate simulations relevant to HEDP science and IFE, we need to know material properties—transport coefficients, opacity, equation of state and atomic structure—in diverse and extreme conditions across disparate length and time-scales. This need for fundamental materials science and materials tunability/performance at extreme conditions cross-cuts fundamental HEDP and FST (*e.g.* point to blanket materials section). How do we tune materials properties to sustain plasma interface conditions, radiation damage, *etc*. without knowing the physics of transformation and rule book for degradation and damage? Key studies in fundamental HEDP materials at relevent or surrogate conditions for FST will provide novel, *in situ* spatio-temporally resolved measurements of damage. These new measurements are enabled by a number of the above listed DPS recommendations, in particular the MEC, LCLS multi-PW upgrade and, LaserNetUS, (can say or point to these sections, *etc.*)

The pursuit of this foundational understanding uniquely bridges our community with planetary modeling, exoplanets, and condensed matter physics: IFE targets, initially at ambient conditions, make transitions through warm dense material states to hot, dense plasma. Understanding fundamental changes in atomic and electronic structures under compression and heating requires novel experiments and new modeling techniques. At 10–100 Mbar pressures (created by shock waves or isentropic compression), there are significant uncertainties in the rates governing phase transitions on pico- to femtosecond timescales, the equation of state, the conditions at which phase boundaries occur, and the response of materials to gradients in temperature and electromagnetic fields. Moreover, in the ultra-dense, strongly coupled, degenerate plasma regime, the macroscopic manifestations of quantum effects might lead to novel, stable materials not yet realized by any other science discipline.

**Expert Groups**: HEDP and GPS

### *Recommendations*

**Support research to understand the transport properties (thermal, particle diffusion, viscosity, radiation transport, and nuclear reaction rates) of strongly-coupled plasmas**.

Well-diagnosed, university-scale laboratory experiments can make precise measurements of transport rates. Examples include non-neutral plasmas, ultracold neutral plasmas, dusty plasmas, and pulsed power devices. Understanding these systems requires the development of theory and modeling capabilities to describe the influence of strong coupling and magnetization on transport rates in classical strongly coupled plasmas, as well as the combined influence of degeneracy in dense plasmas. Non-equilibrium physics in warm dense matter during the solid-to-plasma transition must also be investigated (*e.g.* measure and model the dynamical processes of the warm- and hot-dense matter with femtosecond temporal resolution).

**Support research to understand the phase diagram, atomic and electronic structure of strongly-coupled and possibly degenerate plasmas across the pressure/temperature/applied magnetic field diagram through coupled theory, simulation and experimental studies.**

Novel heated and compressed HED states can be created using advanced experimental facilities (lasers, pulsed power, and XFELs), which require precision equation-of-state and transport property measurements to pressures of ~100 GPa to ~100 TPa. Without these benchmark platforms our community will not have credibility in claiming that we understand the fundamental physics well enough to move forward to explore new physics, new regimes, and novel states. We need to measure and model ionization of matter in ultra-dense states, where screening and/or Pauli exclusion forces are strong, to realize novel (potentially useful) states of matter such as high-temperature superconductors, transparent electrides, and at extreme temperatures at which even heavy elements can be stripped of most electrons. Also, to Identify signatures (*e.g.* line broadening) that can be used to reveal fundamental electronic and ionic structure and serve as diagnostics for laboratory plasmas.

**Study the dynamics of multi-qubit entangled states in laser-diagnosed, strongly-coupled, pure-ion plasma crystals as possible systems for quantum computation and/or quantum simulation.**

Trapped ions are a contending approach to quantum computation and quantum simulation. Such systems may exhibit quantum supremacy when they include more than 30-40 quantum bits. Penning traps can confine hundreds of ions in a crystalline plasma state, providing a proof of principle demonstration. Further developments in experiment, theory and simulation will be required to realize the potential of this promising approach. This is an interdisciplinary research field as much of the physics of laser manipulation and quantum computation relies upon AMO science, while the physics of ion traps, decoherence, and the strongly correlated interactions between ions relies upon plasma science. This is an area in which FES may explore partnerships to translate the plasma physics advances to quantum computation. This may be fostered by the ongoing DOE-wide initiative in quantum information science, and is an example topic where FES science can provide high-impact contributions to the broader National Quantum Initiative. Other opportunities where FES science contributes to, and benefits from, quantum computation are described further in the theory and computation cross cut section (CC-TC).

## DPS-E: Create and Explore Antimatter Plasmas

Most plasmas deal with the interaction of positively charged ions and negatively charged electrons. Some exotic plasma situations can introduce antimatter, such as the positively charged "positron," the antiparticle of the electron. In such plasmas, the antimatter components can collide and release light, energy, and/or new particles into the plasma. For example, on Earth, high-intensity lasers can be used to interact with plasmas to induce pair-production (production of both electrons and positrons). Indeed, it is possible to create such plasmas through the interaction of a number of exotic particle sources (accelerator rings, nuclear reactors, radioactive isotopes) with neutral plasmas. In space, exotic neutron stars known as "magnetars" are associated with extreme magnetic fields and are also believed to involve significant antimatter interactions. Because antimatter is challenging to understand, these plasmas may offer some unique opportunities to learn more about both antimatter and plasma physics. Questions that drive this area include: (1) Can trapped antimatter plasmas provide stringent tests of the fundamental symmetries of nature? (2) What are the properties of an electron/positron pair plasma? (3) Due to the equal mass of each charge carrier, a trapped electron–positron plasma is expected to exhibit much simpler properties, such as a vastly simplified CMA wave diagram, than a traditional electron–ion plasma. In this sense, it represents the "hydrogen atom" of plasma physics. Can this simplified system be created and provide high-precision tests of plasma theory? The FES role in this field has been to support advances in basic plasma physics, often in the context of a multinational collaboration engaged in this research, as well as fundamental research in nonneutral plasma properties and techniques at several smaller U.S. institutions. The support of nonneutral plasma physics as a viable and vibrant subfield of plasma physics has been and remains critical to advances in antimatter research.

**Expert Groups**: HEDP and GPS

### *Recommendations*

**Use antimatter plasmas to study the CPT and gravitational symmetry between matter and antimatter, and to investigate the properties of exotic atoms and molecules in which an electron is replaced by an antiproton.**

The study of matter-antimatter symmetry addresses our deepest understanding of nature. It may also shed light on a major mystery of modern physics, the Baryon asymmetry. One of the most powerful targets for this study is the antihydrogen atom: the simplest stable antimatter system. As antihydrogen is synthesized by mixing positron and antiproton plasmas, advances in nonneutral plasma physics have been, and remain, critical to the field. The antihydrogen synthesis rate is determined by the parameters of these plasmas, and has increased from a few—at best—antiatoms per day to thousands of antiatoms per day as our ability to understand and control the constituent plasmas has improved. Experimentally, the necessary advances have included the development of a full suite of appropriate plasma diagnostics, plasma compression, plasma stabilization, collisional, radiative, evaporative, and expansion plasma cooling, positron capture, and plasma-based ECR magnetometry. Theoretically, the advances have included improvements in our understanding of plasma expansion, heating mechanisms, and the antihydrogen synthesis processes.

**Use studies of positron–molecule attachment and lifetimes to investigate fundamental atomic processes and material properties.**

Annihilation of positrons on atoms and molecules is a fundamental process that tests our understanding of QED and many-body processes in atomic physics. In addition, positron scattering off of atoms and molecules using low-temperature positron beams has allowed the first state-resolved measurements of vibrational excitation and electronic excitation of molecules and improved scattering and collisional ionization cross-sections. Positron probes have also been extensively employed to characterize materials via positron annihilation lifetime spectroscopy (PAL). Current capabilities include the ability to measure the concentration and size distribution of voids in materials as well as measure the elemental composition at surfaces. These already-useful capabilities can be improved and extended through higher accuracy (*e.g.* more mono-energetic, and/or spin-polarized) positron beams. The creation and manipulation of such beams is a subject of current research in trapped nonneutral plasma physics.

**Create and probe an electron–positron pair plasma from light using an ultra-intense laser.**

Electron-positron pair production, as predicted by QED theory, offers the possibility of a direct transformation of light into matter. The advent of ultra-high power lasers and advances in gamma-ray sources opens the possibility to experimentally realize theoretical predictions to explore this new light-matter interaction regime. This regime is challenging because one needs not only high photon energies to surmount the production threshold, but also high photon densities to overcome the smallness of the cross-section and achieve an appreciable yield. Laser-driven gamma-ray sources are the key to overcoming these challenges. Several promising approaches have been proposed that rely on such sources. As the energies of particles and intensity of EM fields are increased, a new possibility for producing pair plasma arises, through a cascaded production process of electrons, positrons, and high energy photons. These cascades come in two types. The first is the shower-type cascade, where the initial particle energy is repeatedly divided between the products of successive Compton and Breit-Wheeler processes and typically happens in the collision of a high energy particle beam with an intense laser pulse. The second is the avalanche-type cascade, where the EM field both accelerates and causes QED processes. In this case, the number of particles grows exponentially, fueled by the energy transformation from the EM field into electrons, positrons, and high energy photons. Colliding multiple laser pulses at one spot provides an optimal field configuration. This field configuration is also advantageous for producing copious amounts of high energy gammas. These types of experiments require the development of ultra-intense lasers, particle accelerators, and gamma-ray sources.

**Trap an electron–positron pair plasma in a well-diagnosed laboratory environment to test foundational plasma physics.**

An electron positron pair plasma has many interesting properties (in theory) that have not been tested in experiments. For example, the predicted electromagnetic wave properties in plasmas consisting of equal mass positive and negative charged particles is very simple in comparison to plasmas in which ions are much more massive than electrons. There are also significant differences for electrostatic waves such as the ion acoustic wave, and cross-magnetic field particle transport should be significantly altered by the strong modification of drift wave fluctuations in such a plasma. Electron-positron plasmas can be created at high density (but in small regions) through pair production in an ultra-intense laser field; or at low density in a large region through their admixture of electron and positron nonneutral plasmas. Each method has its own advantages and disadvantages, and both should be pursued. There is also interest in creating a Bose-Einstein condensate consisting of positronium atoms (a Hydrogen atom in which the proton is replaced by a positron). A Positronium gas can be formed by interaction of cold positron beams with material surfaces. The critical temperature of such a condensate is roughly a thousand times higher than that of a Helium condensate at the same density.

## *DPS-2: Understand the Plasma Universe*

We live in a plasma universe. This is not readily apparent, because the Earth is part of the less-than-one-percent of the visible universe that is not a plasma. However, the transition to the plasma universe begins only about fifty miles above the Earth's surface, where the thin upper atmosphere is ionized into plasma by the Sun's ultraviolet radiation in what we aptly call the ionosphere. Further out lies the magnetosphere, a plasma where the Earth's magnetic field shields the surface and life on it from the fast-moving plasma called the solar wind, which emanates from the Sun. The Sun is a massive sphere of plasma, powered by the release of fusion energy at its core. This energy finds its way to us as light and heat, and without it, there would be no life to contemplate its own existence in the plasma universe. Being both nurtured by and sheltered from the effects of plasma, we are free to investigate the rest of the plasma universe. This universe is a spectacularly varied and multi-scaled subject that touches every aspect of our place in the galaxy and beyond—from the heliospheric termination shock and interstellar space to accretion disks around black holes, astrophysical jets, quasars, and active galactic nuclei.

Fortunately, the same fundamental plasma science applies across all of these scales in the universe, including those created on Earth. Laboratory experiments are a growing and powerful tool for understanding the plasma universe by offering unprecedented control of plasma environments and precision measurements, which complement and are not available to traditional approaches such as spacecraft missions and telescope observations. There exist major scientific opportunities that deserve urgent support for a drastic expansion of the U.S. plasma space astrophysics program in the next decade. Understanding the plasma universe through laboratory experiments, theory, and computational modeling is a critical component in fulfilling the DOE FES's mission of stewarding discovery plasma science. The extremely diverse nature of space and astrophysical plasmas offers unique opportunities to critically test and expand our understanding using plasma physics knowledge over a wide range of scales, boundaries, conditions, and geometries.

Inherent in many of the objective-specific recommendations below is the fact that it is not necessary to recreate the exact parameters of plasmas observed in space and astrophysics (either in laboratories or simulations) in order to advance understanding of those plasmas. It is possible to study aspects of properly posed physics questions that can be used to rigorously test theories inspired by observations or *in-situ* spacecraft measurements. Historically, our understanding of naturally occurring physical problems can best be advanced by attacking such problems with varied, complementary tools.

Ten fundamental plasma processes or effects have been identified that play crucial roles in space and plasma astrophysics, and need to be investigated and understood. In a random order, they are:

- Magnetic dynamos
- Magnetic reconnection
- Plasma turbulence
- Collisionless shock waves
- Plasma and photon transport properties
- Wave-particle interactions
- Atomic and chemical processes under extreme conditions
- Coupling of multiple physical processes over multiple scales
- Effects of boundaries and interfaces between plasmas, and
- Physics of flowing plasmas.

These processes or effects occupy a substantial part of the base of plasma physics knowledge. Our capability to understand and predict space and astrophysical plasma phenomena critically depends on the maturity of our understanding of these processes and models built on them.

**Programmatic Recommendations**

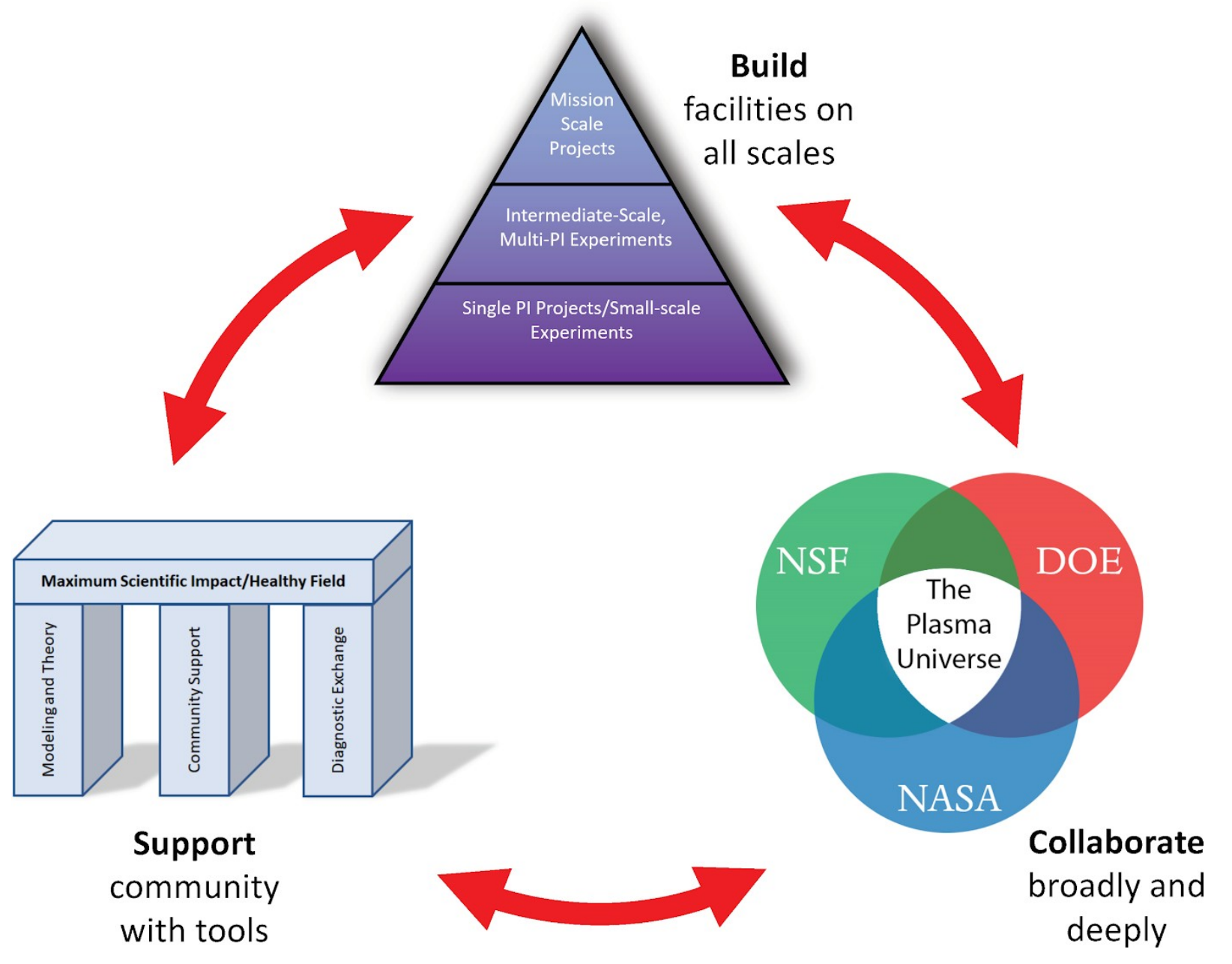

The first programmatic recommendation is to synergize three main components: to **build** facilities on all scales to provide access to plasma conditions relevant to space and astrophysics; to **support** communities with the tools that they need to perform research on these facilities; and to **collaborate** with relevant units of funding agencies, especially NASA and NSF, where space and astronomical observations are made. We recommend building an intermediate scale general plasma science facility to investigate the physics of solar wind plasma in addition to supporting and continuously upgrading facilities on a broad range of scales and topics. We recommend establishing a new MagNetUSA and utilize LaserNetUS to maximize scientific impact of existing facilities and ongoing investments through sharing research tools and supporting a vibrant and collaborating user community. **DOE science centers** can be an effective and flexible platform to form collaborations on the time-critical topics of interests.

The recent or ongoing high-profile NASA missions such as Magnetospheric Multiscale (MMS), Parker Solar Probe (PSP), and the European Space Agency (ESA) Solar Orbiter, and NSF initiatives such as Event Horizon Telescope, Multi-Messenger Astronomy, and Daniel K. Inouye Solar Telescope (DKIST) motivate timely laboratory study of their underlying fundamental plasma processes, such as magnetic reconnection, collisionless shocks, plasma turbulence, and plasma physics under extreme conditions. In addition to the existing DOE/NSF partnership on plasma science, we recommend forming **a new DOE/NASA partnership** for rapid progress in target areas of space missions. Such partnerships can take forms of jointly funded centers or programs, and should be mutually beneficial.

These three components are closely interconnected and interdependent: laboratory facilities provide well-controlled and well-diagnosed research platforms to study space and astrophysical plasma processes, while these facilities are used by a well-supported vibrant community, in close collaboration with space plasma physicists and astrophysicists who explore the plasma universe. All three of these components need to be well-funded and well-coordinated in order to make rapid progress in understanding the plasma universe.

Additionally, fundamental theory is needed to translate scales between laboratory plasma experiments and space/astronomical data. Terrestrial plasma laboratories are necessarily small in physical scale and less energetic. The second overarching programmatic recommendation in Understanding the Plasma Universe calls for supporting **theory, numerical modeling, and associated data analyses**. Numerical modeling is critical for understanding experimental results and interpreting observations. Given the current landscape of high-performance computing at the cusp of exascale, and the development of novel numerical and data analysis methods, the time is ripe for their deployment and support at a commensurate funding level with respect to laboratory experiments and space/astronomical observatories. The broad parameter range that the plasma universe spans necessitates support for fluid, particle-in-cell, continuum Vlasov-Maxwell, hybrid, atomic, and molecular codes, under frameworks facilitating interactions with experiments and observations in a unified, open-source fashion.

## DPS-F: Understand plasma interactions between the Sun, Earth, and other objects in the solar system

The interaction between our terrestrial neighborhood and the outside plasma universe is a prime area for exploring plasma phenomena from both a purely scientific point of view, as well as one that directly impacts the survival of life on Earth and our exploration of the universe. For example, energetic particles from the Sun can enter the Earth’s protective magnetosphere through various plasma mechanisms; these particles can be hazardous to communications and national security satellites as well as space travel. Large fluctuations in the Earth’s magnetic field caused by the impact of coronal mass ejections (severe geomagnetic storms) can destroy energy infrastructure by inducing high voltages in power grids. Estimates of the societal and economic impact of a severe geomagnetic storm to the United States are the equivalent of ten simultaneous hurricane-Katrina-level natural catastrophes. This is but one of a multitude of plasma interactions between the Sun and other objects in our solar system. Although by no means exhaustive, a list of other community supported topics and outstanding questions are provided in the three recommendations of this strategic objective. An overarching recommendation is for increased cross-fertilization of ideas and support through formal memoranda of understanding between DOE and other U.S. government agencies such as NASA and NOAA. A successful example of an existing cooperative agreement is the NSF/DOE partnership in plasma science and engineering. This recommendation is intended to be consistent with the National Academy of Sciences’ 2007 report, *Plasma Science: Advancing*

*Knowledge in the National Interest*, in which such partnerships are not intended “to replace or duplicate the plasma science programs in other agencies.” It is here recommended that the strengths, interests, and technical expertise of each agency be brought to bear in synergistic approaches to investigate or support the investigation of common science problems.

**Expert Groups**: HEDP and GPS

## *Recommendations*

### Support studies of fundamental plasma processes at or near the Sun

Studying the interactions between the Sun with the rest of the solar system begins by understanding the Sun itself. The Sun is our nearest star and is required for life to exist on Earth, yet basic processes such as energy transport within and near the Sun are not well understood. For example, the Sun’s atmosphere (the corona) is more than 300 times hotter than its surface. How does energy flow underneath the surface? What are the dominant mechanisms that transport energy to the corona and heat it? Is it due to damping of Alfvén waves launched on the solar surface or due to a large number of small flares driven by surface convection? Dangerous magnetic storms and large solar flares are the result of stored magnetic energy being released via impulsive instabilities or magnetic reconnection. How does the magnetic energy get stored and triggered impulsively for massive Coronal Mass Ejections or CMEs? Answering such questions represent areas of opportunity for research by an expansion of the NSF/DOE partnership investment, as well as partnering with NASA, for plasma scientists specializing in experimental, theoretical, and simulation approaches to these questions.

### Support our understanding of the fundamental plasma physics of the solar wind

The next step in the Sun-solar system interaction problem is to understand the fundamental properties of the solar wind plasma that connects the Sun to the rest of the solar system. Basic questions about the solar wind remain unanswered. How is the solar wind able to accelerate to very high speeds against solar gravity? How does the solar wind maintain its temperature via collisionless dissipation despite the wind’s rapid expansion? How do plasma waves and instabilities redistribute energy from very large scales to scales small enough to heat this plasma? In recent decades, much progress has been made by laboratory experiments in conducting detailed studies of phenomena that complement measurements by rockets and satellites within the confines and at the boundary of the Earth’s magnetosphere. However, the realm of the solar wind has been largely unexplored in the laboratory due to the lack of an experimental device in the world that is capable of producing the demanding plasma parameters required to do so. This represents a blind spot in the community’s ability to deeply understand the fundamental plasma physics of the vast majority of the solar system and the medium through which plasma waves and energetic particles travel from the Sun to the Earth. In the era of missions such as the Parker Solar Probe and the Solar Orbiter, the U.S. has a timely opportunity to create a global leadership role in exploring plasmas relevant to the solar wind in a laboratory facility. This goal can be realized by deploying current technologies to achieve laboratory plasma conditions that reproduce a critical subset of dimensionless parameters necessary to study the physics of the solar wind as well as other solar and astrophysical phenomena.

**Support research into the interactions of the Sun with planetary magnetospheres and unmagnetized systems**

The third component of the plasma interactions between the Sun and our solar system is to understand the endpoints of the Sun's plasma and its effects on planetary magnetospheres and unmagnetized systems throughout the heliosphere. While the specific interactions and subsequent triggered plasma processes are myriad, (hence beyond the scope of being listed here explicitly) the following are meant to represent the broad range of topics and outstanding physics questions in need of further investigation and support:

- The physics of energetic ion scattering by waves is pertinent to both planetary magnetospheres and magnetic fusion energy devices. How are the Earth's radiation belts populated by energetic electrons and ions from the solar wind; and, what are the physical mechanisms by which both high and low frequency plasma waves naturally depopulate these belts?
- What role do plasma waves play in structuring the earth's ionosphere, which can (for example) affect signals from global positioning satellites?
- How do three-dimensional and particle kinetic effects impact magnetic reconnection in the Earth's magnetotail and dayside magnetopause?
- At Earth and elsewhere in the solar system, how do flow- and gradient-driven instabilities grow in the magnetopause and/or plasmapause and how do they contribute to plasma transport, energy and momentum coupling, wave generation, and turbulence there?
- How do unmagnetized objects like our Moon interact with the solar wind, including phenomena such as dust dynamics on the Moon's surface and the wake behind the Moon?
- What are the properties of magnetized collisionless shocks, such as the Earth's magnetospheric bowshock in the solar wind? How do such shocks form in the first place and how do they evolve over time?

As with other recommendations within this strategic objective, it bears re-emphasizing that projects and missions by NSF and NASA provide much of the inspiration for research in these areas. However, further partnerships with those agencies and DOE can encourage collaboration between scientists traditionally funded separately by these agencies in order to achieve a more complete understanding of these problems.

## DPS-G: Understand the origin and effects of magnetic fields across the universe from star and planet formation to cosmology

The most prominent effect of plasma physics across the universe is the presence of magnetic fields on almost all scales at which plasma also exists. Understanding the origin of these magnetic fields is a grand challenge problem. This includes the origin of the Earth's magnetic field and its evolution, as well as other planets and their moons in our solar system. Further afield is the solar dynamo, which is regularized to a distinct and still mysterious 22 year cycle, one of many dynamic behaviors of stellar dynamos. Even farther out are galactic dynamos, which take a completely different form of dynamics with substantially larger spatial scales and longer temporal scales.

The effects of magnetic fields have been long recognized outside our heliosphere from the study of the interstellar medium of the Milky Way, where magnetic fields constantly interact with thermal and nonthermal plasmas within and in between the molecular clouds. The planetary and solar structures that form from these clouds require multi-scaled plasma processes, from the interaction of electrons, to dust grains, to the evolution of accretion disks. We must understand several fundamental plasma mechanisms to explain these phenomena.

Our view of the cosmos must be seen through the lens of a plasma universe. Our understanding of this plasma fundamentally alters our understanding of the evolution of the cosmos. Traditional laboratory astrophysics has made significant contributions to our understanding of the plasma universe but much more needs to be done to keep up with ever expanding astronomical databases and their analyses. Further studies of these effects must be supported.

**Expert Groups**: HEDP and GPS

### *Recommendations*

**Support further understanding of the origin of the planetary magnetic fields, stellar dynamos, and the origin of magnetic fields on galactic and cosmological scales.**

Outstanding questions include: How does the geomagnetic field evolve in time and is it going to flip its direction soon? Why do different planets and their moons have different magnetic fields? Where do the Sun's small-scale magnetic fields, known as turbulent dynamo, come from? A critical question is whether a universal mechanism exists for all of these dynamo processes or whether each dynamo has its own specific conditions and outcomes. Many aspects of these dynamos should be further studied in the laboratory under well-controlled experimental setups, in addition to the opportunities in exploring the state and dynamics of the Earth's metallic inner/outer core as well as other planets' cores.

**Support studies of magnetic field effects during formation of stars and planets (including exoplanets) in accretion disks and stellar jets.**

The discovery of exoplanets has opened an exciting area of astronomy and plasma astrophysics. Outstanding questions include: What are plasma properties at the edge of atmospheres and the interiors of planets in order to determine exoplanet evolution and habitability? What are the plasma properties and transport as a function of distance from the host star, the host star brightness, planet mass, and planetary magnetosphere? More broadly, what is the role of plasma and magnetic fields in accretion disks where stars and planets are being formed? How are magnetic fields generated to facilitate angular momentum transport in accretion disks, how are they maintained against dissipation, and how do plasmas convert gravitational energy to radiation during the accretion process? What is the role of magnetic fields in the collimation, stability, and radiation of stellar jets over long distances? Many of these processes, such as angular momentum transport in accretion disks and jet launching and stability, have been studied successfully in the laboratory, and further investigations into these and other fundamental processes should be supported.

Dusty plasmas are ubiquitous in space, including the interstellar medium, molecular clouds, and protostellar-protoplanetary disks. Outstanding questions include how do dust grains interact with ambient gas and plasma regarding their charging, breakup and magnetization, and how do they coalesce to grow into planetesimals and eventually into planets (and exoplanets)? The synthesis and coagulation of dust particles to form large structures can and has been studied by laboratory dusty plasma experiments. Filamentary instabilities that are excited in dusty plasmas have been studied in laboratory experiments, and it is possible that they are relevant for mechanisms of planetesimal formation and should be studied further by experiments and simulations.

**Support further studies of atomic and molecular spectroscopy in astrophysical environments.**

Important tasks include identifying, interpreting, and predicting a large number of atomic and molecular spectral lines from astronomical observations. Understanding how atomic and molecular spectral lines are affected or modified is critical for astrophysics and cosmology research when viewed through the lens of large expanses of plasma, which can range from the collisionless fully ionized inter-galactic medium to the collisional weakly ionized molecular clouds in the Milky Way. Continued support should be provided in these areas of laboratory astrophysics.

## DPS-H: Understand the causes and consequences of the most energetic, extreme, and explosive phenomena found in the cosmos

Many objects in the universe, from the crushing pressures of planetary cores to the intense fields of magnetized neutron stars or the event horizon of black holes, challenge our knowledge of how plasmas and particles, including nuclear and atomic physics processes, behave under extreme conditions. Laboratory investigations of these phenomena thus provide rare opportunities to test the robustness of plasma physics and to expand its frontiers into uncharted territories. Plasma mechanisms allow for some of the most energetic acceleration events possible in the universe (such as ultra-high energy cosmic rays), far beyond any powerful man-made accelerators. However, the mechanisms for these events, which include the most luminous sources (such as active galactic nuclei) and the most powerful explosions (such as gamma ray bursts) known in the universe, are not well understood. Material under extreme pressures like those in the interiors of white dwarfs or Jovian planets can transform into new plasma states like warm dense matter or exhibit exotic phase transitions, which can only be studied in detail in laboratory experiments. Such plasma conditions also affect nuclear physics and can shed light on stellar dynamics and nucleosynthesis models relevant to the big-bang and the abundances of elements in the cosmos. Plasmas subject to intense fields can further exhibit new atomic physics, such as the appearance of novel spectroscopic features that could be used to study astrophysical objects. A solid knowledge base is needed to understand the multiple-scale physics of these plasmas, including under extreme conditions, in order to build reliable models to explain and predict astrophysical observations.

**Expert Groups**: HEDP and GPS

### *Recommendations*

**Support research to assess the mechanisms by which particles are heated and accelerated to some of the highest energies observed in the universe.**

Cosmic rays are known to be accelerated to ultra-high energies by plasmas from exploding stars (supernovae), extragalactic jets, and gamma ray bursts (GRB). However, it remains unknown both which processes (for example collisionless shocks, magnetic reconnection, turbulence) are responsible for these energetic particles, and how these processes accelerate and heat particles to extreme energies. Experiments can recreate these processes in the laboratory with appropriately magnetized and scaled conditions, allowing for well-controlled, reproducible, and well-diagnosed studies of particle acceleration and heating relevant to astrophysics.

**Support further understanding of plasma and atomic physics under extreme fields and densities, from planetary cores to black holes.**

Material at the center of planets and stars is subject to extreme densities, temperatures, and pressures, resulting in exotic states of matter. Likewise, plasmas around compact objects such as black holes are subject to intense fields and pressures. Critical to understanding these conditions is knowledge of equations of state, opacities, and particle and energy transport, which are difficult to observe in astrophysical systems. Conversely, these extreme conditions may manifest new spectroscopic features that can be observed. Laboratory experiments with high-power lasers and pulsed-power can now generate plasmas and matter in intense electromagnetic fields, under extreme pressures, and subject to strongly non-equilibrium and non-thermal conditions. Such experiments are essential for understanding the conditions under which non-Maxwellian, non-LTE codes are valid, for deriving material equations of state, and for accurately modeling plasma parameters.

**Support further studies of how plasmas affect and moderate nuclear reaction rates, nucleosynthesis, and abundances in the universe.**

Being able to study stellar-relevant plasmas is essential to understanding the rates of astrophysical nuclear reactions, which in turn are needed for stellar dynamics and nucleosynthesis models. Plasma physics experiments can thus contribute measurements of great value to improve our understanding of these nucleosynthesis processes, including big-bang and stellar nucleosynthesis. Additionally, the unique properties of plasmas or plasma-generated beams of particles or photons are potentially a powerful tool for producing better nuclear data for astrophysics.

## DPS-3: Create Transformative Technologies

Plasma science plays a critical role in enabling many of the technologies in our modern society. Controllable plasma chemistry, advanced plasma source design, and fundamental understanding of the nature of plasma surface interactions are a few areas of needed advance in basic plasma science to continue to advance applications. HEDP science is also positioned to make broader contributions to society that will be enabled through advanced source development for bright sources of particles and photons that will advance not only basic plasma science understanding but also areas of medicine, energy, and manufacturing. For example, plasma assisted manufacturing contributes to a broad range of industries including aerospace, microelectronics, defense, and energy. Plasma science enables fabrication of devices on the atomic scale, and plays a role in over a quarter of the hundreds of process steps taken to fabricate the ubiquitous electronic devices that drive our nation's economy. Plasma science enables advanced coatings and surface treatments that reduce the weight of commercial aircraft and make materials impervious to degradation due to chemical or environmental exposure. Laser plasma interactions provide unique bright radiation sources that could be employed in many technological areas like non-destructive evaluation of aircraft components or civil engineering structures. Laser produced tin plasmas are already being used to manufacture the latest generation of computer chips, a multi-hundred billion dollar a year industry. These contributions have been enabled by bridging fundamental plasma science with other science domains and application needs to make plasma science a foundational part of American leadership in a plurality of industries.

Today, plasma science is positioned to make substantive contributions to a growing portfolio of applications and industries critical for growing national leadership in existing applications and establishing leadership in new technologies. To realize this potential, continued advances need to be made in plasma science, specifically in the areas of plasma based chemistry production, in plasma-based radiation sources and integration of plasma science into new multidisciplinary efforts. Through this effort, plasma science has the potential to play a key role in ending cancer, feeding an ever growing national and global population, and growing national leadership in high technology manufacturing.

Specific disruptive technology themes have been identified where plasma science has a clear path to make substantive contributions to the growth of the U.S. economy and the well being of her citizens. These themes serve two roles. The first is to present research focus areas in basic plasma science that require both a plasma-centered emphasis that falls within the FES charter and the interdisciplinary collaboration necessary to translate these advances to domains spanning biology, material science, and environmental science. The second is to motivate and provide justification for this basic science research by presenting highly impactful applications of these efforts that will improve both societal well being and national technology infrastructure. It is critical that basic science and engineering challenges in plasma science as well as interdisciplinary areas that intersect plasma science with other areas be addressed. Collectively these advances will be used to develop disruptive technologies that will have a profound impact on the economy and society. FES is in a position to take actions that will realize opportunities spanning the Department of Energy mission and beyond in the areas of energy, medicine, agriculture, transportation infrastructure environmental stewardship, and advanced manufacturing. We have defined a set of strategic objectives and concrete recommendations that will advance plasma science understanding to contribute to these critical areas.

The U.S. programs should invigorate low temperature plasma research and key areas within HED plasma science at universities and national labs into a sustained program to develop enabling disruptive technologies based on efficient plasma generation techniques, the understanding of the resulting plasma conditions, and complex plasma-surface interactions. Universities are well suited to develop the basic understanding necessary to develop new technologies, for developing these technologies at lower Technology Readiness Level (TRL) and for workforce development. National labs are well suited to characterize and evaluate higher TRL technologies for industry and provide a comprehensive study of the plasmas and complex plasma-surface interactions that are the basis of most of the enabling disruptive technologies. Together, universities and laboratories working in these areas present a compelling space for public-private partnerships with pathways for scientific discovery and deployment of viable solutions to substantive challenges facing our country and planet in the coming decades.

**Increase support for single-PI-scale research projects**

Low temperature plasma science is a fast-moving field of research. It is characterized by the potentially short time of development from concept to engineering devices. Furthermore, many of these advances do not require large facilities and are aptly carried out in laboratories led by individual PIs. This fast pace of innovation is well served by support for a broad range of single-PI-scale research grants. Small-scale grants also foster innovative new ideas, and continued growth of the field, by enabling new PIs to enter through frequent grant solicitations.

There is a strong argument that has been advocated for in this community planning process that the DOE should have a well-funded, dedicated, national program for low-temperature plasmas. The NSF/DOE partnership has been the traditional mechanism for funding such research. However, this partnership program is oversubscribed as it is tasked with supporting research across the entire spectrum of plasma science and engineering. This limits the growth of low-temperature plasmas, and the industries that it enables. Potential avenues to address this would be for the DOE to increase support for LTPs through targeted DOE solicitations, the existing partnership, and through pursuit of additional multidisciplinary partnerships.

**Foster public private partnerships through recommendations made in the Enabling Technology cross cut**

Public/private partnership is a recurring theme across many science drivers identified in this report, spanning magnetic fusion, HEDP, and general plasma science. The plasma science community working in the area of disruptive technologies have historically built strong and extensive partnerships with industry to move plasma science forward, particularly in the area of integrated circuit fabrication, but also in aerospace and textile industries. It is recommended that FES leverage the experience of the LTP community in fostering partnerships with industry to accelerate the maturation of public-private partnerships spanning the entire FES mission.

## DPS-I: Develop plasma-based technologies that contribute to a stable national energy infrastructure

The contributions that plasma science can make to the nation's energy infrastructure are broad and extend beyond the goal of fusion energy production. Plasma-based technologies can utilize the unique combination of energetic particles, radiation, and chemistry derived from plasma discharges to replace industrial processes that have a substantial dependence on finite resources and high carbon emission. As fossil fuels are displaced, the world's energy infrastructure will continue to evolve toward more electricity generation, particularly from renewable and nuclear sources. Plasma-based technologies can enable a future based on these electricity sources. The primary basic plasma science challenges that need to be addressed to advance this theme center on advancing our understanding of reactive chemistry formation in non-equilibrium systems and the interaction between the plasma state and an array of novel material forms. The fundamental relationship between the non-equilibrium conditions typical to plasma discharges and the chemical reaction pathways that will enable carbon-free industrial processes needs to be advanced. By advancing the understanding of these coupled processes the key deliverable of controllable chemical selectivity that is central to the successful deployment of plasma technology into the energy infrastructure may become possible. Additionally, the complexity of the plasma boundary, particularly with regard to the diversity of material forms that will be required to directly interact with the plasma state (liquids, aerosols, particulate, catalysts, nanoscale structures, *etc.*), will require a re-examination of the plasma material interface that adequately describes these interfacial processes.

**Expert Groups**: GPS

***Recommendations***

**Support research to study plasma-driven physical and chemical pathways for selective processing of materials to displace carbon-generating industrial processes.**

Engineering advanced chemical reactors will require the development of plasma sources that can target certain chemical reaction pathways of interest. However, there are basic science questions that are unanswered that limit the ability to do this. Examples of the fundamental unknowns that govern these processes include incomplete knowledge of species production pathways, interaction cross sections, and energy distributions of both charged states and bound atomic and molecular states. Furthermore, these processes are usually closely connected with the type of plasma source. Advancing the understanding of the coupling from the characteristic time and length scales that define plasma systems to the time and length scales that define material and chemical processes, will enable the development of potentially transformative new technologies.

**Support interdisciplinary, multiple-PI science centers that enable a scope of work that extends beyond basic plasma science to capture plasma interaction with energy-system-relevant metrics.**

Advances in plasma science to enable adjacent technologies require integration of multiple science domains, particularly to deepen our fundamental understanding of how plasmas can be better controlled and how plasmas interact with other material phases beyond bulk solid materials that have typically been employed. Plasma science research integrated with chemistry, material science, and more broadly with industrial-scale energy intensive systems will accelerate deployment of plasma science and technology in this area.

**Support research to advance understanding of the interaction between plasmas and the wide diversity of materials of relevance to advancing the energy infrastructure.**

The outcomes enabled by plasma-based technologies not only depend on the processes occurring within the plasma, but also on what can be extracted from the plasma. For example, plasma reactors may produce a chemical of interest to an energy application, but that chemical must be extracted from the plasma in order to be useful. Delivering either chemical or physical products relies on understanding the interaction between the plasma and the boundary. For example, if the boundary is a liquid, one may need to understand the processes of dissociation of a certain radical as it leaves the plasma and enters the liquid. Modern industrial processes are often associated with the interaction of plasmas with complex materials (liquids, aerosols, particulates, catalysts, nanoscale structures, *etc.*). Understanding this complex interaction will require significant advances in our fundamental understanding of plasma material interfaces.

**Explore partnerships with other agencies within DOE to support these interdisciplinary goals. These include DOE-BES, DOE Fossil Fuels, DOE Nuclear Energy, and ARPA-E.**

This is a highly interdisciplinary driver, spanning energy production, storage, transportation , and distribution. Additionally, it extends into energy intensive industrial processes such as chemistry production and fuel reforming, introducing plasma-based alternatives that present a viable green alternative for many chemical production processes. Key to this science driver is the establishment of interdisciplinary broad research efforts that are structured to combine plasma science with chemistry and material science efforts. Collaboration between OFES and other DOE divisions with a direct energy infrastructure mission such as Nuclear Energy, fossil fuels, and renewable energy would provide interdisciplinary structure that could accelerate advances in this area.

## DPS-J: Develop plasma-based technologies that enable advanced manufacturing

Advances in plasma science have made the United States a leader in the technology and manufacturing sectors that are the cornerstone of the information age. Moving forward, this capability should be expanded to sustain and grow this competitive advantage, particularly with regard to substantive efforts underway in other countries such as China that seek to displace U.S. leadership. This demands advances in basic plasma science that further our understanding of physical and chemical processes that drive manufacturing to a level that enables advanced manufacturing controls such as machine learning and advanced process control. Extending plasma capabilities in processing bulk material surfaces to new material interfaces such as nanoparticles, liquids, bio-inspired materials, and atomic-scale topologies will align plasma science with advanced manufacturing roadmaps in a plurality of industries serving the nation’s high technology sector. These include, but are not limited to, high performance computing and quantum information science, incorporation of advanced manufacturing techniques into plasma science, such as machine learning, and artificial intelligence. Enabling “design for process” concepts to reverse engineer plasma sources for a given task will further accelerate the time to market for these plasma based manufacturing techniques. Advances in plasma-assisted advanced manufacturing are diverging in scale. On one side, the advance of plasmas at near atmospheric pressures have enabled processing of new materials on a bulk scale as well as processing of new materials that are not vacuum hardened. For example, advances in plasma assisted manufacturing for large area laminate surfaces for the aerospace industry, UV resistant and super-hydrophobic fibers for the textile industry, and processing of bio-inspired materials continue to make advances toward substantive contributions in these manufacturing process flows. One the other side, advances in plasma chemistry formation and control have demonstrated the potential to contribute to manufacturing processes at the atomic and molecular scale. For example, atomic and molecular scale assembly of materials and structures through plasma assisted atomic layer deposition (ALD) and molecular layer deposition (MLD) processes, synthesis of nanoscale particles, and increased diversity in material accessibility for design of systems in the microelectronic (now more appropriately defined as nanoelectronic as manufacturing processes extend to dimensions less than 100 atoms across) systems that have reduced power consumption of devices and enabled expansive integration of devices and sensors in the internet of things. Across this spectrum there are overarching challenges where fundamental science and engineering advances are required to realize sufficient process rates and process control to move this technology to commercial volumes.

**Expert Groups**: GPS

***Recommendations***

**Support research to advance understanding of plasma-generated chemical, energetic, and directional selectivity and control over advanced manufacturing processes.**

Advanced understanding of the formation and control of selective and anisotropic chemistry will be required in order to mimic the historic pace of Moore's Law and move into new manufacturing paradigms. The requires the translation of selectivity to highly-controlled material compositions and structures across a plurality of materials spanning liquids, particles, aerosols, and complex non-planar topologies. It will also require the advance of experimental and computational efforts to produce a level of predictable computationally-assisted design that is comparable to engineering tools in adjacent physical domains such as heat transfer, fluid mechanics, and mechanical forces. Currently, there is a recognized level of refinement that is required to capture the coupled physical, chemical, and material domains with sufficient precision to predictably design systems that employ plasma based systems for advanced manufacturing. Even in the semiconductor industry, the most mature and successful industry that plasma science can point to as having made substantive contributions to advances in high tech manufacturing, the systems that are vital for further advances that rely on plasma technology still require a level of iterative design and build that is not needed in other areas of semiconductor manufacturing such as lithography, bulk material synthesis, or ion implantation. By advancing the underlying science of plasma based systems incredible increases in efficiencies, even in established manufacturing technologies such as semiconductor device fabrication, can be realized while also contributing to new advances in manufacturing across a plurality of high technology sectors. This should include both university and laboratory engagement. It is likely best served by several single-PI-scale grants.

**Support interdisciplinary, multiple-PI science centers to understand plasma's role in advanced manufacturing flows including material synthesis, removal, modification, functionalization, and advanced manufacturing modalities**

Precision control of manufacturing processes requires high-level understanding of the basic science that drives that specific process. For example, precise control of laser welding processes requires inputs spanning radiative transport, heat transfer, and material science to enable weld processes that continue to make advances in manufacturing to this day. Plasma assisted manufacturing requires a similar effort, both in existing plasma based manufacturing processes and those that are envisioned to make substantive contributions to the national economy in the coming decades. Advancing fundamental understanding of physical and chemical processes in plasma systems is a vital first step in integration of intelligent manufacturing controls to unit manufacturing processes and process flows where the plasma state plays a critical role. To this end, convergent efforts spanning plasma science, advanced manufacturing control methodologies, and specific manufacturing processes are needed. This requires collaboration between universities, national laboratories, and industrial partners.

**Form partnerships with DOE-BES, ARPA-E, and DOE's Advanced Manufacturing efforts.**

This is a highly interdisciplinary driver, spanning energy production, storage, transportation, and distribution. Additionally, it extends into energy intensive industrial processes such as chemistry production and fuel reforming, introducing plasma-based alternatives that present a viable green alternative for many chemical production processes. Key to this science driver is the establishment of interdisciplinary broad research efforts that are structured to combine plasma science with chemistry and material science efforts. Collaboration between OFES and other DOE divisions with a direct energy infrastructure mission such as BES, ARPA-E, and DOE's Advanced Manufacturing efforts would provide interdisciplinary structure that could accelerate advances in this area.

## DPS-K: Develop plasma-based technologies that improve the physical well being of society

As a steward for plasma science, DOE-FES is uniquely positioned to lead the advance of science that can make substantive contributions to the physical well being of society with contributions spanning medicine, agriculture, and environmental science. This field is interdisciplinary, but the desired outcomes center on fundamental plasma science challenges, particularly in areas that rely upon process selectivity, which require advancing the underlying knowledge of the physics and chemistry that drive reactive plasma systems. It is therefore imperative for the LTP community to develop an understanding of the role of the plasma in relevant processes and interactions. An example where this understanding is critically important is plasma-based engineering of biological processes. This area converges plasma physics, plasma chemistry and plasma engineering with biology. It focuses on the interaction of the reactivity produced by low temperature plasmas, usually at atmospheric pressure, with soft biological matter (*e.g.* liquids, cells, tissues, food, plants, agricultural products). LTPs provide a unique, rich environment of reactive oxygen species, reactive nitrogen species, charged particles, photons, and electric fields. One of the unique features of plasmas compared to other sources of reactivity is the ability to very rapidly change the reactive species production pathways, thereby enabling feedback systems that customize in real time the reactivity delivered to objects of interest. As such adaptive plasmas for medical, agriculture and perhaps material science applications represent emerging areas that require research support. Key to advancing plasma science in these areas is enabling collaboration across entities outside of the energy infrastructure sector such as medicine, agricultural science, and environmental science.

These can be achieved through public-private partnerships or inter-agency research initiatives with entities such as NIH, FDA, USDA, and EPA.

**Expert Groups**: GPS

***Recommendations***

**Support university-scale projects to advance understanding of plasma interactions with bio- and enviro-inspired materials.**

Many of the fundamental science questions associated with the interaction of plasmas with soft matter can be addressed by small-scale single to few PI-led investigations. Universities provide a natural place to conduct this research not only because it is the appropriate scale, but also because it is interdisciplinary. Every university has biology, chemistry, physics and engineering departments, and many have medical colleges. This proximity can foster interdisciplinary research. The future of this field will also rely on training students in multiple disciplines, which is naturally done at universities.

**Support interdisciplinary, multiple-PI science centers that enable a scope of work that extends beyond basic plasma science to capture plasma interaction with living systems, natural resources, and diverse ecosystems.**

Of the three disruptive technology objectives outlined here, plasma systems for agriculture, environment, and medicine has arguably the greatest need for interdisciplinary efforts, as it spans science domains that have had little interaction over the last several decades and stand to have the greatest substantive advances through interdisciplinary study. These efforts should include pathways for the evaluation of plasma technology in medical devices and scalable systems that can provide tangible solutions on the scale of agriculture, environmental remediation, and human well being and include expertise in biological system response, ecosystem responses to new technologies, and evaluation of next generation agricultural systems for irrigation, waste management, nutrient control, and distribution networks.

**Explore partnerships with NIH, USDA, FDA, and EPA as well as engagement with DOE's current efforts in addressing the energy-water nexus.**

As mentioned previously, interdisciplinary research is vital to advance plasma science in this area. DOE-OFES has established a strong foundation in plasma science. In order to enable this foundational work to contribute to disruptive technologies outside of the primary OFES deliverable of achieving fusion energy, collaborations with adjacent agencies directly tied to these technology domains should be pursued. Without this, plasma interaction with these disparate material systems, vital to advance these technologies, falls between agency level funding priorities.

## DPS-L: Develop plasma-based technologies that provide secondary sources and other new capabilities, to benefit fundamental science, industry, and societal needs.

One of the most promising applications of HED plasmas, with major potential impact for science, industry and society, is the realization of bright compact sources of high-energy particle beams and photons. The practical benefits of such sources are numerous and compelling; not only can advanced radiation sources be used to probe and create novel HED plasma states, but they also have potential for impacting societally important areas like the fabrication of the future generations of computer chips to medicine. HED plasmas offer several possible methods to realize promising sources covering photon energies spanning from the extreme ultraviolet to gamma rays. In the case of laser-driven photon sources, these methods include exploiting relativistic phenomena in overdense plasmas, radiation from relativistic electrons accelerated in the plasma wakefields, and coherent and incoherent line radiation and continuum radiation resulting from atomic processes in dense laser-created plasmas and Z-pinches. Similarly, different approaches have been demonstrated or proposed for the generation of intense beams of energetic ions, neutrons and positrons. The realization of useful sources will require the development of higher efficiency laser drivers, an increase in the high repetition rate of the lasers, new targets, and new diagnostics.

**Expert Groups**: HEDP

### *Recommendations*

**Support the development of laser-driven plasma sources of high energy photons and particle beams**

Important societal needs and fundamental science studies require a new generation of more powerful, efficient, and compact sources of high energy photons and directed particle beams. Plasmas are efficient sources of high energy photons ranging from the extreme ultraviolet to gamma rays. They can be tailored to generate both bright coherent beams and powerful incoherent radiation. Plasmas can also generate and accelerate particles to form particle beams of unprecedented flux and energy. New laser technology for secondary sources should be developed in a coordinated effort amongst agencies. Such sources could reach presently un-obtainable parameters and open a path to new science and unique solutions to societal needs.

# Fusion Science and Technology

Fusion energy science is now sufficiently mature to warrant the mission to construct a fusion pilot plant capable of the production of net electricity. The transition to a mission-driven program for fusion energy is motivated by steady advances in plasma science, progress at major new international facilities such as ITER and W7-X, rapid advancement in computational and modeling capabilities, and burgeoning investment from private industry. Here we describe a community-driven plan to embrace exciting new research opportunities in fusion science and technology (FST) that are required to realize the goal of fusion energy. Throughout the CPP, the content for this chapter was primarily developed by community members through the Magnetic Fusion Energy (MFE) and Fusion Science and Technology (FM&T) topical area activities with additional contributions from the High Energy Density Physics (HEDP) topical area. This plan reflects the strong agreement among the community that research in this area should be driven by the mission to enable economically competitive fusion energy in the United States, in order to address the urgent issues of energy sustainability and security. The recommendations made here are broadly consistent with the recent National Academies Burning Plasmas report, and we support the recommendations of that report for the U.S. to remain a partner in ITER and to begin a science and technology program leading to the construction of a fusion pilot plant (FPP) that would operate as early as the 2040s. The community recognizes that significant additional investment in fusion materials and technology is needed, as these areas are relatively under-developed, apply to nearly any plausible FPP design, and likely set the timescale on which any FPP could be successful. We recommend establishing a multi-institutional, multidisciplinary program for exploring FPP designs, together with industry, to drive and integrate the latest scientific innovations, identify the critical cost drivers of an FPP, and inform research priorities accordingly (FST-PR-A).

Our research plan for FST is driven by three major themes we call *Science Drivers (SDs)*: 1) Control, Sustain, and Predict Burning Plasmas; 2) Handle Reactor-Relevant Conditions; and 3) Harness Fusion Energy. Each Science Driver represents a unique area of scientific inquiry, but they are interlinked, and all three must be accomplished to achieve fusion energy. In this strategic plan for FST, we have identified eight Strategic Objectives (SO) and five FST Program Recommendations (PR) that each relate to one or more of the Science Drivers. Among these objectives and recommendations, the community identified four research areas to be of highest priority. Expanded research and development efforts are required to develop plasma-facing components capable of withstanding reactor-relevant conditions (FST-SO-A) and structural and functional materials that can withstand FPP neutron fluxes (FST-SO-B). Additionally, blanket and tritium technology should be aggressively pursued (FST-SO-C). Increased emphasis in these areas of fusion technology must also be accompanied by a robust research program that allows for the completion of the tokamak physics basis (FST-SO-D) and the realization of FPP-relevant plasma conditions.

Remaining a full partner in ITER remains the best option for U.S. participation in a burning-plasma experiment (FST-PR-B). New initiatives by private interests to achieve burning plasmas also hold great promise, although with greater risk, and have galvanized a new generation of scientists and engineers. The U.S. fusion program should embrace these initiatives and seek to support and utilize new private facilities to advance fusion science and technology, where possible (FST-PR-C, FST-SO-H.4).

However, ITER and other planned/existing facilities will not be able to fully address the high heat flux and neutron fluence conditions that will be present in a fusion power plant. To address these areas and enable the design of a pilot plant that projects to an economically viable fusion power plant, new programs and facilities will be needed. The highest priority new facilities needed for rapid progress towards an FPP are a fusion prototypic neutron source (FST-SO-B.2) and a high-power-density tokamak facility for developing and testing divertor solutions (FST-SO-D.2). In parallel to these facilities, targeted investments should be made in programs to provide critical new research capabilities and enhance U.S. leadership. This includes expanding programs to develop suitable materials for FPP-relevant plasma facing components (FST-SO-A), including completing and operating MPEX, and to advance the science and technology required for functional fusion blankets (FST-SO-C). These programs are needed to develop solutions that are critically important to most fusion reactor concepts.

Confinement is known to be a primary driver of the cost of an FPP, and many important questions regarding confinement physics remain. While tokamaks currently represent the leading candidate for an FPP, a robust program is needed to complete the tokamak physics basis for an FPP, utilizing ITER, DIII-D, NSTX-U, and other domestic and international facilities (FST-SO-D). Examples of critical issues that need to be resolved include disruptions, power handling, sustainment, and core-edge integration. In parallel, the success of W7X and significant recent advances in theory and modeling capabilities motivate the design and validation of innovative quasisymmetric stellarator concepts, which represent an area of U.S. leadership (FST-SO-E) and could offer an attractive path to an FPP. This should be done in addition to a renewed program to develop select alternative MFE confinement concepts (FST-SO-H.3). The U.S. currently has a significant world lead in Inertial Confinement Fusion due to the large investment by the National Nuclear Security Administration (NNSA). This should be leveraged to build an Inertial Fusion Energy (IFE) program that offers a distinctly different and potentially attractive path to fusion energy (FST-SO-H.1-2). When determining support for alternative confinement concepts, scientific and technical feasibility of the concept as a pilot plant, as well as potential to reduce the cost of a fusion pilot plant, should be major considerations.

The U.S. fusion program must distinguish itself from those of its international partners by focusing on developing scientific and technological solutions that will enable fusion energy that is economically viable in the U.S. market. The commercialization of fusion power will require licensing of plants, and these activities must start immediately and connect with the design of an FPP (FST-SO-G). Our program should encourage and prioritize innovations that have the potential to substantially lower the cost or accelerate the realization of fusion power (FST-SO-F), while closely engaging with industry to develop promising technologies at lower cost. Research should be broadly based to allow for the exploration and utilization of new, potentially transformative technologies that could allow us to reach our vision more quickly.

Rapid scientific and technological innovation will require new programs and facilities. The community recognizes that designing and constructing major new facilities may not be possible without progressively redirecting resources from existing facilities. Given the possibility of constrained budgets, there is significant support among the community to pivot resources from existing facilities to fund new programs and facilities, if necessary, so that new facilities can be operational within ten years or less. The resources and research programs of existing facilities should immediately evolve to reflect the priorities of this plan. Any such transition must be mindful of the workforce needs and impacts associated with diverting operations budgets to construction.

To better understand the views of the community on the prioritization of elements within this plan, a prioritization assessment exercise was performed at the CPP-Houston workshop. Building off of the format of the MFE+FM&T Knoxville workshop, presentations, small group breakout discussions, reporting back, and eventually polling were all used to gauge the community's prioritization of 23 FST program elements. These elements, which map directly to one or more of the recommendations within this strategic plan, were assessed by CPP-Houston attendees against the criteria of *importance to mission*, *urgency*, *impact of investment*, *using innovation to lower cost,* and *U.S. leadership and uniqueness*. Ordering of the Strategic Objectives within this report is meant to reflect the rough prioritization of the Objectives captured through community input and discussion at the MFE+FM&T Knoxville and CPP-Houston workshops. It is important to note that this ordering does not imply that all recommendations in Strategic Objective-A are higher priority than the recommendations in all other Strategic Objectives. A detailed description of the polling and prioritization processes, a map of the program elements to recommendations in this report, charts summarizing the polling of the FST program elements, and a discussion on the robustness of the polling results can be found in Appendix A.

## *Vision Statement*

Our vision is for fusion energy to be a major source of safe, economical, and environmentally sustainable energy in time to address critical energy and security needs of the U.S. and the world.

## *Mission Statement*

Establish the basis for the commercialization of fusion energy in the U.S. by developing the innovative science and technology needed to accelerate the construction of a fusion pilot plant at low capital cost.

## Definition of a Fusion Pilot Plant

The following three deliverables are likely required to demonstrate that fusion has the potential to be a safe, economical energy source in the U.S., and therefore define the minimum mission scope for a fusion pilot plant (FPP). These deliverables are consistent with the recommendations of the 2018 National Academies report.

- Produce net electricity from fusion
- Establish the capability of high average power output
- Demonstrate the safe production and handling of the tritium, as well as the feasibility of a closed fuel cycle

# Values

The Strategic Plan reflects our values:

1. Prioritize research most important to the *FPP Mission*
2. Act with *Urgency* to address energy security and sustainability
3. Embrace a *Culture of Innovation and Diversity*
4. Maintain *Flexibility* to benefit from innovation
5. Establish a firm *Scientific Basis*
6. Aspire to *U.S. Leadership*
7. Build and strengthen *International Collaboration* where beneficial
8. *Engage All Stakeholders*, including Labs, Universities, and Industry

# Science Drivers

The objectives and their associated recommendations in the strategic plan for Fusion Science and Technology are driven by underlying scientific questions that are encapsulated in a set of three overarching "Science Drivers" (SDs):

1. Control, Sustain, and Predict Burning Plasmas
2. Handle Reactor Relevant Conditions
3. Harness Fusion Power

In the subsequent sections outlining the Objectives of this Strategic Plan, an explicit linkage is provided that connects the Objective back to the relevant Science Driver(s). The description of each of the Science Drivers follows.

## SD1: Control, Sustain, and Predict Burning Plasmas

A critical need in the quest for fusion energy production is the ability to control and sustain a burning plasma. This requires establishing scenarios for maintaining high performance in a burning regime and preventing damage associated with transient events, through the development of tools to predict, avoid, and mitigate such events. Burning plasmas, in which the heating is primarily due to the energy released from fusion reactions, pose challenges to stability and control that are not fully accessible by present experiments and for which significant uncertainty exists. Therefore, projecting to this regime with confidence requires the development of theory and modeling tools, carefully validated against experiments, that are capable of predicting all the important aspects of plasma behavior, using both reduced models and integrated simulations spanning alpha particle physics, transport and confinement, stability, boundary layer physics, and plasma material interactions. This also requires making advances in diagnostic techniques to measure the relevant plasma quantities with the needed spatial and temporal resolution, which must be robust in a burning plasma environment and have long-term survivability (SD2: Handle Reactor Relevant Conditions). The plasma is the energy source in a fusion reactor, so it must be integrated with all the support systems and ex-vessel components in order to Harness Fusion Power (SD3).

## SD2: Handle Reactor Relevant Conditions

A fusion pilot plant will produce heat, particle, and neutron fluxes that significantly exceed those in present confinement facilities, and new innovative approaches to overcoming this challenge are required. These intense conditions affect all regions of the device in distinct ways, including the plasma-facing components (PFCs), structural materials, functional materials, magnet materials, diagnostic materials, and ex-vessel components. For the plasma facing components, solutions must combine advancements in plasma physics and improvements in a material's ability to withstand the sustained heat and particle fluxes of an FPP. This should be recognized as a shared burden between science and technology, benefiting from an integrated strategy that develops and tests multiple approaches. Critical physics gaps include finding solutions for controllable, detached divertors; establishing a predictive understanding of divertor heat loads; developing methods to accommodate or avoid damaging edge transients; and understanding SOL transport and material migration. These are all necessary steps to developing the linked requirements for in-vessel components such as radio-frequency (RF) antennas and first wall and divertor PFCs. Many of these engineering and design issues are independent of confinement concept, and progress can be made with dedicated test stands and exposure facilities, coupled with advancements in theory and modeling.

In a fusion pilot plant, high fluxes of 14 MeV neutrons produce unique effects in materials which are presently poorly understood. A scientific understanding of how materials properties evolve and degrade as a function of neutron energy and fluence is needed to be able to safely predict the behavior of materials in fusion reactors. Even those components not directly exposed to high fluxes of neutrons may be affected by secondary aspects including gamma rays, decay heat, or radioactive dust that are generally not a concern for surrounding systems in today's experimental reactors. Additionally the effects of corrosion, material compatibility, joints, and tritium trapping must be understood in order to construct a fusion pilot plant. Consideration for the manufacturing tolerances, maintenance, repair and lifetime, including disposal, need to be considered at an early stage. Demonstration solutions for Handling Reactor Relevant

Conditions (SD2) must both ensure sufficient survival of PFCs and integrate near-edge plasma conditions that allow for a Controlled, Sustained Burning Plasma core (SD1), which ultimately requires dedicated confinement facilities to validate integrated modeling and simulation. All the materials in the device must withstand the intense conditions, while having tritium retention levels compatible with techniques to Harness Fusion Power (SD3).

### SD3: Harness Fusion Power

Interlinked with a burning plasma and materials that can withstand fusion reactor conditions are all the key systems required to capture the power, breed fuel, and ensure the safe operation of the reactor. Before a device is constructed, materials and components must be qualified, and a system design must ensure the compatibility of all components. Just as the plasma and materials in a fusion reactor will need to advance beyond today's capabilities, the balance of plant equipment, remote handling, tritium breeding, and safety systems will also require significant advances. Connected to all of these systems will be the measurement and diagnostic systems needed to ensure the safe operation of the device and safe work environment for the site personnel, including surveillance program for materials, stress and motion detectors for components, tritium detectors, and radiation detectors. All the systems in a reactor must work in concert with each other, so all the systems needed to Harness Fusion Power must help achieve SD1 to Control, Sustain, and Predict Burning Plasmas. Additionally, the systems for breeding and extracting tritium, the detectors and safety systems, the power conversion systems, and all other components will be exposed to harsh conditions including fusion neutrons, high temperatures, corrosive media, and more, so all these components of the reactor must also be designed to accomplish SD2: Handle Reactor Relevant Conditions.

## *Strategic Objectives and Recommendations*

### FST Strategic Objective A: Demonstrate solutions for managing high heat and particle loads sufficient to design plasma-facing components (PFCs) for a fusion pilot plant

*The unmitigated power and particle exhaust expected in a fusion pilot plant cannot be sustained by present materials used in confinement devices. Presently operating or planned non-confinement facilities can recreate heat and particle fluxes prototypical of mitigated levels expected for FPP scenarios, allowing for off-line development of candidate materials sufficient to warrant testing in existing or future confinement devices. This work should be complemented by developing a validated predictive modeling capability to assist in identifying PFC solutions that extrapolate to a fusion pilot plant. The research described below will advance the technological readiness level (TRL) of actively cooled solid PFCs, the present leading candidate, to the point where they could be chosen for use in an FPP design. In parallel, R&D will be pursued to raise the TRL of liquid metal (LM) PFC approaches that may provide an*

*alternate PFC solution for an FPP. A decision point to proceed with either solid PFCs or LM PFCs for the baseline operational phase of an FPP is targeted to occur between the conceptual and detailed FPP design stages (as determined by design studies described in [FST-PR-A]) and drives the need for a continuous and staged research and development program of both concepts in the near-term. Figure FST-SO-A.1 illustrates the staging of these research tasks. The research described in this Strategic Objective was recognized by the community as amongst the highest priorities for the development of an FPP throughout the community planning process. The development of PFCs is a key element for the FPP Mission and is an urgent task as the timeline to develop and qualify materials has historically been long. Furthermore, given the potential for materials to last longer (and/or be replaced quickly) and be compatible with a higher fusion power density (and therefore directly impact plant size and capital costs), it seems clear that the FST-SO-A challenges are of the highest priority for moving towards an FPP design.*

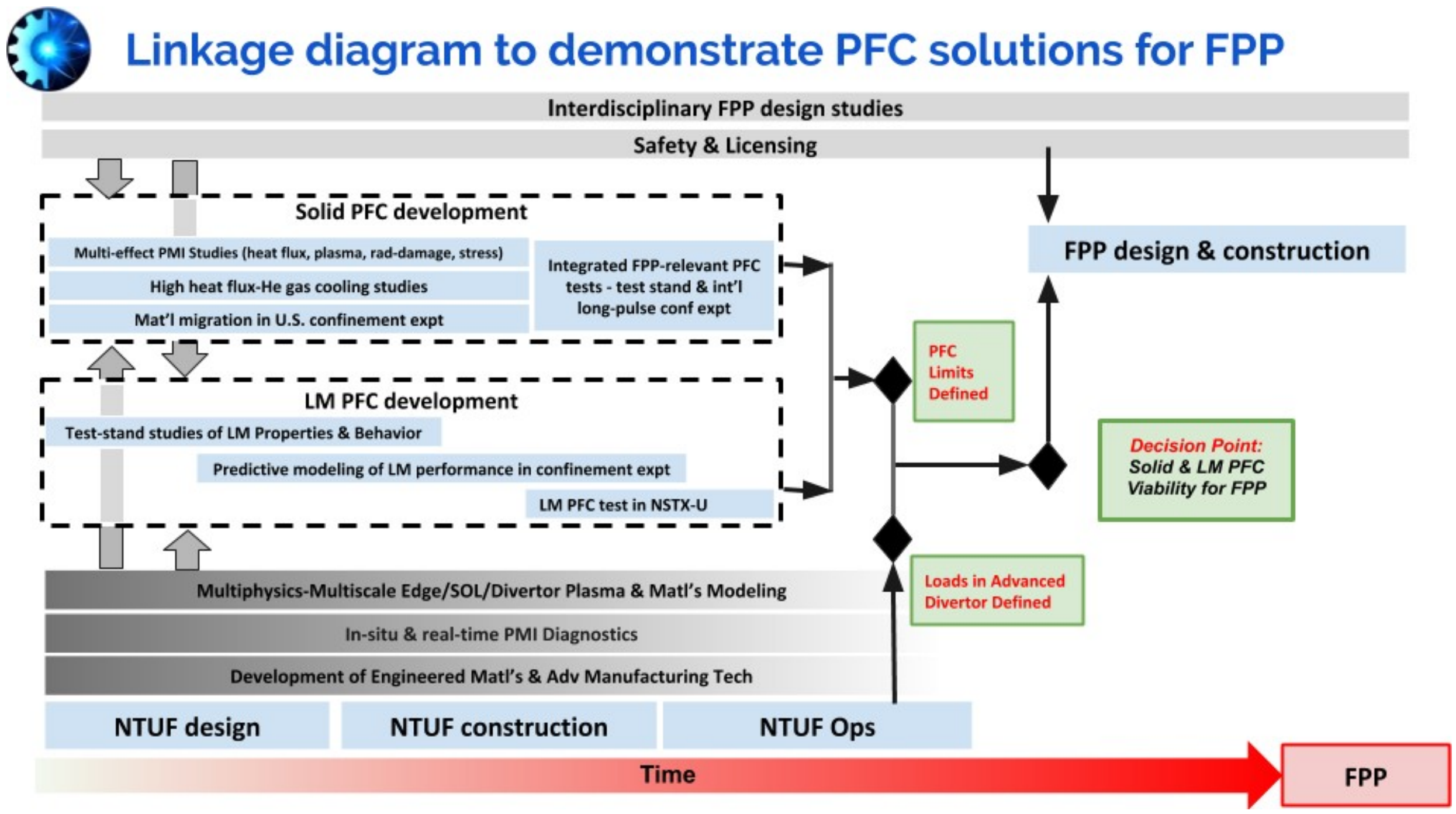

Figure FST-SO-A.1: Linkage diagram for FST-SO-A

**Relation to Science Drivers**:

Handle Reactor Relevant Conditions: The achievement of a self-heated burning plasma within an FPP inherently results in the generation of a significant flux of unburned fusion fuel, helium fusion ash, and intense heat flux that impinge on the PFCs. The PFC system, comprised of the first wall (FW), divertor target, and radio-frequency (RF) antenna protective structures, must be designed to withstand these fluxes and operate for an extended period of time without failure while also being bombarded by the energetic neutrons produced by the fusion reactions. The PFC system will face extreme challenges in an FPP, including heat transfer and removal, material damage and transmutation due to plasma and neutron irradiation, tritium fuel retention leading to possible concerns of inventory control, safety, and self-sufficiency, remote maintenance and repair. The PFC system is also intimately linked to the structural and functional materials that support them, the blanket system that allows the fusion energy to be harnessed for useful purposes, and the radio-frequency launchers that heat the plasma and help control plasma currents. The FST-SO-A as described below provides a staged science-based R&D program organized around four (4) tasks that can lead to a PFC system that enables the future development of a credible FPP design.

**Expert Groups**: FM&T-Plasma Material Interaction and High Heat Flux; FM&T-Fusion Materials; MFE-Boundary and Divertor Plasma Physics; Plasma and Material Interaction; FM&T-Measurements and Diagnostics; FM&T - Magnets and Technology

1. **Improve our understanding of plasma material interactions in solid materials at FPP relevant conditions and demonstrate new actively cooled solid-material PFC solutions**

   This task can be addressed in the near-term by upgrades to existing linear plasma exposure devices (*e.g.* TPE, PISCES, *etc.*) and by the completion of the MPEX device in the mid-2020s. These linear plasma devices need to be used to study PMI in plasma-facing armor materials under combined high plasma-ion flux irradiation, high heat flux, and neutron irradiation (or energetic ion beams as a neutron surrogate) at FPP-relevant material temperatures. These experiments will then provide insight into the impact these multi-effect processes have on the critical FPP issues of tritium retention and inventory, Tritium Breeding Ratio (TBR), and plasma-facing armor material property degradation (FST-SO-B). In parallel, the performance limits of the actively cooled PFC system comprised of plasma facing armor, actively cooled substrate and armor/substrate joining technology under high heat flux with FPP-relevant cooling is needed to enable prototyping of actively cooled solid-PFC component designs. The latter issues require the design and deployment of a dedicated non-plasma high heat flux test stand, extended to include a He gas cooling capability, that can test component level PFC designs by the mid-2020s to inform a decision point by mid-2030s on solid PFCs for an FPP.

   Demonstration of an adequate PFC system lifetime and exploration of the influence on FPP confinement scenarios are also critical issues that require a dedicated research effort. Given the limited information on material migration and slag management provided by linear plasma experiments, this issue requires targeted experiments on existing confinement devices. Experiments in confinement devices can determine first wall (FW) charge-exchange (CX) neutral flux, composition, and energy spectrum; ion flux to the FW; and the resulting FW erosion rates and resulting redeposition/co-deposition occurring elsewhere. When coupled with mapping of low- and high-Z impurity transport, redeposition, co-deposition, and material migration rates in FPP-relevant plasma exhaust scenarios, these experiments will permit validation of material migration models (FST-SO-D.5, FST-SO-E). These models will then allow predictions of FW and divertor lifetimes, in-vessel fuel inventory evolution, and impact on tritium self-sufficiency.

The next generation of confinement devices either proposed as part of this CPP process or now nearing operation will also be valuable platforms to evaluate the performance of PFCs. For example, NTUF, which is likely to start operations with solid PFCs, will allow exposure of divertor materials to, and integration of components with, high power load operation. When and if off-line tests are successful, NTUF could allow for testing of first wall material candidates, integration of/with gas cooled components and hot wall concepts. This experience will be essential to demonstrating the viability of solid PFCs for an FPP (PMI-HHF EG Strategic Block). However, testing of actively cooled PFC technologies in this platform will be limited if a short-pulse NTUF concept is pursued. Future facilities will also serve as platforms for research into material migration and slag management, but again will be limited to lower fluences. Overall, NTUF will have a critical role to play in PFC research, and the specific testing capabilities will be determined as the NTUF design is advanced.

In addition to developing domestic capabilities, the U.S. should continue existing international collaborations that provide access to facilities with different capabilities, such as higher fluxes (*e.g.* MAGNUM-PSI) that can better mimic the conditions expected near the strike-point at the divertor. The community is also aware of new international facilities that may come on-line in the near-future and that can complement the domestic program (*e.g.* JUDITH2). Gas cooling and related technology, such as using He and operating confinement devices with hot walls, is also an area where the U.S. has had, and could reclaim, leadership. Many international partners are ramping down their research in this area and focusing their efforts in water cooling instead. World-wide, experimental facilities that look into gas cooling systems exist, but either run at lower pressures (*e.g.* at KIT) or lower temperatures (in Korea) than what the community has identified as necessary for the FPP mission. Other ideas for cooling technologies and coolant materials do exist, mainly in the context of blanket research (FST-SO-C).

As these material and PFC studies mature and are used to validate predictive models for PFC performance, it will be necessary to use these results to advance the readiness of an actively cooled solid PFC divertor and FW for an FPP, targeting He-gas cooled designs as a potential area for U.S. leadership. First steps towards this goal can be completed through the use of the test stands mentioned above combined with experiments on international linear plasma test stands. Looking beyond such work, collaborations should be pursued on long-pulse confinement devices, both stellarators (FST-SO-E.3) and tokamaks (FST-SO-D.5) located in the E.U. and in Asia, to provide opportunities for U.S. teams to contribute in areas like PFC material development, diagnostics, design/modeling support, and component manufacturing in partnership with U.S. industry. In areas where collaborations provide limited experience, high-heat flux test stands and plasma exposure facilities can be used in conjunction with detailed engineering design studies and remote handling tests to raise the TRL of actively cooled PFC systems on a timeline consistent with FPP scoping studies (FST-PR-A).

Finally, recognizing that it is unlikely that existing materials will provide adequate PFC system performance, it is imperative to initiate and sustain a program for the development of new, innovative solid materials that will form the basis of the solid first wall armor, solid divertor targets and the liquid metal PFC substrates through techniques such as advanced manufacturing, nano-engineered materials, material by design, virtual engineering, use developments from aerospace and other high heat flux applications, *etc.* (FST-SO-B.1) As computational techniques such as machine learning become more easily available, accessible, and efficient, many of these innovative approaches also become applicable to a wider range of fields. Historically, both within and outside fusion program research activities, innovation and computational material research have been areas of U.S. leadership. The PFC development program presented here would allow the fusion program to take advantage of these strengths and thereby maintain this leadership, and in general push for a greater role of the U.S. in world-wide PFC research.

**2. Improve the readiness of liquid metal (LM) plasma facing materials and test slow-flow and/or fast-flow PFC concepts on confinement facilities**

This near-term recommendation is composed of several research tasks. First, it is recommended to use existing and proposed domestic non-plasma LM test-stands (*e.g.* LMX, FLIT, *etc.*) and small-scale plasma experiments (LTX-$\beta$, HIDRA) to determine the relevant LM properties and behaviors needed to inform design of PFCs in confinement device experiments. These properties and issues include (but are not limited to) the effect of plasma exposure on LMs; free-surface flow stability of LMs in relevant magnetic field geometries; LM-substrate interactions such as corrosion/erosion in the presence of plasma, high temperature, and irradiation; LM evaporation, and other loss mechanisms; and tritium retention and removal in LM PFC candidates. Second, these results need to be used to predict losses from the LM at FPP relevant FW temperature; LM vapor penetration into the divertor and main plasma chamber; and impacts on plasma operations, plasma performance, and machine safety using a combination of small-scale experiments and initial models. In parallel with these domestically-focused efforts, the program should support collaborations with teams investigating LM PFCs on international confinement devices, linear plasma facilities and test stands. Further information on the biggest challenges that LMs face as PFCs, as well as the properties most important to investigate for this role, can be found the results of previous studies (*e.g.* FESS) and in the Strategic Block of the PMI-HHF Expert Group. The resolution of key LM science questions is essential to determine feasibility of LMs as an alternative to solid PFCs in a timely manner.

If the results of these studies show that LM PFCs still appear to be a promising alternate approach to solving the PFC challenge, then the community should carry out an upgrade to an existing domestic confinement facility to use LM PFCs, targeting NSTX-U (FST-SO-D.3). In concert with such a confinement experiment, theory and modeling programs should be expanded to complement this effort and develop a predictive capability for the self-consistent impact of LM-based PFCs on plasma confinement and, at the same time, predicting the evolution of the LM properties as a result of the plasma interaction. These studies should be prioritized to generate the information necessary to determine if LM PFCs will be viable in an FPP. Based on the results of the above LM PFC R&D effort along with the identified performance limits of solid-material PFCs, a LM PFC upgrade could be considered for NTUF to provide the necessary FPP design validation data.

Overall, liquid metal PFC research has been an area of U.S. leadership. There is a tenuous basis for this leadership in fast flowing systems that has been acquired via collaboration with FLiLi system implemented in EAST. However, capillary porous systems—first developed in Russia—are being incorporated into European confinement experiments (*e.g.* FTU and TJ-II). Therefore, despite collaborations with international facilities (*e.g.* MAGNUM), and with U.S. expertise being tapped internationally (*e.g.* by the COMPASS-U and, potentially, ST40 teams), this leadership is now at risk for slow-flow systems. Successful completion of the roadmap recommended here would allow the U.S. to regain and then maintain its leadership.

3. **Advance the integration of full-physics and reduced material models and edge/divertor plasma models to permit validated prediction of PFC performance under FPP conditions, in coordination with FST-PR-A.2**

In parallel to the above activities, it is crucial to continue to advance and validate the multi-physics, multi-scale theory and modeling capabilities (FST-PR-D.2) necessary to support this R&D plan by comparison with single-effect and multi-effect PMI experiments. In particular, it will be important to be able to predict armor material evolution under simultaneous plasma and energetic neutron irradiation—including near-surface morphology, bulk material transmutation and defect production and evolution—and effects on material properties, fuel retention, diffusion and trapping, all while operating at high temperature and under high mechanical stresses. Furthermore, it will be key to develop predictive models of FW and divertor erosion, redeposition, co-deposition and material migration coupled together with the concomitant edge/SOL/divertor plasma conditions and then validate against dedicated experiments on confinement devices. The resulting validated models will then be used to project performance of the PFC system under FPP conditions and, ultimately, to make a determination of the risks and benefits of choosing a solid-material PFC vs. LM PFC design for the FPP. This effort will require leadership-class computing and advanced algorithms to enable use of these models in FPP design efforts. This should be supported, for example, through including this as a topical area in regular open SciDAC calls.

The community recognizes that theory and modeling are areas of high impact of investment. For instance, in areas such as development of new materials, computational research coupled with and guiding experiments can be significantly more cost effective than purely experimental development. Furthermore, modeling of materials and plasma-material interactions is recognized by the community as an area of current U.S. leadership, especially for large scale models and integrated systems, and that we should strive to maintain.

4. **Develop and deploy *in situ* and *ex situ* materials characterization tools in both off-line test stands and plasma simulators as well as in confinement experiments to permit more rapid evaluation of PFC system performance and behaviors (FST-PR-F.1)**

The most important and impactful *in situ* and *in operando* (*i.e.* while plasma is operating, also called ‘real-time’) PMI diagnostic capability needs include developing the ability to characterize material properties such as surface composition and morphology evolution, heat transfer properties of bulk solid materials and liquid flow (either LM or coolant), and mechanical properties of solid PFC and LM substrate materials under single-effect (*e.g.* plasma exposure, high heat flux, or displacement damage) and multi-effect (*e.g.* plasma exposure with high heat flux and displacement damage) conditions. Such work in both off-line PFC testing facilities and confinement facilities will accelerate the rate of discovery and lead to the development of detailed understanding of the underlying PMI science and of the implications on PFC performance. Furthermore, improving existing *post mortem* materials analysis tools and incorporating capabilities onto existing major fusion facilities to track surface composition and structure evolution dynamically *in situ* (if possible), would also have high impact. Finally, linking such surface and materials diagnostics with plasma measurements of *e.g.* main plasma edge/SOL density, temperature and flow velocities; impurity atom and ion composition, distribution, flows and transport processes and with measurements of material erosion, redeposition and co-deposition, and material transport would lead to the ability to quantitatively validate models and predict these processes that are critical to PFC system performance.

The U.S. currently leads the world effort to develop *in situ*, “real time” PMI diagnostics. A strong and focused program will not only allow to maintain the leadership in this area of innovation, but also greatly contribute to leadership in PMI research.

## FST Strategic Objective B: Determine the structural and functional materials that will survive under fusion reactor conditions

*The scientific understanding of materials degradation under fusion conditions must be advanced, and materials must be developed to withstand the harsh operating conditions, in order to successfully design, license, and operate any fusion pilot plant, independent of confinement concept. Many of the conditions are orders of magnitude more extreme in an FPP than in today's fusion experiments and fission reactors. Key parameters include the neutron energy spectrum, neutron flux and fluence, He and H transmutation, operating temperature, particle flux, heat flux, corrosion effects, stress, and magnetic field. All in-vessel and ex-vessel structural and functional materials need to be designed to handle their respective reactor conditions. Functional materials include those for breeding tritium (breeder material, neutron multiplier, tritium permeation barrier coating), flow channel inserts, materials for diagnostics,* etc. *Programs are needed to develop the high temperature structural materials design criteria and to support the licensing of a reactor, which will require new experiments or facilities. There are numerous opportunities to innovate in materials to allow advanced pilot plant designs, such as additive manufacturing, modern computational materials science, and machine learning. At this stage, the advances that can be made in materials will benefit all confinement concepts of an FPP including magnetic fusion energy (MFE) and inertial fusion energy (IFE). The recommendations to accomplish this strategic objective are to test and develop materials, conduct fission and fusion neutron irradiations of materials, and develop high temperature structural design criteria. These activities must coordinate with the FPP design exploration (FST-PR-A) and the licensing and remote maintenance strategy (FST-SO-G).*

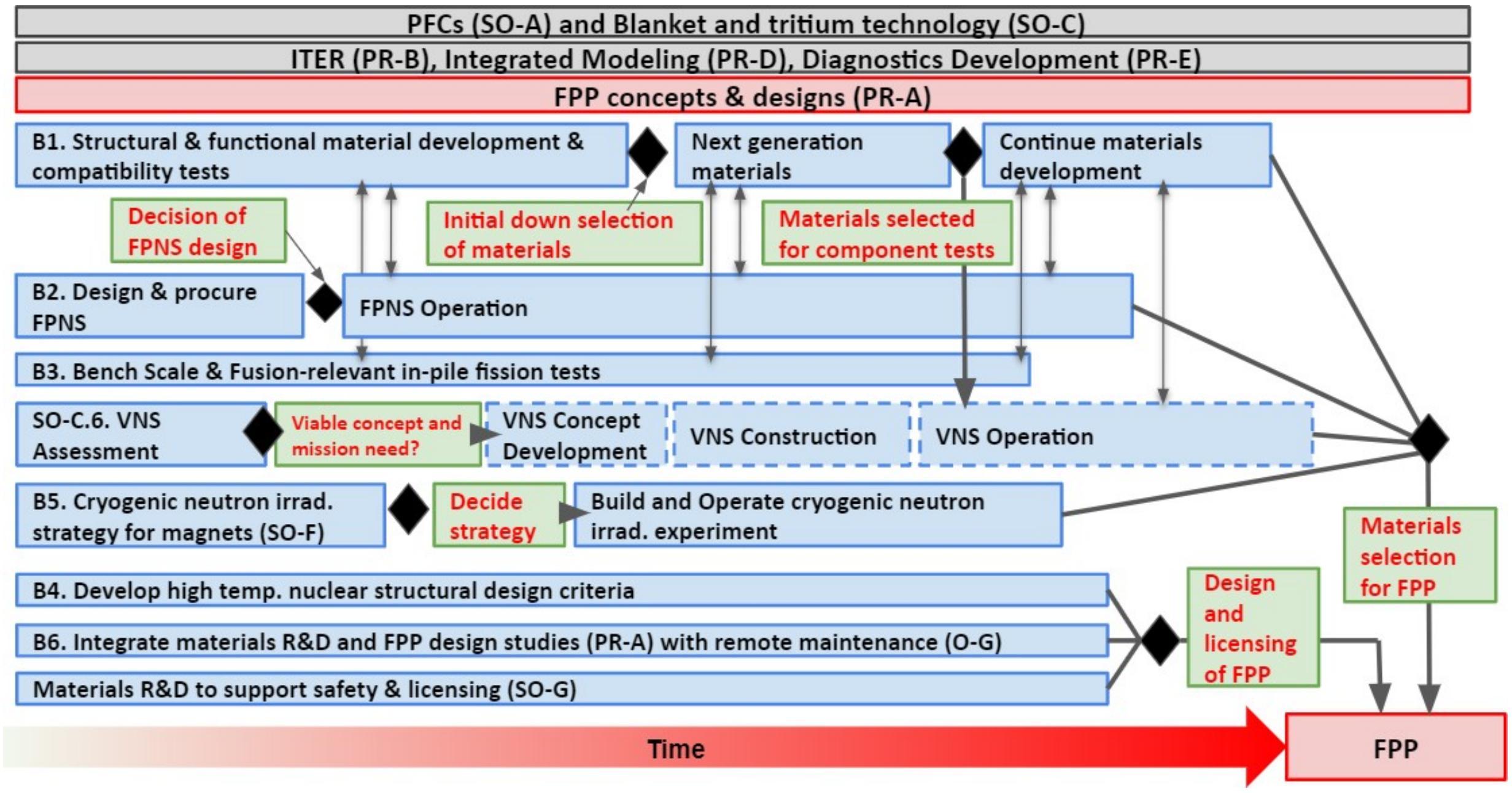

Figure FST-SO-B.1: Linkage diagram for FST-SO-B

**Relation to Science Drivers**:
Handle Reactor Relevant Conditions: Today's materials cannot survive for long in the harsh environment expected for an FPP. In an FPP, structural and functional materials must survive several orders of magnitude higher fluence of fusion neutrons and particles than in today's devices. Material solutions must be discovered to realize an FPP.

**Expert Groups**: MFE-Boundary and Plasma and Material Interaction; FM&T-Fusion Materials; FM&T-Blanket, Tritium, and Systems; FM&T-Plasma Material Interaction and High Heat Flux; FM&T-Measurements and Diagnostics

1. **Expand the fusion materials program to develop structural and functional materials that will survive the conditions in a fusion reactor**

   Innovative solutions for structural and functional materials are needed to realize an FPP. Material degradation is caused by neutron irradiation, coolant corrosion, high temperature, and helium and tritium embrittlement. In addition, the effect of neutron irradiation on tritium permeation and retention in structural materials and plasma facing components need to be considered. Compact FPP designs, which might be more economically attractive, require developing materials that can withstand higher neutron fluxes and fluences. This recommendation encompases experiment, theory, and simulation needed to find such solutions. Examples of opportunities for innovation include, but are not limited to: development of innovative materials; development of advanced manufacturing capability for fusion materials including novel geometries, joining dissimilar materials, and integrating diagnostics; utilization of modern materials science methods including computational materials discovery to design innovative alloys and composites; and utilization of machine learning techniques to enhance radiation damage characterization, simulations, and materials design. This program is urgently needed because it will inform the design of key multiple-effects fusion technology and energy facilities including the blanket component test facility (BCTF) described in (FST-SO-C.5), volumetric neutron source (VNS) described in (FST-SO-C.6 and FST-SO-B.4), and an FPP.

2. **Immediately design, construct, and operate a Fusion Prototypic Neutron Source (FPNS)**

   The energy spectrum of fusion neutrons is significantly different than that of fission neutrons. Consequently, the effects of atomic displacement damage, solid transmutation products, and helium gas production on material performance and degradation are distinctly different between fusion and fission radiation environments. Current progress of fusion materials research is inhibited by the lack of a suitable D-T fusion neutron source anywhere in the world. The fusion materials community evaluated this need through a workshop in 2018 and determined the base requirements of such a facility include a damage dose rate > 10 dpa/year and helium-to-damage ratio of ~10 appm/dpa in reference ferritic/martensitic steels. The details of the community findings are in the workshop report [FPNS 2019]. An FPNS is intended for studying fundamental properties of materials which can be achieved with a small irradiation volume of ~50 cc; it is not intended for obtaining engineering data. There are several viable candidates for producing fusion neutrons for an FPNS. The detailed design should begin immediately to allow for operation in approximately 5 years. An FPNS would help downselect materials for an FPP. Additionally, an FPNS can further research objectives for blanket and tritium science. Results from FPNS experiments can be used to validate theory and simulation. The proposed FPNS facility would differ from international facilities, including IFMIF-DONES, by being operational on a shorter timescale at significantly lower capital cost and thus enabling a more aggressive timeline for an FPP. Operation of this device would position the U.S. as a world leader in fusion materials testing. In addition to data collection from FPNS, which will focus on small scientific sized samples (not engineering data) and single effects, there will be a need for integrated volumetric neutron testing. The strategy for such tests is discussed in FST-SO-C.6.

3. **Carry out in-pile fission irradiation testing of fusion relevant materials**

Lacking fusion irradiation facilities, we currently rely on irradiation testing using fission reactors to build an understanding of how fusion-relevant material properties vary as a function of neutron energy, dose rate, solid transmutation rate, and gas transmutation rate. Spectral tailoring of a mixed energy spectrum neutron irradiation by thermal neutron shielding can be done, in order to approximate the fusion neutron spectrum. In addition, fast neutron irradiations can be relevant for large areas of a fusion reactor that are farther away from the first wall. Therefore, while the complete understanding of the fusion neutron irradiation effects on structural materials can only be addressed with fusion neutron environments (including the FPNS and future burning plasma environments), the need for in-pile neutron irradiation will remain essential. In-pile irradiation may also be useful for blanket component mockup testing (FST-SO-C) given the large volume of irradiation that is available as compared to FPNS. Nevertheless, given that material damage, transmutation, and gas production in a fusion environment is different from that in fission, experimental results from in-pile fission irradiation testing need to be closely coupled with modeling for predicting material behavior in fusion environments. In the near term, this recommendation would provide the ability to downselect candidate materials for further study using a FPNS. Research is already existing in this area and can be expanded immediately without the need for new facilities. The U.S. is already a leader in this area and should continue ongoing international collaborations in this area, for example those with Japan and the E.U.

4. **Develop high-temperature nuclear structural design (HTNSD) criteria that builds on the ITER Structural Design Criteria (ISDC)**

   Fusion design criteria will be significantly different and more complex than those used for fission reactors or ITER, and thus need a suitable design methodology. These differences include transmutations, unique components, gradients in solid structures, unique safety concern factors, complexity of material systems, and operating temperatures. Therefore, in order to design an FPP, it is necessary to develop HTNSD criteria, related to creep, fatigue, and corrosion behavior of structural materials. Today, the U.S. has not established HTNSD criteria for designing an FPP, and no such effort exists internationally. This recommendation is also particularly critical for design qualification and licensing purposes of an FPP, which both require a long-term effort. To prevent delay on the FPP licensing, this recommendation should begin in the near term and in parallel with other activities. Data collected in recommendations FST-SO-B.1, 2, and 3 is input for the HTNSD criteria, and the development of the HTNSD criteria would also iteratively feedback to guide activities in those recommendations. This recommendation should build on the ITER structural design criteria (ISDC) and leverage opportunities with "Generation IV" fission reactor R&D.

5. **Initiate a near term activity that determines the strategy for a cryogenic neutron irradiation experiment for magnet materials (FST-SO-F)**

   High field and high temperature superconducting magnets represent a breakthrough technology that could lead to a compact FPP. The development and unirradiated testing of superconductor magnets is detailed in FST-SO-F. Superconductor performance is strongly dependent on the quality of the crystal structure and radiation damage will determine the service lifetime and economics of the superconducting magnets. Currently, using existing fission irradiation facilities, magnet materials can be neutron irradiated at ambient or higher temperatures to start to understand their behavior under irradiation. However, superconductors' response to neutrons while at their typical operating temperature will be significantly different than their behavior during high temperature irradiation, but very little data on cryogenic irradiation properties of superconductors exists. There are currently no facilities in operation in the world to irradiate superconductors at cryogenic temperatures, and there are no capabilities to test superconductors during irradiation; these represent critical knowledge gaps to being able to design and operate an FPP. The first step of this recommendation is to determine the strategy for cryogenic neutron irradiation of superconducting materials, which could be accomplished with a purpose build experiment in a fission facility or future fusion neutron facility. The viability and cost of these options should be evaluated and one option selected. Once the irradiation strategy is in place, this recommendation will establish the scientific understanding of superconducting magnet materials, joints, and insulation under neutron irradiation at relevant service conditions.

6. **Integrate the fusion pilot plant design (FST-PR-A) and materials development with the remote maintenance equipment and strategy (FST-SO-G)**

   The behavior of components in a fusion reactor is complex and design-dependent, requiring research into the critical design-dependent phenomena that might lead to failure. Integration among FPP design, materials research, and remote maintenance strategy is needed. Materials development will support the design studies. Near-term initiation of FPP engineering design activities will provide useful prioritization and scheduling information for numerous fusion materials and technology R&D activities. The current Fusion Energy System Studies program (*e.g.* ARIES, FNSF) are not robust enough to inform materials prioritization and need to be expanded as recommended in FST-PR-A. Functional materials used in diagnostics and remote maintenance equipment will need to be designed to handle the high doses and contamination, for the fusion core, near-core and any transportation route/cask, and the hot cell.

## FST Strategic Objective C: Develop the science and technology necessary to breed, extract, and safely manage large quantities of tritium

*There is strong community consensus that a fusion pilot plant needs to demonstrate closure of the tritium fuel cycle,* i.e. *demonstrate a tritium breeding ratio (TBR) greater than one. This includes both the "inner" cycle of plasma fueling, exhaust, and reprocessing, and the "outer" cycle of tritium breeding and extraction. Because of the inherently low burn fraction of tritium in the plasma and the need to progress to high availability operation, the tritium throughput in the inner cycle will greatly exceed any prior experiment. New technologies need to be developed to support such operation while minimizing the scale-up required of the tritium plant and the increased tritium inventories that this implies.*

*Tritium breeding blanket technology is at a low TRL (~1-2). There is broad community consensus that substantial increases in effort are necessary to advance tritium breeding blanket technology, and significant risk associated with not doing so given the necessity for achieving tritium self-sufficiency in an FPP. Some foundational developments are needed in advance of integrated design and testing of potential breeder blanket concepts. These include the development of different "functional" materials (lithium-bearing solids or liquids that breed tritium) and compatible structural materials that will survive the fusion nuclear environment. As soon as these have developed to the point at which a credible down-selection in blanket concepts can be made, a campaign of progressively integral testing is required to achieve the high confidence in the operation of the blanket required for deployment in the FPP. This includes large-scale non-nuclear component testing in a flexible Blanket Component Test Facility (BCTF), potentially followed by nuclear testing in a Volumetric Neutron Source (VNS), which would integrate the same effects but in a prototypic nuclear environment, and demonstrate actual tritium breeding and extraction. The feasibility, cost, and schedule associated with options for component-scale irradiation testing (ITER TBM, VNS, early FPP operation,* etc.*) first need to be assessed. As the TRL is raised on these technologies, there*

*would be decision points for pursuing different types of blankets, tritium systems, and fueling systems that would be specific to design decisions for an FPP. However, in the near term, all the below recommendations, except Recommendation 8 for pellet fueling, will equally benefit any fusion concept including tokamaks, stellarators, inertial fusion energy, and alternate concepts. Recommendation 8 would benefit both tokamaks and stellarators as well as potentially other magnetic confinement concepts.*

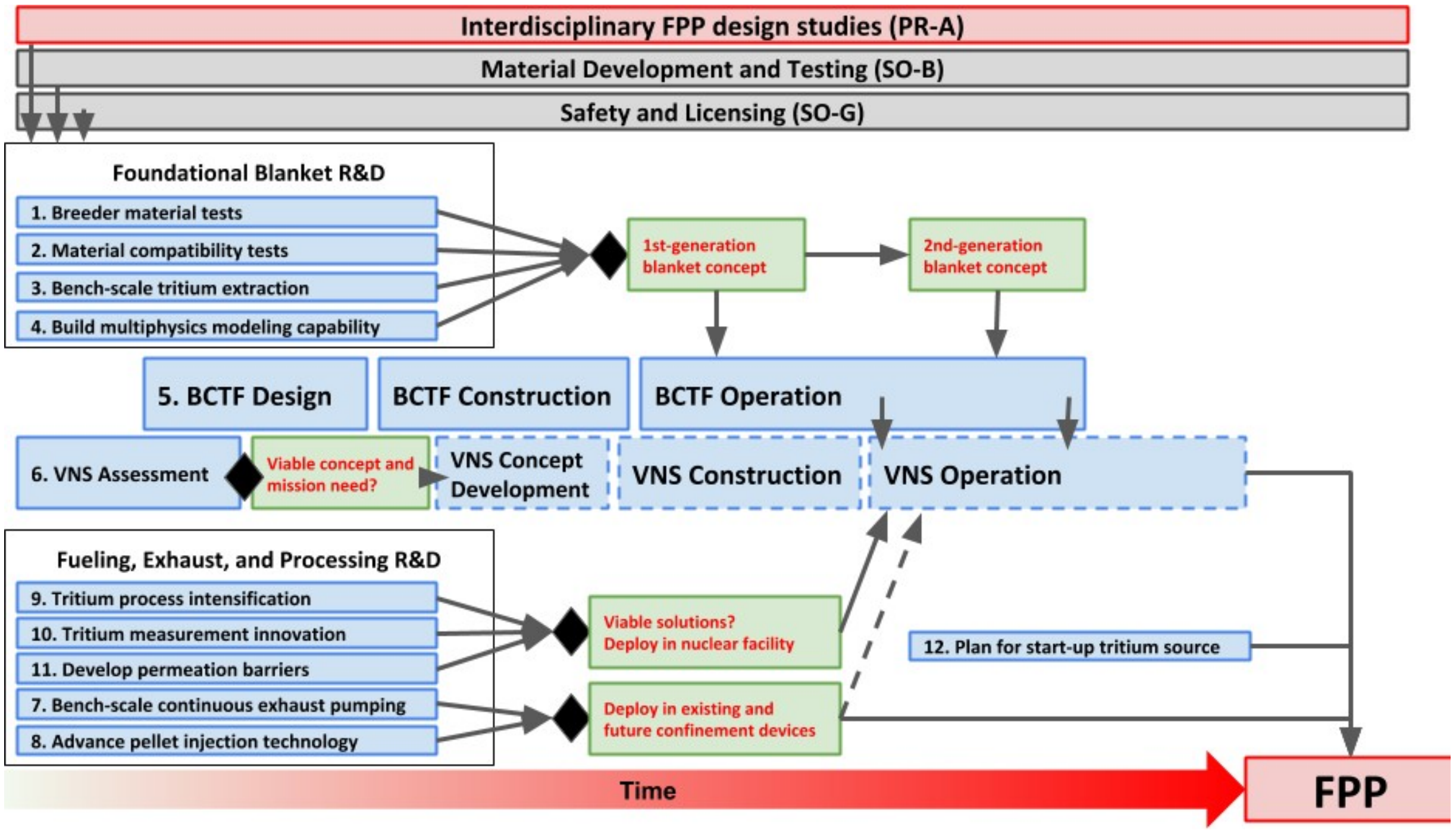

Figure FST-SO-C.1: Linkage diagram for FST-SO-C

**Relation to Science Drivers**:

Harness Fusion Power: This is the primary purpose of the blanket and tritium handling systems that safely recover tritium and process it for use as fuel. In addition to its central role in producing the tritium needed to fuel the fusion reactor, the blanket must also remove most of the energy produced, and is therefore the heart of the power conversion system.

Handle Reactor Relevant Conditions: Along with the divertor, the blanket and integrated first wall must survive the most difficult conditions of the fusion energy environment, including high incident heat and 14 MeV neutron fluxes. These create high temperature and stress gradients in materials that are experiencing radiation damage at a high rate.

**Expert Groups**: FM&T-Blanket, Tritium, and Systems; FM&T-Fusion Materials; FM&T-Magnets and Technology; FM&T-Measurements and Diagnostics

1. **Initiate small-scale tests for a variety of functional breeder blanket materials to advance blanket concept designs**

   Given the low development level of breeder concepts in the U.S., it is recommended that multiple breeder concepts be advanced in parallel until the knowledge base is sufficient to make a down-selection between these; design constraints (*e.g.* for a compact device) may facilitate an earlier decision. For solid breeders, innovative (*e.g.* advanced or additively manufactured) porous ceramic designs should be pursued; for these, the critical issues are identification of the optimal ceramic compound, fabrication, and determination of mechanical, heat transfer, and tritium retention properties. Irradiations in a fission reactor will be required, or in FPNS (FST-SO-B.2) if spectral effects are important. For liquid breeders such as liquid metals or molten salts, in addition to measurement of the fundamental physical (including tritium transport) properties, the critical issues are understanding magnetohydrodynamic flows in the presence of large temperature gradients, and how these influence heat, tritium, and corrosion product transport, for which the operation of forced convection loops will be required. Because the blanket is also the heart of the power conversion system, blanket concepts that support high temperature operation (and therefore high plant efficiencies) should be pursued. Blanket material/concept development should also seek to minimize tritium inventories and unwanted tritium permeation. While blanket concepts developed in the U.S. will likely differ significantly from those developed internationally, collaborations should be pursued where there are commonalities in the underlying materials and transport phenomena.

2. **Support testing of compatibility between breeder and structural materials required for a viable integrated design**

   Existing candidate structural materials may have significant compatibility issues with solid or liquid breeders at high temperature, which may be exacerbated by high magnetic fields or high radiation environments, leading to erosion of structural materials, transport of activated corrosion products, and reduced component lifetime. Compatibility of blanket materials and related issues such as liquid metal embrittlement need to be investigated in small scale experiments, including natural convection loops. Compatibility issues identified through such work should be mitigated via additional materials development (FST-SO-B), with input on potential impacts on radioactive waste generation and related safety issues (FST-SO-G).

3. **Construct bench-scale experiments to test tritium extraction concepts and transport in breeder and structural blanket materials**

   Tritium bred in the blanket must necessarily be extracted and re-used as fuel. Efficient tritium extraction is necessary to reduce circulating tritium concentrations, which drive unwanted tritium permeation. Lower tritium inventories therefore reduce reliance on permeation barriers or other mitigation systems, and so efficient extraction supports tritium accountancy and safety in the FPP. The vacuum permeator is a leading concept for fast-flowing liquid breeders; initial concept development requires that this and/or other concepts be tested in a bench-scale forced convection loop to understand the relevant tritium transport phenomena. Such a facility can also be used to investigate tritium retention and transport phenomena in solid breeder or structural materials. Prototype extraction system development and testing in a BCTF and VNS or an in-pile fission loop must follow to qualify the component for FPP operation.

4. **Develop models and a multiphysics modeling capability to enable integrated blanket designs**

Computer modelling will play an increasingly important role in fusion design and technology, where the complexity of the physical processes involved, and the highly interconnected nature of systems and components call for support from sophisticated and integrated computer simulation tools (FST-PR-D). Models for physical phenomena critical to blanket design such as neutronics, structural mechanics, electromagnetics, fluid dynamics, magnetohydrodynamics, heat transfer, tritium and corrosion product transport, and radiochemistry, validated with data obtained in the preceding experiments, need to be unified in a multiphysics modeling framework. This should facilitate "loose" coupling where sufficient and "strong" coupling (parallel execution with continuous multi-directional data transfer between codes) where necessary, and be capable of exploiting high performance computing, necessary to meet the computational demands imposed by complex geometries and phenomena such as high magnetic fields.

5. **Design, construct, and operate a Blanket Component Test Facility (BCTF) to perform *non-nuclear* testing of integral scale blanket components**

As soon as it can be sufficiently informed by the fundamental blanket research outlined above, work should begin on the design and construction of a BCTF, and on the design and fabrication of blanket prototypes based on the selected concept(s) to be tested in the facility with industry involvement. The purpose of the BCTF is to provide for large component-scale testing in an environment that is as prototypic as possible without neutrons or radioactive materials. Excluding the latter provides for a more flexible test facility with greatly reduced safety consequences associated with component failures. The facility should therefore integrate all non-nuclear features of a fusion nuclear facility simultaneously in its operation, including the use of actual breeder/coolant fluids at prototypic temperatures, pressures, and flow rates, surface and surrogate volumetric heating and associated large temperature and stress gradients, magnetic field, vacuum conditions, coolant ancillary systems (heat exchangers, purification systems, *etc.*), and simulated breeding by injection of hydrogen and helium (and subsequent extraction). The facility should operate over long periods to simulate the lifetime expectations for the blanket in an FPP (*e.g.* 10 days pulses, with 0.5 day dwell, and a total service of 2.25 years). The facility will provide integral effect data that will additionally serve to validate the multiphysics models recommended above for use in FPP blanket design.

6. **Identify a strategy for component-scale irradiation testing, *e.g.* in a volumetric neutron source (VNS), that fits within our timeline for the FPP**

   While most design evolution and integration issues should be resolved in the non-nuclear Blanket Component Test Facility (BCTF), this will not reveal any integral effects arising from the fusion radiation environment and the presence of tritium, and unknown such effects may arise more immediately than the anticipated long-term material evolution due to radiation damage. Furthermore, volumetric heat and tritium generation must be simulated in some way in the BCTF, and may differ in the true radiation environment in important ways. Low fluence component-scale irradiation testing will reveal design issues that were not revealed by the non-nuclear BCTF environment. Additionally, information is ultimately needed about the long-term evolution of materials in a high fluence environment as it impacts cost through availability. These two mission needs could be accomplished in one or more facilities. Because of the significant risk associated with installing a blanket in the FPP without any knowledge of such effects, a means for component-scale irradiation testing is highly desirable.

   The ITER Test Blanket Module (TBM) program addresses the low-fluence mission for international programs, but the U.S. is not a member of the TBM program. Opportunities for collaboration with other TBM partners should be explored, but we may need different blanket concepts to accomplish a low cost FPP, and it appears unlikely that a U.S.-led TBM could still be deployed in ITER. TBM data is not openly available to partners in the way ITER data is, and furthermore it is unclear whether data could be obtained on a timeline consistent with our FPP. Regardless of interaction with ITER, the U.S. also needs a strategy for high fluence evaluation of blanket components. A Volumetric Neutron Source (VNS) could provide both low- and high-fluence component-scale irradiation testing. Historically, VNS referred to a tokamak that demonstrated high fluence and high availability, but several alternate concepts have been proposed more recently, including the gas dynamic trap, shear-stabilized Z-pinch, and beam target fusion generator. Because there exists some uncertainty regarding the feasibility of such alternate concepts and the cost and time required to develop them, it is recommended that a near-term scoping activity be undertaken that would outline requirements for component-scale irradiation testing (*e.g.* low vs. high fluence missions) and critically assess options (ITER TBM, VNS, early FPP operation, *etc*.) for achieving these.

7. **Initiate bench-scale tests of plasma exhaust pumping technologies that enable continuous operation and facilitate Direct Internal Recycling (DIR)**

   Because D-T burn fractions are inherently low, fuel has to circulate through the fueling and exhaust systems many times before it is burned in the plasma. This combined with a need for continuous operation in an FPP implies an increased throughput in the tritium plant, resulting in much larger tritium inventories. In order to reduce this, continuous pumping systems should be pursued that can separate the D-T mix from other exhaust gases and send it directly to the fuel injectors, bypassing the tritium plant and isotope separation systems. Such technologies include the super-permeable metal foil pump, which both pumps and separates hydrogen isotopes from other gases, as well as continuous cryopumping, in concert with other means for removing impurities such as neon.

8. **Advance pellet injection technology to meet the fueling and disruption mitigation needs of a fusion pilot plant**

   Pellet injection technology has been demonstrated at the proof of principle level for D pellet fueling. D-T pellet technology has been demonstrated by firing from an injector in a lab setting, but never deployed on a confinement device, and there are remaining challenges to realizing reliable D-T pellet injection for long pulse operation. Shattered Pellet Injection (SPI) was developed for the ITER Disruption Mitigation System (DMS) and is being implemented on many machines to support the design of the system for ITER. Significant further research is needed to improve the reactor implementation of the technology. SPI disruption mitigation research should continue to include alternate concepts such as those employing low-Z powders, and investigate the feasibility of high speed systems needed for high time response systems. The above technologies should be advanced through deployment in existing and future devices including ITER, W7-X, JT-60SA, NTUF, I-DTT, and other international devices to support dedicated research [FST-SO-D.4].

9. **Support other innovations in tritium processing technology that reduce the size, cost, and tritium inventory of the tritium plant**

   An FPP will need to exhaust and process tritium at a rate one to two orders of magnitude faster than anything that has been attempted previously and will require an enormous amount of chemical processing equipment. It will also require continuous 24/7 operation which also has not been attempted previously with tritium at this scale. If direct internal recycling is not developed successfully due to the challenges associated with removing impurities at such high processing rates, process intensification and ingenuity will be required to reduce building footprint (capital cost) as well as process system holdup and inventory. Process intensification that closely couples heat and mass transport with chemical reaction has the potential to both increase throughput of the Thermal Cycling Absorption Process (TCAP) isotope separation system and decrease its size. Modifications to such systems that can accommodate the necessary high processing rate should be pursued.

**10. Support innovation for measuring tritium inventory**

Strict tritium management will be required for an FPP to meet regulatory controls and site limits, ensure compliance with safety bases, and to maintain continuous facility operation. Previous operational experience with tritium in fusion facilities has been limited to TFTR and JET, where the total throughputs were on the order of 100 g. Tritium accountability for these machines was laborious and had numerous difficulties, and in the case of JET, led to restrictions of the D-T program. An FPP will have a tritium throughput of several kilograms a day and will have the additional complexity of tritium breeding. Thus, it is not possible or desirable to scale these accountability techniques to an FPP. Similarly, ITER's accountancy process will require the collection of all the tritium into hydride beds for calorimetry, which is not feasible for a continuously operating pilot plant. An ideal analytical method for tritium would be accurate (>99%), fast (seconds or less), in-line (no waste or cross-contamination), reliable and inexpensive – such a technique does not currently exist. This presents a challenge to safety and licensing, but it also limits tritium processing concepts. For example, the direct internal recycle process will be constrained by the ability to quickly and accurately measure tritium concentrations in the recycle loop. Techniques will be needed for measuring tritium concentrations in liquids such as PbLi, Li, or FLiBe.

**11. Perform research and development for tritium permeation barriers and capture technologies to significantly improve tritium confinement**

Current state of the art technologies employed by ITER for confinement and tritium cleanup will cost hundreds of millions of dollars and take up large spaces (10K's square feet). Also, currently ITER will be licensed to release up to 6000 curies per year for normal operations and up to 18,000 curies for years in which open vessel maintenance is required. Current known U.S. NRC limits are roughly two orders of magnitude below these limits so significant progress is needed to ensure that an FPP can be affordable and can meet emissions requirements to obtain a license. Tritium permeates all materials to various degrees, so advanced materials and confinement barriers are an absolute necessity as the large components of the torus could accumulate a significant tritium holdup, further increasing accident and safety basis consequence. Once it permeates, tritium also has significant damaging effects on all classes of materials which needs to be further characterized for materials being considered for plasma facing components. One method that has been suggested to reduce tritium permeation is the use of barrier coatings. Materials such as SiC, Al, or other materials are known to have low permeability to hydrogen isotopes and creating coatings of permeation barriers could make a significant difference in tritium emissions. However, demonstration of uniform coating methods as well as durability of the permeation barriers and materials in the operating environment will be important to adopting that strategy.

12. **Develop a comprehensive plan to provide the start-up tritium inventory for a Fusion Pilot Plant**

    While continued operations of an FPP will be fueled by tritium bred in the blanket, a significant initial start-up inventory is needed that depends on the rates of fueling, exhaust, breeding, extraction, processing, and retention in the machine. There is a limited non-defense related supply of tritium in the world, and by the time an FPP is ready to start up there will be even less tritium available due to decay and use by ITER. Perhaps 1 – 2 kg of tritium is needed, and this will be extremely expensive no matter how it is obtained. A comprehensive plan needs to be developed to address a supply chain for this tritium.

## FST Strategic Objective D: Advance the tokamak physics basis sufficiently to design a low cost fusion pilot plant

*The Mission laid out in the strategic plan requires innovative new approaches to optimize the tokamak in order to reduce the capital cost of a fusion pilot plant. A number of gaps in the tokamak physics basis need to be closed in order to confidently design a tokamak FPP and to achieve this mission. These include advancing our understanding of energetic particles and burning plasma physics relevant to a high-fusion-gain FPP; transport and stability physics needed to enable high-average-power output; and scrape-off-layer, divertor and plasma-material interaction physics required for a plasma exhaust solution in an FPP environment. Critical issues must also be addressed to integrate improved understanding into operational scenarios for an FPP. Achieving this objective requires a comprehensive, multidisciplinary science program that uses existing and planned domestic and international facilities, takes full advantage of our ITER partnership and leverages new opportunities through private-public partnerships, increases confidence in validated theory and modeling, and recommends a key investment in a new tokamak facility to begin operations between 2025-2030. The suggested capabilities of this facility, established through previous community-wide activities [PMI 2015, Transients 2015, Nat. Acad 2018] and this Community Planning Process, are specified to allow timely closure of physics gaps relevant for tokamak divertor solutions and their integration with potential FPP operating scenarios. This suggested new facility is referred to in this report as the new tokamak user facility (NTUF), but we strongly urge FESAC and the community to consider alternative names.*

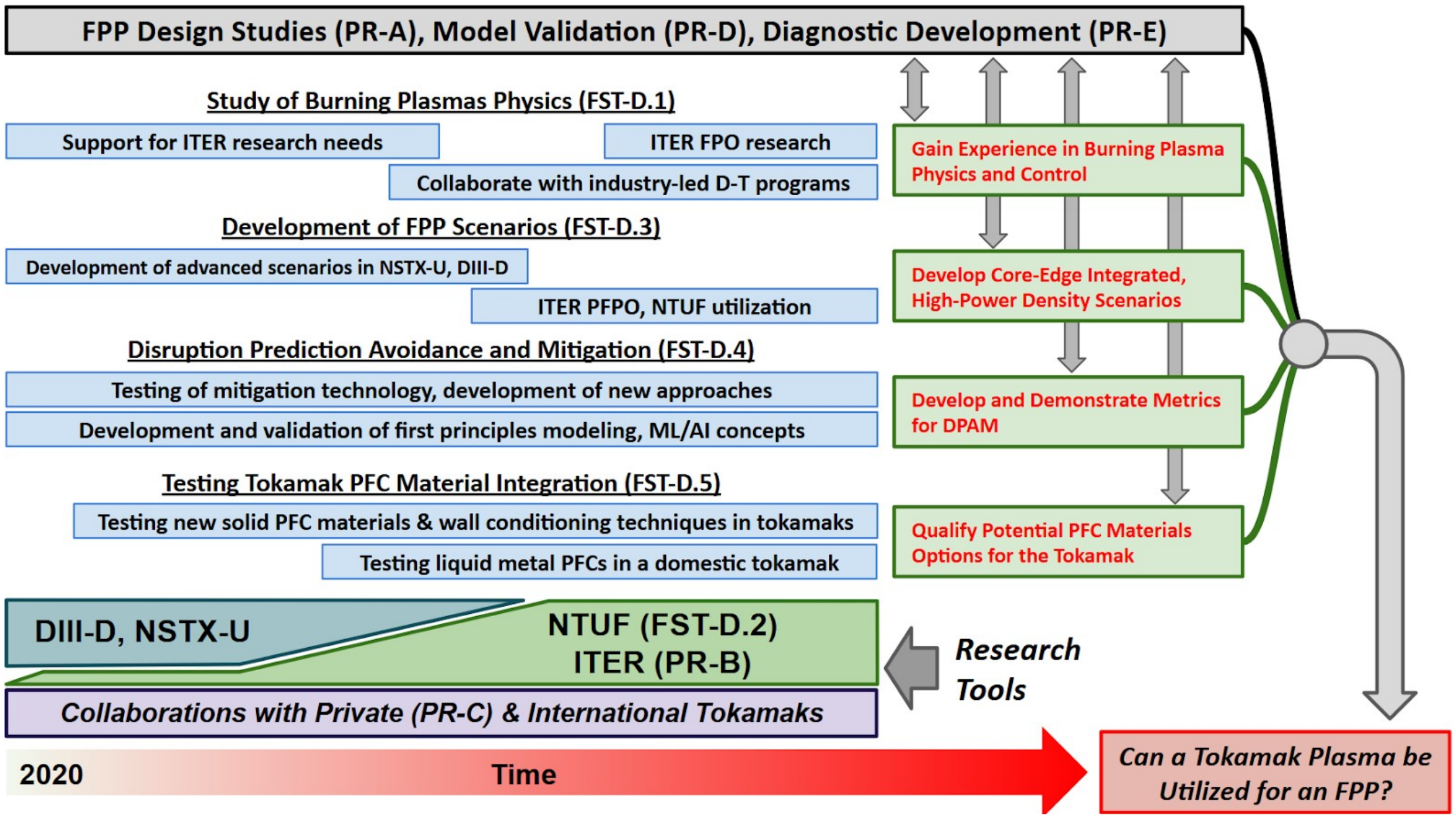

Figure FST-SO-D.1: Linkage diagram for FST-SO-D

**Relation to Science Drivers**:

Control, Sustain, and Predict Burning Plasmas: Understanding how to control, sustain and predict the behavior of a burning plasma is fundamental for safe and sustained electricity production from deuterium-tritium (D-T) fusion. Through the operation of ITER, along with the broadest array of current and planned facilities worldwide, the tokamak has the opportunity to develop the most comprehensive physics basis of any MFE concept. Control of tokamak plasmas requires the development of techniques to share multiple actuators so that the plasma can be initiated, sustained for the desired duration without deleterious instabilities, and safely ramped down and restarted upon demand. Techniques to diagnose tokamak plasmas are mature and are used to validate physics-based models and demonstrate sophisticated real time control algorithms, but these successes must be extended to burning plasmas.

Handle Reactor Relevant Conditions: The strategy for handling the high levels of heat and particle fluxes that will be encountered in a tokamak-based pilot plant will be driven by testing innovative plasma divertor concepts and heat flux mitigation approaches at reactor-relevant conditions. By completing experiments on existing facilities, leveraging upcoming experiments, and designing, constructing and operating a new tokamak that will achieve reactor levels of power and particle exhaust, we will improve our ability to develop physics-based exhaust strategies for the fusion pilot plant.

**Expert Groups**: *MFE-Boundary and Divertor Plasma Physics, Plasma and Material Interaction; MFE-Transport and Confinement; MFE-Energetic Particles; MFE-Transients; MFE-Scenarios;*

*MFE-Global Context and U.S. Leadership; FM&T-Plasma Material Interaction and High Heat Flux; FM&T-Magnets and Technology; FM&T-Measurements and Diagnostics*

1. **Leverage all opportunities to access, prepare for, and study burning plasma physics for validated extrapolation of self-consistent alpha heating, the physics of alpha particle driven instabilities, and helium ash removal for optimization of the FPP**

   Execute programmatic recommendation (FST-PR-B), which includes fulfillment of U.S. commitments to construct and operate ITER and establishment of a U.S. team to fully participate in ITER during pre fusion power and fusion power operation. This enables the U.S. to gain critical knowledge to inform the FPP design process, to develop and utilize knowledge of technology and diagnostics in a fusion nuclear environment needed for FPP control, and to complement science from NTUF to inform compact tokamak pilot plant scenarios. In addition, the U.S. should position itself to partner with industry to perform experiments on privately funded D-T tokamaks, such as SPARC and ST-F1, which are planned in the mid/late 2020s, contingent on demonstration of HTS technology for fusion magnets. These efforts should include modeling and simulation for operational scenarios in compact experiments with significant power from alpha heating, and on the development of diagnostics for science and control in a high flux, low fluence fusion nuclear environment. Research on DIII-D, NSTX-U and collaboration on international facilities should be used to help resolve issues identified in the ITER research plan, targeting areas of U.S. leadership such as ELM control, disruption avoidance and mitigation (FST-SO-D.4), energetic particle physics, and the development of control strategies for high performance burning plasmas.

2. **Establish the capability to test tokamak divertor solutions at conditions typical of an FPP that can also be integrated with FPP operating scenarios by designing and constructing a new tokamak facility**

   This new tokamak user facility (NTUF) must have the flexibility to investigate innovative tokamak divertor solutions, encompassing long-legged concepts and PFC material options, at heat and particle fluxes that are at the same scale as those projected for the pilot plant as well as the ability to simultaneously achieve core plasma energy confinement and bootstrap current that project to a high-average-power output, net-electric pilot plant. This simultaneous achievement of high power density across a range of core plasma scenarios and divertor solutions represent combined capabilities that cannot be obtained through international collaboration or a major upgrade to an existing facility. Without NTUF, there would be significant risk in the integrated exhaust scenarios when designing a tokamak-based FPP, and teams (FST-PR-A) would need to rely on extrapolations of modeling validated in regimes at parameters far away from pilot plant conditions. These proposed capabilities for NTUF represent a prioritization made by the community within this planning process, and a more detailed outline of the motivation for this facility, how it relates to community input and its context within the worldwide program is given in Appendix B. By combining these capabilities, NTUF would represent an opportunity to establish U.S. scientific leadership, leverage advances in theory and modeling (FST-PR-D) and motivate the development of a fusion workforce (Cross-cut WF). NTUF could provide a platform to demonstrate and drive technology development within this plan, such as PFCs (FST-SO-A), H&CD actuators (FST-SO-F) and diagnostics (FST-PR-E). To help inform the case for Mission Need for this facility, FES should support combined physics and engineering teams to establish conceptual engineering design and cost estimates for facilities with the capabilities mentioned above. Results from NTUF are expected to be necessary to inform decisions for near-term, industry-driven FPP approaches as well as to contribute to the techno-economic analysis of the FPP deliverables outlined in FST-PR-A. To be compatible with these timeline requirements, FES should establish the Mission Need for this new facility no more than one year following acceptance by FESAC of this recommendation, and support detailed design and construction so that the completion of NTUF allows for research output to begin before the end of the decade. While the existing physics basis is sufficient to proceed urgently with the Mission Need and conceptual design activities, existing facilities should be used to inform the completion of the detailed design of NTUF. These activities should include development and qualification of innovative and efficient current drive approaches on DIII-D, as well as development of techniques to avoid or mitigate edge localized modes.

3. **Develop the science basis for candidate scenarios that project to high-average-power on FPP utilizing domestic and international facilities, transitioning this research to NTUF and ITER**

An FPP will need candidate scenarios that project to high-average-power with integrated power exhaust solutions. These scenarios encompass both (a) the plasma and neutral physics that support solutions for the power and particle exhaust through the low collisionality, opaque pedestal as well as (b) the core-edge integration physics that enables high-average-power FPP operation. This recommendation combines the breadth of expertise and knowledge in tokamak physics that exists within the U.S. community, while focusing and prioritizing the work to help achieve our overall mission and integrating with ongoing design studies in FST-PR-A. Scenarios in compact FPP concepts have unique requirements related to establishment, sustainment and confinement of high absolute plasma pressure core, integrated with boundary solutions compatible with high power density. By utilizing DIII-D, international collaborations, and through completion of the NSTX-U Recovery Project and utilization of NSTX-U, the U.S. should develop and advance high-confinement, high-non-inductive-fraction scenarios suitable for long pulse and steady-state operation. A key activity is to use the unique capabilities of NSTX-U to study the aspect ratio optimization of the tokamak, extending the scaling of confinement and stability to lower collisionality. The research activities at DIII-D can support the development of the physics basis for FPP scenarios through utilization of advanced, innovative and flexible H&CD actuators (FST-SO-F). Additionally, the excellent diagnostic coverage on DIII-D can begin to answer critical questions on pedestal and boundary operational space for FPP designs as well as challenge predictive tools for boundary plasmas. Through physics model validation on DIII-D, NSTX-U, and international collaborations, the U.S. should further develop power exhaust scenarios, including extending divertor science and pedestal physics towards reactor-relevant conditions and investigating the compatibility of core/mantle radiation with high-confinement plasmas.

While existing domestic facilities can answer important open scientific questions, outlined in many recent community reports [PMI 2015, Transients 2015, ReNeW 2009], allowing us to maintain timely progress toward an FPP, completion of this objective will also require access to new facilities. As NTUF becomes better defined through FST-SO-D.2, a clear plan to transition from existing facilities to NTUF and ITER, as well as possible collaborations with other international and private facilities, should be established. The community supports a transition strategy in which existing facilities are not decommissioned before NTUF is near operation; that ensures continuity of experience and knowledge; and that allows existing domestic devices sufficient resources to complete essential mission objectives. As ITER moves to first plasma, a U.S. team should be prepared, using all available collaboration mechanisms, to participate in ITER during pre fusion power operation (PFPO) to obtain the non-burning plasma physics results that can help inform FPP design scoping. The critical research areas include disruptions (further emphasized in recommendation 4 below), scalings of confinement and pedestal behavior with gyroradius-normalized machine size, parametric dependencies of the SOL heat flux width, ELM suppression from 3D fields and core-edge integration with an opaque scrape-off layer.

4. **Advance multiple methods for disruption prediction, avoidance, and mitigation to inform FPP design**

Finding solutions to overcome the disruption challenge is critical to establishing the readiness of the tokamak at the scale of a fusion pilot plant, and a strategy that advances multiple prediction, avoidance, and mitigation (PAM) approaches is necessary. New or improved concepts need to be explored, testing the technology (FST-SO-C.8) and physics in all available domestic and international experiments. First-principles physics model development should be continued, leveraging SciDAC programs and complementary experimental validation efforts. Use of machine learning and artificial intelligence approaches to predict and avoid disruptions should be continued, leveraging improvements in available diagnostics (FST-PR-E) and high performance computing (FST-PR-D). Later in the 2020s, NTUF, ITER and other facilities accessible to the U.S. community, both international and domestic, should be utilized to demonstrate disruption prediction, avoidance, and reliable mitigation at high plasma current and energy density. Participation in the testing of the shattered pellet mitigation system should be emphasized as a priority for the ITER Pre-Fusion Power Operation phase. Results on PAM should be coordinated with FPP conceptual design efforts (FST-PR-A) and metrics of success should be developed.

5. **Test tokamak-specific integration of new PFC materials to reduce extrapolation uncertainties such as material migration and identify potential plasma facing materials for FPP designs**

   While the program outlined in FST-SO-A will work to develop and qualify potential solid and liquid metal plasma facing materials and components, there will remain tokamak-specific challenges that will impact their ultimate performance in an FPP. Existing domestic facilities, targeting DIII-D, and opportunities through collaboration on international long-pulse devices should be used to test advanced solid PFC materials such as SiC or new W-alloys. This should build on successful demonstrations of these materials using test stands and draw on experiences of qualifying new materials outside of fusion where possible. Following test-stand demonstrations to raise the TRL of liquid metal PFCs, discussed in FST-SO-A.2, an existing domestic tokamak facility, targeting NSTX-U, should be upgraded to use liquid metal PFCs and to explore their integrated effects on the plasma and to validate modeling. These opportunities should build on investments in fundamental science of plasma material interactions, open up opportunities for U.S. leadership, and be completed in a timely manner, in order to allow follow-on research using NTUF and to support FPP design studies (FST-PR-A). Future FPP operations will require more comprehensive solutions than existing or near-term devices for managing the consequences of sustained plasma material interactions. Using available tokamaks, strategies and concepts for active wall conditioning and material flow-through PFC solutions should be developed and tested. NTUF and burning plasma experiments will have access to more FPP-relevenant plasma edge conditions and research should include improving measurements of the neutral and plasma wall flux necessary to estimate sources of main-chamber erosion and results of material migration. These data will be needed as input for developing strategies for slag management, which should be completed in coordination with remote maintenance and RAMI activities (FST-SO-G).

## FST Strategic Objective E: Advance the stellarator physics basis sufficiently to design a low cost fusion pilot plant

*The stellarator has unique features that make it attractive as a fusion energy reactor. It is intrinsically steady state, can operate at high plasma density to achieve high gain while potentially relaxing plasma exhaust constraints, has relatively benign responses to magnetohydrodynamic instabilities, avoids current-driven disruptions, and has low recirculating power needs. These benefits all provide an opportunity to develop a net electric pilot plant at low capital cost. The stellarator is unique among magnetic fusion concepts in that it does not*

*require the plasma to provide critical aspects of its confinement. Rather, as the confining magnetic field in a stellarator is generated from external magnets, there is considerable flexibility to develop configurations optimized for performance, reliability, and simplicity. For example, the concepts of quasi-symmetry and quasi-omnigeneity have emerged to solve the traditional weakness of the stellarator, poor neoclassical transport at low collisionality, a prediction that has been validated experimentally in HSX and W7-X. With these advances, stellarators now demonstrate confinement properties and scaling that project to an attractive fusion pilot plant.*

*In the ensuing decades since the designs of HSX, W7-X, and NCSX, there have been considerable advances in theoretical understanding of stellarator confinement physics. In particular, quasi-symmetric configurations represent an approach distinct from those taken by the international community and offer unique opportunities to optimize for reduced turbulent and energetic particle losses (in addition to reduced neoclassical losses) while simultaneously integrating with novel plasma exhaust configurations. This is an area where the U.S. is recognized as having world leadership. In addition, enhanced computational tool development, advances in coil technology and design, and the availability of new enabling technologies (such as additive manufacturing techniques) provide opportunities to design and build simplified stellarator concepts.*

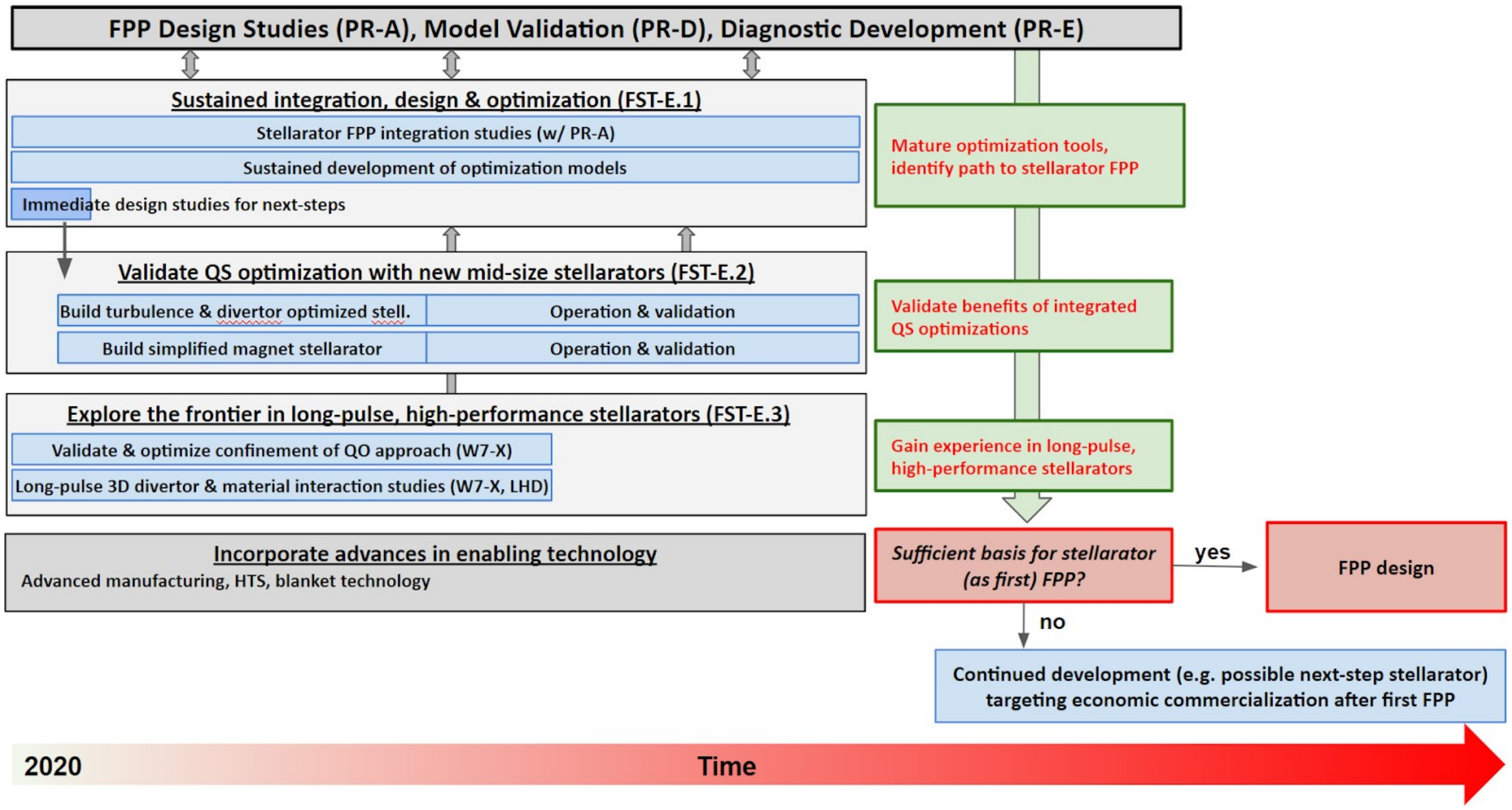

Figure FST-SO-E: Linkage diagram for FST-SO-E

**Relation to Science Drivers**:

Control, Sustain, and Predict Burning Plasmas + Handle Reactor Relevant Conditions: The stellarator is intrinsically steady-state and requires no significant active control. The stellarator's physics properties are largely dictated by external 3D magnetic field geometry. With significant recent advancement in 3D physics understanding and improvement in computational tools, the stellarator is ready to demonstrate improved confinement properties based primarily on theoretical prediction. As the stellarator is intrinsically steady-state, it also provides a unique opportunity for studying material erosion, redeposition, and migration on long timescales.

**Expert Groups**: MFE-Boundary and Divertor Plasma Physics; MFE-Plasma and Material Interaction; MFE-Transport and Confinement; MFE-Energetic Particles; MFE-Transients; MFE-Scenarios; MFE-Global Context and U.S. Leadership; FM&T-Plasma Material Interaction and High Heat Flux; FM&T-Magnets and Technology; FM&T-Measurements and Diagnostics

1. **Expand and sustain an integration, design, and optimization effort to identify candidate stellarator configurations that scale to an FPP**

   The U.S. stellarator program should employ and continue to develop the theory, modeling and optimization tools (FST-PR-A, FST-PR-D) needed to identify and refine possible low cost FPP configurations that incorporate recent U.S. advances in optimizing for reduced neoclassical, turbulent, and energetic particle losses, improved MHD stability, compatible divertor configuration and magnet design simplicity. The activity should also be used in the near term to coordinate and finalize optimization and design of new domestic stellarator facilities that are proposed to validate these U.S. innovations (FST-SO-E.2). Through continuously incorporating advances in theory and computation, results from domestic experiments and international collaboration (FST-SO-E.3), and advances in enabling technologies, this activity will clarify the pathway to a low cost stellarator pilot plant. This should include identifying key decision points such as if and when additional stellarator experiments beyond those proposed in FST-SO-E.2 would be required.

2. **Design, construct, and operate one or more stellarators to test innovative optimizations and demonstrate performance that projects to an economically competitive pilot plant**

   The construction of new mid-size stellarator experiments is essential to validate recent U.S. advances in optimization using the approach of quasi-symmetry. These new experiments must be initiated in the near term in order to make timely decisions on the viability of the stellarator for an FPP. Two devices would be required to explore the two related yet distinct approaches of quasi-axisymmetry (QAS) and quasi-helical symmetry (QHS). Whereas both configurations have similar neoclassical transport physics, geometric differences can impact equilibrium and microinstability properties with the implication that QAS and QHS are predicted to have different global stability and turbulent transport properties. Further, QAS configurations can be found at smaller aspect ratio than QHS. The HSX program has tested the hot electron physics of QHS, while no experimental facility to test QAS physics exists. There is a crucial need to test the hot ion, high beta properties of a quasi-symmetric stellarator to validite the physics basis required for FPP. The proposed facilities will pursue the distinct optimizations of both quasi-symmetric approaches, while also incorporating unique technology innovations. Possible capabilities of these new facilities have been discussed through community-wide activities [Stellcon 2017]. The proposed mid-scale quasi-helically symmetric (QHS) device is targeted to demonstrate for the first time turbulent transport reduction by design of the 3D magnetic field, in addition to good energetic particle confinement, while simultaneously integrating a novel non-resonant divertor configuration. The technical readiness to proceed with the design and construction of a mid-scale QHS stellarator is high and can start immediately [Stellcon 2017]. A small scale quasi-axisymmetric (QAS) device is proposed to build simpler stellarator coils using permanent magnets while maintaining optimization towards low turbulent transport and improved MHD stability.

3. **Validate core physics and investigate steady-state divertor and plasma exhaust solutions in long-pulse, high-performance optimized stellarators**

   Collaboration with the international stellarator community enables access to high performance, long pulse stellarator physics. The international stellarators LHD and W7X rely on different optimization approaches than quasi-symmetry, providing complementary research to broadly validate the physics that forms the basis of the optimization models. The early success of W7-X marks a new era for stellarator research and the U.S. should continue as a central partner in exploring the properties of the quasi-omnigenous configuration. Furthermore, the U.S. should also pursue collaborative research opportunities to explore the physics of, and possible solutions for, steady-state divertor and plasma material interactions enabled by the long-pulse superconducting stellarators (FST-SO-A). International collaboration can provide high scientific return on investment as both LHD and W7X are operational.

## FST Strategic Objective F: Innovate the magnet, heating, and current drive technology needed to reduce the pilot plant capital cost

*Innovations in the technology that will enable low-cost magnetic fusion energy will be needed, particularly in the areas of magnets, heating, and current drive. Innovations in magnet technology, where the U.S. has a strong leadership role, can open the pathway to achieving high magnetic fields that can reduce the size of confinement devices and eventually a fusion pilot plant. Development of joints for superconducting magnets, for example, could radically alter how plants are designed and maintained by enabling access to components inside the toroidal field coils and by facilitating vertical remote maintenance. Radiofrequency (RF) power must provide practical solutions for high-efficiency heating and current drive with location control to access and sustain burning plasma conditions in a reactor. The use of advanced manufacturing techniques can lead to innovative and cost-effective approaches to the development of components otherwise too complex to manufacture with traditional techniques, including launchers that can operate in a fusion nuclear environment. Innovations and*

*optimizations in magnets and RF launchers, sources, transmission, and plasma/wave interactions benefit multiple confinement concepts including tokamaks, stellarators, and many alternate configurations. The use of public-private partnerships will play an important role in advancing this area. In addition, it is anticipated that the needed test facilities in this area can provide a workforce development opportunity as an entry point into the field. Collaboration with universities and international partners, including ITER, will help reduce costs and duplicated effort. Testing of new launcher concepts in confinement devices can provide valuable feedback for the advancement of innovative design concepts. Note that additional areas of enabling technologies are covered in other parts of the report, including pellet injection (FST-SO-C.8), first wall materials (FST-SO-A), and structural materials (FST-SO-B).*

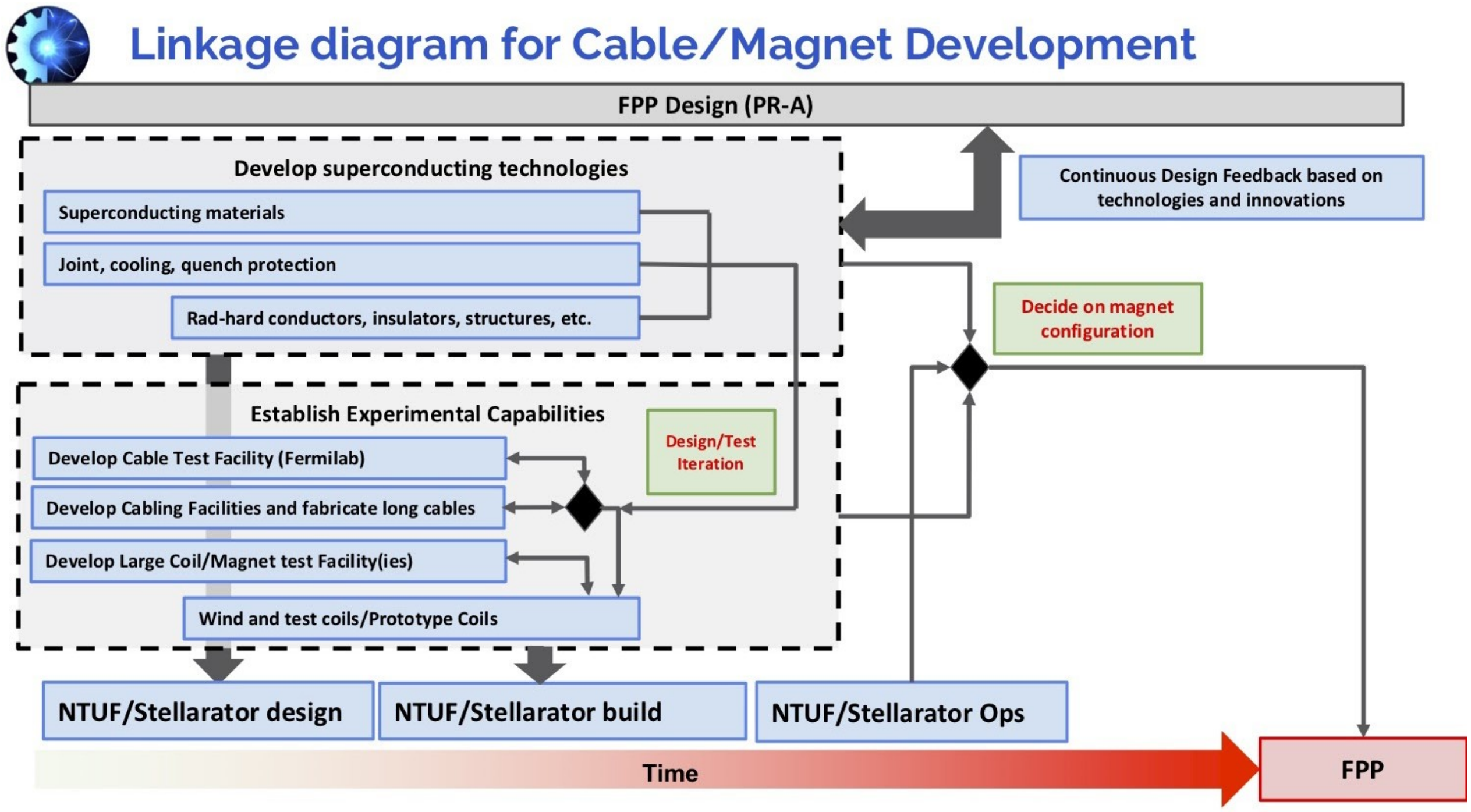

Figure FST-SO-F.1: Linkage diagram for FST-SO-F

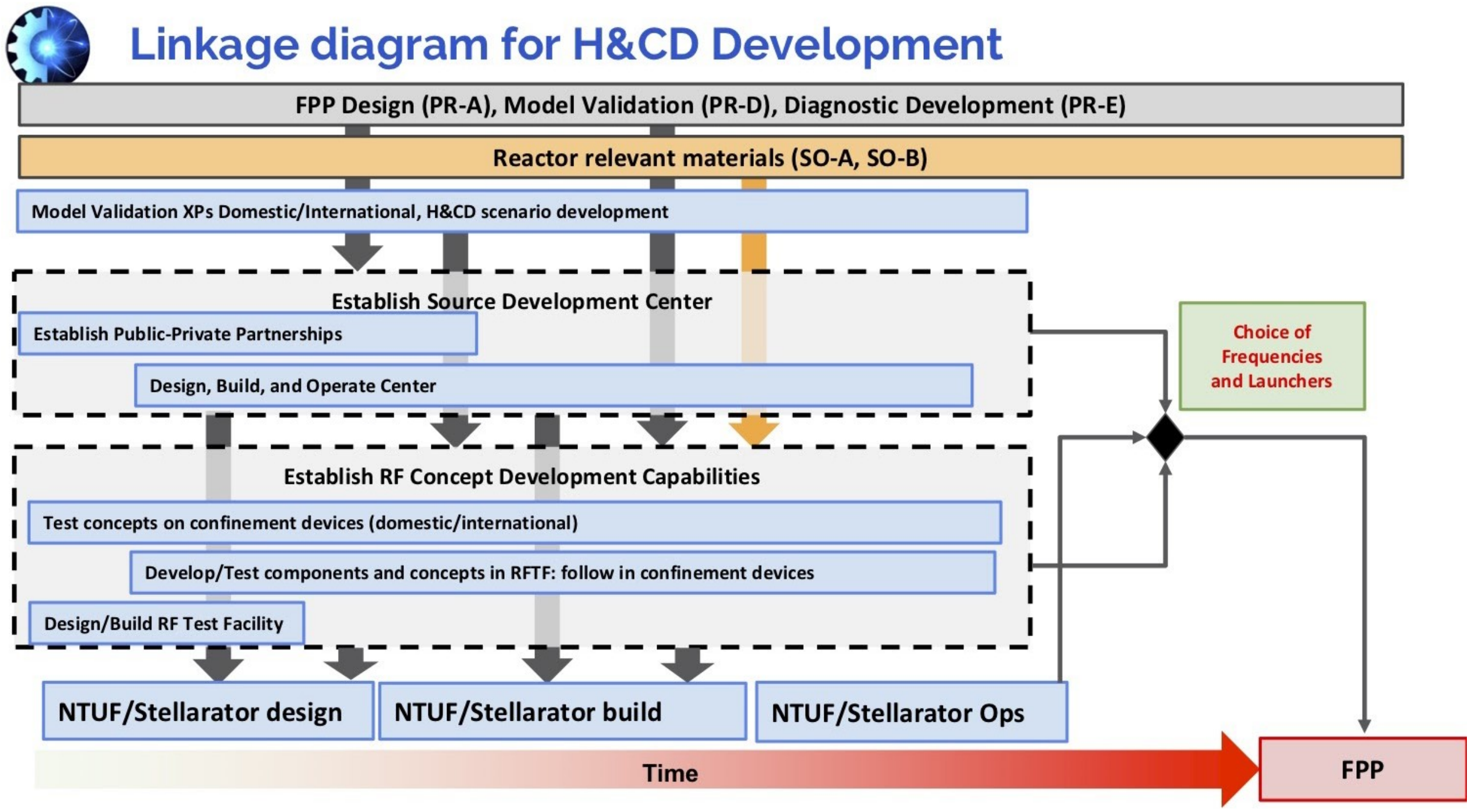

Fig. FTS-SO-F.2

**Relation to Science Drivers**:

Control, Sustain, and Predict Burning Plasmas: This objective provides necessary technology that will enable the success of a reactor concept and practical operating scenarios. High-field magnets are needed to confine the plasma, and innovations that allow for higher magnetic fields and other technologies will define the confinement concept. The electrons and ions will need to be heated to achieve plasma burn, and current drive will be needed for some reactor concepts. Modeling coupled with validation experiments on current and future confinement devices will confirm the scenario development physics for the considered RF heating and current drive methods. Innovations in the sources and launchers are required to provide the needed steady-state solutions.

Handle Reactor Relevant Conditions: Magnets and RF launchers will need to operate in a fusion nuclear environment. The magnet support structure, superconductors, and insulators must survive the neutron and radiation flux. The RF launchers will be part of the first-wall and must be constructed out of materials that are able to operate at elevated temperatures in the high-heat and neutron/radiation flux environment. Innovations will be required in launcher-compatible plasma-facing materials, structural materials, and required cooling technology.

**Expert Groups**: *MFE-Energetic Particles; MFE-Scenarios; FM&T-Magnets and Technology; FM&T-Plasma Material Interaction and High Heat Flux; FM&T-Fusion Materials*

1. **Establish the experimental capabilities required to develop and test high field magnets and cables**

   Various facilities are needed to test the anticipated advances in high magnetic field technology for magnetic fusion energy. Development of prototype cable technology is a high priority, and the availability of a recently announced cable test facility at Fermilab will help with both public (multiple DOE offices) and private cable development. The establishment of a large-bore magnet test facility is needed to demonstrate prototype fusion magnets at up to 20 T. Coordination with private industry in this area will be essential to the success of this endeavor, as numerous industry partners and DOE offices are currently involved in many of these efforts.

2. **Integrate achievable magnet technologies into the FPP multiphysics design tools**

   The practical performance capability and anticipated advances in magnet and cable development must be coordinated with the design teams for the next generation of confinement devices and FPP concepts (FST-PR-A). Innovations in multiphysics design tools (FST-PR-D) can push innovations in magnet configurations that otherwise would be too costly or time consuming to pursue for different confinement concepts.

3. **Continue to develop and test high-current, high-field cable technologies through the FPP conceptual design phase**

   A variety of magnet-related technologies will need to be advanced for the specific requirements of next generation fusion devices using superconducting materials. Radiation damage effects in cable conductors, insulation materials, and magnet structural materials need to be determined and understood so that effective mitigation approaches and their design consequences can be accurately assessed. Advances in joints could revolutionize superconducting magnet design, assembly, and remote handling (FST-SO-G.4). Development of advanced cooling methods and quench protection will be needed. Work needs to be coordinated with industry to advance progress and reduce potential duplication of effort.

4. **Establish the capability to develop and test RF launcher concepts compatible with operating in a fusion nuclear environment**

   There is a need to develop steady-state, high-heat-flux nuclear compatible RF launcher components/concepts, since many existing launcher concepts are not compatible with this environment and will not meet the requirements of an FPP. These needs can be met and the research gaps closed by development of components/concepts using a new RF Test Facility and by testing on confinement devices as available through FST-SO-D, FST-SO-E and FST-SO-H activities. The new facility will also be needed to validate near-field RF-plasma interaction models that can be used to predict the impact of wave-interactions in the scrape-off-layer (SOL) on future confinement devices. A dedicated non-confinement facility is needed to provide better diagnostic access and operational flexibility/availability compared to a confinement device. This facility will reduce risk and raise the maturity of the design of launchers and diagnostics prior to implementation on a confinement device, since confirmed concepts will be necessary for successful operation of future devices and component failure on a confinement device cannot be tolerated. The facility will need the appropriate magnetic field and relevant plasma parameters to simulate the SOL environment, and such a capability would be unique to the U.S. compared to other RF test facilities worldwide. It is anticipated that innovative advanced manufacturing techniques (FST-SO-B.1) will be essential to the realization of launchers needed to operate in the first-wall environment, and the concepts will be tested at high RF powers in the magnetized plasma environment of the facility. When possible, testing of launcher concepts on confinement devices should be done to confirm high-power long-pulse compatibility. First-wall compatible launcher materials will be tested and qualified at high plasma heat-flux in a separate device, such as MPEX (FST-SO-A.1).

5. **Establish a new RF source R&D center to develop the RF technology needed for a fusion power plant**

   A new RF source R&D center is needed to develop the RF generators and power supplies needed to provide high-efficiency steady-state heating and current drive. Unprecedented levels of power at extended pulse lengths will be required in an FPP. The TRL for high-power steady-state generators in the electron cyclotron range of frequencies (ECRF) is low, particularly for the 200+ GHz gyrotrons needed for a high-magnetic field confinement device. Improving the efficiency of ECRF, ion cyclotron range of frequencies (ICRF), and lower hybrid range of frequencies (LHRF) sources above the existing level (~50%) will reduce recirculating power requirements for an FPP, and very long pulse testing will establish source lifetime limits in the intended mode of operation. Public-private partnerships will be essential to the success of this endeavor (FST-PR-C).

6. **Support the development of reactor-relevant heating and current drive scenarios through experiments that support model development and validation on both current and future confinement devices**

   The validation of heating and current drive models on current experiments needs to be supported to advance operating scenarios needed for a future reactor. The RF SciDAC effort is a particular strength for the U.S. community, and can be expanded to support this area, both domestically and internationally. Code development benchmarked by experiments (through FST-SO-D, FST-SO-E, FST-SO-H, and FST-PR-B activities) and RF diagnostics (FST-PR-E) should be utilized to produce a full predictive understanding of RF launchers in a next step device. Scenarios that use RF heating and current drive methods to reduce the requirements on the central solenoid need increased emphasis in the tokamak approach. It is particularly important to develop high efficiency heating and current drive methods that are reactor relevant and operate on the desired location in the plasma. While the use of negative neutral beams may be complementary to RF systems, collaborations with international partners should be pursued if needed. Note that the model validation area will be coordinated with similar activities in FST-PR-D.

## FST Strategic Objective G: Develop the balance of plant technology, remote handling and maintenance approach, and licensing framework necessary to ensure safe and reliable operation of the fusion pilot plant

*When moving from plasma experiments of today to an FPP, there are numerous challenges beyond the plasma and the fusion core. These challenges include the overarching need to ensure safe and reliable operation, the balance of plant equipment (BoP), remote handling and maintenance, and the licensing approach. Licensing an FPP and procuring its BoP equipment are long-term items, and some constraints related to safety, complexity, or cost can render unworkable an otherwise promising FPP design.*

*An electricity-producing fusion nuclear device such as an FPP will need to be licensed. An agency outside the fusion program,* e.g. *the NRC, grants the license. Public safety is the primary issue for licensing. Pathways for release of tritium or other activated products to air, ground and water beyond the site boundary are of particular concern. The fusion program has a large and integral role in the licensing process because the fusion program provides the modeling, design rules, and materials data that propagate through models of accident scenarios to develop a credible well-documented safety basis for a license. The near term action items for the fusion community in this area focus on generating the data that will eventually support the safety basis for a license.*

*While we should pay attention to developments in BoP equipment in fission and non-nuclear applications, fusion has higher operating temperatures, different working fluids, and unique considerations like tritium safety. Heat exchangers, turbines, and other BoP equipment must be developed specifically for an FPP to use the heat transfer fluids of helium (He) and/or lead lithium (PbLi).*

*In an FPP, almost all maintenance must be performed remotely using robotic systems to handle radioactive materials. These robotic components must function while exposed to difficult-to-shield gamma rays accompanied by significant decay heat, and work with components that are contaminated with tritium, activated dust/debris, and material deposited on their surfaces.*

*The strategies for Reliability, Availability, Maintainability, and Inspectability (RAMI) in an FPP significantly impacts the design of its in-vessel components. Also, the power conversion system and the tritium processing plant have high cost and high impacts on the FPP mission. Early*

*confirmation that the strategies will work is needed to mitigate costly later redirection that would delay the safety basis and licensing discussions for a decision-to-build. The R&D approach is to use early small-scale tests for this purpose. The development of appropriate data is linked to many of the recommendations in [FST-SO-C](#) (tritium/blankets/fueling/pumping), [FST-SO-A](#) (heat/particle-loads), [FST-SO-B](#) (materials) and the aspect of design and systems integration and the safety basis in [FST-PR-A](#) (design activity), [FST-PR-B](#) (ITER) and [FST-PR-E](#) (diagnostics).*

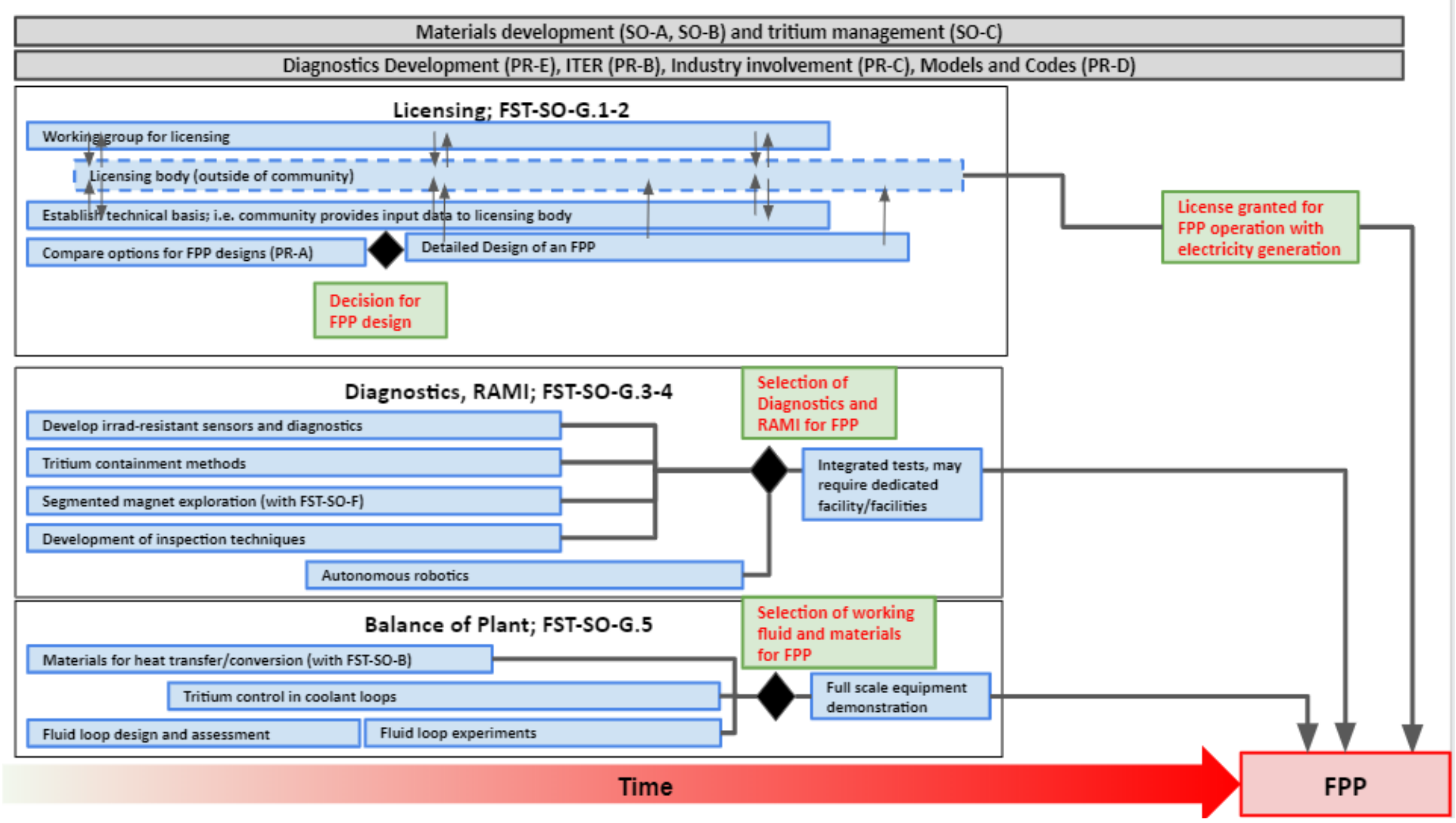

Figure FST-SO-G.1: Linkage diagram for FST-SO-G

**Relation to Science Drivers**:
Harness Fusion Power: Transitioning from a fusion science program to a program aimed at commercializing fusion energy will require the advancement of numerous systems needed to Harness Fusion Power. Extracting the fusion energy and interfacing with the grid will be critical and have unique challenges as compared to current commercial energy sources. We need a strategy for licensing an FPP, and an efficient R&D path for high-cost high-impact systems (RAMI and power conversion).

**Expert Groups**: FM&T-Magnets and Technology

1. **Start a working group to develop a licensing approach for fusion**

   A fusion pilot plant that produces significant power and electricity will also possess hazardous inventories of tritium and activation products. It will require a licensing process to ensure safe operations, protection of the public, and sufficient proliferation resistance. No regulations in the U.S. deal specifically with licensing of fusion reactors. Nor is it clear that existing regulations, written for fission reactors and other types of radiological facilities, are directly applicable to fusion reactors. To mitigate the risk of a long lead time for licensing an FPP, a licensing working group should be established that engages national laboratory experts, gathers lessons learned from ITER (FST-PR-B) and any other relevant licensing experience (*e.g.* criteria for selecting licensing basis events), and engages NRC regulators and industry and public stakeholders to outline the necessary licensing framework for fusion. The working group should become a standing committee with some group members included in the FPP design studies (FST-PR-A). While the fusion community does not grant a license, it serves as an important resource to the regulating body as a strategy for licensing fusion is developed. There is precedent and an important need for the community experts engaging with the NRC; examples of such interactions include the recent exploration of licensing of medical isotope facilities and molten salt reactors.

**2. Establish a technical basis for fusion reactor safety and licensing**

This activity provides technical data for the safety basis for licensing. Starting limited work in target areas as soon as possible has a high payoff for relatively modest investment. Moreover, an early look can direct R&D toward more easily licensed designs. Innovations may lower cost, for example, by shrinking plant size, lowering tritium inventory and improving safety, and in this way speed and ease the path for licensing.

As described above, the fusion community does not do the licensing itself, but there is a large body of scientific research that the fusion community must undertake to make licensing an FPP possible. This research would be completed in coordination with the design exploration activity in FST-PR-A, the licensing working group in Recommendation 1, and the licensing body. The licensing body itself does not perform the research, so we, as the ones applying for the license, must generate all the information related to safety and operations. We must prove to the licensing body how we came to these conclusions about our design and its behavior under various scenarios, and how we qualified materials, components, and everything else in the plant. There will be iterative interactions with the licensing body because they will give feedback on areas where more data is needed, and then that research must be undertaken. It is necessary to have detailed designs for any licensing assessment, so that happens later in the timeline, but the data gathering relevant to the licensing should begin immediately because much of it has long lead times (for example, materials qualification).

The shorthand term of "establishing a technical basis," as used in the title of this recommendation and on the sequencing diagram, actually covers several areas of scientific research including developing materials properties and component databases; developing high temperature structural design criteria (HTSDC);

developing design rules and ASME/other code development; and analyzing safety, accidents, hazardous materials, and energy sources in an FPP. The materials properties database and HTSDC are covered in more detail in FST-SO-B; they can begin immediately and interact with the licensing working group in Recommendation 1 in an iterative way. It is important to note that materials themselves are not licensed, so even if materials are chosen for an FPP that have been previously used in other licensed nuclear facilities in the U.S., those materials would have to undergo additional analysis and testing specific to their use in an FPP. The fusion community and FES would have a very natural role in this research in a parallel way to how materials research is currently performed at national labs, and to some extent universities, to support the licensing applications of commercial fission and medical isotope facilities. Similarly, as part of analyzing safety, potential accidents, and the tritium inventory and control for an FPP, there is a need for accurate fundamental properties, chemistry, and interaction data for tritium (for example tritium permeation rates in materials under relevant conditions). This aspect of the research would fall under FST-SO-A and FST-SO-C with coordination with the licensing working group in Recommendation 1, and can be initiated immediately.

Licensing an FPP will require quantification of radioactive material inventories and their location. This, in turn, requires experiments and verified and validated tools for neutronics, activation analysis, and plant-scale transport analysis of tritium, activated corrosion products, activation products in the form of dust, *etc*. Although tritium is perhaps the largest single safety factor specific to an FPP, additional potential hazards come from the combination of radiological (tritium, activated dust/fume), chemical (liquid metal, molten salt), toxicity (beryllium, lead), and thermal (afterheat). The fusion community must establish the scientific and technical basis for understanding potential hazards and their controls in normal and off-normal scenarios. The U.S. has some leadership in licensing (e.g. MELCOR code) and long-established expertise in safety and tritium handling.

3. **Develop new, specialized sensors and diagnostics for in- and ex-vessel survey**

New sensors, diagnostics, and integrated data acquisition for remote maintenance, and integrated plasma core and plant control are necessary to ensure worker safety and comply with licensing requirements. Early on, we need to identify gaps. Moreover, an early look at areas outside fusion to anticipate progress can direct R&D toward technology and innovations that may lower cost.

Many traditional instrumentation methods will not work in a fusion environment due to high magnetic fields, radiation, service temperature, and contamination. There is also a hierarchy of needs based on their lifetime, the severity of the operating environment and when the prospect of likely solutions must emerge. The gap analysis noted above should engage appropriate expertise, become part of the ongoing design activity (FST-PR-A) and have a strong link with diagnostics development (FST-PR-E). We should leverage experience outside fusion in controlling complex systems and in related areas such as neural networks, autonomous learning, *etc*.

Additionally, developing the architecture and software for an overall plant control system is a critical area for plant safety, and software and automation are important. For an FPP, the data acquisition system (DAS) and logic network (software) for plasma core and overall plant control (coupled) must digest a huge amount of real-time data and translate this into a useful control system. This includes raw data from sensors plus signals that confirm that the sensors and systems are operating properly. A deployment plan will be needed with later confirmation that systems are operating properly during shake-down at each stage (startup with H and/or D, then later D-T operation).

4. **Establish strategies for remote calibration, alignment, maintenance, and replacement of components**

Early on, we should investigate and anticipate advances in robotics and measurement that can simplify RAMI and pursue a design approach that looks to innovation in design to reduce the potentially large costs related to RAMI. A small effort in this area in the near term will help avoid major delays or redesigns of an FPP.

The fusion nuclear environment will require advancements in the design of components that are compatible with remote maintenance. The ongoing design activity (FST-PR-A) should carefully consider the RAMI approach from the start because of its strong impact on the design of in-vessel systems with the potential to simplify systems. A radical example is jointed superconducting magnets to enable vertical assembly/disassembly, which increases the ease of remote maintenance and makes blanket sector designs more flexible (FST-SO-F). A strategy and methods for re-welding or re-connection of irradiated material in a high radiation environment needs to be developed with special attention paid to reduced concentration and release of tritium. The early studies should target areas with high impact on cost and evaluate the likelihood of successful developments, and the ongoing design studies should be used to guide R&D investment. Then and later we should leverage developments in autonomous and intelligent robotic systems in the public and private sector, along with international experience (*e.g.* JET, ITER) for fusion facility remote handling and maintenance in areas where conditions prevent human workers. This links to FST-PR-B (ITER) and FST-PR-C (Industry).

Intermediate work will likely require a suite of small facilities to develop and test equipment for remote sensing and component handling. Later, due to the high cost of the equipment for RAMI and huge impact of any shortcomings in performance, we anticipate that full scale demonstrations of this equipment will be needed for the decision to build an FPP.

5. **Carry out conceptual design and small scale tests of balance of plant equipment**

Specialized BoP equipment is needed for an FPP due to fusion's higher operating temperatures, different working fluids, and unique considerations like tritium safety.

We should start advancing heat exchanger technology by compatibility tests in fluid loops with materials that can operate at high temperature with He and PbLi coolants. PbLi is corrosive to many materials, especially at high temperatures, and no material yet identified works for a long-lived PbLi heat exchanger. Developing/adapting He gas turbines and compressors is another need, which should leverage related activities where possible (for example, work within NASA). Again, initial small-scale tests are useful to confirm productive directions for the R&D path in combination with outreach toward approaches to Brayton cycle power conversion and high temperature heat exchangers used in fission and aerospace. Solutions for the BoP must address tritium control, activation product transport, impurity cleanup, temperature ranges for operation, and effects of radiolysis. These near-term activities have connections with the design activity (FST-PR-A), blanket development (FST-SO-C) and materials development (FST-SO-B).

In the near and intermediate term, we anticipate the need for smaller facilities to develop and test some BoP equipment (*e.g.* gas turbines and heat exchangers). Due to the high cost of the BoP equipment for an FPP and the high impact of poor performance on its mission, we anticipate that full scale demonstrations of this equipment will be needed for the decision to build an FPP.

## FST Strategic Objective H: Develop alternative approaches to fusion that could lead to a lower cost fusion pilot plant, utilizing partnerships with private industry and interagency collaboration

*Magnetic fusion energy (MFE), magneto-inertial fusion energy (MIF/MTF), and inertial fusion energy (IFE) offer distinct paths to the commercialization of fusion energy. While the tokamak and stellarator MFE configurations are relatively mature, and a focus of this report, neither is ready to meet the requirements for a FPP. Other MFE configurations, commonly referred to as "alternate concepts," offer potential for significant reduction in engineering complexity and therefore lower cost, often in exchange for increased challenges in plasma confinement or stability. Some alternate MFE concepts require no plasma current and are therefore inherently steady-state. Others have moderate or zero toroidal field, reducing magnet requirements. Elimination of auxiliary heating is another benefit associated with some alternate concepts. Similarly, IFE concepts offer potential advantages through the separation of driver technology and target implosion, with tradeoffs in complexity associated with intense driver-target interactions, high repetition rate, and requirements for the first wall. Compact, pulsed MIF/MTF concepts that rely on magnetic insulation and compression of the plasma have characteristics intermediate to MFE and IFE.*

*This strategic objective aims to develop select alternative IFE and MFE fusion concepts as part of an innovative program that builds on existing U.S. leadership. Many of the efforts now*

*supported by industry are based on concepts different from the tokamak. Federal support for fusion research by DOE-ARPA-E also embraces alternate concepts. Leveraging and coordinating research across all sectors is one goal of this strategic objective. Doing so will inject innovation and scientific breadth that benefits all approaches. Research on alternate concepts develops science and technology that helps advance tokamak and stellarator research and could contribute to FPP. Also, many of the next steps for the development of alternate concepts are innovative experiments at the small-to-intermediate scale, creating excellent opportunities for workforce development.*

*Within this strategic objective, Recommendations 1 and 2 address the need for an Inertial Fusion Energy (IFE) program, with a focus on maximizing target gain and the efficacy of the driver, while lowering cost. This is distinct from the National Nuclear Security Administration's (NNSA) national-security-oriented mission. The main purpose of the NNSA program is advancing and delivering inertial confinement fusion capabilities in support of the Stockpile Stewardship Program (SSP). Additionally, the current DOE-FES Program provides support for the basic plasma science at HEDP conditions. This program will leverage existing NNSA capabilities, attract new scientists to these missions, and increase the return on current and future U.S. investments. Given that additional requirements in energy gain and efficiency are necessary to achieve IFE, combined with the concomitant uncertainties in physics, success will only be achieved through the continuity, sustainment, and growth of a coordinated multi-disciplinary research community involving multiple institutions that can prioritize research important to energy production. Recommendation 3 seeks to establish rigorous criteria and metrics for evaluating alternate concepts, and recommendation 4 directly supports partnerships with private industry, and other agencies. All four recommendations are critical to the implementation and impact of this objective.*

**Relation to Science Drivers:**
<u>SD1: Control, Sustain, and Predict Burning Plasmas</u>: The ultimate goal of alternative concepts is to explore novel methods for creating and confining burning plasmas other than the tokamak or stellarator. It is necessary to develop such concepts to sufficient technical readiness so that their viability can be determined. Alternative confinement concepts may also enable scientific and technological progress that could benefit many approaches to fusion energy.
<u>SD2: Handle Reactor Relevant Conditions:</u> A number of alternative concepts are proposed as the basis for neutron sources needed to develop fusion materials.

**Expert Groups**: HEDP topical area, MFE-Transport and Confinement

1. **Establish a program that can pursue requirements for IFE**

   A credible approach to IFE requires an integrated design with high gain, an efficient driver, and targets that are relatively inexpensive. This motivates the development of concepts that are simple and/or robust, which are not a focus of the national-security-oriented National Nuclear Security Administration's (NNSA) Inertial

Confinement Fusion (ICF) Program. (The primary mission of the NNSA stockpile stewardship program emphasizes high yield rather than high gain.) This could include novel approaches to compression, heating, and confinement, such as fast ignition, shock ignition, and implosions that are magnetized. To address conditions relevant to a pilot plant, it will also require a better understanding of beam propagation and target injection in environments that are hostile, and the response of the first wall to intense nuclear and x-ray radiation (FST-SO-B, FST-SO-A). Data is needed outside the NNSA parameter space in beam-plasma interactions and/or current loss, equation-of-state, opacity, nuclear processes, and material interactions, with and without magnetic fields. Assuming use is made of existing facilities, diagnostics, and computational resources, in close partnership with the NNSA, and other agencies, the potential for progress is significant. At the same time, it is also necessary to understand the integrated science of heavy ion beam accelerators and transport systems, high-bandwidth and deep-UV lasers, and pulsed power, with greater utility to IFE. Existing capabilities are constrained in terms of cost, reliability, operability, and peak and average power, and advances in HEDP, the science and technology of particle beams, solid-state lasers, excimer lasers, and pulsed power provide exciting new opportunities.

2. **Advance technologies relating to the driver, target fabrication, and diagnostics and modelling that could speed development of IFE**

Complementary to scientific understanding (FST-SO-H.1), the viability of IFE requires advances in driver technology, developing cost-effective methods for target fabrication, and advances in diagnostics and modeling that address reactor-relevant conditions. The product of target gain and driver efficiency must approach or exceed 10 to project to an energy producing pilot plant. Also, the driver must have a path to operating at high repetition rates (0.1 - 10 Hz) at a reasonable scale-of-capital, with little or no need for maintenance. (For perspective, the fusion-scale facility at LLNL operates at ~$3 \times 10^{-5}$ Hz, and has to refurbish its optics on a rolling basis.) This motivates research into technologies including heavy-ion beams, broadband lasers, Raman and Brillouin amplifiers, plasma optics, and other concepts that advance the basic feasibility of IFE. The goal (in all cases) should be to reduce the scale and complexity needed for a cost-effective pilot plant. As all concepts in IFE are tied to the quality, quantity, and cost of target(s), this should include research into new fabrication techniques, such as additive manufacturing, after key physics are demonstrated. Lastly, these efforts should include support for diagnostics (FST-PR-E) and modeling (FST-PR-D) that function with real-time data, analysis, and feedback, at high repetition rates, and advances in Quantum Information Science and Machine Learning. An IFE activity within the Department of Energy's (DOE) Office of Science can leverage capabilities developed by the NNSA, but these tools have restrictions on their use, and we support a review of requirements and needs for open design capabilities consistent with national security.

3. **Establish a staged, three-tier program to develop promising alternative magnetic fusion energy configurations using rigorous evaluation and metrics**

   A three-tier strategy is proposed to provide a staged development approach for alternate MFE concepts. Tier 1 is focused on validating the physics central to a concept's basic viability. Energy confinement is a typical focus at this stage. Tier 2 is a transition to the development of self-consistent solutions to confinement, sustainment, stability, and boundary interface. There should be enough maturity at the end of Tier 2 to make informed assessments of fusion reactor potential, including capital cost. Tier 3 aims at the demonstration and validation of integrated solutions to confinement, sustainment, stability, boundary interface, and control. All stages, including Tier 1, must be supported with sufficient experimental, measurement, and modeling capabilities to provide clear answers to basic questions on scientific and technical feasibility. A set of rigorous metrics must be applied both to enter concept development and to evaluate progression to subsequent stages. Validated models that describe fundamental behavior and performance, including scaling parameters, must emerge not later than Tier 2. Successful progression through the staged approach implies a concept could be considered for a pilot plant demonstration that could be a major component of the national program or target or development by private industry.

4. **Leverage private industry and interagency investments (*e.g.* DOE ARPA-E) in alternative fusion approaches through collaborations in theory, modeling, measurement capabilities, and technology transfer**

   There are significant opportunities for beneficial public-private partnerships in alternative fusion approaches. Research being conducted on alternates in the private sector benefit the government-funded program by absorbing risk while continuing to develop confinement concepts with potentially reduced engineering costs but at a lower TRL. On the other hand, as private ventures tend to focus on the development of a single fusion concept on a focused track, they can stall on technical problems that might be easily solved at a more resource-rich national laboratory. For this reason, private ventures can greatly benefit from a relatively modest investment from the government. From the private industry perspective, these investments would ideally come in the form of focused, modular grants to address a specific problem, and there are existing models which could be expanded upon, like the Innovation Network for Fusion Energy (INFUSE). Programmatic recommendation FST-PR-C describes an overarching strategy for coordinated research encompassing all approaches to fusion (tokamaks, stellarators, alternates, IFE, MIF/MTF) supported by the federal government and the private sector.

# *Fusion Science and Technology Program Recommendations*

## FST Program Recommendation A: Establish a multi-institutional, multidisciplinary program to develop fusion pilot plant concepts

*Fundamental to this strategic plan is the recognition that advances in plasma understanding and technology enable consideration of new economically attractive fusion pilot plant (FPP) concepts. In order to improve the ability of the U.S. fusion program to dynamically respond to innovations, a coordinated program is needed which continuously evaluates emerging developments to identify new research priorities and opportunities based on their implications for FPP concepts. This multi-institutional, multidisciplinary program will develop potential FPP concepts using validated predictive modeling tools as well as commercially available engineering tools, coordinating with private and public stakeholders to identify FPP deliverables, and performing techno-economic analysis of these concepts in order to help inform research priorities.*

1. **Initiate and sustain a coordinated FPP conceptual studies program which brings together a broad cross-section of experts from across the domestic fusion program and private industry, in order to inform U.S. fusion research needs and priorities as advances in scientific understanding and technological innovation are achieved**

   This program would support developing a variety of conceptual designs for FPP options based upon best current scientific understanding and integrated modeling tools. These options should span a range of confinement configurations (FST-SO-D, FST-SO-E, FST-SO-H) and operating scenarios, and be updated following advancements in relevant modelling capabilities informed by experimental validation activities and technologies. Participation in the program should be broad and open to the entire U.S. fusion community, enabling diverse sets of institutions and researchers to pursue innovative approaches to developing economically attractive FPP concepts. It is explicitly recognized by the fusion community that these studies are intended to help clarify and inform research needs and priorities, and do not represent a commitment on the part of DOE to further develop or construct any particular FPP concept. Without implementation of this program, it is unlikely that advances in understanding and technology achieved by U.S. investments in fusion energy will be adopted and leveraged in a timely fashion, resulting in a slower and more expensive FPP development path.

2. **Integrate predictive plasma and material modeling capabilities ([FST-PR-D]) with the industry-standard engineering tools that will be required to carry out FPP costing studies and conceptual design, ensuring a direct pathway to incorporate innovations in science and technology enabled by this strategic plan**

   To enable the success of recommendation FST-PR-A.1, it is vital that the full range of the U.S. fusion community's world-leading predictive modeling capabilities (enabled by FST-PR-D) be fully utilized in developing FPP concepts. These plasma and material modeling capabilities must be connected to state-of-the-art commercial engineering tools in order to perform holistic assessments of FPP facility concepts. Considerations of safety and licensing requirements and needs identified in FST-SO-G should be integrated into these joint modeling capabilities. In addition to the development of these physics and engineering capabilities, improved costing algorithms and tools should be developed to support necessary techno-economical analysis of potential FPP designs. Without implementation of this recommendation, U.S. investments in science, technology, and advanced computing cannot be easily adopted or leveraged by FPP design concepts. These modeling and costing tools should be made available to private industry though partnership models outlined in FST-PR-C.

3. **Establish and engage the appropriate expertise needed to conduct techno-economic analysis of potential FPP concepts within the U.S. fusion program**

   Determining the economic attractiveness of FPP concepts requires growing expertise in techno-economic analysis within the U.S. fusion community. In particular, it is widely recognized that improved costing algorithms and tools are needed for accurate assessments of fusion facilities, including the FPP studies entailed within FST-PR-A.1 above. The experts trained and recruited through this effort would be responsible for leading the development of those tools. Without such an effort, the ability of the U.S. fusion community to confidently perform accurate cost-benefit analyses of FPP concept innovations will remain limited.

4. **Conduct FPP mission scoping, engaging with public and private stakeholders to identify an optimal set of deliverables for the fusion pilot plant, considering multiple confinement concepts and operating scenarios**

   These discussions should establish both relative priorities of FPP deliverables, such as those suggested in this plan here [FPP-Defintion], and quantitative values to be met. Potential deliverables would include targets for wattage of net electricity production, how long power generation should be sustained, and fluence levels achieved. Possible staged approaches to FPP construction and operation should also be evaluated, in addition to licensing and regulation considerations developed by [FST-SO-G]. These deliverables should be periodically revisited to ensure that the FPP concepts remain well-aligned with a dynamically evolving U.S. energy market. Not developing these deliverables in close collaboration with all relevant public and private stakeholders, or allowing them to remain static over long timescales, risks setting the U.S. fusion program on a path which develops FPP concepts that will not be economically attractive.

## FST Program Recommendation B: Participate fully in ITER to advance our capability to predict, control, and sustain a burning plasma and to obtain the critical science and technology input needed to design a fusion pilot plant.

*This recommendation echoes the National Academies recommendation that the U.S. should remain an ITER partner to gain experience with a burning plasma at a power plant relevant scale while benefiting from the shared cost through international partnership [Nat. Acad. 2018a Appendix H]. Access to burning plasma with a significant fraction of alpha particles, the dominant heating source in a burning plasma, is an essential step in completing the tokamak physics basis [FST-SO-D] and is required for validation of the theory, simulation, diagnostics, and tools needed to optimize and achieve successful, high performance operation of an FPP. ITER will be used to test methods for control of reactor plasma confinement and stability, plasma interactions with first wall materials, and fusion power output. Critical research during ITER pre fusion power operation (PFPO, scheduled 2028-2034) is needed to inform FPP design scoping [FST-PR-A]. ITER is an important step in understanding the optimal application scheme and advancing pellet injection technology for fueling and disruption mitigation [FST-SO-C.8], and participation in the testing of the shattered pellet mitigation system is a priority [FST-SO-D.4]. Operation of ITER is now on the horizon with First Plasma expected in 2025. As an ITER partner, the U.S. receives full benefit from ITER but provides only a fraction of the cost. The ITER project is on track with execution of project scope through First Plasma now at over 65 percent, and machine assembly officially scheduled to start in April 2020. We are now in a position to begin reaping the benefits of our investment in ITER. These recommendations are essential to ensure that U.S. credibility is maintained, ITER remains on schedule, and remaining risks are resolved to enable ITER to meet or exceed its performance goals.*

1. **Fulfill commitments to ITER construction and operation**

   The U.S. should fund agreed-upon in-kind and cash contributions needed to construct and operate ITER on a schedule compatible with the ITER facility plan. In the near term this includes fully funding ITER Sub-Project 1 (First Plasma) and Sub-Project 2 (Post-First Plasma). This is essential for maintaining U.S. credibility within the global fusion community and to ensure successful completion of this vital experiment.

2. **Develop a U.S. workforce program for domestic and onsite ITER participation**

In order to ensure access to burning plasma knowledge and critical research needed to inform the FPP, the U.S. urgently needs to establish a team that will enable full participation in ITER during systems commissioning, pre fusion power and fusion power operation phases. The U.S. should create a centralized, coherent framework to develop the longer-term workforce needed to effectively participate in ITER experimental planning and capitalize on the U.S. return on investment. A clear plan to transition the workforce from existing facilities to ITER and new facilities, such as NTUF, must be developed through FST-SO-D.2 and as existing facilities complete essential mission objectives. The U.S. should utilize available collaboration mechanisms and take advantage of opportunities to engage workers in existing ITER activities. Immediately, the U.S. can establish a flexible, rapid response approval mechanism allowing U.S. institutions to take advantage of the ITER Project Associates (https://www.iter.org/jobs/IPA) scheme. DOE should clarify U.S. participation in the IPA process and implement a mechanism to ensure broad involvement by universities, industry, and national laboratories in ITER contracts and operation. Enabling collaborations between universities and ITER could also provide long-term and stable funding paths to grow new faculty lines in fusion and plasma science [CC-WF.B.7].

3. **Utilize existing facilities and funding models to support ITER readiness through start of Pre-Fusion Power Operations**

The ITER Research Plan (IRP) identifies a number of priority research questions which require best possible answers prior to embarking on the PFPO campaign, anticipated within the next decade. Contributions to needed R&D from existing world facilities are coordinated through the International Tokamak Physics Activity (ITPA). The U.S. should remain engaged in the ITPA organization and increase participation as needed in order to optimize contributions from existing U.S. facilities toward ITER. Extra priority should be given to ITPA activities relevant for an FPP. In the near term, DIII-D should be used to help resolve outstanding ITER scenario issues identified in the IRP. For R&D tasks in which a concerted effort is required, the U.S. should use open, competitive Funding Opportunity Announcements to establish new multi-institution research teams.

4. **Leverage ITER involvement to inform decisions for an FPP**

Preparation for ITER operation and eventual lessons learned from participation in operation are expected to resolve questions for an FPP. The U.S. needs to establish and execute a process to obtain and understand the technology development for the wide range of engineering systems on ITER, generally produced by other Domestic Agencies, as guaranteed by the ITER agreement. Many of these systems are already completed or near completion, so gaining this technology knowledge can begin immediately. Immediate engagement is needed for long lead-time FPP elements. The R&D results emerging from U.S. participation in ITER will include:

A. *Tritium handling, safety, and licensing components*

While relevant information from the ITER Test Blanket Module program may be limited for FPP design [FST-SO-C.6], the U.S. can gain from ITER R&D activities for tritium fueling, exhaust, processing, and inventory [FST-SO-C], providing opportunity to translate to an early technical basis for design, manufacturing, commissioning, and start-up operation of large-scale FPP tritium handling. The U.S. is expected to provide the Tokamak Exhaust Processing System for ITER Sub-Project 2. The FPP high-temperature nuclear structural design (HTNSD) criteria will be developed through extension of the ITER Structural Design Criteria [FST-SO-B.4]. ITER participation would advance U.S. modeling capabilities to determine licensing and balance-of-plant requirements in order to assess overall FPP performance, safety margins, and costs [FST-PR-D.3, FST-PR-A.2]. Participation enables expanded verification, validation, and uncertainty quantification activities for a full range of models and workflows to study parameters and operating regimes not accessible on U.S. domestic facilities [FST-PR.D.5]. Lessons learned from ITER will help develop a licensing framework [FST-SO-G.1] and strategies for remote handling and maintenance [FST-SO-G.4].

B. *Science informing tokamak pilot plant scenarios*

The proposed NTUF device [FST-SO-D.2] is ultimately required for demonstrating the feasibility of the U.S. path toward an economical FPP, as the high absolute plasma pressure core and high power density boundary is in a different regime than ITER. However, U.S. research during ITER PFPO is needed to provide key data on other critical gaps to help inform FPP design scoping. This includes the performance of disruption avoidance and mitigation technology supplied by the U.S. [FST-SO-D.4]; ELM suppression from 3D fields; scaling of confinement, pedestal, and scrape-off-layer physics; and control strategies for high performance burning plasmas. ITER PFPO model validation will be used to develop power exhaust scenarios [FST-SO-D.3], energetic particle physics for predictive integrated modeling [FST-PR-D], and reactor-relevant heating and current drive [FST-SO-F.6]. Experience from ITER can be leveraged to develop needed RF generators and power supplies [FST-SO-F.5].

C. *Knowledge of diagnostics in a fusion nuclear environment for control purposes*

The U.S. is expected to deliver seven instrumentation systems for Sub-Projects 1 & 2. Not all of the strategies developed for ITER diagnostics may be applicable to an FPP, but engaging directly with relevant ITER work begins to solve a number of challenges with diagnostic and instrumentation survivability [FST-PR-E.2], design, and commissioning challenges [FST-PR-E.3, CC-MD].

## FST Program Recommendation C: Deploy various models of public-private partnerships to develop technology at a lower cost and move towards fusion commercialization

*Partnering with private industry to advance fusion development is a necessary step on the path to commercialization of fusion energy within the U.S. Building from examples in commercial space, fission, and other industries, public-private partnerships will provide private capital to augment public research through cost-shares, facilities, and capabilities that may not otherwise be developed within the public program, and transfer technology to be used in the private sector. Including private industry with the public effort helps to provide urgency and alignment with future markets, introduces an opportunity to take bigger risks, and drives commercial solutions.*

1. **Develop, utilize, and expand programs for public/private partnerships that leverage lessons learned from DOE and other areas of the federal government**

   The U.S. should make the Innovation Network for Fusion Energy (INFUSE) a permanent FES applied research program to support industry's investment in fusion energy. INFUSE should be expanded and adapted based on input from the FY19 pilot program and mechanisms should be developed for direct University participation. FES should also coordinate with ARPA-E to support necessary follow-on FES research from ARPA-E funded activities and assess the potential for alternate concepts to form an approach to a fusion pilot plant. The U.S. should exploit the ability of government-funded science to be completed in partnership with privately funded fusion companies. Using peer-reviewed proposals, FES should develop best practices for collaboration of public researchers on private facilities to further the public scientific research needs (*e.g.* using the FRC at TAE as a Discovery Plasma Science and SPARC as a burning plasma collaborative facility). FES should explore the use of cost-share programs, as was done for the NASA COTS and DOE SMR programs, focusing on reactor technologies and prototype demonstrations prioritized within the Strategic Objectives. Finally, FES and industry should look back at lessons learned from early (70's and 80's) participation of private industry in fusion as well as PPPs in the early development of fission, both domestically and abroad.

2. **Dialogue should be fostered between government researchers and experts in the private sector (*e.g.* venture capital, finance, private industry, *etc.*). The private sector should be recognized as a key stakeholder when developing goals for FES programs.**

   Part of the motivation of the U.S. program to push towards a low capital cost fusion pilot plant is the recognition that this would be more attractive to industry for the eventual commercialization of fusion power plants. The dialogue between government researchers and experts in the private sector must continue throughout the development of the FPP. This continued interaction with all stakeholders will be key for advancing fusion power, especially as the fusion program embraces the energy mission and develops designs and cost estimates for a fusion pilot plant (FST-PR-A).

## FST Program Recommendation D: Develop and utilize a hierarchy of validated models for predictive integrated modeling, by continuing the partnership between FES and ASCR, expanding capacity computing infrastructure, and utilizing advances in computational architecture and capability.

*Advancing multi-scale, multi-physics theory and modeling capabilities in a diversity of topical areas is necessary to predict the complex interactions between numerous plasma, material, and engineering processes that will occur within a fusion pilot plant (FPP). These advances are required to extrapolate with confidence from present experiments to the fundamentally new physics regimes typical of any FPP. These fundamental advancements also form the basis for the models that must be used to develop FPP designs in FST-PR-A. There is a hierarchy of approaches required, spanning from high fidelity simulations to faster reduced complexity models. This hierarchy of models must be integrated to represent physics processes of the entire plasma, describing dynamics from the plasma core to the first wall. This "whole plasma model" capability must then be extended, to whole device modeling which integrates plasma dynamics with the divertor, wall and blanket behaviors, and then finally to whole facility modeling which also integrates balance of plant, costing, and licensing concerns to the site boundary, to support FPP design and costing activities in FST-PR-A. Carrying out the necessary range of modeling and simulation work needed to realize fusion energy in a timely and cost-effective manner will require continued close partnership between FES and ASCR to fully leverage current and future computing resources, including U.S. investments in exascale computing. This work is vital to accelerating the pace of fusion energy development, by both helping to focus research priorities and providing a means of risk management when extrapolating from current-day experiments to FPP conditions.*

1. **Support advancing fundamental scientific understanding of fusion-relevant plasma and material physics through theoretical and computational exploration to enable innovation and new conceptual solutions**

   Support for fundamental theoretical investigations of fusion plasma and material physics questions is central to this strategic plan. More concretely, these studies provide the foundation for building the scientific understanding needed to improve our predictive modeling capabilities. For instance, there are a variety of research gaps where our fundamental physics understanding is limited, such as how edge localized modes are suppressed by resonant magnetic perturbations, what controls the L-H power threshold, what limits the maximum achievable pressure in stellarators, and how plasma facing component and structural material properties are degraded by plasma and neutron irradiation. Developing an improved fundamental understanding of these questions is vital for building predictive physics-based models that can be extrapolated to future regimes with confidence. Without this foundational work, the risks entailed in extrapolating current results to future facilities are greatly magnified, and our ability to innovate efficiently is greatly reduced. One example of how fundamental theoretical research can directly enable transformational innovations and approaches to fusion energy can be seen in the optimization of stellarator transport and confinement detailed in FST-SO-E.

2. **Develop hierarchies of validated predictive models suitable for timely design and optimization of future facilities**

   FES should support increased collaboration between theory, computation, and experiment through means such as SciDAC projects. The goal of these collaborations should be to build, verify, refine, and validate predictive models of all fusion-relevant plasma and material processes, at a variety of fidelities and computational costs. This activity is vital, for example, to deliver validated energetic particle transport models for predictive, integrated modeling, which are needed to understand and quantify the consequence of fast-particle interactions with instabilities for current drive, thermal profiles, wall heating, *etc*. As identified in the 2015 Integrated Simulation workshop [Integrated Simulations 2015], these model hierarchies should include both high-fidelity and a range of reduced models, with reduced model predictions verified against analytic theory and high-fidelity simulation results. Innovations in data analysis and machine learning should be used to complement traditional model development approaches. Without building and validating these hierarchies, our ability to accurately design and optimize FPP concepts in a timely and resource-efficient manner will be limited, our confidence in extrapolating beyond current facilities and plasma regimes will not be warranted, and the risks in the development of FPP concepts will be magnified.

3. **Develop physically rigorous and computationally robust model integration methods to enable predictive whole-facility modeling and optimization of FPP concepts**

   Developing these integration methods will require participation from experts in analytic theory (to ensure the approaches taken are physically rigorous) and computational scientists (to ensure the algorithms used are computationally robust). This effort should prioritize improving the ability of the FPP conceptual study program (FST-PR-A) to make timely decisions which incorporate advances in understanding. Beyond integrating U.S. leadership capabilities in physics and engineering modeling together, ITER participation and international collaboration should be used to advance U.S. modeling capabilities for topics such as balance-of-plant and licensing requirements. Implementation of this recommendation is critical to building accurate modeling tools capable of describing the complex interplays and trade-offs of various physics phenomena and engineering choices inherent to any FPP concept. Without such a capability, our ability to correctly assess overall FPP performance, safety margins, and costs will be strongly limited.

4. **Invest in computational infrastructure and software engineering needs to enable optimal utilization of current and future high performance computing platforms**

   Continued investments in the development of computing frameworks will provide a variety of beneficial returns to the community such as improved data management and analysis, easier interfaces to experimental data, code couplings and benchmarkings with experiments, code couplings, *etc*. Consistent with this recommendation as well as the general cross-cutting recommendations to improve code reproducibility and accessibility, compatibility with common data standards and formats such as IMAS to support code benchmarking and expanded accessibility should be encouraged. Support is needed for dedicated software engineers, who will be tasked with ensuring the continued functionality of "legacy" production codes on new architectures, and building standardized software libraries that minimize the overhead of new code development. Finally, investments in capacity as well as capability computing resources are needed to continue advancing our modeling capabilities and sustain U.S. leadership. Without these investments, U.S. leadership in predictive modeling of fusion plasmas will be at risk, along with our ability to efficiently utilize the significant DOE investments in advanced computing platforms.

5. **Support expanded verification, validation, and uncertainty quantification activities for full range of needed models and workflows.**

   Verification and validation with rigorous uncertainty quantification remains an essential requirement for confidence in predictive models, particularly as they will likely play crucial roles in the licensing process. Future validation efforts should focus on regimes that are relevant to burning plasma conditions. These efforts should also take full advantage of ITER participation as well as collaborations with international and private facilities to study parameters and operating regimes not accessible on U.S. domestic facilities. Validation platforms can and should include "test-stand" level facilities as well larger confinement devices. Increased emphasis should also be placed on the development of synthetic and virtual diagnostics in collaboration with the work outlined in FST-PR-E, particularly those needed to support the operation of nuclear facilities such as the FPP. Without these activities, confidence in our ability to accurately design FPP concepts and extrapolate to FPP regimes will be low, and risks in the operation and licensing of an FPP will be magnified.

## FST Program Recommendation E: Establish a program for developing diagnostics, measurement, and control techniques that can be used in a reactor environment.

*Advances in diagnostics, instrumentation, data handling and interpretation are broadly needed for plasma science, fusion materials research, and power plant operation. Development and implementation of plasma science diagnostics are needed to provide sufficient data for model validation and the prediction of burning plasma behavior. Diagnostic advancements for fusion materials studies are needed to improve the understanding of the interaction of the fusion environment with materials at a fundamental level. Power plant operation and control will rely on new diagnostics and instrumentation to monitor proper operation of the facility, including in-vessel and ex-vessel systems and sensors for data processing and automated real-time decision making. These sets of plasma diagnostics and engineering instrumentation may evolve over the operational stages of an FPP. For example, some systems may be needed in the early phase to confirm adequate operation during commissioning of the plant. Others will need the robustness to survive in the nuclear environment during extended operation. The intent of this Programmatic Recommendation is to accelerate progress and increase readiness of diagnostic systems for a fusion reactor.*

This set of recommendations emphasizes diagnostic needs specific to magnetic fusion and technology programs. Broader aspects of plasma diagnostic development needed to advance plasma science, to develop the tokamak physics basis, to close gaps, and to validate models are addressed in further detail in the Cross-cutting section of this report.

1. **Develop critical *in situ* and combined effect diagnostics for fusion materials research and plasma science needed to validate models, which includes new capabilities on existing confinement devices as well as on smaller "lab-scale" experiments**

   This objective focuses on materials and scrape-off layer plasma characterization techniques needed to advance the fundamental science underpinning the fusion technology systems highlighted in this report. As a starting point, the program should target efforts to improve the scientific productivity of existing materials and surface analysis tools for fusion-specific problems. Here, a relatively modest investment of effort to improve data analysis algorithms or detector efficiency could help realize significant advancements in measurement spatial resolution, sensitivity, and throughput. Critical measurement R&D needs for specific fusion technology areas must also be considered. For example, within the blanket and tritium fuel cycle topic, this includes fundamental research on liquid metals, including development of techniques to characterize flows. In addition, new methods of characterizing corrosion would enable materials compatibility testing to be achieved on much shorter timescales than is currently possible. From the perspective of plasma-material interactions, there are large uncertainties associated with fundamental quantities such as tritium trapping energies, recombination rates, and transport parameters. Much of the challenge arises due to difficulties with characterization due to limitations of existing techniques. This has been a significant obstacle to modelling how surfaces evolve in a plasma environment, and completely new measurement approaches are likely needed to precisely quantify these parameters. To supplement smaller-scale test stands, we recommend the incorporation of improved materials analysis capabilities into existing major fusion facilities (both linear plasma and toroidal confinement devices) to track surface composition and structure evolution (*i.e.* using *in vacuo* sample transfer). It is also imperative to establish new diagnostics to complement larger-scale facilities that have been proposed to study "combined effects", such as concurrent ion beam damage and plasma exposure, or concurrent high heat flux and plasma exposure.

Existing tools for *post mortem* characterization provide exquisite detail into material structure and composition (with resolution down to atomic scale) and should continue to be used and refined for fusion-specific applications. Available techniques capable of time-resolved measurements, tracking surface evolution and the dynamic response of materials to plasmas are much more limited. A robust program to develop novel *in situ* diagnostics for probing dynamic effects of plasmas on materials (surface composition/structure) is also suggested. Finally, it is essential to upgrade scrape-off-layer diagnostics to improve measurements of incident particle fluxes and heat loading.

2. **Initiate the R&D needed to solve diagnostic survivability challenges (materials & electronics) imposed by the nuclear conditions expected throughout a fusion pilot plant facility**

Existing plasma diagnostics and engineering instrumentation will be severely challenged by the harsh environment of burning plasmas. Because the lead-time associated with completing the required materials and component testing is on the order of 10 years or more, it is recommended that a R&D program be started immediately to develop an understanding of the associated degradation mechanisms so that suitable countermeasures can be developed. This includes characterizing the real-time and long-term impacts of neutron-, ion-, and gamma-induced damage to insulating materials, and encompasses radiation-enhanced conductivity, radiation-induced electrical degradation, thermal conductivity degradation, and mechanical strength degradation. The effects of radiation on refractive optics, optical fibers, viewports, feedthroughs, supporting structure, and diagnostic mirrors must also be considered and suitable alternatives/countermeasures must be devised. It is recommended that testing of relevant diagnostic materials be initiated as a near-term priority using existing fission neutron sources. As available test stands come online, materials testing is recommended using a fusion prototypic neutron source (FST-O-B), within larger scale component testing to be performed in a volumetric neutron source (FST-O-C).

R&D on these topics has been underway as part of the ITER diagnostics program. While not all of the strategies developed for ITER diagnostics may be applicable to an FPP, much of the fundamental work is still highly relevant. To take full advantage of the progress that has been made by the international community, it is recommended that the U.S. should engage ITER on these topics directly (FST-PR-B).

3. **Develop nuclear environment compatible plasma diagnostics and engineering instrumentation needed for control and safe operation of an FPP and benchmark these new instruments on available facilities**

Keeping in mind the challenges associated with a burning plasma environment, this recommendation focuses on developing a suite of new techniques that can survive in a nuclear environment. The topic of radiation-hardened engineering instrumentation and plasma diagnostics is a required area for innovation. However, to ensure that these new techniques are available in time to be implemented on an FPP, these activities must be approached with urgency, given the typical lead-time of ~10 years required for diagnostic development. As a starting point, it is necessary to establish the diagnostics needed for anticipated operating modes (*e.g.* using the ITER classification: machine protection and basic control, measurements for advanced control, and performance evaluation physics), leveraging from lessons learned in ITER diagnostic design and commissioning (FST-PR-B). Potential gaps where required time, spatial resolution, or durability may not be adequate must then be identified. Following completion of these steps, the program should develop new techniques to replace plasma diagnostics and engineering instrumentation which are otherwise incompatible with the fusion environment. This may require development of new radiation-resistant materials and rad-hardened instrumentation. Furthermore, manufacturing and design approaches to improve diagnostic integration, *in situ* calibration, or replacement will need to be established in concert with the design of remote maintenance systems (FST-O-G). The use of synthetic diagnostics may be helpful in performing quantitative assessments of existing plasma diagnostics as well as in the design of new techniques. Where possible, these systems should be validated on existing confinement devices or subsequent intermediate-scale facilities.

The needs of plasma diagnostics and engineering instrumentation must be considered at the outset of the FPP design activity. The FPP may start with extensive diagnostics as the machine is initially being tested. It is important to keep in mind that the diagnostics will occupy space needed for tritium breeding blankets. To accomodate adequate tritium breeding, diagnostics may need to be reduced to a minimum set needed for control and machine protection. It will be necessary to have diagnostic port design and layouts that account for the impacts on blankets and other components surrounding the plasma, as well as regulatory requirements. Strategies for remote calibration, alignment, maintenance, and replacement will need to be developed to minimize personnel exposure. In addition to the plasma diagnostics, the scope must be established for the required engineering instrumentation and tooling required to assess damage to activated components and analyze components removed from the nuclear facility. Finally, algorithms and control systems will be needed to manage and analyze the large amount of data produced by the various sensors. This includes automated techniques for real-time control and safe operation of the facility.

4. **Develop advanced control techniques to maintain high-performance burning plasmas without disruptions or other major excursions**

   The plasma diagnostic and control techniques for any future pilot plant will have to be advanced well beyond what is in use on existing devices and what is planned for ITER. The control systems must be made as robust as possible, since any deterioration of plasma control could lead to loss of confinement (potentially leading to disruptions), posing a significant risk of damage to the internal components and surfaces of the machine itself. Improvements are needed to ensure that the reactions of the control systems in response to transient events are as fast as possible. The control system must also enable operation of the device near its limits, where high-performance plasmas are needed to maximize power output. The functions of the control system will likely need to be accomplished while relying on a limited set of plasma diagnostics, given the challenges associated with their survivability in a nuclear environment described above. It will be essential to develop methods to quickly extract information on the state of the plasma from this limited set of plasma diagnostics. Control of the plasma will also rely heavily on robust models of how the plasma responds to available actuators.

# Cross-Cutting Opportunities and Recommendations

Fusion and plasma science spans a diverse range of topics and goals, from understanding plasma dynamics in the universe to providing technological advances that will improve human health and well-being, including limitless fusion energy to power our future civilization. The two large areas of FST and DPS represent the DPP-CPP organizational areas of General Plasma Science (GPS), High Energy Density Plasmas (HEDP), Magnetic Fusion Energy (MFE) and Fusion Materials and Technology (FM&T). Several areas of research having common importance to this wide breadth of fusion and plasma science were chosen for strategic focus on cross-cutting opportunities: Theory and Computation (TC), Measurements and Diagnostics (MD), Enabling Technology (ET) and Workforce Development (WF), including Diversity, Equity and Inclusion. The DPP-CPP discussions on these cross-cut areas aimed to identify scientific and technological opportunities that are largely held in common across all topical areas, and to identify organizational or strategic frameworks that advance or leverage common areas of interest or need among the topical areas for the purpose of advancing fusion and plasma science broadly. Attention was also given to identifying research methods and tools in neighboring disciplines outside of fusion and plasma science that would advance science and technology broadly through coordinated research activities. As such, the recommendations given below represent opportunities to advance fusion and plasma science broadly if carried out in coordination within FES and in coordination with adjacent fields of science and technology that have similar needs and opportunities. In these cross-cut areas, FES should encourage coordination within the fusion and plasma science realm and in cooperation with other government offices, agencies and the private sector to maximize these cross-cutting opportunities.

## *Cross-cut TC: Theory and Computation*

*Theory and computation provides the basis for interpreting experimental observations and transforming those observations into physical understanding. This understanding is manifested through the development of a hierarchy of models, ranging from the formulation of governing equations and their solutions via analytic ("pencil and paper") theory, to direct numerical solution of those governing equations via computer, to semi-empirical scaling relations that combine insights from theory with calibration parameters drawn from experimental measurements. These approaches are complemented by purely empirical models of physical phenomena, such as widely-used confinement scaling laws in fusion energy derived from regression analysis of experimental observations or neural networks and other machine-learning–based approaches. The process of testing a model's accuracy and fidelity is often referred to as verification and validation (V&V), and this process is central to any robust research program. While analytic theory and small-scale computational analysis form the foundation of the plasma and fusion communities' modeling capabilities, large-scale simulations using advanced high-performance computing platforms play an ever-increasing role in interpreting and predicting plasma and fusion system behavior. Plasma theory and modeling have historically been an area of U.S.*

*leadership, and maintaining this strength will require continued investment in theory and modeling research, computational infrastructure, and workforce development. Three recommendations are identified below to enable the U.S. to maintain leadership in this vital component of plasma physics and fusion energy research. Each recommendation is viewed as a vital and equally important component of the strategic plan's approach to ensuring a robust theory and modeling program that underpins the full range of research and development in both the FST and DPS areas.*

## Recommendations

1. **Support a broad spectrum of verification and validation activities throughout the entire FES program, to enable the development of physical understanding and foster innovation**

   Verification and validation (V&V) are the means by which we test the fidelity and accuracy of the models that embody our understanding of plasma and material physics. Carrying out this V&V research requires dedicated support for a broad spectrum of activities, from analytic theory to large-scale simulations on exascale computing platforms. Beyond analytic theory's central role in developing the new models and conceptual frameworks which enable us to improve our understanding of physical phenomena, analytic theory is vital for defining the test and use cases around which many V&V studies are built. Therefore continued robust support for fundamental theoretical research is an essential foundational component of a healthy research program and portfolio. Complementary to this sustained support for analytic theory is the need to continue investments in computational infrastructure and accessibility (described in greater detail below), to enable the robust V&V of the computational models that increasingly embody our understanding of physical phenomena. Beyond the specific recommendations below, it is of paramount importance to ensure that FES-supported researchers continue to have access to computational resources and tools appropriate for carrying out V&V studies. In many cases, such as uncertainty quantification studies, a higher premium may be placed upon ensembles of simulations at moderate resolution, rather than a single simulation of the highest possible resolution. These approaches are sometimes referred to as capacity vs. capability computing workflows. Ensuring support for capacity computing workflows on future platforms in addition to the traditionally emphasised capability approach is therefore essential to continued time- and resource-efficient V&V research within the FES portfolio.

2. **Harness growth in advanced scientific computing tools to improve fundamental understanding and predictive modeling capabilities**

   The efficient use of advanced and high-performance computation has long been a hallmark of the plasma and fusion research activities supported by FES. To sustain U.S.

leadership in this area and enable continued productive computational modeling, it is important for FES to take a broad and balanced approach to future investments in computational infrastructure. Continuation of the close partnership with the Advanced Scientific Computing Research (ASCR) program of DOE is recommended, to enable the plasma community to take full advantage of current and future computing platforms, including coming exascale machines. Towards this end, it is also recommended that FES increase support for software engineers to partner with FES-funded code development efforts. These engineers are needed to keep key production codes running efficiently on new computing platforms, and to provide tools such as standardized software libraries that can lower the time and financial costs of both developing new codes and transitioning existing codes to new architectures. This recommendation reflects the reality that the complexity of software development for these new platforms (*e.g.* those making heavy use of GPUs) has increased greatly, and building efficient, scalable codes for them requires specialized expertise in computer science as well as plasma and material physics. Growing this expertise within the FES portfolio represents a specific workforce development need and opportunity, particularly since it provides an opening to connect with disciplines and researchers outside of traditional plasma physics and fusion energy programs. As such, it is an important component of the workforce development activities detailed in CC-WF.

It is also recommended that FES maintain support for a broad spectrum of computing needs and workflows, recognizing in particular the vital roles both capacity and capability computing play in advancing the frontiers of our understanding. In particular, both FES management and the community as a whole should be proactive in considering and communicating the full breadth of actual computing needs when specifications and requirements for new high-performance computing platforms are being identified. It is likely that many other research communities have similarly broad computing needs, and opportunities for greater interaction with other programs (both within and outside of DOE) on this topic should be pursued. For instance, opportunities to learn from other communities such as high-energy physics on how cloud computing resources can be efficiently utilized as complements to traditional DOE-supported computing platforms. Finally, it is recommended that FES take a measured approach towards balancing investments in new areas such as machine learning and quantum computing with ongoing research programs using established computing platforms and algorithms, which in many cases have been optimized for production use through significant previous DOE investments.

3. **Support improvements to the accessibility, interoperability, reproducibility, and user-friendliness of FES-fundeded codes and their outputs**

Given the critical role computational modeling now plays within the FES research portfolio, and the opportunities for innovation enabled by new modeling capabilities, it is important to ensure that FES-funded computational tools are reliable and broadly usable by the plasma and fusion communities. It is therefore recommended that FES encourage open source development of current and future codes as feasible (recognizing limits due to *e.g.* export control restrictions), as well as implementation of best practice methodologies such as the use of version control software. Continued support for the development and use of common data standards and computing frameworks is also recommended. In order to increase the accessibility of FES-funded codes and tools, it is recommended that FES increase support for hands-on training and interactions between code users and developers, through means such as visiting scholar programs and increased travel support. Complementing these activities should be increased support for, and requirements on, accurate and timely code documentation.

Finally, it is also recognized that the growth in advanced computing, and its role within the FES program, motivates new approaches to data management and analysis. For example, the largest and highest-resolution simulations are of necessity limited in number, and the computing resources needed to perform them obtained through highly selective (and time-consuming) proposal processes, such as the INCITE and ALCC programs. It is therefore recommended that FES begin identifying ways to make the datasets generated by these simulations more widely accessible to the community, to enable a broader cross-section of researchers to use the data in new and innovative ways. For instance, these data sets could be used to help guide development of new theoretical approaches, verify and validate reduced models, or form parts of the data sets needed for building neural nets. This recommendation draws from the observation that, much like the data obtained from large experimental facilities, these simulations, and the computers used to perform them, represent significant financial investments on the part of DOE, and therefore the simulation data should be made similarly available to the broader community. In implementing this recommendation, FES is encouraged to examine how other fields such as climate modeling or astrophysics and cosmology have made large simulation datasets publicly accessible.

## Cross-cut MD: Advance the development of Measurement and Diagnostics techniques for plasma science and fusion energy

*New developments in diagnostics, instrumentation, data handling and interpretation are broadly needed to advance plasma science, fusion materials research, and fusion energy studies. The diagnostics should push beyond the boundaries of what is possible with existing techniques in terms of spatial, spectral, and temporal resolution. These developments are intended to be strongly coupled with modelling efforts (CC-TC) for which validation data is crucial for an accurate prediction of plasma behavior. For example, diagnostic advancements for fusion materials studies are needed to improve the understanding of the interaction of the fusion environment with materials at a fundamental level. Advanced diagnostics are also required to diagnose a plasma and to characterize particle and photon beams generated in ultra-intense*

*laser-matter interactions. Many of these diagnostics must be robust to survive high-radiation environments during extended operation. The intent of this Measurements and Diagnostics (MD) cross-cut is to accelerate progress and increase readiness of diagnostics for a fusion reactor and for plasma-science applications. The high-level objectives for the MD cross-cut process are to: 1) Identify cross-cutting diagnostics critical to the scientific discovery of plasmas relevant to the different topical areas; 2) Encourage and shape initiatives that address cross-cutting opportunities among the topical areas; 3.) Identify diagnostics in disciplines outside FES that could advance the research activities of plasma science and fusion energy; 4) Identify organizational and strategic frameworks that advance common areas of interest.*

1. **Pursue advances in diagnostics development, including innovations that provide access to a wider range of plasma parameter space and enable survivability in extreme environments**

   Many existing diagnostics need to be advanced in terms of their spatial, spectral, and temporal resolution in order to allow new discoveries in plasma and fusion science. Additionally, new developments of advanced diagnostics are needed to enhance our fundamental understanding of plasmas in general and take us to the next level in our efforts of developing a fusion energy source. Innovation is needed to measure plasma, material, and component parameters in extreme environments, such as in an FPP. These diagnostics can be categorized as those that rely on particles and photons emitted from the plasma itself, those that rely on external probes, or a combination of both. Some existing diagnostics that need advancement include, but are not limited to:

   ● <u>Neutron spectrometry (*e.g.* for diagnosing alpha transport and heating).</u> Diagnosing alpha transport and heating of fuel ions is essential for understanding the science of ignition and burning plasmas at the NIF, SPARC and ITER. Discovery Plasma Science may also benefit from this. This can be done by measuring the alpha knock-on high-energy tail in the neutron spectrum. The challenge of implementing this technique is to define a system that can measure this high-energy tail that is $10^{-3}$ or smaller, depending on the plasma conditions, than the primary-neutron component. The diagnostic must have sufficient energy resolution and sensitivity to measure a very small component in the neutron spectrum, while at the same time be relatively insensitive to background.

   ● <u>High-resolution x-ray spectroscopy of high-Z elements.</u> High-resolution x-ray imaging spectroscopy of impurities is critical for providing important information of ion temperature, electron temperature and toroidal/poloidal rotation velocity in MFE plasmas. For DPS and HEDP, high-resolution x-ray spectroscopy will provide important information on line/edge shapes and shifts for validation of atomic-physics models, and collisional-radiative/radiation-transport theories and codes. The needed instrumentation advancements include improved spectral resolution, broader spectral coverage, and higher time resolution, ideally with small port access requirements and small equipment footprint.

- Optical Thomson Scattering (OTS). OTS will play a critical role in studying collisionless shocks, particle acceleration, instabilities and magnetic reconnection in laboratory astrophysical plasmas, and for studies of the dynamics of under-dense hohlraum plasmas in ignition experiments at the NIF. In the context of MFE plasmas, this technique will be essential for studies of the spatial distribution of electron temperature and density, as well as the impact of non-thermal heating on the electron velocity distribution. It might also be an important method for studies of confined alphas. The overarching challenge of OTS is the need to measure a tiny signal in the presence of a large background; each new implementation of OTS requires advances in laser technology, detector capability, and/or optical design. For MFE applications, increases in OTS measurement rep rate are needed to study dynamic processes such as disruptions and instabilities.

- Plasma-material interface diagnostics. How material composition and structure evolve during high-flux plasma exposure is a critical problem throughout fusion and plasma science. A major obstacle to understanding these effects has been the limitations of existing characterization techniques, given the high magnetic and electric fields present in a plasma discharge. Improvements to existing forms of surface analysis (improving spatial resolution and chemical sensitivity), as well as the development of new in-situ diagnostics will greatly enhance our understanding of the fundamental mechanisms governing plasma-induced surface evolution.

- High-resolution gamma-ray spectroscopy for studies of fast particles. This technique has been identified as an important technique for MFE plasmas as it will provide essential information about the fast ion behavior and relative density profiles of fuel and impurity ions. For HEDP, the technique has been discussed for probing alpha transport, mix and compression. The major challenge of implementing this technique is to find a design that is sensitive enough to measure very small signals in the presence of large backgrounds, mainly hard x-rays.

Additional techniques identified as needing advancement are optical and x-ray laser beam probes, ion beam diagnostics (*e.g.* Thomson parabolas), ultra-fast diagnostics such as x-ray streak cameras, and micro/nano probes to better capture kinetic-scale physics. An issue facing many diagnostic implementations is the need for synthetic diagnostic data that mimics the characteristics of real data. This synthetic data, which could be generated by large-scale simulations, provides scenarios for diagnostic development and testing. It will also be required to train machine-learning analysis processes.

The detection materials and electronics in many these diagnostics must be made resistant to damage and malfunction caused by high levels (both in terms of fluence and flux) of radiation. This can be achieved by shielding, improved detector materials, and/or advanced analysis techniques. In the context of FM&T and MFE, this applies to a large range of diagnostics that must be developed to monitor: the vacuum vessel, thermal shield, magnetic field coils, in-vessel control coils, in-vessel structures, first wall and divertor, tritium blanket, auxiliary heating, gas feed, and in-vessel support and remote control. Having these systems will be critical for measuring and controlling the MFE machine performance for the entire lifetime of the device, especially as the allowable port space for diagnosing fusion reactors is potentially reduced. The U.S. should leverage the knowledge gained from the diagnostic systems which are being developed and deployed for ITER nuclear operation, especially through renewed funding support for U.S. ITER diagnostics. For DPS and HEDP, this recommendation applies to diagnostics used to primarily diagnose plasmas. Detectors that are immune to electromagnetic pulse interference and energetic particle bombardment are needed to diagnose plasmas generated by intense lasers, as well as plasmas being investigated in the radiation environment of space.

2. **Support the generation of atomic, molecular, nuclear, and spectroscopic data that meets needs in multiple plasma disciplines, support the generation of analysis tools, and establish central databases and best practices for data management**

There are profound advantages to generating critical data and establishing central data repositories. It meets needs that are shared across plasma science areas; centralizes the storage and maintenance of data; increases data quality and accuracy; reduces the number of updates and redundancies; and generates a higher return on investment. These generation efforts and databases should include:

- High-quality atomic, molecular, and spectroscopic data
- Nuclear data
- Open-source software ecosystem for plasma research
- List of calibration facilities and capabilities
- Best practices for storing data

Atomic, molecular, spectroscopic, and nuclear data enables many useful diagnostics and is a fundamental part of plasma modeling. While much progress has been made, a structured collaborative effort between the plasma community and atomic, molecular, and nuclear physics community will facilitate tremendous progress in all areas of plasma science. Feedback from the plasma modeling and diagnostic communities is critical to this effort, and will lead to identification of gaps and refinement of atomic and nuclear data, and benchmarking of the derived coefficients. These databases are needed in the U.S. and internationally; supporting maintenance of these databases in the U.S. would encourage U.S. leadership in this area.

3. **Establish a forum to better guide diagnostic work across the topical areas**

   There is often a disconnect between the diagnostic communities across the different topical areas. The seeming chasm in parameter space ($10^{12}$ in time, $10^{11}$ in density) often prevents community members from seeking insight or ideas from the other topical areas, but in fact there are many commonalities. There are several existing, partially-overlapping, domestic and international diagnostics forums, *e.g.* the High Temperature Plasma Diagnostics topical meeting, the International Conference on Plasma Science, the Laser-Aided Plasma Diagnostics workshop, and the APS-DPP conferences. It is proposed that a standing task force of expert diagnosticians (~10 people) from the different topical areas be formed to leverage and coordinate the expertise across the topical areas. This group will discuss and identify diagnostic techniques that are ripe for transfer between topical areas. Establishing this forum in the U.S., supported with funding by the U.S. DOE, will position U.S. scientists to be leaders in emerging diagnostic technologies for the international fusion energy and plasma science efforts.

## Cross-cut ET: Enabling Technology

*The realization of fusion energy and the advancement of plasma science will require advances in adjacent technology areas such as power delivery, advanced materials, advanced manufacturing techniques, and algorithms for modelling, control, and optimization. Although more broadly stated here to capture the cross cutting nature of this effort between FES and DPS, these enabling technology categories directly relevant to the fusion energy effort were summarized in the 2018 panel report delivered to FESAC titled "Transformative Enabling Capabilities" [TEC 2018]. In that report, advanced algorithms, high critical temperature superconductors, advanced materials and manufacturing, and tritium fuel cycle systems were specifically stated, and are aligned with one or more of the topic areas stated here. This cross cut seeks to advance these mission critical areas with recommendations that will strengthen the domestic supply chain, integrate enabling technology efforts into FES and DPS facilities at all scales, and identify FES and DPS mission needs, options, and timelines with regard to enabling technology development.*

*There are many examples of enabling technologies that cut across multiple efforts in the fusion energy and plasma science mission. High critical temperature superconductors can provide not only magnetic confinement for fusion energy, but also help to replicate astrophysical conditions in terrestrial laboratories and advance accelerator technology. Plasma facing materials are a ubiquitous issue across all fusion energy and plasma science efforts, and advancing the*

*development of materials that are resilient to the conditions formed by the plasma state will broadly advance the field. Advances in algorithms for predictive modelling, optimization, and control of complex systems will become a powerful tool in magnetic confinement and improve non-recoverable engineering cost and operating efficiencies of existing and future industries enabled by plasma technology.*

*Below are recommendations that leverage: existing facilities and facility networks; structuring of future facilities and pilot experiments; public/private partnership; and increased efficiencies by leveraging both facilities-level infrastructure and small to medium scale experiments. We further recommend identifying and prioritizing gating technologies that have low technology readiness and thereby present the greatest potential delays to mission deliverables in the next 5-10 years.*

## Recommendations

1. **Create programs to support public-private partnerships across the full breadth of fusion energy and plasma science. Create and support "Plasma Science Technology Networks" to motivate private partnership engagement and collaboration and strengthen the domestic supply chain for components enabling plasma technology**

   Public-private partnerships have been successful for accelerating technology readiness and commercializing enabling technologies. For example, previous public-private partnerships in low-temperature plasmas have been fruitful through semiconductor industry lead consortia such as the Semiconductor Research Corporation and Sematech as well as agency driven programs such as the NSF Industry/University Collaborative Research Center program. A FES-driven program that promotes industry helps to provide both the production of fundamental, enabling technologies as well as the needed drive towards commercialization. A good example of an existing DOE program that meets this need is INFUSE. This program should be broadened to further engage all sectors, including universities and other private companies that provide enabling technologies for fusion energy and plasma science, and all FES focus areas. Also, it needs to be structured so that the partnerships can evolve as the TRL levels for these industries grow from developmental to commercial. The mechanisms for engaging with this program should be simplified to encourage broader participation. Diversification of partnership frameworks to include consortia, direct research collaboration with laboratories and universities, and SBIR/STTR proposals should be released that involve cross-cutting technology opportunities. This program should establish a mechanism to test systems developed for non-fusion applications that may fulfill an FES enabling technology need. For example, material science advances for space vehicles may be applicable to the FES PMI effort and if applicable, mechanisms for extending technology to the FES mission should be available. Finally, this program should identify commercialization pathways for enabling technology supply chains outside of fusion energy and plasma science applications in order to diversify markets and provide more stability for these industries. Many of the technologies developed in FES have broad application in other DOE and government offices, so interagency partnerships should be supported (DPS PR-Collaborate).

Partnership programs should be structured to grow and strengthen the supply chain for these enabling technologies. Several critical technology areas have non-existent or shrinking domestic supply chains, and the relatively nascent private fusion technology supply chain viability will largely depend on streamlined engagements with laboratories and universities as well as development of supply chain networks similar to those found in the defense supply chain industry. As an example, several current enabling technology domestic supply chains that the HEDP community rely on for laser glass, optics, gain materials, capacitors, gyrotrons, and vacuum electronics have reached a critical point that jeopardizes U.S. leadership in this area.

2. **Broadly support advanced materials and manufacturing**

The performance of materials under the extraordinary conditions they are subjected to in fusion energy systems and basic plasma science experiments is one of the largest technical barriers for both the fusion energy mission and basic science discovery. For example, the construction of an FPP still requires advances in structural materials, high critical temperature superconductors, resilient plasma facing materials, and materials capable of withstanding extraordinary levels of power transmission as well as power dissipation. Improving the fabrication of these components is critical for improving the precision and reproducibility of experiments as well as reducing the cost of fusion energy systems. For example, advanced techniques such as additive manufacturing can enable new experiments and measurements with features such as dissimilar material joints, unique geometries with complex internal structures, and embedded sensors. Additionally, additive manufacturing can provide more economical fabrication of critical components for both FST and DPS systems. Materials and manufacturing includes the conceptualization, synthesis, and test of materials that meet FES mission needs and expands the range of operation for laboratory based plasma discovery. A broad multidisciplinary effort in material synthesis, modeling, measurement, and fabrication will close the technology gap for multiple expert area objectives.

3. **Leverage small- and mid-scale experiments and facilities to develop transformative enabling technology**

Small to mid-scale experiments and facilities focused on the research and development of enabling technology can provide an entry point to develop the workforce needed to advance plasma science and technology (CC-WD). Compared to large-scale research facilities, smaller-scale experiments/facilities are often more flexible, provide hands-on work, offer easier diagnostic access, and often provide the chance for combined experimental and theoretical research. Students and early-career staff are typically part of a small but focused research team that encourages close interactions and collaboration. Examples include university experiments and the RF Test and Source Development Facilities described in FST-SO-F. Additionally, leveraging small and mid-scale experiments enables the evaluation of low TRL technologies and develop the workforce in plasma science adjacent areas that address enabling technology needs, contributing to the growth and strengthening of a domestic enabling technology supply chain. Calls should include single-PI and interdisciplinary multi-PI projects to advance low-TRL opportunities.

4. **Build enabling technology development and evaluation into pilot plant design, DOE facilities, and collaborative networks**

   Future facilities should incorporate pathways for engaging with enabling technology experts in private industry, the national laboratory complex, and academia to develop and evaluate critical enabling infrastructure. Integration of enabling technology development should be a specific evaluation metric for future infrastructure proposals. Infrastructure level support that enables testing, advancing technical readiness, and deployment of enabling technology should be decoupled from "basic science" challenges, similar to infrastructure programs supported through NSF (MRI, SCSC, *etc.*) and DOD (DURIP).

5. **Initiate a program to identify those enabling technologies that are mission critical but are currently at low TRL or have no known solution**

   Both the fusion energy mission and plasma science effort have multiple enabling technology needs. Many of these needs are at lower TRL, and some challenges have multiple technical solutions. A well structured program should be initiated that will identify enabling technology needs for fusion energy and plasma science. This program will periodically provide an overview of the state of these technologies that include expected TRL level and competing technologies that can achieve the same end goals. An example of such a periodic reporting structure is the "International Technology Roadmap for Semiconductors" that has been published under a variety of names annually from 1992–2015. The ITRS identified critical enabling technology in materials, manufacturing, and design over a ten year cycle, identifying multiple pathways to achieve the device performance scaling defined by Moore's law, and was instrumental in moving the industry on an unprecedented pace of doubling product performance every 18 months for almost 25 years. A similar effort for fusion energy and plasma science enabling technology that draws experts from the laboratories, academia, and industry to develop and periodically update a technology roadmap for enabling technology in fusion energy and plasma science will be a valuable resource that will help the field develop data-driven timelines and deliverables gated by emerging technologies.

# *Cross-cut WF: Develop a Diverse and Inclusive Workforce for Fusion and Discovery Plasma Science, Engineering, and Technology*

*The fusion and plasma science community is preparing to enter a new phase of expansion and development, which will require significant growth in our workforce. Increasing diversity and inclusion in the fusion and plasma science community are essential components of any workforce development program, and these advancements will only be made possible through deliberate effort to reach out to and engage underrepresented groups. In addition, significant consideration must be given regarding how to improve the climate of our community to be more inclusive of all groups. A multi-stage, comprehensive workforce development program is urgently required to meet our future needs, beginning with public outreach and education to increase awareness of fusion energy and plasma science careers, along with the creation of opportunities for undergraduates through internships, apprenticeships, and academic projects, and retaining the workforce with expanded job opportunities. Universities are an essential link in the chain of workforce development, but also serve a larger purpose in providing unique and independent contributions to research that promote a diversity of thought and innovation that need to be maintained. Opportunities for university leadership in the plasma/fusion research community must be expanded to incentivize the addition and retention of faculty, university scientists, and technical workers. The increase in job opportunities presented by private enterprises can provide a powerful new tool for enticing new entrants to the fusion community as well as new avenues to retain our experienced personnel. Obtaining trained personnel to carry out the development of new facilities and technology programs will require expanded focus beyond the typical plasma physics development path and into a wider array of engineering/technology disciplines. The recommendations in this section were developed from community input in the forms of initiatives/white papers, group discussion at workshops and conference calls, and five focus group discussions. The recommendations are divided into three categories focusing on: A) Diversity, Equity, Inclusion (DEI); B) Workforce Development Tools; C) Community Outreach and Engagement. The categories and recommendations within are interlinked and of equal importance.*

A supplementary appendix that contains additional details and background information regarding the recommendations found in this section can be found here.

## Recommendations

### Category A: Embrace diversity, equity, and inclusivity to attract the broadest array of talent and diversity of thought

This category provides recommendations based on our Statement on Diversity, Equity, and Inclusion, which can be found here.

**Recommendation A-1: Engage Diversity, Equity, Inclusion (DEI) Experts to Advise our Community and Develop Assessment Tools**

*i.* We acknowledge that we are not subject matter experts in issues of climate and inclusion. *We urgently recommend that DOE supports efforts to bring in outside subject matter experts to study the social climate within fusion energy/plasma physics fields, and then take action on expert recommendations to ensure the psychological safety of our community in this process.* There is precedent for this in our field demonstrated by AIP's report "The Time Is Now: Systemic Changes to Increase African Americans with Bachelor's Degrees in Physics and Astronomy," which DOE can use as one evidenced-based standard of practice.

ii. Develop methods/techniques for performing regular assessments of our community to ascertain the effectiveness of these efforts.

iii. Committee of Visitors (COV) assigned to review the DOE FES program should include a review of what efforts are being made to monitor and encourage diversity among funding recipients and DOE facilities.

**Recommendation A-2: Implement New/Updated Policies and Codes of Conduct to encourage DEI:** While the community is working to engage outside experts and wait for their assessments, DOE must undertake immediate actions to establish a culture of respect for our current and future FES community. Policy actions for our current community derived from evidence-based standards of practice include:

i. Requiring cultural competence, unconscious bias, and bystander intervention training for program managers and principal investigators. Recommendations on how to increase the effectiveness of training should be utilized from the 2018 National Academies report on *Sexual Harassment of Women* including in-person rather than online training and that includes active participation with other trainees. The National Academies report also identified difficulties in understanding how federal agencies deal with cases of sexual harassment after reviewing their public websites, which can create an unnecessary barrier for a victim to report incidents of misconduct. DOE Office of Science has general information on how incidents of harassment can be reported on their website. DOE FES should investigate how they can more directly assess reports of harassment in their program and ensure that all individuals supported on federal funds (not only the PIs) understand their rights and abilities to report harassment. This is another topic that would greatly benefit from advice from an outside expert in order to ensure the adoption of best practices.

ii. Codes of conduct (*e.g.* APS, IEEE) should be clearly articulated at technical meetings to reinforce to session leaders and attendees how to maintain respectful conversations. Individual institutions should be encouraged to adopt these codes of conduct within their own groups and suggest improvements.

iii. Consider the adoption of double anonymous peer review of proposals. Examples can be found from other federal agencies including NASA.

**<u>Recommendation A-3:</u> Incorporate consideration and promotion of Diversity, Equity and Inclusivity efforts as an integral aspect of the review process for institutions seeking federal funding from DOE OFES:** The DOE should establish a component addressing DEI efforts through applications to their Funding Opportunity Announcements (FOAs) as part of the "Program Policy Factors." How the DOE implements such a component can happen in a variety of ways ranging from requiring additional statements or documents in the application package to integrating DEI considerations throughout the main project narrative. An effective model of the latter is the required integration of Broader Impacts in all National Science Foundation (NSF) proposal submissions. Broader Impacts (which can include DEI issues) are required to be incorporated in and evaluated throughout the main project narrative. The former approach could be modeled after current additional required documents such as the Data Management Plan found in NSF proposal packages as well those for the DOE. An additional DEI Statement document would not necessarily be an integral part of the project, but at a minimum would be required for consideration for funding. In both cases, the peer reviewers and program managers should take an applicant's description of DEI efforts into their overall evaluation and reward proposals for quality attempts and intentions for addressing these issues. While such a measure is not guaranteed to enact changes within the research community itself, it will force potential PI's to articulate how these issues might be addressed within their own research programs.

**<u>Recommendation A-4:</u> Create an accessible environment for all members of our community:** In order for students, scientists, and engineers with disabilities to fully participate in discovery plasma and fusion energy sciences, the environment must be made accessible to them. DOE facilities should prioritize the removal of accessibility barriers such as entrances blocked by one or two steps and signage that is not compliant with legal standards such as those set by the Americans with Disabilities Act (ADA). Websites and databases created with DOE support should strive to be screen reader (text-to-speech) compatible and otherwise consistent with ADA standards in order to be broadly accessible. Conferences supported by DOE should have an accessibility policy in addition to a code of conduct.

**<u>Recommendation A-5</u>: Increase funding opportunities for underrepresented groups**: Efforts should be made to expand recruitment pools (geographically, fields of study, types of institutions, *etc.*) and identify currently underrepresented areas. New funding opportunities should be considered for these underrepresented groups. These could include: graduate or postdoctoral fellowship aimed at women and minorities (*e.g.* Rackham Merit Fellowship Program is a public university program to provide financial assistance to students from underrepresented groups; DOE funding for Solar Foundation targeting workforce development for veterans and underserved communities); expand undergraduate internship opportunities for women, minorities, and students without traditional plasma training; small funding grants available to women, underrepresented minorities, and early-career scientists.

**<u>Recommendation A-6:</u> Create Parental Leave Policies:** Family leave policies should be more uniformly applied to research institutions across the nation. The recent enactment of the Family and Medical Leave Act (FMLA) “provides up to 12 weeks of paid parental leave in connection with the birth, adoption, or foster care placement of a child for employees covered by FMLA provisions applicable to Federal civilian employees.” All institutions receiving funds from federal agencies must strive to better conform to a 12 week paid leave policy program. If/when members of research teams supported by DOE FES funds are in need of parental leave (such as during circumstances described in the FMLA), DOE FES should allow funded PIs to continue to financially support their personnel for up to 12 weeks of paid leave at their existing salary/benefit level. In addition, DOE FES should work with PIs to adjust milestones/deliverables accordingly to accommodate research team members who take family leave in order to avoid penalizing research programs that enact family leave policies. Research institutions should make parental leave policies and information transparent and easily accessible. Furthermore, flexible working hours and telecommunication options should be supported, including remote presentation options for conferences. Institutions should also ensure that private lactation space is available (*i.e.* apply the federal law under section 7 of the Fair Labor Standards Act).

## Category B: Increase the pathways to fusion starting at the undergraduate level, and create more opportunities for participation of technical workers from other fields

Additional details on recommendations B-1 through B-5 including timelines and sample budgets can be found here.

**Recommendation B-1: Establish Student Design Competitions**: The goal of the student design competitions is to show that the realization of fusion as an energy source encompasses many disciplines and contributes broadly to national science and technology goals. A series of student design competitions can be established to target topical areas identified in the FESAC Report on Transformative Enabling Capabilities[2]: advanced algorithms, advanced materials and manufacturing, and high-temperature superconductors. These FES driven student competitions can begin as joint competitions between FES and organizations with established design competitions (*e.g.* NASA) to leverage existing university applicant pools. The topics of the student projects should be designed to reach a wide variety of universities, colleges, and community colleges and should not rely solely on an institution having an established fusion or plasma physics program. Additionally, a significant fraction (30%) of the project can be scored on community outreach, especially to the K-12 grades, to promote the student team’s work and the overarching program goal of realizing commercial fusion energy.

[2] R. Maingi, et. al., Summary of the FESAC Transformative Enabling Capabilities Panel Report, Fusion Science and Technology, 75: 3 (2019)

**<u>Recommendation B-2:</u> Develop Flexible Post-Undergraduate Education Options**: More formal post-undergraduate educational programs can be established to provide either Capstone Certificates (9 credits in a concentrated area) or Master of Engineering (30 credits) degrees. Both of these options can be offered online so that they will be accessible to a wider range of students, including those who are working full-time at institutions around the world. Additional certificate programs can be developed to address specific knowledge gaps for individuals in other fields to transition into fusion or engineering. DOE should consider the development of apprenticeship programs, potentially in coordination with universities, to allow for the training of specialized workers for positions including technicians, computer science, and electricians. Successful models have been developed internationally including by the United Kingdom Atomic Energy Authority (UKAEA) that could be studied and possibly emulated.

**<u>Recommendation B-3:</u> Employ scientists and engineers with BS/MS degrees at FES facilities:** The two large MFE user facilities, DIII-D and NSTX-U, routinely receive many more experimental proposals than available run time. In an effort to reach full utilization of these facilities, the workforce can expand to employ more scientists and engineers with Bachelor's and Master's degrees. Rather than fill measurement support/analyst roles with PhD-trained staff, which can underutilize their expertise, or Ph.D. students, which can result in year-to-year variations in quality, these roles could be filled by, long-term, trained technical staff at MS/BS level, creating a near-term opportunity to broaden the fusion workforce. Oversight could still be provided through collaborators, including opportunities for student training, but the responsibility for the day-to-day operations would be transferred to the host, improving its ability to deliver a 'Walk-In/Walk-Out' experience for the user. Online education options mentioned above can be used to fill knowledge gaps for the incoming workforce.

**<u>Recommendation B-4:</u> Create Private-Public BS/MS development program:** This program is similar to the ideas expressed above, but is designed to engage the BS/MS workforce prior to graduation. Upon graduation, individuals in the development program will gain valuable knowledge through rotations at universities, national laboratories, and private companies. As an example, an individual graduating with an electrical engineering degree can consider rotations that consist of: diagnostic development, remote handling, and power systems design. The costs can be split during the rotation period between private and public funds. In addition, students could concurrently pursue further education through an online graduate degree or certificate during their rotations, which would allow them to combine the fundamental theory with the hands-on practice of the material. A more focused approach with a participant working at one institution could be similar to the Graduate Scheme (http://culhamgraduatescheme.com) program utilized at the Culham Center for Fusion Energy.

**<u>Recommendation B-5:</u> Establish Private-Public fellowship program:** A Private-Public Fusion Energy graduate fellowship or Private-Public Fusion Energy Postdoctoral Fellowship can be established to target gaps identified in FESAC Transformative Enabling Capabilities report as well as the Workforce Development report[3]. As part of the Graduate Fellowship program, awardees would take an internship during their 4th year of graduate study at a private company. This can be something similar to the existing DOE Office of Science Graduate Student Research (SCGSR) Program with the addition that students may conduct part of their thesis research at a private company. Similarly, a Postdoctoral fellow can split time between privately and publicly funded programs. These positions could be used to transfer knowledge between institutions - such as the application of advanced algorithms to machine control or the use of a diagnostic on multiple machines. ORISE has significant experience managing student internship and fellowship programs and could consider expanding its portfolio to embrace private-public collaborative opportunities.

**Recommendation B-6: Create a coordinated paid summer internship program** for undergraduate students that would position students in internships at universities, national labs, and private companies throughout the country. Students would participate in an introductory workshop/course where fusion and plasma science are introduced and its many challenges identified. These workshops could be organized by experts from both the Discovery Plasma Science and Fusion Science & Technology communities. This would be a chance for them to get to know about the big picture of plasma science and fusion, the local subset of the community, and each other. The workshop could be held at different places each year (universities, national labs, companies) and would bring experts from different fusion and plasma science fields to speak to the summer cohort. Considering that fusion and plasma science are not taught at many undergraduate institutions, this would be the first introduction to many of the students to these topics. This could be modeled in part after the PPPL NUF/SULI One-Week-Course[4], but with a broader emphasis beyond plasma physics. After the workshop, students would go to their summer institutions to continue their internship. Their summer work can be presented in poster or oral form locally and it can be uploaded to a central website so that their work can be publicized more broadly, unless the work deals with sensitive intellectual property and cannot be made public. Afterward, the program would fund the students' attendance and participation at a relevant meeting (APS-DPP, SOFE, ASME, etc.) This model is very similar to that of the National Undergraduate Fellowship (NUF).

**Recommendation B-7: Cultivate and increase faculty tenure lines at universities and colleges through faculty development grants and collaborative opportunities with National Laboratories:** As a means to address the continued decline in the number of university faculty lines in the areas of plasma and fusion science, DOE should work with universities to identify opportunities for increasing the number of joint appointments between universities and national laboratories (*e.g.* ORNL, JLab, LBNL, PPPL, etc.) where appropriate to the lab's mission scope and the university department's needs. Important lessons can be learned from the High Energy Physics community to develop lasting collaborations and joint faculty positions between universities and national/international facilities, such as has been

[3] FESAC, Final Report on Fusion Energy Sciences: Workforce Development Needs (2014).
[4] https://suli.pppl.gov/2019/course/

done between CERN, Argonne National Laboratory, and the University of Chicago. Opportunities to develop more lasting collaborations between universities and ITER, facilitated through DOE national laboratories, could help provide long-term and stable funding paths that encourage universities to invest in new faculty lines in fusion and plasma science. The DOE should also implement faculty development grants awarded to a department looking to expand their plasma science faculty or to develop a plasma program at a college or university without one. A model for this type of award is the Faculty Development in the Space Sciences awarded through the Geophysics Directorate of the NSF. The grant provides funding to a department for a new faculty member to be hired and full academic support for the first five years, after which the college or university takes over continued employment of the new line. In particular, this grant is aimed not only at departments with established space science programs, but to institutions looking to expand their research portfolio. A fusion or plasma science-based faculty development program could be similarly used for encouraging the expansion of the relatively low number of plasma faculty in colleges and universities throughout the country.

## Category C: Increase literacy of plasma science and fusion energy and improve student involvement in related degrees/employment opportunities by developing a community outreach network

**Recommendation C-1:** **Support New Public-Facing Website:** U.S. fusion community and FES should support the creation and management of a centralized website that would serve as the public face of the U.S. fusion community. The website would feature resources for K-12 teachers/students, internship opportunities, jobs in fusion, fusion news, etc., and would be curated and managed by a committee composed of members of the U.S. fusion community, from diverse fields and diverse institutions, including private fusion companies. This committee would be tasked with compiling and curating the content and maintaining it up to date. An institutionally and thematically diverse committee will help identify a broad range of internships/resources/jobs highlighting the multidisciplinary nature of fusion energy and plasma science research. FES support for this initiative is critical for its long-term impact: Having FES give its “blessing” gives the website legitimacy within the community and outside of it. It would also help recruit collaborators and committee members. Monetary support from FES would help make a more professional website (by hiring web designers), and help keep it up to date, both content-wise and aesthetically.

**Recommendation C-2:** **Support Pre-College Outreach by the Plasma Science and Fusion Community**: A substantial gap in current workforce efforts is the development of plasma science and fusion energy *(PS&FE)-specific* pre-college outreach activities that will increase literacy and improve student involvement in related degrees/employment opportunities. To address these pressing needs, we recommend the establishment of a Plasma Network for Outreach and Workforce (Plasma-NOW). Plasma-NOW will consolidate the various outreach efforts in our community under a single umbrella network that will facilitate the exchange of ideas within the PS&FE community and maximize their effectiveness on student/public involvement with PS&FE-related programs. To ensure rapid results, we recommend the establishment of a ***DOE OFES-sponsored committee*** that will conduct a survey on best outreach practices available in the PS&FE community as well as identify how successful STEM practices can be adapted to the specific needs of the PS&FE workforce development. The outcome of this effort will result in a ***report and a strategic plan*** for PS&FE pre-college outreach, which will outline the vision and proposed structure of Plasma-NOW.

DOE OFES endorsement of Plasma-NOW will encourage serious input from the PS&FE scientific community. Additional funding support will enable the committee members to conduct field studies and engage the PS&FE community in the proposed activities. DOE can further support these efforts by encouraging submissions of proposals to the open solicitation call by PIs interested in contributing their time and educational resources to support the Plasma-NOW efforts. Plasma-NOW will be organized following the example of similar initiatives in other STEM fields, such as the highly successful NISEnet. A proposed structure for Plasma-NOW is shown in Fig. CC-WF.1.

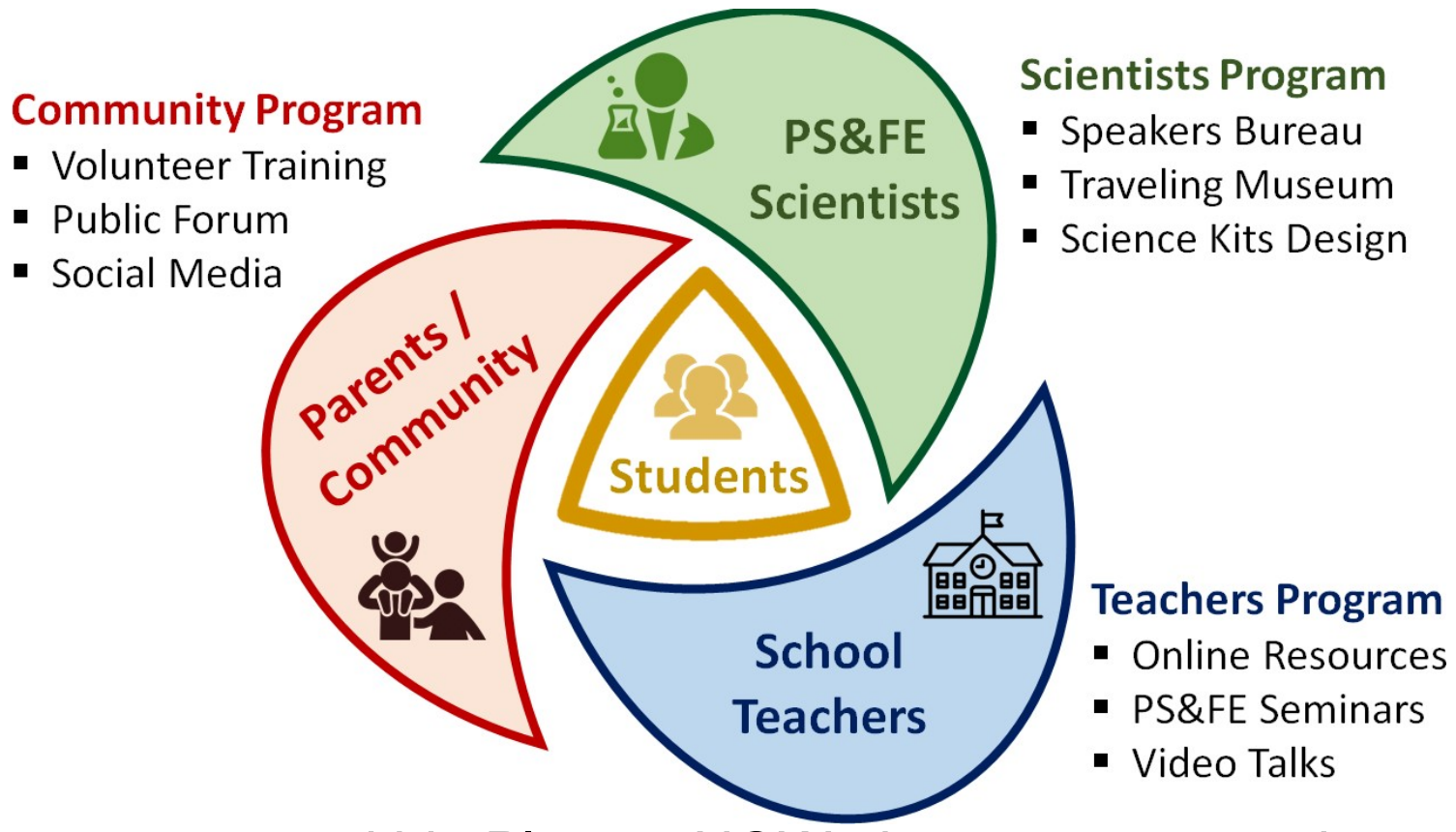

Figure CC-WF.1: The programs within Plasma-NOW aim to engage students both directly through contact with scientists, as well as indirectly through teachers and community programs.

# Appendices

## *Appendix A. The Prioritization Assessment Criteria Applied to Fusion Science and Technology Program Elements*

### Prioritization Assessment Criteria

At the CPP-Houston meeting, the attendees discussed and applied Prioritization Assessment Criteria (PACs) to help prioritize different elements within the Fusion Science and Technology (FST) program. These criteria were inspired by the Principles, Values, Metrics, and Criteria Working Group in the 2017 U.S. Magnetic Fusion Research Strategic Directions community workshops. A revised set of principles were introduced to the CPP process at the 2nd Joint Workshop for MFE and FM&T at Knoxville, and these principles were adopted as the FST Values. The PACs, defined below, were developed from a subset of these Values by the MFE and FM&T program committees prior to the CPP-Houston meeting. The CPP-Houston attendees considered the PACs during discussion of the strategic plan in breakout groups. The PACs were defined as the following:

***1. Importance to FPP Mission***
How essential is the research enabled by this facility or program for ensuring the success of the FPP?

***2. Urgency***
How critical is it that this facility/program is started (or continued) immediately to enable a timeline of an FPP by the 2040s?

***3. Impact of Investment***
Does this facility/program likely provide significant scientific or technological progress relative to the investment?

***4. Using Innovation to Lower Cost***
Does this facility/program take advantage of new innovation not previously utilized by the fusion program that could potentially lead to a lower cost pilot plant?

***5. U.S. Leadership and Uniqueness***
Would the facility or program provide unique capabilities or make the U.S. a leader in areas that are required for the commercialization of fusion?

## Process at the CPP-Houston Meeting and Analysis

In an attempt to assess the community's views on the relative importance of these criteria in determining prioritization, the PACs were ranked by the attendees at the CPP-Houston meeting via polling software. "Importance to Mission" was ranked the most important PAC to consider when performing prioritization, with "Urgency," "Impact of Investment," "Innovation to Lower Cost," and "U.S. Leadership and Uniqueness" following in order of importance. The FST strategic plan has a number of objectives and recommendations, and to obtain more detailed information, each Strategic Objective was subdivided into at least two program elements, creating a total of 23 elements shown in column A of Table A.1 The approximate mapping of each element to the recommendations in the FST plan is shown in column B of Table A.1. Following the completion of each CPP-Houston plenary presentation, breakout group discussion, and plenary discussion, the attendees were asked to give a 1-5 score (with 1 low and 5 high) for each program element for each of the PACs. The polling results are shown in columns C to H in Table A.1.

The FST-SO-H strategic objective was developed by the CPP Program Committee at a Writing Retreat that followed the 2nd Joint MFE and FM&T Workshop in Knoxville and the 2nd HEDP Workshop in Menlo Park. Inertial Fusion Energy (IFE) was one of the high priority HEDP areas of research identified at Menlo Park, and a community letter in support of IFE research was submitted to the CPP process, signed by more than 100 individuals. To create a strategic plan with the two broad themes—FST and DPS—the IFE strategic objects were merged into the FST plan and organized in FST-SO-H together with new recommendations on alternative MFE configuration research, which also received a community letter of support. The CPP participants in the HEDP topical area were therefore not involved in the development of the PACs, and they did not participate in discussion or polling at the Knoxville workshop. As a result, while IFE is an important element in the strategic plan moving forward, polling on IFE is not included in Table A.1.

The total attendance at the CPP-Houston meeting was approximately 180 people. The self-identified primary associations with the four topical areas were 52% MFE, 19% FM&T, 18% HEDP, and 11% GPS. Two of the five days of the CPP-Houston meeting were divided into parallel FST and DPS sessions. On these days, the MFE and FM&T registrants primarily attended the FST sessions, and the HEDP and GPS registrants primarily attended the DPS sessions. Typical polls collected ~120 responses, with some extending up to ~150. The polling for all program elements was conducted with the FST discussion participants at Houston. Polling on the elements related to FST-SO-H included a larger group and was open to all attendees of both the FST and DPS community at CPP-Houston.

**Table A1. Houston Attendee Polling of FST Program Elements by Prioritization Assessment Criteria (PACs)**

| A | B | C | D | E | F | G | H |
|---|---|---|---|---|---|---|---|
| | | Mission | Urgency | Impact of Invest | Innovation | US Leadership | Ave. of Col C-G |
| Fusion Prototypic Neutron Source | FST-SO-B.2 | | | | | | |
| Blankets Sci & Tech Program | FST-SO-C.1,2,4 | | | | | | |
| Tokamak Disruptions Program | FST-SO-D.4 | | | | | | |
| Burning Plasmas | FST-SO-D.1 | | | | | | |
| Materials Development Program | FST-SO-B.1,3,4,5 | | | | | | |
| Magnets | FST-SO-F.1,2,3 | | | | | | |
| New Tokamak User Facility (NTUF) | FST-SO-D.2 | | | | | | |
| PFC Material Integration Program | FST-SO-D.5 | | | | | | |
| Solid PFC Program | FST-SO-A.1,3,4 | | | | | | |
| Tritium Fueling & Exhaust Program | FST-SO-C.3,7,10,11,12 | | | | | | |
| Tokamak Scenarios Program | FST-SO-D.3 | | | | | | |
| PPP and Intragency MFE Alternates | FST-SO-H.4 | | | | | | |
| Liquid PFC Program | FST-SO-A.2,3,4 | | | | | | |
| QS stellarator | FST-SO-E.2 | | | | | | |
| QS Stell Opt +international collab | FST-SO-E.1,3 | | | | | | |
| Blanket Component Test Facility | FST-SO-C.5 | | | | | | |
| ITER Team | FST-PR-B.2,3 | | | | | | |
| Three Tier MFE Alternatives Program | FST-SO-H.3 | | | | | | |
| RF Test Stand + Scenarios | FST-SO-F.4,5,6 | | | | | | |
| Volumetric Neutron Source | FST-SO-C.6 | | | | | | |
| RAMI and Balance of Plant | FST-SO-G.3,4,5 | | | | | | |
| Licensing | FST-SO-G.1,2 | | | | | | |
| Inertial Fusion Energy (IFE) | FST-SO-H.1.2 | Polling Data Not Presented Due to Parallel Development of IFE in HEDP Topical Area (See Main Text) | | | | | |
| **Max Value** | | **4.73** | **4.57** | **4.46** | **4.39** | **4.40** | **4.35** |
| **Min Value** | | **2.81** | **2.82** | **2.94** | **2.25** | **2.52** | **2.94** |

The color of each cell in Columns C-G of Table A.1 reflects the average value the community assigned to each program element on a 1 to 5 scale. Each column, C-G, represents a different criteria and has the same color scheme applied. This color scheme is chosen such that the highest value in the column is indicated by a **dark red** color, and the **dark blue** indicates the lowest value in that column. The color scale spans from dark blue (indicating low values) to dark red (indicating high values) with white in the middle, between the highest and lowest values (indicated on Table A.1 below each column). The numerical data that was used to generate this plot has been removed to address concerns raised by members of the community. Column H in Table A.1 represents the value for each program element averaged over all 5 prioritization assessment criteria (an average of Columns C-G). The color scheme used in Column H is the same as is applied to columns C-G. The order of program elements is determined by sorting the chart by column H, the average value.

## How Polling Data Was Used in This Process

All of the strategic objectives (SOs) and program recommendations (PRs) in this report are viewed by the community as important for achieving our stated missions. The order of the SOs in the FST section of this report is intended to roughly reflect the community's assessment of the relative priority that should be assigned to each SO. This ordering was proposed before the CPP-Houston meeting based on a combination of polling results, verbal discussion, and written feedback generated primarily from the MFE / FM&T Knoxville meeting. Polling data was useful information in determining this ordering insofar as it gave clear, quantitative evidence that many of the high-priority elements enjoy broad community support that cuts across the demographics and institutional affiliations of those who were polled. The feedback from attendees at CPP-Houston confirmed that the ordering of SOs is an appropriate reflection of community consensus.

The PACs discussed at CPP-Houston and described above are intended to provide more fine-grained information regarding the community's view of the merits of various program elements, as related to the stated mission to pursue a low-cost FPP. The intent is that this information, in combination with other information in this report, continued community discussion, and cost estimates for these elements that will be developed later, will provide a basis for prioritization and sequencing within budget scenarios. It is not intended that the polling results presented here should be used as the sole basis for any funding or sequencing decisions, nor has this data been used as the sole basis for determining or indicating priorities within this report.

### Robustness of the Results

The results displayed in Table A.1 appear to be robust. The elements there are ordered by an average across PACs, and this ordering is generally consistent with the ranking of the PACs (*e.g.* with elements with high "Importance to Mission" also tend to have high average values). The detailed CPP-Houston polling presented in Table A.1. qualitatively confirms polling data obtained from the MFE+FM&T Knoxville workshop. The attendees at the Knoxville and Houston meetings differed significantly, and yet both workshops yielded strongly consistent results. Therefore, the results appear largely independent of which cross-section of the MFE & FM&T community was polled. Most importantly, the elements having the highest average values (taken across all PACs) also had small standard deviations. This data, taken together with consistent polling done at the Knoxville workshop, indicates that there is strong community consensus that these program elements should be high priorities in FES.

## *Appendix B. Assessing the Needed Capabilities for the New Tokamak User Facility*

The combined efforts of both Advocacy Groups and Expert Groups have laid out a strong basis of support for the U.S. fusion program requiring a new tokamak facility. These efforts reinforce and advance the process that began with the community input at Austin and Madison in 2017. This resulted in the National Academies panel recommending to either upgrade an existing facility or establish a new facility to accomplish a pre-pilot plant research program to demonstrate sustained, high-power density magnetically confined plasmas. Through feedback from the FST community, there is strong support for the recommendation that a new tokamak, rather than an upgraded facility, is necessary to accomplish our goals. The intention of such a "new tokamak user facility" (NTUF) is that, in conjunction with ITER and other accessible tokamaks, we can sufficiently close the critical gaps related to plasma exhaust and core/edge integration in planned FPP operating scenarios, investigated through FST-SO-D.3. Here, closing gaps represents the sufficient completion of the physics basis necessary to design an FPP, and does not imply that research into the diverse areas of confinement physics is finished, similar to community progress following the 1999 and 2007 publishing of the ITER Physics Basis and Update, respectively. Based on community discussions it is expected that no other new domestic tokamak facility beyond NTUF will be required before designing a tokamak FPP. This assumption links the scope of NTUF not only to present day capabilities and those anticipated from planned international and non-governmental fusion programs, but also to the assumptions of the FPP mission goals. These goals are outlined in the Definition of a Pilot Plant, and we acknowledge that they are likely to evolve as high priority work under FST-PR-A is completed. For the tokamak, strategies for producing net electricity at low capital cost are anticipated to increase power density and exacerbate an already significant power and particle exhaust challenge, which in turn increases the likelihood that the needed divertor mitigation will affect the pedestal and core plasmas. Even without the detailed design of an FPP, there is a pre-existing motivation from the 2015 Workshop Reports for a facility that can be used to explore and optimize exhaust solutions that can also simultaneously demonstrate suitable core performance. Results from a facility like NTUF are needed to distinguish *potential* FPP operational scenarios from those that are likely to be *achievable*. NTUF's research output is a necessary input into FPP scoping (FST-PR-A), and NTUF's design can begin without additional input from FPP design activities. The broad mission space of potential FPPs is already known from numerous power plant and pilot plant studies (*e.g.* ARIES, EU-DEMO, ARC, CAT, ST-PP, *etc.*), and these provide reasonable estimates of fusion pilot plant heat and particle fluxes. Potential solutions, spanning a range of divertor geometries and materials are known and were considered by the community in 2015, leading to the recommendation for a Divertor Test Tokamak. What remains is demonstrating that these proposed divertor solutions are compatible with the pedestal and core plasmas that are required for sustaining high power density operation of a tokamak-based FPP.

By considering these constraints, the following two capabilities are recommended for NTUF to satisfy. These represent a down-selection and prioritization driven by community input.

- (CAP-A) The flexibility to investigate innovative tokamak divertor solutions, encompassing long-legged concepts and PFC material options, at heat and particle fluxes that are at the same scale as those projected for the pilot plant.
- (CAP-B) The ability to simultaneously achieve these divertor solutions at core plasma energy confinement and bootstrap current fractions that project to a high-average-power output, net-electric pilot plant.

This simultaneous achievement of high power density across a range of core plasma scenarios and divertor solutions represents combined capabilities that cannot be obtained through international collaboration or a major upgrade to an existing facility. Community feedback confirms that capabilities to *"demonstrate reactor-relevant heating and current-drive actuators that utilize minimal, fusion nuclear compatible approaches for feedback control"* and *"test solutions for managing divertor and main-chamber PFC material migration and erosion for pulse lengths and power densities prototypical of a pilot plant and demonstrate sustained, disruption* *free operation"* are lower priority to include in NTUF than CAP-A and CAP-B. This prioritization is motivated by the cost and schedule implications that would prevent progress in other areas of the FST plan as well as likely delay progress on research obtainable with a more limited scope NTUF that focus on [CAP-A] and [CAP-B]. Thus while a broad array of design options still need to be considered for NTUF, the deprioritization of the high duty factor capability is expected to allow both copper and superconducting magnetic coil based designs for NTUF to be considered. It is expected that NTUF will not use tritium, as the goal is to complete urgent research objectives utilizing CAP-A and CAP-B at lower cost and reduced complexity compared to devices that can pursue burning plasma physics as outlined in FST-SO-D.1.

These capability statements and the material included in this Appendix would provide important input into DOE's crafting of formal Mission Need statement, as outlined in Section 4.d in DOE G 413.3-17 "Mission Need Statement Guide."[5] As outlined in FST-SO-D.2, we also need a near-term, FES-funded activity to scope these capabilities into pre-conceptual facility designs. Rather than characterizing NTUF by engineering metrics, *e.g.* size, field, aspect ratio, heating power or cost, community discussions focused on the capability gaps which drive the first stage of the Critical Decision process within DOE. While pre-conceptual studies completed as part of the Initiative submission process have outlined potential NTUF design options, we emphasize here and in FST-SO-D.2 that near-term resources are needed to encourage multiple institutions to advance pre-conceptual design investigations in an open, collaborative manner.

*Support for Capabilities from Community Initiatives and Expert Group Strategic Blocks*

A comprehensive argument for the need for new tokamak facility is laid out in [Buttery - 2019] where it refers to the vital need to find self-consistent solutions for sustainment of high pressure core plasmas that are compatible with pedestal, scrape-off layer, wall and divertor power handling. The Initiative notes that investigating single issues in separate edge or core focused facilities *'will not give the required predictive capability to project to [FPP], adding more steps to the fusion path'*. This strongly motivates the combination of [CAP-A] and [CAP-B], and rather than attempt to summarize the arguments we suggest reviewing that Initiative for specific details. A similar message emphasizing integration is outlined in [Menard - 2019] Goal 2, noting that for high-performance core plasmas with high edge power density, *"Such integration appears very challenging with existing and planned facilities and may motivate a new [NTUF] facility"*. It is assumed in all of the capabilities, but specially worded in [CAP-B] that there is not a fundamental restriction that would prevent using NTUF to explore scenarios that would extrapolate to a pulsed, high-duty factor FPP as outlined in [Mumgaard - 2019]. A device aimed at [CAP-B], properly equipped with sufficient measurements, would advance Understanding Plasma Transport, as outlined in the Transport and Confinement Expert Group's Strategic Blocks, but would fail to address Demonstrating Integration due to the NTUF's lack of access to burning plasma core. Similarly, [CAP-B] would address multiple areas within Program Elements

[5] *https://www.directives.doe.gov/directives-documents/400-series/0413.3-EGuide-17-admchg1*

1-3 in the Strategic Blocks developed by the Scenarios Expert Group, but would not fully close these gaps because of the absence of alpha heating.

Strong support for a [CAP-A] is given in [Canik - 2019], where the arguments are laid out for the need to focus on the long-legged divertor concept due to the gap in available facilities worldwide that can study such concepts at high power density. Indirect support for [CAP-A] is also emphasized in [Buttery - 2019] which proposes to absorb mission scope of development and qualification of the divertor solution in the new facility. In general, three of the capabilities are to some degree supported by [Kuang - 2019], which invokes the 2015 PMI Workshop Report findings for the need for a U.S. led Divertor Test Tokamak, for which the capabilities correspond closely to the Priority Research Directions (PRD), *e.g.* CAP-A/PRD-B and CAP-B/PRD-E. The recommendation for a facility that combines [CAP-A] and [CAP-B] is within the four Strategic Blocks developed by the Boundary/PMI MFE Expert Group. Similarly the combined capabilities address similar themes of core-edge integration that were raised in the Scenarios and Transients Expert Groups' Strategic Blocks.

Despite not supporting high duty factor operation, NTUF could still contribute to reducing risks for FPPs by narrowing the uncertainty in main-chamber erosion rates, driven by the reactor relevant conditions required for [CAP-A], and be a platform for prototyping innovative solutions for management of large volumes of eroded material ('slag management') that are compatible with high performance plasmas required for [CAP-B]. In [Abrams - 2019] the materials research that can be accomplished in a short-pulse, high-power density device consistent with NTUF is outlined, which could allow NTUF to deliver science under FST-SO-D.5 even without a dedicated capability on material migration enabled by high duty factor operation. In [Abrams - 2019] and [Unterberg - 2019] there is an emphasis on the need for high temperature walls, ≥ 600 degC, which could be a design feature to manifest from [CAP-A] as the pre-conceptual design studies recommended in FST-SO-D.2 are completed. Together, [Unterberg - 2019], [Stangeby - 2019] and [Abrams - 2019] all raise a similar concern about the large volumes of eroded wall material build up and its potential to prevent continuous FPP operation. The short pulse NTUF as described above can investigate, although not fully demonstrate, innovative solutions to this challenge. Additionally, ITER and other long-pulse international facilities will offer windows into the nature of the erosion problem, although they may not be able to investigate innovative approaches such as the proposed 'flow-through' low-Z coatings, dedicated divertor design through wall contouring or even remote maintenance through advanced robotics. Due to the expected cost and time requirements of high duty cycle intermediate facilities, demonstration of slag management solutions may not be fully addressed before an FPP, similar to the full integrated fusion nuclear science mission.

It is also expected that NTUF can be equipped to use liquid metal PFCs, captured as one of the PFC material options in [CAP-A], which may provide a solution for some erosion and slag management issues, and can provide an opportunity to inform an important baseline FPP design decision as mentioned in FST-SO-A. In [Goldston - 2019], a plan for maturation of the liquid lithium PFC technology is laid out that first engages test-stand and existing confinement facilities like NSTX-U. In [Menard - 2019], the basic strategy is outlined for establishing Mission Need for a device that can close the science and technology gaps between present devices and the FPP, similar to the one being made here for NTUF. An important capability of a new facility was identified to be the ability to test all ranges of divertor and first wall options including both slow-flow and fast-flow liquid metals. Also emphasized in [Menard - 2019] is the need to consider these types of options early in the device design to ensure physics and engineering requirements can lead to a viable design. For the NTUF on the timeline mentioned in FST-SO-D.2, this motivates the need to mature the LM technology quickly, as suggested in FST-SO-A.2, to ensure future testing in NTUF is possible, without delaying the baseline NTUF design that is expected to use solid PFCs.

In the end, the argument from [Buttery - 2019] remains that if the candidate scenario is not investigated in an integrated manner (excluding burning plasma physics, which will be investigated through ITER and private ventures, FST-SO-D.1), then more steps are added on the path to the pilot plant. A single step facility with the capabilities described above, combined with a burning plasma experiment like ITER, should allow for tremendous progress in the tokamak physics basis. Nevertheless, phasing the operational stages of NTUF will be important in delivering the science output required for FPP designs (FST-PR-A) in a timely manner, consistent with commissioning of the increasingly complex portions of the facility. For example the divertor physics mission can be explored initially, expanding in scope to the integrated scenarios. In this respect, it may be feasible to equitably meet the scientific needs of multiple Initiative groups on the same facility over time.

*Comparison to Present DOE Facilities and Planned Capabilities Beyond DOE*

[CAP-A]: This capability requires a facility with both toroidal field strength and power density within proximity of those expected in an FPP. While both MAST-U and TCV will have access to long-legged divertors, they lack the necessary field strength and power density to meet this capability. The Italian DTT is expected to approach the pressure and power density requirements, but the configuration for the long-legged divertor is not the baseline configuration [DTT Green Book]. There are present predictions for lower current Super-X configurations in that device, and likely restrictions in heating power due to the large outer gap. Not having a domestic facility with the CAP-A capability risks delaying the sufficient completion of the physics basis to enable timely FPP construction.

[CAP-B]: The I-DTT will have the power density to explore the divertor solution, but will not support the advanced tokamak scenarios in its baseline configuration [DTT Green Book]. DIII-D, JT-60SA and KSTAR can explore advanced scenarios but not at sufficiently high power density to qualify the scenario for a compact FPP. ITER arguably would be able to investigate the steady-state, AT regimes (depending on the needed bootstrap current fraction), but not until after 2040. It also is limited in what H&CD actuators it will use, focusing on electron cyclotron (EC), ion cyclotron (IC) and neutral beam (NB), preventing evaluation of potential improvements in LH or Helicon at high power density. Moreover, ITER will not have the ability to investigate a flexible divertor solution, so if the standard lower single null (LSN) solution will not work for FPP,

then we receive that information too late to feed into FPP. Similar arguments broadly apply to conditions expected for SPARC. While there is sufficiently high power density, there may be insufficient divertor flexibility to develop solutions for FPP, and limited ability to investigate high confinement with high bootstrap current fraction.

As mentioned above it is the combination of these capabilities that would make NTUF unique and world leading. This is highlighted further in Figure 1, where its intended operating space is sketched along axes of divertor heat flux and average pressure and metrics of normalized core plasma performance relative to existing and future devices. It is argued that to progress to an FPP for a tokamak we need to move diagonally within this space, reaching high core performance at high pressure and mitigate high divertor power density. Different FPP operational scenarios, such as high-duty factor pulsed or long-pulse steady-state, will differ in the exact end point. ITER is intended to be placed between the existing devices and FPP, while also focusing on D-T scenarios which NTUF will not. The near-term larger-scale, D-D facilities like JT-60SA and I-DTT will tend to operate closer to the top left and bottom right, respectively. With NTUF we should be able to dramatically reduce the size of the step, for aspects of both plasma physics and fusion technology, between ITER and FPP. As outlined in FST-SO-D.2, physics and engineering teams will be required to provide further detail on just how NTUF fills this space as well as allow the cost and schedule impact to be understood.

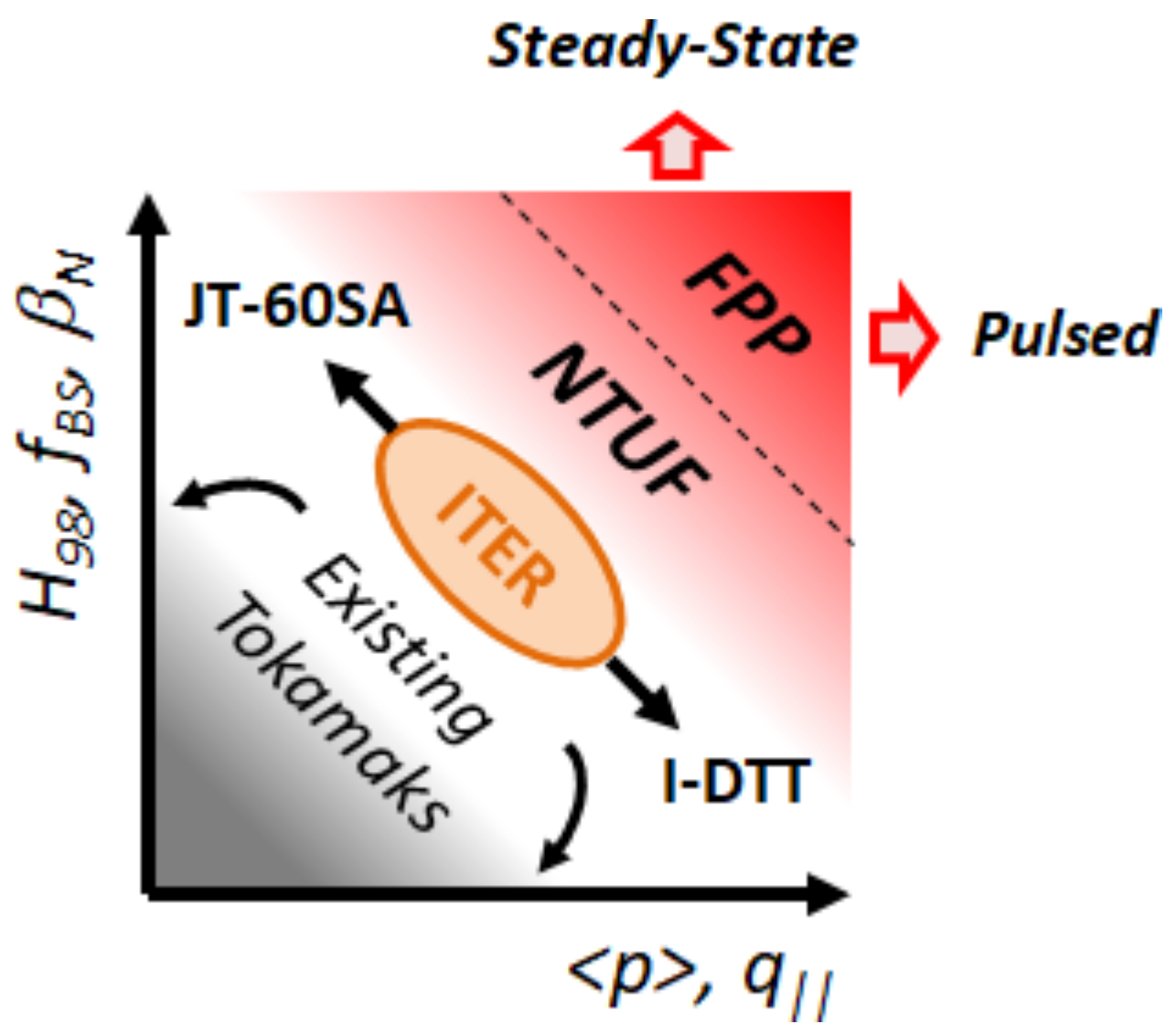

*Figure A.1: Diagram showing the intended placement of NTUF on axes of absolute parameters such as divertor heat flux and average pressure and normalized core plasma performance relative to ITER, FPP and existing and near-term devices.*

<u>References</u>

# *Appendix C. Focus Groups*

The DPP-CPP wanted to reach out to groups who may not be as vocal at the workshops, may have difficulty attending, and/or may have separate concerns that are not as easily voiced in the big workshops. Thus, we hosted ~1.5 hour discussions, moderated by Laurie Moret and with Lauren Garrison as the notetaker, with each of these groups:

- Women
- Underrepresented minorities
- Graduate students
- Early career scientists and engineers

Currently, the percentage of women and non-white individuals in the field of fusion and plasma science is significantly less than their percentage in the U.S. population. Looking forward, GenZ is projected to have ~50% white and ~50% all other races [Brookings 2018]. If we are not actively recruiting women and non-white individuals, we will be ignoring ~75% of the future potential workforce. The percentage of women and underrepresented minorities in science does not simply get better with time—it requires action. Diversity and inclusion is not about filling a quota or being nice. Diversity and inclusion (across all aspects of personal characteristics and diversity of thought) is the best strategy for solving challenging problems—like developing fusion energy. This advantage of diversity is shown in many studies, such as this article in Forbes which states, “Diversity is a key driver of innovation and is a critical component of being successful on a global scale.”

The following invitation to participate in the focus groups was sent to the DPP-CPP google group mailing list and to many other relevant mailing lists:

-----

**We need you! Please sign up to participate in a focus group about the community driven strategic plan for the DOE Office of Fusion Energy Sciences.**

Throughout 2019, the American Physical Society Division of Plasma Physics Community Planning Process (APS-DPP-CPP) is developing a strategic plan for the DOE Office of Fusion Energy Sciences.

There are numerous ways that everyone can give feedback and participate in the process, such as attending one of the workshops (see the full list of events here: https://sites.google.com/pppl.gov/dpp-cpp/ ). In addition to the workshops and webinars listed on the website, we are seeking additional targeted feedback from individuals who are less likely to attend our workshops. If you self-identify with any of these groups: women, graduate students, early career scientists and engineers, or underrepresented minorities and you are in a career or field of study related to any of these topics: Magnetic Fusion Energy, Fusion Materials and Technology, High Energy Density Physics, or Discovery Plasma Science, please sign up to participate in a focus group session. Signups will be on a first come basis, with the caveat that we may need to limit slots allotted per institution so that we can hear from a broad group and keep each session small. However, additional times may be scheduled if the below options fill up, so please let us know if you'd like to participate and aren't able to sign up during one of the times below.

+Monday November 4; 1:00-2:30 pm EST; focus group for Women

+Monday November 4; 2:30-4:00 pm EST; focus group for Graduate Students

+Wednesday November 6; 9:30-11:00 am EST; focus group for Early Career Scientists and Engineers

+Wednesday November 6; 11:00 am-12:30 pm EST; focus group for Underrepresented Minorities

**To sign up, please fill out this short form by November 1 with your information and desired focus group session**
https://docs.google.com/forms/d/e/1FAIpQLSfAyYVV7eCEG-21hT4jzB5JD-rIxOmPqDE8xNRqMgWD8v853A/viewform.

**Input from everyone is vital to the success of this effort.**

Sincerely,

Lauren Garrison, on behalf of the DPP-CPP chairs

--------------

Participation in a focus group discussion was open to anyone who self-identified with the topic group. We did not allow advocates or managers who interacted with but were not themselves a member of a demographic to participate or observe. We limited participants to one of the sessions even if they identified with multiple groups both to allow more participants and because the same questions were asked in each discussion group. Sign ups were limited to 12 participants per session and because of response an additional discussion time for Early Career scientists and engineers was added and took place November 7, from 11:30am–1:00pm EST. Despite more sign-ups, participation in each discussion session ranged from 4–8 participants.

We gathered feedback in three areas:

1. About the process of the DPP-CPP—have people been involved, have they been heard, what can we do better, *etc.*

2. About the strategic plan we are developing and about OFES—what sorts of things/facilities/action items do they most want in the strategic plan, how do they interact with OFES now and how could it be better, *etc.*
3. About their current situation and outlook for the field—questions about the atmosphere in their current position, outlook for careers, discrimination/barriers to their careers, and opinions about the U.S. fusion science and technology program in a global context, *etc.*

The same script of questions about these three topics was used for each discussion group, but naturally different participants were more enthusiastic about different questions so all discussions were not identical. Notes were taken during the calls, but no attribution was made and no personally identifying details were used for any statements reported out of the focus groups. All the highlights reported here were echoed by many participants across many or all of the different discussion groups. That is not to say that every participant would agree with every statement made here. It is important to remember that the focus groups were very small by design to allow for full and supportive conversations of sometimes difficult topics. That means that even ideas that were supported by many in the focus group discussions do not necessarily represent the consensus view of that segment of the community or the community as a whole. The ideas that appeared to come out strongly from multiple focus groups were presented at the MFE+FM&T Knoxville Community Workshop in November 2019. Then, those ideas originating in the focus groups were treated as additional input to the CPP for discussion in the wider community. Some of the ideas that originated in the focus group discussions were embraced by the wider community and became part of this report. For example, some ideas for recommendations in the Workforce Cross-Cut area had their seed in the focus groups, and the idea of urgency was adopted across the Knoxville Workshop and adopted into the Mission Statement for FST.

Highlights and common themes from the focus group discussions:

1. Urgency
   Every focus group mentioned climate change and an urgent need to develop fusion energy as part of the solution. Younger people are keyed in to the climate issue and are going into the fusion field because they see it as a way to have an impact. There is a risk of losing this talent if we do not proceed quickly enough towards fusion energy. There was a general sentiment that the leaders in the fusion field are not aware of the need for urgency or are not taking it seriously enough.
2. Excitement
   There was a great deal of enthusiasm about the field of fusion and plasma science. Most participants have a positive outlook on their career in this field going forward. Many benefits to working in plasma science were mentioned including applications to energy, climate, health, advanced materials, and much more.
3. CPP Strategic Plan
   Everyone was excited about their own area of research and wanted it to continue and be represented in the CPP strategic plan. However, after the series of questions, a key takeaway from all the focus groups was that any sort of large new facility would be exciting to bring in new people and help unite the existing community. The participants felt that having the community agree on building something soon, even if it was not related to their personal research, would be the best way forward. They wanted more of a focus on innovation and things that will shake us out of the current stagnation. Additionally, for the FST section of the plan, materials and technology should be emphasized.

4. FES
   While the audience for the CPP Strategic Plan is broader than FES, it is an important component and a section of the questions in the discussion asked people's opinions of the current FES research portfolio and ideas for the future. It was clear that many people in these focus groups, especially graduate students and early career scientists and engineers, were not familiar with what FES does or how they could engage with FES. Funding decisions and proposals are often done at a high level by the leaders at an institution, so graduate students and early career people did not feel they had any knowledge of or input to the process. FES and leaders/mentors at each institution could take action to educate students and early career scientists and engineers about what FES does, what funding opportunities are available, and how to apply for grants.
5. Discrimination, unconscious bias, harassment, and barriers to advancement
   Focus group participants shared some experiences of explicit harassment or discrimination. There were many more instances of unconscious bias, off color comments, and uncomfortable situations. Unfortunately, nearly all focus group participants had either had a negative experience themself or knew of a colleague's negative experience in the field. Although the focus groups were discussing discrimination based on gender, age, race, and numerous other identifiers, the trend of the discussion matches with the findings of the NAS study [Nat. Acad. 2018b], which described the situation for sexual harassment of women with an iceberg metaphor. The tip of the iceberg is the small number of very serious incidents that would require disciplinary or legal action, but under the water are the vastly more numerous but individually less serious infractions. The pervasive harassment can be just as damaging as a single instance of sexual coercion. Participants in the focus groups felt that unconscious, institutionalized gatekeeping in this field tends to reward people that "look like" the majority. These microaggressions, uncomfortable comments, and other smaller incidents add up to create an uncomfortable atmosphere that causes people to leave the field.
6. Mentoring
   Most people in the focus groups felt they were receiving good mentoring, but almost all of this mentoring was informal. Because much is informal, some women and minorities feel at a disadvantage. Also, there are unique challenges for mentoring when many in our field are working for one institution but located at a different facility. The field is always evolving, and some early career people are working in new areas and have trouble finding mentors in their area at their institution. Because the field is hopefully expanding and starting new initiatives, this issue of early career scientists and engineers needing to change research topics and not being able to find mentors in their new areas was seen as something that would become a larger issue as we move forward.

7. Graduate Students
   Graduate students expressed some of the same concerns as the other focus groups but have some additional unique aspects. Focus group participants in this area discussed that they sometimes feel looked down upon and are not always taken seriously. Mental health challenges need to be considered; it was brought up by focus group participants that graduate students (in all science, not just plasma) are at a significantly increased risk for these issues compared to others of their same age. This discussion point in the focus groups agrees with findings; for example, Evans et al. saw over six times the rate of depression and anxiety in graduate students compared to the general population. Many schools do not have a plasma program, so graduate students would benefit from more learning and training opportunities at other universities and national laboratories. Reaching out to students at universities without a dedicated plasma program could be important for growing the workforce in the field.
8. Suggestions for improving diversity and inclusion
   The focus group participants had many ideas for how to improve conditions in the field.
   - Leaders in the field (possibly all PIs) should be required to take training on diversity issues, unconscious bias, cultural competence, bystander intervention, or similar.
   - There should be small funding grants available to women, underrepresented minorities, and early career scientists. This would help break out of a cycle of name recognition to attract new people to the field.
   - We all need to be more proactive and sensitive about recruiting in areas and regions which we do not normally target.
   - There should be a plasma physics graduate or post-doctoral fellowship aimed at women and minorities.
   - We should expand undergraduate internship opportunities for women, minorities, and students without traditional plasma training.
   - There was a desire to see acknowledgement from leaders that diversity matters.
   - To change anything, this effort has to be shared among all people, not just the under groups themselves.
   - FES could require a broader impacts section for their grants, similar to NSF.

## Appendix D. Glossary of Acronyms

| | |
|---|---|
| ADA | Americans with Disabilities Act |
| AI | Artificial Intelligence |
| ALCC | ASCR Leadership Computing Challenge |
| AMO | Atomic, Molecular, and Optical |
| APS | American Physical Society |
| ARC | MIT concept for fusion reactor |
| ARIES | Advanced Reactor Innovation and Evaluation Study |
| ARPA-E | Advanced Research Projects Agency—Energy |
| ASCR | Office of Advanced Scientific Computing Research |
| ASME | American Society of Mechanical Engineers |
| BCTF | Blanket Component Test Facility |
| BES | Office of Basic Energy Sciences |
| BoP | Balance of Plant |
| COMPASS-U | EU tokamak facility under facility in Czech Republic |
| COTS | Commercial Orbital Transportation Services |
| COV | Committee of Visitors |
| CPP | Community Planning Process |
| CX | Charge Exchange |
| D-T | Deuterium-Tritium |
| DAS | Data Acquisition System |
| DEI | Diversity, Equity, and Inclusion |
| DIII-D | Tokamak facility at General Atomics |
| DIR | Direct Internal Recycling |
| DKIST | Daniel K. Inouye Solar Telescope |
| DOD | U.S. Department of Defense |
| DOE | U.S. Department of Energy |
| DPP | Division of Plasma Physics |
| DPS | Discovery Plasma Science |
| DURIP | Defense University Research Instrumentation Program |
| EAST | Experimental Advanced Superconducting Tokamak (facility in China) |
| ECRF | Electron Cyclotron Range of Frequencies |
| ELM | Edge Localized Mode |
| ESA | European Space Agency |
| ET | Enabling Technology |
| EU-DEMO | EU concept for demonstration fusion reactor |
| FES | DOE Office of Fusion Energy Sciences |
| FESAC | DOE Office of Fusion Energy Sciences Advisory Committee |
| FLiLi | Flowing Liquid Lithium |
| FLIT | Flowing Liquid Torus |
| FMLA | Family Medical Leave Act |
| FM&T | Fusion Materials and Technology |
| FNSF | Fusion Nuclear Science Facility |
| FOA | Funding Opportunity Announcement |
| FPNS | Fusion Prototypic Neutron Source |
| FPP | Fusion Pilot Plant |
| FRC | Field Reversed Configuration |
| FST | Fusion Science and Technology |
| FTU | Frascati Tokamak Upgrade (facility in Italy) |
| FW | First Wall |
| GPS | General Plasma Science |
| GPU | Graphics Processing Unit |
| GRB | Gamma Ray Burst |
| H&CD | Heating and Current Drive |
| HED | High Energy Density |
| HEDP | High Energy Density Physics |
| HEDLP | High Energy Density Laboratory Plasma |
| HIDRA | Hybrid Illinois Device for Research and Applications (facility at U. Illinois) |
| HSX | Helically Symmetric Experiment (facility at U. Wisconsin—Madison) |
| HTNSD | High Temperature Nuclear Structural Design |
| HTS | High Temperature Superconductor |
| HTSDC | High Temperature Structural Design Criteria |
| ICF | Inertial Confinement Fusion |
| ICRF | Ion Cyclotron Range of Frequencies |
| I-DTT | Italian Divertor Test Tokamak |
| IEEE | Institute of Electrical and Electronics Engineers |
| IFE | Inertial Fusion Energy |
| INCITE | Innovative and Novel Computational Impact on Theory and Experiment |
| INFUSE | Innovation Network for Fusion Energy |
| IPA | ITER Project Associates |
| IRP | ITER Research Plan |
| ISDC | ITER Structural Design Criteria |
| ITER | International tokamak facility under construction in France |
| ITPA | ITER Tokamak Physics Activity |
| JET | Joint European Torus (EU facility in UK) |
| JT-60SA | JT-60 “Super Advanced” tokamak (facility under construction in Japan) |
| JUDITH2 | Electron beam facility in Germany |
| KSTAR | Korea Superconducting Tokamak Advanced Research (facility in Korea) |
| LHD | Large Helical Device heliotron (facility in Japan) |
| LHRF | Lower Hybrid Range of Frequencies |
| LM | Liquid Metal |
| LMX | Liquid Metal Experiment (facility at PPPL) |
| LPI | Laser-Plasma Interaction |
| LSN | Lower Single Null |
| LTP | Low Temperature Plasma |
| LTX | Lithium Tokamak Experiment (facility at PPPL) |
| MAGNUM-PSI | Linear facility for PMI in the Netherlands |
| MAST-U | Mega Ampere Spherical Torus Upgrade (facility in UK) |
| MD | Measurement and Diagnostics |
| MEC | Matter in Extreme Conditions (facility at SLAC) |
| MELCOR | Engineering code for nuclear systems |
| MFE | Magnetic Fusion Energy |
| MIF | Magneto-Inertial Fusion |
| ML | Machine Learning |
| MMS | Magnetospheric Multiscale satellite experiment |
| MPEX | Material Plasma Exposure Experiment |
| MRI | Major Research Instrumentation Program |
| MTF | Magnetized Target Fusion |
| NASA | National Aeronautics and Space Administration |
| NCSX | National Compact Stellarator Experiment |
| NIF | National Ignition Facility (facility at LLNL) |
| NIH | National Institutes of Health |
| NNSA | National Nuclear Security Administration |
| NRC | Nuclear Regulatory Commission |
| NSF | National Science Foundation |
| NSTX-U | National Spherical Torus Experiment Upgrade (facility at PPPL) |
| NTUF | New Tokamak User Facility |
| NUF | National Undergraduate Fellowship |
| OTS | Optical Thomson Scattering |
| PAL | Positron Annihilation Lifetime |
| PAM | Disruption Prediction, Avoidance, and Mitigation |
| PFC | Plasma Facing Component |
| PFPO | ITER Pre-Fusion-Power Operation |
| PI | Principal Investigator |
| PISCES | Plasma Surface Interaction Experimental Facility (facility at UCSD) |
| PMI | Plasma-Material Interaction |
| PPP | Public-Private Partnership |
| PR | Program Recommendation |
| PS&FE | Plasma Science and Fusion Energy |
| PSP | Parker Solar Probe |
| QAS | Quasi-Axisymmetric |
| QED | Quantum Electrodynamics |
| QHS | Quasi-Helical Symmetric |
| QS | Quasi-Symmetric |
| RAMI | Reliability, Availability, Maintainability, and Inspectability |
| RF | Radiofrequency |
| SCGSR | Office of Science Graduate Student Research |
| SciDAC | Scientific Discovery through Advanced Computing |
| SCSC | South Carolina Statewide Collaboration |
| SD | Science Driver |
| SMR | Small Modular Reactor |
| SO | Strategic Objective |
| SOFE | Symposium on Fusion Engineering |
| SOL | Scrape-off Layer |
| SPARC | MIT/CFS HTS tokamak concept |
| SPI | Shattered Pellet Injection |
| SSP | Stockpile Stewardship Program |
| ST40 | Tokamak Energy spherical torus facility in UK |
| TBM | Test Blanket Module |
| TBR | Tritium Breeding Ratio |
| TC | Theory and Computation |
| TCAP | Thermal Cycling Absorption Process |
| TCV | Tokamak à Configuration Variable (facility in Switzerland) |
| TFTR | Tokamak Fusion Test Reactor (facility formerly at PPPL) |
| TJ-II | Heliac facility in Spain |
| TPE | Tritium Plasma Experiment (facility at INL) |
| TRL | Technology Readiness Level |
| UKAEA | United Kingdom Atomic Energy Authority |
| V&V | Verification and Validation |
| VNS | Volumetric Neutron Source |
| W7-X | Wendelstein 7-X stellarator (facility in Germany) |
| WF | Workforce Development |
| XFEL | X-Ray Free Electron Laser |

# *Appendix E. The Community Planning Process*

(A description of the process will be added in an addendum to this report.)

# *Appendix F. Community Workshops*

## High Energy Density Physics Workshop, July 16-17, 2019, College Park, MD

## Agenda for the HEDP Community Workshop
## July 16-17, 2019

### Tuesday, July 16

**Session 1:** **Plenary session (Introduction and Overview)** Remote Link

9:00 – 9:40 Welcome and Overview of DPP-CPP effort Carolyn Kuranz

**Session 2:** **Plenary session (selected lightning talks)** Remote Link

9:40 – 10:30 Plenary Session: Lightning Talks I
(Hu, Willingale, Feister/Orban, Murphy, Vogman)

10:30 – 10:45 Coffee Break

10:45 – 12:15 Plenary Session: Lightning Talks II
(Gatu Johnsen, Ross, Eggert, Dyer, Herrmann, Fox, Wei, Seidl)

12:15 – 13:15 Lunch break

**Session 3:** **Break-out session I (lightning talks, and start defining initiative quad charts**)

13:15- 14:15 Breakout Session: Lightning Talks

**Fundamentals I** (Hydrodynamics, Magnetized HEDP, Laboratory Astrophysics, Computation I) Remote Link

**Fundamentals II** (Nuclear Physics, Warm Dense Matter & Materials, HED Atomic Physics) Remote Link

**Fundamentals III** (Nonlinear optics and Laser Plasma Interactions, Relativistic HED and High Field Science, Intense Beams and Particle Acceleration, Computation II) Remote Link

**Facilities and Diagnostics** (Laser Facilities, Pulsed Power, X-ray Light sources, Radiation sources, Enabling Technology, and Diagnostics) Remote Link

**Inertial Fusion Energy** (Driver and Reactor Technology and High Yield Target Physics) Remote Link

14:15 – 16:15 Breakout Session: Discussions and prepare summary presentations)

Coffee break at 15:30

**Session 4:** **Plenary session** Remote Link

16:15 – 16:30 Fundamentals I
16:30 – 16:45 Fundamentals II
16:45 – 17:00 Fundamentals III
17:00 – 17:15 Facilities and Diagnostics
17:15 – 17:30 IFE

17:30 Adjourn

## Wednesday, July 17

9:00 Welcome and Overview

**Session 5: Break-out session (finalize quad charts for the different initiatives)**

9:30 – 12:00 (Coffee Break at 10:30)

**Fundamentals I** (Leads: Merritt, Tzeferacos, Schaeffer, Doss) Remote Link
**Fundamentals II** (Leads: Zylstra, Starrett, Fournier, Gleason) Remote Link
**Fundamentals III** (Leads: Willingale, Geddes, Geissel) Remote Link
**Facilities and Diagnostics** (Leads: Rocca, Ditmire, Douglass, Pickworth) Remote Link
**Inertial Fusion Energy** (Leads: Thomas) Remote Link

12:00 – 13:00 Lunch break

**Session 6: Plenary session (present final quad charts and summaries)** Remote Link

13:00 – 13:15 Hydrodynamics
13:15 – 13:30 Magnetized HEDP
13:30 – 13:45 Laboratory Astrophysics
13:45 – 14:00 Computation I (Fluid modeling)
14:00 – 14:15 Nuclear Physics
14:15 – 14:30 Warm Dense Matter & Materials
14:30 – 14:45 HED Atomic Physics
14:45 – 15:00 Computation II (Kinetic Modeling)
15:00 – 15:15 Relativistic HED and High Field Science
15:15 – 15:30 Intense Beams and Particle Acceleration
15:30 – 15:45 Nonlinear optics and Laser Plasma Interactions
15:45 – 16:00 Pulse Power Facilities
16:00 – 16:15 X-ray Light Sources
16:15 – 16:30 Laser Facilities
16:30 – 16:45 Radiation Sources and Enabling Technology
16:45 – 17:00 Diagnostics
17:00 – 17:15 IFE Driver and Reactor Technology

17:15 Wrap-up and assignments

17:30 Adjourn

Joint MFE and FM&T Workshop, Jul 22-26 2019, Madison WI

**Please Note : All submitted initiatives will be discussed in at least one expert group or cross-cut breakout session. The schedule for these sessions will be posted separately.**

**The room assignments for breakout sessions are listed on the breakout agenda**

**Registration is available starting at 7:30 AM Monday and Tuesday mornings in the Atrium of Engineering Hall. Coffee will be available during registration.**

*Talks are linked by the speaker's name (when approved by the author).*
*Initiative papers are linked to the talk title.*

## Monday (July 22nd)

1800 Engineering Hall, UW-Madison Campus (https://www.map.wisc.edu)
https://fusion.zoom.us/j/337056531 (WS)
Scribe notes

| Time | Speaker/Topic | Notes |
|---|---|---|
| 8:20 | Schmitz | Meeting logistics |
| 8:30 | Solomon | Welcome and Meeting Goals |
| 9:30 | Coffee break | |

| | | | | | |
|---|---|---|---|---|---|
| **Theme I : Identifying the Elements for a Balanced MFE Program in the 2020s**<br>Chair : Saskia Mordijck<br>Scribes : Guttenfelder, Petty | | | | **Theme I : Materials and Technology Future Directions**<br>Chair : Paul Humrickhouse<br>Scribes : Tynan, Donovan | |
| *Talks in these sessions will be 12 minutes long with 8 minutes for discussion each* | | | | | |
| 1800 Engineering Hall<br>https://fusion.zoom.us/j/337056531 (WS)<br>Scribe notes | | | | 1610 Engineering Hall<br>https://zoom.us/j/5669851878 (OS-1)<br>Scribe notes | |
| 10:00 | Nazikian | A national initiative to accelerate ITER research | | 10:00 | NAS report and discussion |

| | | | | | | |
|---|---|---|---|---|---|---|
| | | and maximize the US return on ITER | | | | |
| 10:20 | Hill | Realize the Full Potential of DIII-D to Advance Development of Cost-Effective Fusion Power Plants | | 10:20 | | |
| 10:40 | Battaglia | The NSTX-U Facility in the 2020s: Advancing the Physics Basis for Configuration Optimization Toward a Compact Fusion Pilot Plant | | 10:40 | Kessel | A Critical Integration Step for Fusion Blankets, the Blanket Component Test Facility |
| 11:00 | Greenwald | Collaborations On The SPARC Device | | 11:00 | Duckworth | High Field Superconducting Magnet Technology Development for Fusion Devices and Science Missions |
| 11:20 | Bader | A U.S. Intermediate Scale Stellarator Experiment | | 11:20 | Baylor (Shimada) | Integration of the Fueling and Pumping in the Fusion Energy Fuel Cycle |
| 11:40 | Lazerson | International stellarator research in support of a low capital cost pilot plant | | 11:40 | Kolasinski | Facilities Instrumentation for Components Removed from a Compact Pilot Plant |
| 12:00 | McCollam | Reversed-field pinch research toward Ohmic ignition at high engineering beta | | 12:00 | Shimada | In-pile fission irradiation to advance tritium-related material and technology challenges |
| 12:20 | Discussion | | | 12:20 | Discussion | |
| 13:00 | Lunch | | | | | |
| 14:30 | Expert Group Breakout Session – Link to Agenda | | | | | |
| 15:30 | Coffee | | | | | |
| 16:00 | Expert Group Breakout Session – Link to Agenda | | | | | |
| 17:30 | Adjourn | | | | | |

## Tuesday (July 23rd)

| **Theme II: Science and Technology Challenges on the Path to Fusion**<br>*Talks in this session will be 12 minutes plus 8 minutes of discussion each* | | | | |
|---|---|---|---|---|
| | 1610 Eng. Hall<br>Zoom (NH)<br>Scribe notes | 1800 Eng. Hall<br>Zoom (WS)<br>Scribe notes | 2255 Eng. Hall<br>Zoom (OS-2)<br>Scribe notes | 3534 Eng. Hall<br>Zoom (OS-1)<br>Scribe notes |
| Time | **Power Handling**<br>Chairs: Matt Reinke & George Tynan<br>Scribes: Tynan, Winfrey | **Heating and Sustainment**<br>Chair: Rich Magee<br>Scribes: Petty, Collins | **Modeling and Design**<br>Chair: Chris Holland & Brad Merrill<br>Scribes: Ferraro, Lumsdaine | **Materials, Blankets, Diagnostics**<br>Chair: Robert Kolasinski<br>Scribes: Donovan, Zinkle |
| 8:30 | Canik - Initiative | Brookman - Init. | Neilson - Initiative | Pint (Katoh) - Init. |
| 8:50 | Yoda - Initiative | Diem - Initiative | Lyons - Initiative | El-Guebaly - Init |
| 9:10 | Youchison - Init. | Bonoli (Baek) - Init. | Meneghini (Smith) - Initiative | Bohm - Initiative |
| 9:30 | Baldwin - Initiative | Caughman (Pinsker) - Initiave | Snead - Initiative | Ferry - Initi |
| 9:50 | Coffee Break | | | |
| 10:10 | Goldston - Initiative | Ono - Initiative | Kessel - Initiative | Nygren - Initiative |
| 10:40 | Andruczyk - Init | Raman - Initiative | Ghoniem - Initiative | Sowder - Initiative |
| 11:00 | Gray - Initiative | Bongard - Initiative | Ying (Humrickhouse) - Initiative | Parish - Initiative |
| 11:20 | Rapp - Initiative | Sabbagh - Initiative | Zarnstorff - Init. | Hu (Taylor) - Init. |
| 11:40 | Discussion | Discussion | Discussion | Discussion |
| 12:20 | Lunch | | | |
| 13:45 | Expert Group Breakout Session – Link to Agenda | | | |
| 15:30 | Coffee | | | |
| 15:45 | Expert Group Breakout Session – Link to Agenda | | | |
| 17:30 | Adjourn | | | |
| | | | | |

| **Workshop Reception** | |
|---|---|
| 19:00 - 21:00 | Reception for the Joint MFE/FM&T and DPS workshops (cash bar)<br>Memorial Union, Tripp Commons (2nd Floor), UW-Madison campus<br>(https://www.map.wisc.edu) |

## Wednesday (July 24th)

| **Theme III: Cross-cutting Opportunities for the Fusion Energy Sciences**<br>Chair: George Tynan<br>Scribes: Nygren, Hegna<br>*Talks in this session will be 10 minutes plus 5 minutes of discussion each* | | |
|---|---|---|
| 1800 Engineering Hall (held jointly with the DPS workshop)<br>https://fusion.zoom.us/j/337056531 (WS)<br>Scribe notes | | |
| 8:30 | Vay | Integrated ecosystem of advanced simulation tools for plasma modeling |
| 8:45 | Kolasinski (Allain) | High Fidelity Surface Diagnostics for Plasma-Material Interactions |
| 9:00 | Field | Adoption of Advanced Manufacturing for Advancing and Implementing Materials for Fusion Energy Applications |
| 9:15 | Murphy | Equity and inclusion in plasma physics |
| 9:30 | Discussion | |
| 10:00 | Coffee | |
| **Theme IV: The Role of Public/Private Partnership on the Path to Fusion**<br>Chair: Saskia Mordijck,<br>Scribes: Kolasinski, Reinke | | |
| 1800 Engineering Hall<br>https://fusion.zoom.us/j/337056531 (WS)<br>Scribe notes | | |
| 10:30 | Youchison | INFUSE (15 + 5 min) |
| 10:50 | Hsu | ARPA-E (20 + 5 min) |
| 11:15 | Holland | Fusion Industry Association (25 + 5 min) |
| 11:45 | Discussion | |
| 12:15 | Lunch | |

| | |
|---|---|
| 13:45 | Cross-Cut Breakout Session – Link to Agenda |
| 15:30 | Coffee |
| 16:00 | Cross-Cut Breakout Session – Link to Agenda |
| 17:50 | Adjourn |

## Thursday (July 25th)

**Theme V: The Path to a Fusion Pilot Plant**
Chair: Steve Zinkle, Walter Guttenfelder
Scribes: Hughes, Caughman
*Talks in this session will be 12 minutes plus 8 minutes of discussion each*

1800 Engineering Hall
https://fusion.zoom.us/j/337056531 (WS)
Scribe notes

| | | |
|---|---|---|
| 8:30 | Wade | Near-Term Initiatives to Close the Fusion Technology Gaps to a Compact Fusion Pilot Plant |
| 8:50 | Katoh (Snead) | Accelerated Development of Materials as the Enabling Technology for Fusion Energy |
| 9:10 | Egle | The Fusion Prototypic Neutron Source (FPNS), an Affordable, Timely 14 MeV Fusion Neutron Irradiation Facility for Near-term Fusion Material Testing |
| 9:30 | Sutherland | The need for a diverse fusion energy research and development portfolio for the pursuit of economically competitive fusion power |
| 9:50 | Coffee Break | |
| 10:20 | Buttery | A National Research Program to Prepare for a Compact Fusion Pilot Plant by Resolving the Physics of Sustained High Power Density Conditions |
| 10:40 | Menard | Development of Mission Need and Preliminary Design of a Sustained High Power Density Tokamak Facility |
| 11:00 | Merrill | Developing the framework for licensing a fusion power plant |
| 11:20 | Kessel | The Compact Fusion Pilot Plant Mission Definition, Design, and Required R&D Program |
| 11:40 | Gates (Maurer) | The Stellarator Path to a low-cost Pilot Plant |
| 12:00 | Discussion | |

| | |
|---|---|
| 12:40 | Lunch |
| 14:00 | Expert Group Breakout Session – Link to Agenda |
| 15:30 | Coffee |
| 15:45 | Expert Group Breakout Session – Link to Agenda |
| 17:30 | Adjourn |

## Friday (July 26th)

| | | |
|---|---|---|
| **Theme VI: Summary of Workshop Accomplishments and Moving Forward** | | |
| 1800 Engineering Hall<br>https://fusion.zoom.us/j/337056531 (WS)<br>Scribe notes | | |
| 8:30 | MFE and FM&T Expert Group Discussions (same breakout rooms as Thursday afternoon) | |
| 10:30 | Coffee | |
| 11:00 | MFE and FM&T Expert Group Summaries | |
| 12:30 | Co-Chairs | Summary, homework, and the path forward |
| 1:00 | Adjourn | |

## DPS Workshop, July 23-25, Madison, WI

Day 1 – Tuesday July 23, 2019 (***What is the status of Discovery Plasma Science?)***

**Session 1: Day 1 Opening**
**Location: 1227 Engineering Hall, UW-Madison** (https://map.wisc.edu/)
**Chair: John Sarff, Scribe: David Schaffner**
**Zoom connection:** https://zoom.us/j/5100133208

| | |
|---|---|
| 8:30-8:40 | Meeting logistics – Baalrud |
| 8:40-9:00 | Description of Workshop Function, Goals, Outcomes – Baalrud |
| 9:00-9:20 | Open Discussion regarding Function, Goals, Outcomes, Definition of DPS |

**Session 2: Reports on Reports**
**Location: 1227 Engineering Hall, UW-Madison**
**Chair: Hantao Ji, Scribe: Stephen Vincena**
**Zoom connection:** https://zoom.us/j/5100133208
In this first session, a number of speakers will present overviews of recent reports on various aspects of plasma physics related to discovery science.

| | |
|---|---|
| 9:20-10:00 | Report of the *Panel on Frontiers of Plasma Science* – Fred Skiff |
| 10:00-10:40 | Report on *Workshop on Opportunities, Challenges, and Best Practices for Basic Plasma Science User Facilities* – Earl Scime (Remote) |
| 10:40-10:55 | Coffee Break (Atrium of Engineering Hall) |
| 10:55-11:35 | Report on *Enabling a Future Based on Electricity Through Non-Equilibrium Plasma Chemistry* -- Mark Kushner |
| 11:35-12:15 | Report on *Workshop on Opportunities in Plasma Astrophysics* – Ellen Zweibel |
| 12:15-12:30 | Report on NAS Plasma 2020 Progress – Mark Kushner |
| 12:30-2:00 | Lunch |

**Session 3: Group Discussion—Broad Overview of Plasma Science and Discussion of Reports**
**Location: Mechanical Engineering (ME), across the street from Engineering Hall**
*Group 1: ME1270 - High energy laboratory astrophysics, and solar and magnetospheric laboratory astrophysics) - Moderator: David Schaffner*
*Group 2: ME2180 - Low temperature plasmas, plasma surface interactions - Moderator: Steve Shannon*
*Group 3: ME2065 - Single component plasmas, dusty plasmas, laser plasma interactions and theory - Moderator: Steve Vincena*
For the discussion breakouts, participants will split into groups (as listed above). The groups will be given the same set of topics/questions to go through (Group Discussion Report Worksheet). Each group will have a discussion moderator or two whose role will be to keep the discussion moving through the topics and avoiding too many tangents. Each group will also have a discussion scribe to keep minutes of the discussion.

| | |
|---|---|
| 2:00-3:00 | Group Discussion Part 1 (Introductions, start of question discussion) |
| 3:00-3:30 | Coffee Break (Atrium of Engineering Hall) |
| 3:30-5:00 | Group Discussion Part 2 (Remainder of Questions) |
| 5:00-:5:10 | Discussion of Day 2 activities and Adjourn |
| 5:00-5:30 | Group Moderators and Scribes meet to consolidate notes and prepare brief reports, incorporate comments from remote participation (remaining workshop participants can leave) |

**Reception for Joint MFE/FM&T and DPS workshops**

| | |
|---|---|
| 7:00-9:00 | Memorial Union, Tripp Commons (2nd Floor) (https://map.wisc.edu/) |

Day 2 – Wednesday, July 24, 2019 (***Overview of DPS and Joint Cross-Cuts)***

**Session 4: Joint Session with MFE/FM&T—Cross Cutting Opportunities for the FES**
**Location: 1800 Engineering Hall, UW-Madison**
**Chair: George Tynan, Scribes: Nygren, Hegna**
**Zoom connection: https://fusion.zoom.us/j/337056531**

| | |
|---|---|
| 8:30-8:45 | Integrated Ecosystem of Advanced Simulation Tools for Plasma Modeling - Vay |
| 8:45-9:00 | High Fidelity Surface Diagnostics for Plasma-Material Interactions - Kolasinski |
| 9:00-9:15 | Adoption of Advanced Manufacturing for Advancing and Implementing Materials for FES – Field |
| 9:15-9:30 | Equity and inclusion in plasma physics - Murphy |
| 9:30-10:00 | Discussion |
| 10:00-10:10 | Coffee Break (Atrium of Engineering Hall) |

**Session 5: Broad Overview Talks of Discovery Plasmas Science (20+5min)**
**Location: 1227 Engineering Hall, UW-Madison**
**Chair: Steve Shannon, Scribe: Daniel Den Hartog**
**Zoom connection:** https://zoom.us/j/5100133208

| | |
|---|---|
| 10:10-10:40 | Brief Intro and Brief Reports from Group Moderators and Scribes to discuss results of Day 1 Breakout, major themes, conclusions, recommendations, etc. |
| 10:40-11:05 | Troy Carter (UCLA) - Plasma Lab Astrophysics/Basic Plasma Experiment |
| 11:05-11:30 | Bruce Remington (LLNL) - High Energy Lab Astrophysics |
| 11:30-11:55 | Hans Rinderknecht (LLE) - Laser Plasma Interactions |
| 11:55-12:20 | C. Fred Driscoll (UCSD) - Single Component Plasmas |
| 12:20-12:45 | Peter Bruggeman (U Minnesota) - Low Temperature Plasmas |
| 12:45-1:45 | Lunch |

**Session 6: Breakout into Cross-Cutting Groups with Members of MFE and FM&T**
**Location: Engineering Hall**
*Enabling Technology: EH2255*
*Measurements and Diagnostics: EH2309*
*Theory and Modeling: EH2239*
*Workforce Development: EH2345*
The remaining time will be spent in a group breakout session where members of the cross-cutting groups from the three topical areas at Madison will meet. After a brief overview of tasks and goals for the cross cutting groups, the remaining time, until about 5pm, will be free time for cross-cutting groups to organize and discuss without any hard end time.

| | |
|---|---|
| 1:45-2:00 | Introduction to Cross-Cutting Group Meetings |
| 2:00-6:00 (or into evening) | Cross-Cutting Group Meetings |

## Day 3 – Thursday, July 25, 2019 (***Presentation and Discussion of Initiatives)***

**Session 7: Day 3 Opening**
**Location: 1227 Engineering Hall, UW-Madison**
**Zoom connection:** https://zoom.us/j/5100133208

| | |
|---|---|
| 8:30-8:45 | Day 3 Opening Remarks and Description of Sessions |

**Session 8: Initiatives Presentation (10+5min)**
**Location: 1227 Engineering Hall, UW-Madison**
**Chair: Scott Baalrud, Scribes: David Schaffner, Steve Shannon**
**Zoom connection:** https://zoom.us/j/5100133208

*Initiative Presentations Session A*

| | |
|---|---|
| 8:45-9:10 | Carolyn Kuranz (Report of HEDP Meeting) *20+5min |
| 9:10-9:25 | Cary Forest (Stellar Wind Tunnels Presentation) |
| 9:25-9:40 | Fatima Ebrahimi (The Plasma Universe Initiative, Reconnection Presentation) |
| 9:40-9:55 | Eva Kostadinova (Controlling Charging in Dusty Plasmas, Presentation) |
| 9:55-10:10 | Mike Cuneo (Multi-Scale, Multi-Physics Advanced Plasma Hybrid Algorithms, Modeling, and Simulation Presentation) |
| 10:10-10:25 | Philip Efithmion (National Initiative in Low Temperature Plasma, Presentation) |
| 10:25-10:45 | Coffee Break (Atrium of Engineering Hall) |

*Initiative Presentations Session B*

| | |
|---|---|
| 10:45-11:00 | Mark Kushner (Plasma Physics Challenges in Low Temperature Plasma, Presentation) |
| 11:00-11:15 | Greg Severn (Sheath Physics Initiative, Presentation) |
| 11:15-11:30 | Jacob Roberts (Ultracold Neutral Plasmas for Controllable and Precision Plasma Physics, Presentation) |
| 11:30-11:45 | Paul Bellan (Thoughts on Discovery Science, Presentation) |
| 11:45-12:00 | Thomas Schenkel (Quantum Information Science and Fusion Energy Sciences) |
| 12:00-12:15 | Jeroen van Tilborg (Light sources from Laser-Plasma Accelerators, Presentation) |
| 12:15-1:30 | Lunch |

*Initiative Presentations Session C*

| | |
|---|---|
| 1:30-1:45 | Troy Carter (Facility for the Study of Astrophysical Processes: On behalf of Walter Gekelman, Presentation) |
| 1:45-2:00 | David Schaffner (A Large Scale Turbulent Plasma Wind Tunnel: On behalf of Michael Brown, Presentation) |
| 2:00-2:15 | Steve Shannon (Reactive Low Temperature Plasmas: On behalf of Katharina Stapelman, Presentation) |
| 2:15-2:45 | Coffee Break |

**Session 9: Reconvene of Expert Groups or Advocacy Groups for Discussion of Initiatives**
**Location: Mechanical Engineering (ME), across the street from Engineering Hall**
*Group 1: ME1270 - High energy laboratory astrophysics, and solar and magnetospheric laboratory astrophysics) - Moderator: David Schaffner*
*Group 2: ME2180 - Low temperature plasmas, plasma surface interactions - Moderator: Steve Shannon*
*Group 3: ME2065 - Single component plasmas, dusty plasmas, laser plasma interactions and theory - Moderator: Steve Vincena*
In the same working groups from Day 1, we'll gather in these groups to discuss the following topics stemming from the presented and submitted initiatives as well as plan for a second round of initiatives. See the Group Discussion Report Worksheet for a list of specific topics. Group moderators will get any input from remote participation to include in this discussion.

| | |
|---|---|
| 2:45-4:45 | Group Discussion |

**Session 10: Full Workshop Discussion**
**Location: 1227 Engineering Hall, UW-Madison**
**Chair: Scott Baalrud**
**Zoom connection:** https://zoom.us/j/5100133208

| | |
|---|---|
| 4:45-5:30 | Final Full Workshop Discussion, Feedback, Farewell |

## High Energy Density Physics Workshop, Nov 12 - 14, 2019 Menlo Park, CA

**Agenda for the 2nd HEDP Community Workshop**
**SLAC National Accelerator Laboratory, Menlo Park, CA**
**November 12-14, 2019**

**Please see end for list of remote links.**

## Tuesday, November 12

| | | |
|---|---|---|
| 9:00 | Gathering (Panofsky Lobby, SUSB B53) | |
| **Session 1:** | **Introductory plenary session (Panofsky Auditorium, SUSB B53)** | |
| 9:30 | Director's welcome | |
| 9:35 | Overview of the workshop | C. Kuranz |
| 9:45 | FES Perspective | K. Akli |
| 10:15 | Coffee break (Panofsky Lobby, SUSB B53) | |
| 10:30 | Review of 1st workshop (10 min each) | |
| | 1) Atomic Physics | A. Safronova |
| | 2) High-Intensity Laser Plasmas | C. Geddes |
| | 3) Hydrodynamics, Magnetized HED, and Lab Astro | D. Schaeffer |
| | 4) Inertial Fusion Energy | C. Thomas |
| | 5) Laser Facilities and Radiation Sources | T. Ditmire |
| | 6) Measurements and Diagnostics | L. Pickworth |
| | 7) Nuclear Physics | A. Zylstra |
| | 8) Pulsed Power Facilities | J. Douglass |
| | 9) Warm Dense Matter | A. Gleason |
| 12:00 | Lunch break [Group Photo!] (lunch on your own, e.g., SLAC Cafe' or local options) | |
| **Session 2:** | **Tent-Pole initiatives plenary session (Panofsky Auditorium, SUSB B53)** | |
| 13:30 | Overview of the 5 tent-pole initiatives: | |
| | 1) LasernetUS | T. Ditmire |
| | 2) Frontier HED Science on the LCLS MEC Instrument | A. Gleason |
| | 3) Magnetized HED Science and Pulsed Power Technology | D. Schaeffer |
| | 4) Radiation and Particle Sources | J. Rocca |
| | 5) Inertial Fusion Energy Science and Technology Program | C. Thomas |
| **Session 3:** | **Tent-pole initiatives breakout session (5 breakout rooms in B53)** | |
| 14:30 | Discussion of the 5 tent-pole initiatives. | |
| | | |
| 15:30-15:50 | Coffee break (Panofsky Lobby, SUSB B53) | |
| | | |
| **Session 4:** | **Review plenary session (Panofsky Auditorium, SUSB B53)** | |
| 17:30 | Presentations of the discussions of the tent-pole initiatives from Session 3. | |
| | | |
| 18:30 | Adjourn | |

## Wednesday, November 13

| | | |
|---|---|---|
| 9:00 | Gathering (Kavli Auditorium Lobby, B51) | |
| | | |
| **Session 5:** | **Morning plenary session (Kavli Auditorium, B51)** | |
| 9:30 | Summary of previous IFE reports | S. Finnegan |
| 9:45 | The Brightest Light Initiative | T. Ditmire |
| | | |
| **Session 6:** | **Thematic initiatives breakout session (3 breakout rooms in SUSB B53)** | |
| 10:00 | Discussion of the 3 thematic initiatives: | |
| | 1) Frontier HED Science | |
| | 2) Innovative HED Technology | |
| | 3) Inertial Fusion Energy and Technology | |
| | | |
| 10:30-10:50 | Coffee break (Kavli Auditorium Lobby, B51) | |
| | | |
| 12:00 | Lunch break (on your own, e.g., SLAC Cafe' or list local options) | |
| | | |
| **Session 7:** | **Review and cross-cut plenary session (Kavli Auditorium B51)** | |
| 13:30 | Presentations of the discussions of the 3 thematic initiatives from Session 6. | |
| | | |
| 14:30 | Overview of cross-cutting areas: | |
| | 1) Measurement and Diagnostics | J. Frenje |
| | 2) Theory and Computation | A. Arefiev |
| | 3) Enabling Technology | T. Ditmire |
| | 4) Workforce Development | S. Finnegan |

| | |
|---|---|
| 15:15 | Coffee (Kavli Auditorium Lobby, B51) |

**Session 8: Cross-cut breakout session (4 breakout rooms for each cross-cut in SUSB B53)**

| | | |
|---|---|---|
| 15:30 | Discussion of the cross-cutting components of the initiatives. | |
| 17:30 | 1 hour SLAC Tour (shuttles from Kavli over to Far Experimental Hall) | |
| 18:00 | Adjourn | |

## Thursday, November 14

| | | |
|---|---|---|
| 9:00 | Gathering (Kavli Auditorium Lobby, B51) | |

**Session 9: Cross-cut plenary session (Kavli Auditorium, B51)**

| | | |
|---|---|---|
| 9:30 | Presentations of the discussions of cross-cuts from Session 8. | |
| 10:30 | Coffee break (Kavli Auditorium Lobby, B51) | |
| 10:50 | Moderated discussion of the HED strategic vision for FES | |
| 12:30 | Lunch break (on your own, e.g., SLAC Cafe' or list local options) | |

**Session 10: IFE Plenary session (Panofsky Auditorium, SUSB B53)**

| | | |
|---|---|---|
| 14:00 | IFE town hall | |
| 15:00-15:20 | Coffee break (Kavli Auditorium Lobby, B51) | |

**Session 11: Plenary session (Panofsky Auditorium, SUSB B53)**

| | | |
|---|---|---|
| 16:30 | Concluding remarks | C. Kuranz |
| 17:00 | Adjourn | |

## Joint MFE and FM&T Workshop, Nov 18-22, Knoxville, TN

**Please Note. The agenda of the workshop is going to depend on the progress made each day. This agenda is subject to change at any time.**

For a detailed description of what topics are covered in each overarching and strategic objectives, see the draft PC plan:
https://drive.google.com/file/d/1pe0cH7mKfw0EIAmwJZJdTXHhhLOvxu6c/view

**Remote Participation**
https://tennessee.zoom.us/j/748036175
Plenary sessions will be broadcast at the Zoom number above. If you would like to participate remotely in discussion sessions, please register here:
https://docs.google.com/spreadsheets/d/1LRhetIODFhYGqjH_xXKIamiJJ9KLHxkmFJtYNkIuo_g

| **Monday** | | | |
|---|---|---|---|
| | | | |
| **Time** | **Topic** | **Presenter/Lead** | **Location** |
| 7:00 | Registration, Breakfast Snacks and Coffee | | Atrium |
| 8:30 | Welcome, Logistics | Donovan | Ballroom-413 |
| 8:50 | Opening Presentation | Garrison | Ballroom-413 |
| 9:30 | Presentation of Draft Strategic Plan | Grierson/Guttenfelder | Ballroom-413 |
| 10:10 | Clarifying Questions on Draft Plan | Sarff | Ballroom-413 |
| 10:30 | Coffee Break | | Atrium |
| 11:00 | Full Group Discussion of Draft Plan | Sarff (Scribe Notes) | Ballroom-413 |
| 12:00 | Lunch | | |
| 1:30 | Discussion of Programmatic Direction and Elements | Howard | Ballroom-413 |
| 1:45 | Breakout Discussion on High Level Plan | All | Breakout Rooms |
| 3:10 | Coffee Break | | Atrium |
| 3:30 | Breakout Discussion on Programmatic | All | Breakout Rooms |
| | Issues/Approach | | |
| 4:40 | Reconvene for Full Group Discussion | Co-Chairs (Notes) | Ballroom-413 |
| 5:20 | Polling on High Level Plan and Programmatic | Co-Chairs | Ballroom-413 |
| 5:40 | Adjourn | | |

| **Tuesday** | | | |
|---|---|---|---|
| | | | |
| **Time** | **Topic** | **Presenter** | **Location** |
| 7:30 | Breakfast Snacks and Coffee | | Atrium |
| 8:30 | Summary of Monday's Discussion | Solomon | Ballroom-413 |
| 8:35 | Overarching A,B,C / D,E,F | Holland / Donovan | Ballroom-413 |
| 9:00 | Breakout Discussion on Overarching A,B,C | All | Breakout Rooms |
| 10:10 | Coffee Break | | Atrium |
| 10:30 | Breakout Discussion on Overarching D,E,F | All | Breakout Rooms |
| 11:40 | Reconvene for Full Group Discussion | Scribe Notes | Ballroom-413 |
| 12:10 | Polling on Overarching | | Ballroom-413 |
| 12:30 | Lunch | | |
| 2:00 | Perspectives from Congress | Adam Rosenberg Staff Director, Energy Subcommittee, House | Ballroom-413 |
| 2:20 | Presentation on NTUF | Reinke | Ballroom-413 |
| 2:50 | Breakout Discussion on NTUF | All | Breakout Rooms |
| 4:20 | Coffee Break | | Atrium |
| 4:40 | Reconvene for Full Group Discussion | | Ballroom-413 |
| 5:10 | Polling on NTUF | | Ballroom-413 |
| 5:30 | Adjourn | | |

| **Wednesday** | | | |
|---|---|---|---|
| | | | |
| **Time** | **Topic** | **Presenter** | **Location** |
| 7:30 | Breakfast Snacks and Coffee | | Atrium |
| 8:30 | Summary of Tuesday Feedback | Ferraro | Ballroom-413 |
| 8:30 | Presentation of Strategic Objectives SO-E, Materials+SO-B Stellartor physics basis | Zinkle / Hegna | Ballroom-413 |
| 9:00 | Breakout Discussion on SO E | All | Breakout Rooms |
| 10:10 | Coffee Break | | Atrium |
| 10:30 | Breakout Discussion on SO-B | All | Breakout Rooms |
| 11:40 | Reconvene for Full Group Discussion | | Ballroom-413 |
| 12:10 | Polling on SO-E,B | | Ballroom-413 |
| 12:30 | Lunch | | |
| 2:00 | Presentation on SO F,G,H, Tritium and Blankets | Humrickhouse / Nygren | Ballroom-413 |
| 2:30 | Breakout Discussion on SO F,G,H | All | Breakout Rooms |
| 4:00 | Coffee Break | | Atrium |
| 4:20 | Reconvene for Full Group Discussion | | Ballroom-413 |
| 4:50 | Polling on SO-F,G,H | | Ballroom-413 |
| 5:10 | Adjourn | | |
| 5:45 - 8:45 | Reception<br>Appetizers + cash bar | | Scruffy City<br>18 Market Square |

| **Thursday** | | | |
|---|---|---|---|
| | | | |
| **Time** | **Topic** | **Presenter** | **Location** |
| 7:30 | Breakfast Snacks and Coffee | | Atrium |
| 8:30 | Wednesday Recap | Garrison | |
| 8:30 | Presentation of SO-C Innovative Tech + SO-D PFCs | Caughman/Lasa | Ballroom-413 |
| 9:00 | Breakout Discussion on SO-C-Innovative Tech | All | Breakout Rooms |
| 10:10 | Coffee Break | | Atrium |
| 10:30 | Breakout Discussion on SO-D-PFCs | All | Breakout Rooms |
| 11:40 | Reconvene for Full Group Discussion | | Ballroom-413 |
| 12:10 | Polling on SO-C + SO-D | | Ballroom-413 |
| 12:30 | Lunch | | |
| 2:00 | Presentation on SO-A Tokamaks Part 2 + SO-I Remote Maintenance | Mordijck/Nygren | Ballroom-413 |
| 2:30 | Breakout Discussion on SO-A + SO-I | All | Breakout Rooms |
| 4:00 | Coffee Break | | Atrium |
| 4:20 | Reconvene for Full Group Discussion | | Ballroom-413 |
| 4:50 | Polling on SO-A + SO-I | | Ballroom-413 |
| 5:10 | Adjourn | | |

| **Friday** | | | |
|---|---|---|---|
| | | | |
| **Time** | **Topic** | **Presenter** | **Location** |
| 7:30 | Breakfast Snacks and Coffee | | Atrium |
| 8:30 | Thursday recap/Successes from the week | Howard/Garrison | Ballroom-413 |
| 9:00 | Breakout Discussion | All | Breakout rooms |
| 10:45 | Reconvene for Full Group Discussion | Moret | Ballroom-413 |
| 11:30 | Next Steps | Ferraro | Ballroom-413 |
| 12:00 | Adjourn | | |

## DPS Online Webinar Series, November 22, 25, and 26

Three webinars representing the three thematic areas of Discovery Plasma Science were held in lieu of an on-site community workshop.
Zoom connection information for all DPS webinars: https://zoom.us/j/346612730

**Create Disruptive Technologies (Friday, November 22 at 2:00 EST)**
Chairs: Steve Shanon and Yevgeny Raitses

**Understand the Plasma Universe (Monday, November 25, 2:00 EST)**
Chairs: Hantao Ji, Steve Vincena, and David Schafner

**Advance the Foundational Frontier (Tuesday, November 26, 2:00 EST)**
Chairs: Dan Dubin and Daniel Den Hartog

## CPP-Houston Workshop, January 13-17, 2020

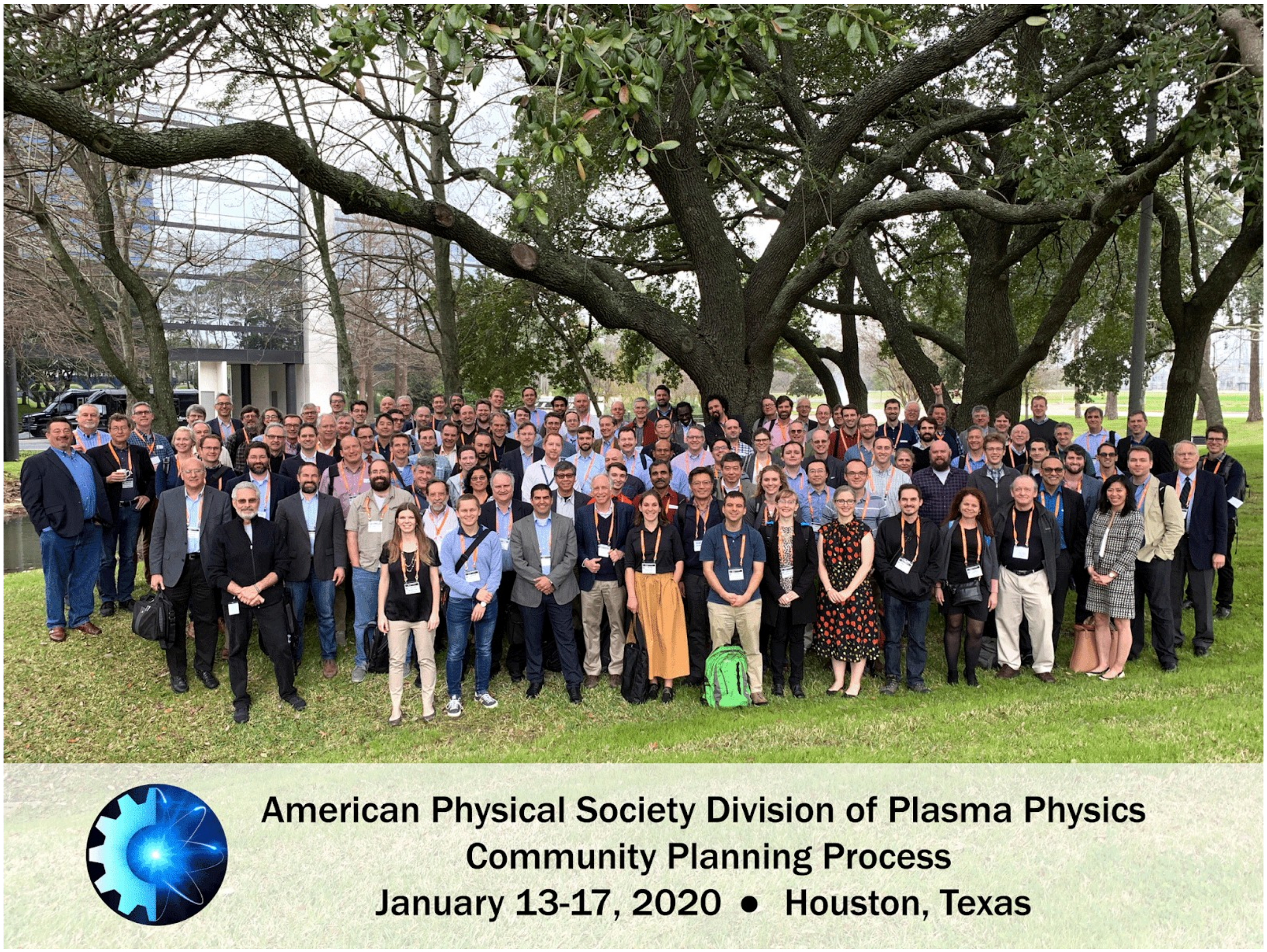

### Week Overview

(Plenary Fusion Science and Technology, Discovery Plasma)

| Monday | Tuesday | | Wednesday | | Thursday | Friday |
|---|---|---|---|---|---|---|
| Plenary | FST | DPS | FST | DPS | Plenary | Plenary |
| **Welcome, Intro, Invited Speaker** | **FST Objectives A, B, G** | **Disc. Plasma Full Group Pres.** | **FST Objectives D, H** | **Disc. Plasma Full Group Pres.** | **Summary of Tues & Weds** | **Summary of Week** |
| **Overview and Summaries of the Plan** | **Breakout Discussions** | **Breakout Discussions** | **Breakout Discussions** | **Breakout Discussions** | **Cross-Cut Presentations** | **Breakout on Final Discussion Topics** |
| | **Breakout Discussions** | **Breakout Discussions** | **Breakout Discussions** | **Breakout Discussions** | **Cross Cut Breakouts** | **Reconvene and Full Group Discussion** |
| **Lunch** | **Lunch** | **Lunch** | **Lunch** | **Lunch** | **Lunch** | **Close at noon** |
| **FES-wide Breakouts** | **FST Objectives F, C, E** | **Discovery Plasma Full Group Pres.** | **Fusion Energy Full Group Pres.** | **Breakout Discussions** | **Cross-Cut Presentations** | |
| | **Breakout Discussions** | **Breakout Discussions** | **Breakout Discussions** | **Breakout Discussions** | **Cross-Cut Breakouts** | |
| | **Breakout Discussions** | **Breakout Discussions** | **Breakout Discussions** | **Full Group discussion** | **Reconvene and Report Back** | |
| **Reconvene Report Back** | **Reconvene and Polling** | **Reconvene for Discussion** | **Reconvene and Polling** | | | |

*Monday 1/13 (FES-Wide Discussion)*

Remote connection via Zoom will be provided here: https://mit-psfc.zoom.us/j/6172534785

| Time | Activity | Presenter | Location |
|---|---|---|---|
| 7:30 | Breakfast and Registration | | Foyer |
| 9:00 | Welcome and Introduction | Solomon | Texas Ball. |
| 9:25 | Overall structure of plan | Ferraro | Texas Ball. |
| 9:45 | Cross-cut Overview | Sarff | Texas Ball. |
| 10:05 | Coffee | | Foyer |
| 10:35 | DPS Overview | Baalrud / Kuranz | Texas Ball. |
| 11:15 | FST Overview | Howard / Garrison | Texas Ball. |
| 12:00 | Lunch (on your own) | | |
| 1:30 | Goals for afternoon discussion | Co-chairs | Texas Ball. |
| 1:40 | Breakout Session | Breakout Groups<br>Discussion Questions | |
| 3:00 | Coffee | | |
| 3:30 | Breakout Session | Discussion Questions | |
| 4:30 | Plenary Reporting from Breakouts | | Texas Ball. |
| 5:30 | End of Sessions | | |
| 6:00 | Welcome Reception-appetizers and drinks provided | | |

*Tuesday 1/14 (Parallel FST / DPS)*

**Fusion Science and Technology**

| Remote connection via Zoom will be provided here: https://mit-psfc.zoom.us/j/6172534785 | | | |
|---|---|---|---|
| **Time** | **Activity** | **Presenter** | **Location** |
| 7:30 | Breakfast | | Foyer |
| 8:30 | Presentation of Prioritization Assessment Criteria: | Howard | Texas Ball. |
| 8:45 | Fusion Energy Full Group Presentation: SO-A,B,G | Lasa / Zinkle / Nygren | Texas Ball. |
| 9:15 | Fusion Energy Breakout Session | Groups / Questions | Breakout Rooms |
| 10:00 | Coffee Break | | |
| 10:20 | Fusion Energy Breakout Session | | Breakout Rooms |
| 11:30 | Fusion Energy Reconvene: Report Back and Polling | Poll setup / Guidance | Texas Ball. |
| 12:00 | Lunch (on your own) | | |
| 1:40 | Fusion Energy Full Group Presentation: SO-F,C,E | Caughman / Humrickhouse / Guttenfelder | |
| 2:10 | Fusion Energy Breakout Session | Groups / Questions | Breakout Rooms |
| | Coffee Break when needed (~10 min) | | |
| | Continue Fusion Energy Breakout Session | | Breakout Rooms |
| 4:20 | Fusion Energy Group Reconvene - Report Back | | |
| 4:50 | Fusion Energy Full Group Polling | | |
| 5:20 | DPS+FST updates and announcements | | Texas Ball. |

**Discovery Plasma Science**

| Time | Activity | Presenter | Location |
|---|---|---|---|
| 7:30 | Breakfast | | Foyer |
| 8:30 | Discovery Plasma Full Group: Instructions on goals for morning breakout sessions | Co-chairs<br>Discussion prompts | Woodbine<br>Breakout rooms |
| 8:50 | Discovery Plasma Breakout Discussion: Split into the 4 DPS Drivers; Session #1 | | Breakout Rooms |
| 10:00 | Coffee Break | | |
| 10:20 | Discovery Plasma Breakout Discussion: Split into the 4 DPS Drivers; Session #2 | | Breakout Rooms |
| 12:00 | Lunch (on your own) | | |
| 1:30 | Discovery Plasma Full Group: Report back from morning sessions; Instructions on goals for afternoon breakout sessions | PC members, Co-chairs | Woodbine |
| 3:00 | Coffee Break | | |
| 3:30 | Discovery Plasma Breakout Discussion : Session #3 | Discussion prompts<br><br>Breakout room assignments | Breakout Rooms |
| 5:20 | Full group discussion and preparation for tomorrow | | Texas Ballroom |
| 5:35 | End of Sessions | | |

*Wednesday 1/15 (Parallel FST / DPS)*

**Fusion Science and Technology**

| Time | Activity | Presenter | Location |
|---|---|---|---|
| 7:30 | Breakfast | | Foyer |
| 8:30 | Fusion Energy Full Group Presentation: SO-D,H | Reinke / Sarff | |
| 9:00 | Fusion Energy Breakout Session | Groups / Questions | Breakout Rooms |
| 10:00 | Coffee Break | | |
| 10:20 | Fusion Energy Breakout Session | Groups / Questions | Breakout Rooms |
| 11:15 | Fusion Energy Reconvene: Report Back and Polling | | |
| 12:00 | Lunch (on your own) | | |
| 1:45 | Afternoon discussion questions and polling questions | | |
| 1:50 | Fusion Energy Breakout Session | Groups / Questions | Breakouts |
| 3:05 | Coffee Break | | |
| 3:25 | Fusion Energy Breakout Session | Groups | Breakout Rooms |
| 4:40 | Fusion Energy Group Reconvene - Report Back | | |
| 5:10 | Fusion Energy Full Group Polling | | |
| 5:25 | End of Sessions | | |

**Discovery Plasma Science**

| Time | Activity | Presenter | Location |
|---|---|---|---|
| 7:30 | Breakfast | | Foyer |
| 8:30 | Discovery Plasma Full Group Presentation: Presentation and discussion of modified organization<br>Link to discussion topics for the day | Co-chairs | Watercourt |
| 10:00 | Coffee Break | | |
| 10:30 | Discovery Plasma Breakout Discussion : All Science Drivers Combined for Discussion #2 | | Watercourt |
| 12:00 | Lunch (on your own) | | |
| 1:30 | Discovery Plasma Breakout Discussion : Combined for Discussion, Principles for Prioritization | | Watercourt |
| 3:00 | Coffee Break | | |
| 3:30 | Discovery Plasma Breakout Discussion : Combined Discussion, Principles for Prioritization | | Watercourt |
| 5:25 | End of Sessions | | |

*Thursday 1/16 (FES-Wide Discussion)*

| **Time** | **Activity** | **Presenter** | **Location** |
|---|---|---|---|
| 7:30 | Breakfast | | Foyer |
| 8:30 | Summary of Tuesday and Wednesday | Baalrud / Ferraro | Texas Ball. |
| 8:50 | Cross Cut Presentations: Theory. & Comp. / Measurement & Diagnostics | Holland / Frenje | Texas Ball. |
| 9:20 | Cross Cut Breakout: Theory. Comp. | Breakout groups / Questions | Breakout rooms |
| 10:20 | Coffee Break | | |
| 10:40 | Cross-Cut Breakout: Measure. Diag. | Breakout groups / Questions | Breakout rooms |
| 11:40 | Cross-cut Report Back and Discussion | | |
| 12:00 | Lunch (on your own) | | |
| 1:30 | Cross-Cut Presentations: Enabling Tech / Workforce Development | Shannon / Donovan | Texas Ball. |
| 2:00 | Cross Cut Breakout: Enabling Tech | Breakout groups | Breakout rooms |
| | | Questions | |
| 3:00 | Coffee Break | | |
| 3:30 | Cross Cut Breakout: Workforce Development | Breakout groups<br>Questions | Breakout rooms |
| 4:30 | Cross-cut Report Back and Discussion | | Texas Ball. |
| 5:00 | End of Sessions | | |

*Friday 1/17 (FES-Wide Discussion)*

| Remote connection via Zoom will be provided here: https://mit-psfc.zoom.us/j/6172534785 | | | |
|---|---|---|---|
| **Time** | **Activity** | **Presenter** | **Location** |
| 7:30 | Breakfast | | Foyer |
| 8:30 | Welcome and agenda | Chairs | Texas Ball. |
| 8:40 | Phase II of the Community Planning Process | Rej | Texas Ball. |
| 9:00 | Breakout Sessions-Executive Summary | Discussion topics | Breakout rooms |
| | Coffee Break (as needed) | | |
| | Breakout Session continues | | Breakout rooms |
| 10:20 | Report back and discussion | | Texas Ball. |
| 11:40 | Closing Statement and Next Steps | Solomon | Texas Ball. |
| 12:00 | Meeting Close | | |

## *Appendix G. Initiatives*

During the community planning process the community produced over 300 new and/or revised initiatives and white papers for consideration. All of these were considered as input to the process and formed the basis for the content in this strategic plan. The deadline for initiatives to be considered at the first round of workshops was July 1, 2019. Many of the initiatives were evaluated by Expert Groups during and after the first round of workshops, which provided the opportunity for valuable feedback and refinement of initiatives. After that, some initiatives were revised to be resubmitted and some new initiatives were solicited for areas that did not have enough coverage. The initiative submission portal was open throughout the CPP. All these initiatives can also be found on the CPP website.

| Date | Title of Initiative | Lead/Contact Author | Link to file |
|---|---|---|---|
| 5/30/2019 | Discussion Group 5 (DG5) summary for NAS Committee on a Strategic Plan for Burning Plasma Research | Derek Sutherland | https://drive.google.com/open?id=1UZPlYOwYeafyP4cwM4NDPtJEf-qc0mAh |
| 6/14/2019 | Development of Flibe (Li2BeF4) Salt Fusion Blankets | Charles Forsberg | https://drive.google.com/open?id=118eLM6rMKk3OCeVTufyJcmVHLI5bkeiO |
| 6/15/2019 | Fast Liquid Metal Program for Fusion Reactor Divertor | Egemen Kolemen | https://drive.google.com/open?id=1yxfxr1TOpaciKUQejSYQ-uyGVhI06-BP |
| 6/16/2019 | Strategic implications of the stellarator for magnetic fusion | Allen Boozer | https://drive.google.com/open?id=1YLEwyew4zHVVV15o7vpTB2iex1BK6NTW |
| 6/19/2019 | Initiative to Develop the ASME Construction Code and Standard for Fusion Energy Facilities, V.0 | William K. Sowder | https://drive.google.com/open?id=1U1-bjfEsH4huMbjhkJODkQaFg8Z48XKo |
| 6/21/2019 | Fast Time-Response Disruption Mitigation for ST and AT Facilities | Roger Raman | https://drive.google.com/open?id=1thQSEI6xjHRweq3wTqRCG-7uTCoYnIBe |
| 6/23/2019 | The Plasma Universe Initiative (PUI) | Hantao Ji | https://drive.google.com/open?id=14Ip1figwHUk22NSJcLmCr7BnagQBALYr |
| 6/24/2019 | Achieving the Ultimate Goal of Radwaste-Free Fusion Through Recycling/Clearance | Laila El-Guebaly | https://drive.google.com/open?id=1XThtUstN2mJK6V9etg0cyaevlgbN7K58 |
| 6/24/2019 | Thoughts on Discovery Science: June 24, 2019 | Paul Bellan | https://drive.google.com/open?id=1cJSincugFJtI9Fehyf32QdxqTUrlJJlz |
| 6/26/2019 | National Fusion Design Initiative | Hutch Neilson | https://drive.google.com/open?id=1VVlf9jLtqD8voRLhkzo7YnCLXAPJCXkJ |
| 6/26/2019 | Working Group 1: Principles, Values, Metrics, and Criteria | Ilon Joseph | https://drive.google.com/open?id=1IqAIXSjxH6svflYKp40LP2bfVg706Spz |
| 6/27/2019 | Laser-Plasma Interactions Enabled by Emerging Technologies | John P. Palastro | https://drive.google.com/open?id=12hLXNTiX_wNLNOsTvwGLoNwzleZsds0s |
| 6/27/2019 | Integrated ecosystem of advanced simulation tools for plasma modeling | Jean-Luc Vay | https://drive.google.com/open?id=1awjbBfFiRNV1hMeIRy73B6vjoHKZEjaQ |
| 6/27/2019 | Frontiers in Nuclear-Plasma Physics | Alex Zylstra | https://drive.google.com/open?id=1GOOeUkse_oGrsCM4dMyphs0YsOKHQC4o |
| 6/28/2019 | Diagnosing fusion burn using neutron spectrometry | Maria Gatu Johnson | https://drive.google.com/open?id=1JRWdiC8txRLuUHBW27SHfJVbtdN7comL |
| 6/28/2019 | Consistent Material Properties | Charles Starrett | https://drive.google.com/open?id=1JW6SfWkRDN-AFApEMLvbyn5B9ZDXVDdX |
| 6/28/2019 | The Materials-Design Interface for Fusion Power Core Components | Mark Tillack & Nasr Ghoniem | https://drive.google.com/open?id=16f7Z3beLETnprBPeUBDLab784aeUYEWj |
| 6/28/2019 | Solving the Challenge of High Divertor Heat Flux using Helium Cooling and Tungsten-Alloy Structures | Minami Yoda | https://drive.google.com/open?id=17w-4oWAA6fLWYeLfzU6T-xRxiuiusOIU |
| 6/29/2019 | The Role of High Repetition Rate Experiments in Advancing HEDP Science | Scott Feister | https://drive.google.com/open?id=1cPL4V36DIpsCRox5ry55wBb4vJVEbC4I |
| 6/29/2019 | Computational predictive capability for fusion LM systems including LM-plasma coupling | Sergey Smolentsev | https://drive.google.com/open?id=11B1jB5lf2KjqIDjU-YOucgzTyqphct9w |
| 6/30/2019 | Accelerated Development of Materials as the Enabling Technology for Fusion Energy | Yutai Katoh | https://drive.google.com/open?id=1v813BpuPsxasARrw8TGo-l0Ho26d-C5A |
| 6/30/2019 | Advanced PFCs for Fusion Reactors | Dennis Youchison and Arnie Lumsdaine | https://drive.google.com/open?id=1oTZedf4DgVbgr6buip9UJwiLcFBE3aXu |
| 7/1/2019 | A National Research Program to Prepare for a Compact Fusion Pilot Plant by Resolving the Physics of Sustained High Power Density Conditions | Richard J Buttery | https://drive.google.com/open?id=1895ECWPfs5C5S0FJxu2cXWzWg4ILYUP9 |
| 7/1/2019 | The Compact Fusion Pilot Plant Mission Definition, Design, and Required R&D Program | Charles Kessel | https://drive.google.com/open?id=1euC_TanGtLBD4OAp0VeYPv46U3itYdh- |
| 7/1/2019 | Plasma Optimization, Preparation of the Plasma Scenario for the Compact Fusion Pilot Plant | Charles Kessel | https://drive.google.com/open?id=1KlzqLGJzuyv3u0p7UyGV9KtGS_Je1IYH |
| 7/1/2019 | A Critical Integration Step for Fusion Blankets, the Blanket Component Test Facility | Charles Kessel | https://drive.google.com/open?id=14yzEycNR4Hj17BEdTEqEyMTKFc4JFJFD |
| 7/1/2019 | The Fusion Energy Systems Studies Integrated Design and Assessments for the US MFE Program, the Early Phases of CFPP Development Toward Industrial Phases | Charles Kessel | https://drive.google.com/open?id=1SdzJWrcL8UPFgzMHhyEfWGjBOwImxLtQ |
| 7/1/2019 | Are High-Entropy Alloys Suitable for Fusion Applications? | Enrique Martinez Saez | https://drive.google.com/open?id=11Od0-9WrYFcAc9aNtH9RAhpF67Itrofx |
| 7/1/2019 | A national initiative to accelerate ITER research and maximize the US return on ITER | Raffi Nazikian | https://drive.google.com/open?id=1qzjUCW-TP25tP1AEJd0OQZXegYqXDuG6 |
| 7/1/2019 | The Fusion Prototypic Neutron Source (FPNS), an Affordable, Timely 14 MeV Fusion Neutron Irradiation Facility for Near-term Fusion Material Testing | Brian Egle | https://drive.google.com/open?id=1qjK95Xc1RfcB4IBsHKEKhXreqkdXyM5r |
| 7/1/2019 | Commonwealth Fusion Systems and the Path to Commercial Fusion Energy | Alex Creely | https://drive.google.com/open?id=1DlcZfzfJLp2ETotn4R-WFhw8wvjeEit5 |
| 7/1/2019 | High Field Superconducting Magnet Technology Development for Fusion Devices and Science Missions | Robert Duckworth | https://drive.google.com/open?id=11CTHs0apxLDwHJwAGhhNKg0HVHRSu2cw |
| 7/1/2019 | Expand Capacity Computing | Brendan Carrick Lyons | https://drive.google.com/open?id=1WDL3Mum1xYQH4k7M4d2YTVbK6TvgpYQs |
| 7/1/2019 | National Fusion Magnet and Conductor Development Initiative | Xiaorong Wang | https://drive.google.com/open?id=1e4KQ7aQIOsUooQXkQ_jY2Q6scRFKGenk |
| 7/1/2019 | The Basis and Potential of Plasma Optics Made from Ion Waves | Robert Kirkwood | https://drive.google.com/open?id=1wI5na5T_a_mV3UWCg5tl2OCSaBq4D5h2 |
| 7/1/2019 | The need for a diverse fusion energy research and development portfolio for the pursuit of economically competitive fusion power | Derek Sutherland | https://drive.google.com/open?id=1rXwryeCv4bjd1HF-jBUnSQudNT6Z0jC0 |
| 7/1/2019 | Plasma-facing materials and components research needs | Chad M. Parish | https://drive.google.com/open?id=1lgVMzSMPo3_VDaGuKDULJhFFl2tCxYt0 |
| 7/1/2019 | Quad Chart | Prof. Chris Orban | https://drive.google.com/open?id=1LPyRYwkdSxT0uvbozzkmTYKEfX5npzn8 |
| 7/1/2019 | Fundamental Understanding of Hydrogen Isotope-Materials Interactions in Fusion Reactors | Xunxiang Hu | https://drive.google.com/open?id=1Wo6FYVN7GOmw35Il1zF_Fq7JtDDKehrz |
| 7/1/2019 | Continued Optimization and Functional Verification of Cellular Solid Breeder for Transformative Solid Breeder Blankets | Alice Ying | https://drive.google.com/open?id=1TZDg8aBvNO_LsjhyYQ6FT8tm6jlMTn-D |
| 7/1/2019 | Quasi-Linear model initiative for fast ion transport | Vinicius Duarte | https://drive.google.com/open?id=1cgBGlFMeEfeIxwFZO8VW7zuLPDtBOFX8 |
| 7/1/2019 | Developing solid-material plasma facing components for fusion reactors with replenishable wall claddings and continuous surface conditioning | Peter Stangeby | https://drive.google.com/open?id=1mA51u4pUGx-n2wBnowncl0CBTRG_X_1J |
| 7/1/2019 | Some implications of recent technology advances on divertor physics performance requirements of DT fusion tokamaks | Peter Stangeby | https://drive.google.com/open?id=1zNwyBgcZH7qogfs6dICeKfX2y9GCY8qa |
| 7/1/2019 | Targeted design of materials for harsh environments in fusion energy applications | Yury Osetskiy | https://drive.google.com/open?id=1s1Ar_79vmwDe2xxad2tiuha0yLlZxuTU |
| 7/1/2019 | Direct Laser Acceleration of electrons by high-intensity laser pulses | Louise Willingale | https://drive.google.com/open?id=1ZQZT5-gnXflUyGyWBV5l4VU6zYiD16ND |
| 7/1/2019 | InteBaylorgration of the Fueling and Pumping in the Fusion Energy Fuel Cycle | Larry Baylor | https://drive.google.com/open?id=1NAaHsvUsMjYTp8i1exFb3883O0tnF6um |
| 7/1/2019 | Direct Laser Acceleration | Louise Willingale | https://drive.google.com/open?id=1m7bV7YJSjhc91633f9aw8Qp7S1HFavZR |
| 7/1/2019 | An Initiative to Establish Power Exhaust Solutions for a Compact Pilot Plant | John Canik | https://drive.google.com/open?id=1H2UX3yNG9vtJLJgduqV0mXxwsbVQOHk1 |
| 7/1/2019 | A Large Scale Turbulent Plasma Wind Tunnel | Michael Brown | https://drive.google.com/open?id=1hgAmn_XQhDsuQ0OZ4fwRxdbFtRMIy6YL |
| 7/1/2019 | Adoption of Advanced Manufacturing for Advancing and Implementing Materials for Fusion Energy Applications | Kevin Field | https://drive.google.com/open?id=1Y3mmWx64oMW08Hp0xfbMBa3q9JHCaHpV |
| 7/1/2019 | The NSTX-U Facility in the 2020s: Advancing the Physics Basis for Configuration Optimization Toward a Compact Fusion Pilot Plant | Devon Battaglia | https://drive.google.com/open?id=11UlAeKQWGQKRE7Pd4cgSkw6s26FJG3l7 |
| 7/1/2019 | Liquid Metal Divertor | Rob Goldston | https://drive.google.com/open?id=1OyDVbKA2CMXu2dsxDn0ObyhIn9sccesX |
| 7/1/2019 | Frontiers in high-energy-density and relativistic physics enabled by EP-OPAL: a multi-beam ultrahigh-intensity laser user facility | Hans Rinderknecht | https://drive.google.com/open?id=18uyw8QJ_c5NAul3XBIRCETg9y-Ema6-m |
| 7/1/2019 | National Fusion Science Undergraduate Internships | Arturo Dominguez | https://drive.google.com/open?id=1CU3ve-ZZsC_c-wdHIqQ8WzDXyRAosVar |
| 7/1/2019 | Innovative X-ray Crystal Spectrometers for High Energy Density Science | Lan Gao | https://drive.google.com/open?id=1x1p7dc9qV5EhT1UdDh_J271_9K0wc4Gd |
| 7/1/2019 | Reliable Long-Pulse Plasma Heating and Current Drive using ICRF | John Caughman | https://drive.google.com/open?id=1BGui4lJQAVcx0ySHqqYPDUcuNyOKHT2X |
| 7/1/2019 | Realistic testing and simulation of synergistic effects in PMI | Tatyana Sizyuk | https://drive.google.com/open?id=16hpTQRN35QeJ_75k9dvFgprKr30XjPBN |
| 7/1/2019 | Controlling charging in dusty plasmas | Edward Thomas, Jr. | https://drive.google.com/open?id=12vgNScxonLXtTqlaqCAYyFelsyBEbi-x |
| 7/1/2019 | Integrated Simulation Tool for Various Effects of Plasma Transients | Ahmed Hassanein | https://drive.google.com/open?id=1CN9_A6Zp0kFQ2UnbbT9eICTxd5N2V4sp |
| 7/1/2019 | Strategy for Advancing the Technical Readiness of Liquid Metal Plasma Facing Components | Travis Gray | https://drive.google.com/open?id=1dBFLR1-4gPue4Rkzk5MWXIqKmi8cDVp_ |
| 7/1/2019 | Addressing critical Plasma-Material Interactions gaps with the new linear divertor simulator MPEX | Juergen Rapp | https://drive.google.com/open?id=1L3WZC08yXLb1umTRqaiJNogGL9Wic0aE |
| 7/1/2019 | The Stellarator Path to a low-cost Pilot Plant | David A. Gates | https://drive.google.com/open?id=1JU-nLrt9ri-dE4yyBT-yie1To2Lb2qUW |
| 7/1/2019 | Plasma-Material-Interaction Challenges and Path towards RF Sustainment of Steady State Fusion Reactor Plasmas | Paul T. Bonoli | https://drive.google.com/open?id=1gMb5paNALVKiZIkGzurOvAShTld90eP4 |
| 7/1/2019 | Computational Modeling of Non-Equilibrium Warm Dense Matter Systems | Tadashi Ogitsu | https://drive.google.com/open?id=1njw3iuf_8xkqovYybYZ_-jwT2l6O465n |
| 7/1/2019 | Establishing an understanding of radiofrequency heating and current drive for a compact pilot plant | Michael Brookman | https://drive.google.com/open?id=1zG8UAL8wskGIjPMsfJQ0G5-7ZxHi xW |
| 7/1/2019 | Towards an Integrated Fusion Design and Materials Development Program | Lance Snead | https://drive.google.com/open?id=17-peak1kXJLd2eOpXEBJ5JOal7GOuyRu |
| 7/1/2019 | Reversed-field pinch research toward Ohmic ignition at high engineering beta | Karsten McCollam | https://drive.google.com/open?id=1HaTtMjGsw9y4V0t1piCiRJkpzsRp5FTy |
| 7/1/2019 | International stellarator research in support of a low capital cost pilot plant | Samuel Lazerson | https://drive.google.com/open?id=1V8t4oWBXa48NLXuApwshj_2XUR4svdkK |
| 7/1/2019 | Efficient X-ray detection at high energies (>10 keV) | Sabrina Nagel | https://drive.google.com/open?id=1d_f6Qxz5kDW4zZiP9_E1chEkp0niyB_Y |
| 7/1/2019 | Plasma Optics for Increasing Laser Performance to Access New Physical Processes | Robert Kirkwood | https://drive.google.com/open?id=1CKfWY_c_D3sMJyfKfwgU8VJrGOe97uW8 |
| 7/1/2019 | A Virtual Integrated Multi-physics, Multi-scale Simulation Predictive Capability for Plasma Chamber Blanket and Fuel Cycle Component Design and Performance Evaluation | Alice Ying | https://drive.google.com/open?id=1FItSBA2tkkAwvyU8fYmfLm_Wfmlo_Bcv |
| 7/1/2019 | The PlasmaPy Project: Building an Open Source Software Ecosystem for Plasma Research and Education | Nick Murphy | https://drive.google.com/open?id=1Jb0dhCDMjXXnK9mbWvewQbQlgpzx5x-Q |
| 7/1/2019 | Enabling scientific reproducibility in plasma research | Nick Murphy | https://drive.google.com/open?id=1BsV_ZxDiSdsOiij8iTRHLHcK0eMweyu9 |
| 7/1/2019 | Equity and inclusion in plasma physics | Nick Murphy | https://drive.google.com/open?id=1svpSBpx_7HbMRStTNCG97Kb20XgvwZuw |
| 7/1/2019 | In-pile fission irradiation to advance tritium-related material and technology challenges | Masashi Shimada | https://drive.google.com/open?id=1kPb92_bDAVS-Y4ZnQs2-I14PGHLx1-T8 |
| 7/1/2019 | Magneto inertial fusion experiments on NIF | John D. Moody | https://drive.google.com/open?id=1LZTXBRYmPRqSuVQ4H8JnRtCO3Nj0h8RM |
| 7/1/2019 | Physics of plasmas in extreme fields | Stepan Bulanov | https://drive.google.com/open?id=1RTVRV2Un2Sr_LRUug1EA3eNqpSfPQqSc |
| 7/1/2019 | Advanced RF Source Technology and Development Center | Stephanie Diem | https://drive.google.com/open?id=1TwlZiM9Ij5FxkL9X0T57I5aLEV_YaGl_ |
| 7/1/2019 | Developing the framework for licensing a fusion power plant | Brad Merrill | https://drive.google.com/open?id=1WIQyCI1t0BADSRasft7Yv4EcH1It04-P |
| 7/1/2019 | Near-Term Initiatives to Close the Fusion Technology Gaps to a Compact Fusion Pilot Plant | Mickey Wade | https://drive.google.com/open?id=1JUhZzZpknHr80xKDN1Vyx-Sd71nRP7Wg |
| 7/1/2019 | Quantum Information Science and Fusion Energy Sciences | Thomas Schenkel | https://drive.google.com/open?id=1V0GOgPUo4k82OJ9FQOjKz1bLaE-jlL6a |
| 7/1/2019 | Physics and applications of ion acceleration driven by high-repetition-rate PW lasers | Carl Schroeder | https://drive.google.com/open?id=1n8MF4V5T2ogwFL_MQBoLAcdK8dH-InZY |
| 7/1/2019 | Advancing Fusion Technology Workforce Development at the University of Wisconsin-Madison Fusion Technology Institute | Tim D. Bohm | https://drive.google.com/open?id=16VeW3KBvEY-zQGk8SEJTLYykm9V1xOUe |
| 7/1/2019 | Realize the Full Potential of DIII-D to Advance Development of Cost-Effective Fusion Power Plants | David N Hill | https://drive.google.com/open?id=17UU6w7a2nCOA4UQmrdkZspCG_IkDfSkW |
| 7/1/2019 | Initiative: Facility for the study of Astrophysical Processes | Walter Gekelman | https://drive.google.com/open?id=16EWEjgZR2RbdTjlv2ARoR_3QOWOc06ud |
| 7/1/2019 | Compatibility Issues for Fusion Energy | B. Pint | https://drive.google.com/open?id=13VUniaDRpRknzLwIZtLTJ8joGtoePY33 |
| 7/1/2019 | Accelerating Fusion Through Integrated Whole Device Modeling | Amitava Bhattacharjee | https://drive.google.com/open?id=1tWqDC0mwUpDfiUwmAE9to9FbFOP54gAq |
| 7/1/2019 | Low-temperature atmospheric-pressure plasmas | Chunqi Jiang | https://drive.google.com/open?id=1MP-8SmsQdNlM6ESJ5vAMHPm_SVldG2jJ |
| 7/1/2019 | Integrated RF Program to Develop Fusion Reactor Relevant Actuators | Masayuki Ono | https://drive.google.com/open?id=1RKZjGmwpbhx9aQmccgfH1MYBvfj0C03y |
| 7/1/2019 | Nuclear reactions and fusion processes in plasmas of varying density and temperature | Thomas Schenkel | https://drive.google.com/open?id=1Zej5ikUdZQD3Gv6UknJn9td_a631QXjl |
| 7/1/2019 | Fusion Nuclear Analysis Advancements for the Fusion Materials and Technology Strategic Planning Process | Tim D. Bohm | https://drive.google.com/open?id=1sZRYq4xEGNy3bsZbj0yS5MiPIQODJ0V1 |
| 7/1/2019 | Microprobe quad chart | Michael Brown | https://drive.google.com/open?id=1z1X4HKD-azf42tCzX6KYHyfXJ_a4BIN3 |
| 7/1/2019 | Federated Data Analysis utilizing AI/ML for efficient collaborative research among big experiment, simulation and personnel resources | C.S. Chang | https://drive.google.com/open?id=1NZvI2q4SbktDqNrWh0bFVCMJ9NmmFexE |
| 7/1/2019 | Utilizing Outreach Activities for Workforce Development | Stephanie Diem | https://drive.google.com/open?id=1SQiXM0-Fiqn9ryVdmDCXgzbZ71UoIOhH |
| 7/1/2019 | Pulsed Power User Facility | Will Fox | https://drive.google.com/open?id=1ixcCCljBORBRUKtRi_Mt-j-7_G8ymrg- |
| 7/1/2019 | Development of Mission Need and Preliminary Design of a Sustained High Power Density Tokamak Facility | Jon Menard | https://drive.google.com/open?id=16nE6P0DiKFlxjdWWAv7MeIBI0TeYWhc_ |
| 7/1/2019 | Advancing toward in-situ plasma facing component surface diagnostics | Sara Ferry | https://drive.google.com/open?id=1jQkKi3LwcyrTlSSltPb7QfnTlXVFCLhT |
| 7/1/2019 | Laser-plasma accelerators: next generation x-ray light sources for high energy density science | Félicie Albert | https://drive.google.com/open?id=17DwddPByzQUtcnPHju_1ue9Qq1x_VtGP |
| 7/1/2019 | GDT_VNS_Initiative | Cary Forest | https://drive.google.com/open?id=13A2RtJZvIseBjXGuoLE3RdXb-pjJaJQN |
| 7/1/2019 | Robust kinetic simulations for improved understanding and predictions | Genia Vogman | https://drive.google.com/open?id=1B_nqQUUxr4k0sq7sQIIk0ZIiq7K3BMR3 |
| 7/1/2019 | GDT Volumetric Fusion Neutron Source | Cary Forest | https://drive.google.com/open?id=19ma5QpflOaAgJQYLI7nQi9Gr78ca8h-A |
| 7/1/2019 | Initiative to Simplify Optimized Stellarators and Test Key Properties | M.C. Zarnstorff | https://drive.google.com/open?id=1EsTw2abKXm8-AqXeZVdxlnMLgG-fwE3q |
| 7/1/2019 | Generation ITER - Exciting opportunities for early career researchers and the US fusion program | F. M. Laggner | https://drive.google.com/open?id=1u9AEpO2J0C1mEJYVtrDIa-F9BlzZtZDX |
| 7/1/2019 | Frontier HED Science on the LCLS MEC Instrument | Andrew MacKinnon | https://drive.google.com/open?id=1Cl0j6OFrmLnP5cDIKGol81UwLFbpFnA- |
| 7/1/2019 | Cusp-confined spheromaks spinning in a hot sheath | Daniel Prater | https://drive.google.com/open?id=1 y0X1NhVtwhxkuV9hUA4CmI7SYOCbsJ |
| 7/1/2019 | LaserNetUS: a high intensity laser network for frontier plasma science in the U.S. | Gilliss Dyer | https://drive.google.com/open?id=1Gjk9JP1M2xub-00X820mtF4PE7Wi5Q6I |
| 7/1/2019 | Cusp-Confined Spheromaks for MFE and Discovery Plasma Science | Daniel Prater | https://drive.google.com/open?id=1CxRuLd-qHs1YHC6ewZ1CQp-qWoCmaV4m |
| 7/1/2019 | Investigations into a Hot Sheath | Daniel Prater | https://drive.google.com/open?id=1WPNgsO8qeKbKyYWVZlCqoWDlqOx32gPP |
| 7/1/2019 | PFC Testing in a Flexible US High Heat Flux Test Facility | Richard E. Nygren | https://drive.google.com/open?id=18QKnIQo1BZ3AKAUGytPGTkGJpK8R5uVe |
| 7/1/2019 | Thermal Management of Plasma Probes | Richard E. Nygren | https://drive.google.com/open?id=1zB-v6LYWjJYuPqZVxKXKEIDfwHev8eft |
| 7/1/2019 | A Near Term Initiative on Advanced Manufacturing in Fusion Applications | Richard E. Nygren | https://drive.google.com/open?id=1A1LinfBXPBWswvxxob50x_k-FUvCn0Zx |
| 7/1/2019 | Smart Tiles | Richard E. Nygren | https://drive.google.com/open?id=1RxM5ZcZhyd8LNfzlMZleWbqrM-F9h42w |
| 7/1/2019 | Negative Triangularity Tokamak as a Fusion Reactor Option | Max Austin | https://drive.google.com/open?id=18AWINb9yRStsn9dlmUQ_hel-gKR_sfYF |
| 7/1/2019 | Neutral H sensor for C-X flux on wall and divertor | Robert Kolasinski | https://drive.google.com/open?id=1cQ-5cMNmIBqkoYM5Ji6CJ-bvXSSHGivi |
| 7/1/2019 | High Fidelity Surface Diagnostics for Plasma-Material Interactions | Robert Kolasinski | https://drive.google.com/open?id=1WtKhfuDCHTg9zaVSP1-AWDOUpoqJmi6i |
| 7/1/2019 | Instrumentation required for monitoring magnetic fusion device and performance | Robert Kolasinski | https://drive.google.com/open?id=1hQ6h8Rec-bRfVoaF0lTSkyRro2M0jxaN |
| 7/1/2019 | Facilities Instrumentation for Components Removed from a Compact Pilot Plant | Robert Kolasinski | https://drive.google.com/open?id=1LXcqYLidLbdNcNLxAnvxVedUOQy00Vyx |
| 7/1/2019 | Smart Pyrometry for Measure Liquid Lithium Surface Temperature | Richard E. Nygren | https://drive.google.com/open?id=1cZkV3KXgthqhDTAD9shNTA60Cyw4Bc-y |
| 7/1/2019 | A National Initiative for Disruption Elimination in Tokamaks | S.A. Sabbagh | https://drive.google.com/open?id=1uKyUJ2c6a7ypxvnpuftSNh2TV1G7zUNg |
| 7/2/2019 | The Plasma Universe Initiative (PUI) - with updated authorship | Hantao Ji | https://drive.google.com/open?id=1PDz4wXoAyJgN7wubAIj0pN3Hr9y9N45M |
| 7/2/2019 | A Diagnostic Plan for FES HEDP Physics | Steven Ross | https://drive.google.com/open?id=1w-ZIOax9kCw6EmZaCSYfOijEpuwriJvP |
| 7/2/2019 | Establishing an understanding of radiofrequency current drive for a compact pilot plan | Michael Brookman | https://drive.google.com/open?id=1torQS_Hk5jOdrIgpakv5DSqtlA3DKN5Y |
| 7/2/2019 | Capturing Appropriate “Strength of Coupling” Knowledge Base for Reactive Low Temperature Plasmas | Katharina Stapelmann | https://drive.google.com/open?id=1_Bi22QChLocAzU7tEHjUqZP2sJQbJiwm |
| 7/3/2019 | Dynamics and Stability of Astrophysical Jets | Chikang Li | https://drive.google.com/open?id=1LZsS-m7-daCf_P1ebq0sqlca5AFecuHl |
| 7/3/2019 | OMFIT: One Modeling Framework for Integrated Tasks | Sterling Smith | https://drive.google.com/open?id=13AW5uqNxKbCE7TyDSxAprBuEpWKRsULY |
| 7/3/2019 | Spectroscopic studies of WDM/HDM using X-ray free electron laser | Hae Ja Lee | https://drive.google.com/open?id=1P2dNnqKc5fliLz1e88F5hMdY3pSTJdSW |
| 7/5/2019 | white paper | Marty Marinak | https://drive.google.com/open?id=1B9dLUBLae4oc_sRHPgD0V9bTbIzR-1_D |
| 7/5/2019 | Improvements in simulation capabilities and necessary steps | Marty Marinak | https://drive.google.com/open?id=1J6tvacszMGe_1eH1zMbTU0gtKAvABUwe |
| 7/5/2019 | Developing a robust temperature diagnostic for HED hydrodynamics experiments | Mike MacDonald | https://drive.google.com/open?id=1h2qigN1y8_ONnD3MNl4mn8n7FSSU3egS |
| 7/7/2019 | Plasma Physics Challenges in Low Temperature Plasma Chemical Conversion for Environment, Biotechnology and Energy (v11 advocate update) | Mark Kushner | https://drive.google.com/open?id=1IHH0VITEtELjYCw2MULuMCGk8E80YUyJ |
| 7/8/2019 | Advanced Radio Frequency Source Technology and Development Center | Stephanie Diem | https://drive.google.com/open?id=1SkYQ-8jSHPREiFthBtRkPl4EjpcYApVt |
| 7/8/2019 | Science and Technologies That Would Advance High-Performance Direct-Drive Laser Fusion | Stephen Obenschain | https://drive.google.com/open?id=1FJeq8jHqDgF4IPYizIlaIk-GCJOQNeYW |
| 7/9/2019 | National Strategy for Liquid Metal PFC Research for Fusion | Daniel Andruczyk | https://drive.google.com/open?id=1040GQbRvun6-Xr7tCK7uBwg89R67I0eR |
| 7/10/2019 | National Initiative in Low Temperature Plasma | Philip Efthimion | https://drive.google.com/open?id=16Z4Z-akXvbPBPuCkM_U3ScbLBPue7pZq |
| 7/11/2019 | Astro2020 papers on inclusion, education, and research infrastructure | Nick Murphy | https://drive.google.com/open?id=1Qhwngm9c5JfFgE_xYiMEQXEHNWYC877m |
| 7/11/2019 | Inertial Fusion Energy Driver Technology | Constantin Haefner | https://drive.google.com/open?id=1yzxdC4a6bCneUx_dySHlxDPfqdcbuaGe |
| 7/12/2019 | Contrained relativistic HED experiments and simulations | Sasi Palaniyappan | https://drive.google.com/open?id=1-DVJGHuhYCMJ9SigiXowLDT3TI7mbkMi |
| 7/12/2019 | Pulsed power development needs: driver technology, workforce, and experimentally validated, multi-scale models & algorithms (seamless integration from PIC/kinetic to radiation-XMHD) | Ryan McBride | https://drive.google.com/open?id=1GN8A6g4xiBcbPs_glEm36K7EjbDxnPZB |
| 7/13/2019 | Benchmarking NLTE Atomic Physics of Open-Shell Ions in Plasmas | Marilyn Schneider | https://drive.google.com/open?id=1joBr7UNXvcc3ZBYFFgSLHw9t0B_lyF1p |
| 7/15/2019 | Advanced Radio Frequency Source Technology and Development Center | Stephanie Diem | https://drive.google.com/open?id=1HwhNptwUQ1_TCx5vIKdMb6gEXOVfK_Xy |
| 7/15/2019 | Quantum plasma | G. Collins | https://drive.google.com/open?id=1c2uvPX9VbA1nvE3HCqn6lilK_qjrOrz9 |
| 7/15/2019 | Full Participation in ITER: The US Fusion Community's First Opportunity to Study a Burning Plasma at High Gain | C.M. Greenfield | https://drive.google.com/open?id=1X16Pyq1U_zCO7NEzhWdIkSTCSqC5AhaP |
| 7/16/2019 | Diagnosing fusion burn using neutron spectrometry v2 | Maria Gatu Johnson | https://drive.google.com/open?id=13LbDJuY-vJpyrQCWy7AuidIbPa_Q9NBz |
| 7/16/2019 | Pulsed Power Science and Technology Development | Ryan McBride | https://drive.google.com/open?id=1l9Fp10ZqdX5cBFc9Oeum_iWqrqwJkrfV |
| 7/16/2019 | Low-Cost, Scalable Power Plants Based on Heavy Ion Fusion | Peter Seidl | https://drive.google.com/open?id=1helCQ4YL4fmR1Vo02KJKp9m9-VTTTIxk |
| 7/16/2019 | Collaborations On The SPARC Device | Martin Greenwald | https://drive.google.com/open?id=1CoajjWPJXk_3Ugtq2nK29MZb6aK7p-TG |
| 7/16/2019 | Benchmarking NLTE Atomic Physics of Open-Shell Ions in High-Temperature Plasmas | Marilyn Schneider | https://drive.google.com/open?id=1VzbyEyBCR6dWOM5yYJsK_DszfBEiJAIp |
| 7/16/2019 | Multi-Scale, Multi-Physics Plasma Hybrid Algorithms, Modeling, and Simulations | Michael Cuneo | https://drive.google.com/open?id=1u0N0z-tguG3KwUTrKBcDeKE-oeG1AlP6 |
| 7/17/2019 | Collisionless SOL initiative_v3 - Richard Majeski | Dick Majeski | https://drive.google.com/open?id=1H6TeCrMi0g0Zx70g1Eq8JKnJRAvUkt62 |
| 7/17/2019 | ECFS Final Statement on the NAS Report on a Strategic Plan for Burning Plasma Research | ECFS | https://drive.google.com/open?id=1P2xGVMeK9Wc7Qy2LCVnVqbVdse2utz9x |
| 7/18/2019 | Tokamak Capabilities Required for Relevant Boundary Model Validation | A.E. Jaervinen | https://drive.google.com/open?id=1wK3H6FUsmL6B4Lm7H6-3axTnJF8vDs9C |
| 7/18/2019 | Ultracold Neutral Plasmas for Controllable and Precision Plasma Physics | Jacob Roberts | https://drive.google.com/open?id=1FtclwWwjNkpn1fxOTNNAGIly95EhK3bu |
| 7/18/2019 | A tri-particle backlighter platform for precision radiography of fields and flows of high-energy-density plasmas | Chikang Li | https://drive.google.com/open?id=1ecvs3rQLh0-iKOGh_wmNopjML9zjP13b |
| 7/19/2019 | Mitigation of scrape-off layer power flow with lithium plasma-facing surfaces | R. Majeski | https://drive.google.com/open?id=1c2tvCGeFbjPHrvchMKxLLqcqqKUgYwTT |
| 7/19/2019 | Reversed-field pinch research toward Ohmic ignition at high engineering beta, v. 2 | Karsten McCollam | https://drive.google.com/open?id=1Ra0NzdOTMpo2xsKJyNVEQPMLKsil4lp1 |
| 7/19/2019 | The Centrifugally-Confined Mirror as a Pathway to Fusion | Ian Abel | https://drive.google.com/open?id=19sF74VCt-oZBaOVgk7oVoZ6R7U2UN2Od |
| 7/19/2019 | Enhance Physics Basis of NTMs -- to Predict, Avoid, Control, Suppress | Eric Howell | https://drive.google.com/open?id=1_3NQLDBhokZVEFU2oqRZ2wNiM3a-1LY0 |
| 7/21/2019 | Innovation Network for Fusion Energy - A Private-Public Partnership Opportunity | Ahmed Diallo | https://drive.google.com/open?id=1nmlUXvi-OKtIm96QcnpYKUPDAvUDTGp6 |
| 7/22/2019 | Advancing Fusion Energy with Predictive Theory-Based Models | Orso Meneghini | https://drive.google.com/open?id=1H6B4MDNL9zFNN5X2k_kTSJ5D9aEeVY8g |
| 7/22/2019 | Light sources from Laser-Plasma Accelerators | Jeroen van TIlborg | https://drive.google.com/open?id=1EFMKDGSexFbB6WLwpVvMmdVkwmRkLdyX |
| 7/24/2019 | Theory-based scaling of SHPD mission metrics | J.M. Park | https://drive.google.com/open?id=1z6qqcZRPAHKQ2AkNhyM8ceVYwZfMrzZS |
| 7/25/2019 | High-energy, high-brightness, compact, tunable X-Ray source based on moderate power laser system | Dmitri Kaganovich | https://drive.google.com/open?id=1RERgjWnYfdGcchq5_xoDVwGd4QsaCbKz |
| 9/11/2019 | IMPORTANCE OF THEORY, COMPUTATION AND PREDICTIVE MODELING IN THE US MAGNETIC FUSION ENERGY STRATEGIC PLAN | Fatima Ebrahimi | https://drive.google.com/open?id=1Z_9cSoLmxR5fCsmAafL1lGK0Nslrky3N |
| 9/17/2019 | Overview of FNST Gaps - White Paper | Richard E Nygrenb | https://drive.google.com/open?id=1fArmqy3fFdTQXS_aziOLwEo_oczKogwQ |
| 9/19/2019 | Joints for Superconducting Magnets – A Game Changer That Can Rank Confinement Concepts and Radically Change Remote Maintenance | Richard E Nygren | https://drive.google.com/open?id=1hE4hZLfPeHypu2Xs92SUxoGOvBVegCYz |
| 9/19/2019 | National Fusion Design Initiative, v2.0 | Hutch Neilson | https://drive.google.com/open?id=1vm_RdCyqu7in0kyTxdEjN1qP2Q8AlO8D |
| 9/23/2019 | Addendum to the Initiative White Paper Collaborations on the SPARC Device | Martin Greenwald | https://drive.google.com/open?id=1BD-RgaKWaW8PAK3SUrJOGD2QtxYRfDSp |
| 9/24/2019 | Construction of a Divertor Test Tokamak (DTT) as called for in the 2015 Plasma Material Interactions Workshop Report | Adam Kuang | https://drive.google.com/open?id=1PwrPgp-oe3ivWuQcY8zEDZQhyDD7mGb4 |
| 9/25/2019 | Computational predictive capability for fusion LM systems including LM-plasma coupling | Sergey Smolentsev | https://drive.google.com/open?id=1R2lmtpawVZOKrK6feW6EbeLbvJyJoLsb |
| 9/26/2019 | Pulsed tokamaks as an early power plant | Bob Mumgaard | https://drive.google.com/open?id=1Mf-PHHpTiUijPjrNvMYTEMbTKbf4w2K3 |
| 9/26/2019 | FINCH: The Fusion Integrated Nuclear Component Hall | Caroline Sorensen | https://drive.google.com/open?id=1FqKyTMQznvZXPXNK13Rigc0ViCHNjbB4 |
| 9/26/2019 | Tools for Today's Fusion Nuclear Science and Technology Development | Kevin Woller | https://drive.google.com/open?id=1ns3eHRn2EUBjBt85wINb1F6gKiKh2AZC |
| 9/26/2019 | PMI STUDIES IN DISPLACEMENT-DAMAGED FIRST WALL AND DIVERTOR TARGET MATERIALS | Matt Baldwin | https://drive.google.com/open?id=1a_JnGkHlfpreurYY2xIymL4OJfZqcwt6 |
| 9/26/2019 | Conferences on equity and inclusion for plasma science | Nick Murphy | https://drive.google.com/open?id=1ylI8nw5WHDMvNLCRM8yYzcpE4w5G2Uxa |
| 9/27/2019 | Revised version. Developing solid-material plasma facing components for fusion reactors with replenishable wall claddings and continuous surface conditioning | Peter Stangeby | https://drive.google.com/open?id=1XrxaMAMRKV3QBelKZPEfDA0_tDpdhSWz |
| 9/27/2019 | 5D and 6D Kinetic simulation of pedestal and scrape-off plasma dynamics on extreme scale computers | C.S. Chang | https://drive.google.com/open?id=1oj5elpv_TF70zD33rfEux6_TzisN4Uha |
| 9/27/2019 | Tokamak Energy: the spherical tokamak route to fusion power using high temperature superconductor magnets | Steven McNamara | https://drive.google.com/open?id=17H2k0hXXKQJ-7xwe0JZBLxyamZYKqTy8 |
| 9/27/2019 | Development of effective disruption mitigation solutions for reactor-grade devices | Nicholas Eidietis | https://drive.google.com/open?id=1UozQbhrh9NF3HjcBvXiWX2c6JvNXXDCI |
| 9/27/2019 | An Integrated Program to Develop Scalable Solenoid-Free Startup Scenarios for Spherical and Advanced Tokamaks | Michael W. Bongard | https://drive.google.com/open?id=1hWDLcUXT91W0GXVkQ8emyXaVErFWcax |
| 9/27/2019 | A proposal for an alternative fusion energy concept research and development program for the pursuit of economical fusion power | Derek Sutherland | https://drive.google.com/open?id=1gsB0MJEbziSfnTNz9j2oiYRnz-uifpEv |
| 9/27/2019 | An Alternatives Program for Magnetic Fusion | Ian Abel | https://drive.google.com/open?id=1HY52f-PwJ7f7d3fGiaAspf3JwgWPRKVz |
| 9/27/2019 | Abridged Pilot Plant Design Initiative | Hutch Neilson | https://drive.google.com/open?id=1wpzUvkXQWABpUNv9ZIzuYRaE79nOTiwC |
| 9/27/2019 | Development of effective disruption mitigation solutions for reactor-grade devices | Nicholas Eidietis | https://drive.google.com/open?id=1camBsJ80kmT19QTDvJLHq_3rHbW6YKQb |
| 9/27/2019 | A Comprehensive US Research Program to Accelerate the Path to Commercial Fusion Energy | CFS and the SPARC teams | https://drive.google.com/open?id=1DXad-JQfazHQIE84RVYdTxAjSa8VM2mX |
| 9/27/2019 | Demonstration of solenoid-free start-up of low inductance plasma for advanced ST or tokamak scenarios using transient Coaxial Helicity Injection | Roger Raman | https://drive.google.com/open?id=1u3hwPDnkcQdf00nu5rmqRAUBHOb064mN |
| 9/27/2019 | Development and validation of models for the divertor heat flux using kinetic first-principles codes | R.M. Churchill and C.S. Chang | https://drive.google.com/open?id=1_521213NOyqXb6SX6Nq1fVTMag6PtVR9 |
| 9/27/2019 | Summary of AI for Science Townhall meetings | C.S. Chang | https://drive.google.com/open?id=10KfgsOxDYbLvOozn-2EX0ZkR6fwg31c9 |
| 9/27/2019 | Advanced Radio Frequency Source Technology and Development Center (revised) | S.J. Diem | https://drive.google.com/open?id=1YJyEa1dBrzFtaIWACchMZUOLJWCaDe6V |
| 9/27/2019 | Utilizing Outreach Activities for Workforce Development via Private-Public Partnerships (updated) | S.J. Diem | https://drive.google.com/open?id=1y6DUcuLgaBxOcrfnTWOuNmZPo_Pprzxh |
| 9/27/2019 | The PlasmaPy Project: Building an Open Source Software Ecosystem for Plasma Research and Education (version 2) | Nick Murphy | https://drive.google.com/open?id=1ASf6PiOHIKAz1tY_GAZm8bFEj_hguf1t |
| 9/27/2019 | Reversed-field pinch research toward Ohmic ignition at high engineering beta, v. 3 | Karsten McCollam | https://drive.google.com/open?id=1tUv4iI5cX3gCo60TqL29DYZ4KCxBUAZE |
| 9/27/2019 | Negative Triangularity Tokamak | Max Austin | https://drive.google.com/open?id=14s3SgfSCFMteRYEGbBoaq8jJytdD_cuB |
| 9/27/2019 | Expand Capacity Computing - Revised | Brendan C. Lyons | https://drive.google.com/open?id=1ogzMdd8mTSKW6tKb703hHrMjJhoNEGeb |
| 9/27/2019 | Revised: IMPORTANCE OF THEORY, COMPUTATION AND PREDICTIVE MODELING IN THE US MAGNETIC FUSION ENERGY STRATEGIC PLAN | Fatima Ebrahimi | https://drive.google.com/open?id=1I7ZqDd07HL4EUrfIOR6d5MSdPm0Uu2LA |
| 9/27/2019 | High Power Microwave Power for ICF and MCF | Andrew Stan Podgorski | https://drive.google.com/open?id=1LBDGp5WJrszJ9At4P6_ge6ylygQxmy7 |
| 9/28/2019 | Supplement to DIII-D Initiative White Paper | David Hill | https://drive.google.com/open?id=1lLzVRygN5XtWHjV8g7yO4ineO0IwTwrp |
| 9/28/2019 | Reversed-field pinch research toward Ohmic ignition at high engineering beta, v. 3.1 | Karsten McCollam | https://drive.google.com/open?id=1cTvR5aBIJD2fLS-rPejiiuUPlDIM_vw2 |
| 9/30/2019 | NSTX-U in the 2020s - v2 and addendum | Devon Battaglia | https://drive.google.com/open?id=1kO6hl3TIL4z5vljCqZokdwVSs3jVo6vx |
| 9/30/2019 | Mitigation of the Impacts of Edge Localized Modes in Tokamaks | T.M. Wilks | https://drive.google.com/open?id=1N93KIBsfsXaYdkxR_1e53vMZ3UvpQcmW |
| 9/30/2019 | Closing gaps in PMI physics and technology in a mixed-material environment using short-pulse tokamaks | Tyler Abrams | https://drive.google.com/open?id=1SHrbcQ-eTx2jQhvnV1pf3P7q0Mw-8B-9 |
| 9/30/2019 | 5D and 6D Kinetic simulation of pedestal and scrape-off plasma dynamics on exascale scale computers, | C.S. Chang | https://drive.google.com/open?id=1o2qK2SMOh-un0kkpPsz2YBDBNY1MGk_W |
| 9/30/2019 | Liquid Lithium DIvertor V2 | Rob Goldston | https://drive.google.com/open?id=1db1DSrUaAYlOSuw2h2QIl8OGpIAsl5HV |
| 9/30/2019 | Outstanding Issues for Solid PFCs in Steady-state Toroidal Devices & A Proposed Long-Pulsed Facility to Tame the Plasma-Material Interface | Zeke Unterberg | https://drive.google.com/open?id=1fTWZg5g1PhJ3cUEuS7c0hiEr-YnNFiv9 |
| 10/1/2019 | Outstanding Issues for Solid PFCs in Steady-state Toroidal Devices & A Proposed Long-Pulsed Facility to Tame the Plasma-Material Interface - V2 | Zeke Unterberg | https://drive.google.com/open?id=1u9Wr1za6U6P8QBu7Z9nCNvaNoPDYYMeb |
| 10/1/2019 | Development and validation of models for the divertor heat flux using kinetic first-principles codes_v2 | R.M. Churchill and C.S. Chang | https://drive.google.com/open?id=13ela6P23g_yPoT_idB-AgdEcvmKIk3bL |
| 10/2/2019 | The PlasmaPy Project: Building an Open Source Software Ecosystem for Plasma Research and Education (version 3) | Nick Murphy | https://drive.google.com/open?id=1V6ywgCT-HSRP8A9h3yvN2e3r9y1iMMQk |
| 10/4/2019 | A U.S. Intermediate Scale Stellarator Experiment | Aaron Bader | https://drive.google.com/open?id=1WDf4Te7rb9hjgDDMQloKu_vvBPPxZIvJ |
| 10/5/2019 | Full U.S. Participation in ITER in the 2020s | Hutch Neilson | https://drive.google.com/open?id=10l-wznZwOzAWVB8F460dYJf4igdFyD1Q |
| 10/7/2019 | International stellarator research in support of a low capital cost pilot plant v2 | Novimir Pablant | https://drive.google.com/open?id=16dpqOj5O7i55Ol6gNxkI479WpqDwl8PC |
| 10/7/2019 | Revised: Construction of a Divertor Test Tokamak (US-DTT) as called for in the 2015 Plasma Material Interactions Workshop Report | Adam Kuang | https://drive.google.com/open?id=1lioVE-eazIZbow4-uVqW-ZlrNY5n5DQU |
| 10/7/2019 | Stellarator Path to a Pilot Plant | David Gates | https://drive.google.com/open?id=1vZn44oFKmc-VgOyA7fsww1Te7XE1jw6P |
| 10/7/2019 | Possible Mission and Viability of a Sustained High Power Density Facility (supplement to Buttery white paper) | Richard J Buttery | https://drive.google.com/open?id=1NKwgcpkcZGA2mbit0HZkcAgQQQwvK4z8 |
| 10/9/2019 | Collisionless SOL initiative_rev | Dick Majeski | https://drive.google.com/open?id=12PVNl4ZtiPFAjZSie59yYs2S7VEVEDot |
| 10/11/2019 | Engineering and Physics Mission for a SHPDX | Rob Goldston | https://drive.google.com/open?id=12RDhVVbeem4ITvlSKToNUeoicN-6I7f- |
| 10/13/2019 | Coupling Near-surface Processes and Bulk Material Performance via Large-scale Modeling and Simulation | Jaime Marian | https://drive.google.com/open?id=107qIB5l-ePZkU36z2vH20HWMFwUXE3Ya |
| 10/17/2019 | The cup of coffee challenge for fusion and the role of ITER - An addendum to the ITER initiative white papers | Raffi Nazikian | https://drive.google.com/open?id=1Tjhb1uScPCbjGNidmvZ1_Wh-LJWBO5IZ |
| 10/18/2019 | Overview of and Reasoning for a Stellarator Strategic Block in the U.S. Fusion Program | Oliver Schmitz | https://drive.google.com/open?id=14mUnyQx7U_N4cDn7nCdiJlJFN1ouhgiO |
| 10/18/2019 | Initiate Collaboration with QST on new Japanese tokamak JT60SA | Matthias Knolker | https://drive.google.com/open?id=1d0vinsN0-32aANLEX7S-nMcRZZn11VjO |
| 10/24/2019 | The Importance of Partnership Between National Facilities and 'Frontiers' Fundamental Science Programs | Richard Buttery | https://drive.google.com/open?id=1Lyrzj--goWND8S3gUMztptP8_xY9fAIN |
| 11/1/2019 | Plasma is turbulence, and turbulence is plasma | Ilya Dodin | https://drive.google.com/open?id=1Jg1sfMbsTbl5vDq1cNMArk_-prHmd6DV |
| 11/1/2019 | Magnetized HED platform for studies of electron and ion magnetization | Arijit Bose | https://drive.google.com/open?id=12GB5BMbVv1us1WsxNvYEQhn3wd1r61KL |
| 11/4/2019 | Non-Thermal Plasmas as Emerging Technologies | David Go | https://drive.google.com/open?id=1HhiSxkhnPXhCFjFgyaMaJCG0Qw4jGKnU |
| 11/4/2019 | The Plasma Genome Initiative | David Go | https://drive.google.com/open?id=1CqFM9lLBmNF6EWwqrO5ORv37ZwEjHZWO |
| 11/4/2019 | Regional Plasma Centers of Excellence | David Go | https://drive.google.com/open?id=13uwgeb0TtoE1UEahWPXa_Q-UeWa9GT3A |
| 11/4/2019 | Meta-stable and exotic material synthesis through exaggerated non-equilibrium plasmas and extreme synthesis conditions | David Staack | https://drive.google.com/open?id=15mlxbMNPndEgJ5VDlby0OrlYW-WcqSFh |
| 11/4/2019 | Short-Pulse, Time-Domain Microwave Power and Power Delivery System for Fusion Version 1 | Andrew S. Podgorski | https://drive.google.com/open?id=1oZXhFP2EIuNv065ECyPIoOmFSi_8EYpo |
| 11/5/2019 | Investigating Cross-Scale Coupling | Paul M. Bellan | https://drive.google.com/open?id=1K0lw5-r6JK7oil_VDLao81EhS84ekXf9 |
| 11/5/2019 | Major Scientific Challenges and Opportunities in Understanding Magnetic Reconnection and Related Explosive Phenomena in Magnetized Plasmas | Hantao Ji | https://drive.google.com/open?id=1SINslXfpvnUA27-nnN4C_sK4eRIN06wr |
| 11/6/2019 | Bulk Materials Synthesized by Nonequilibrium Plasma | Eiljah Thimsen | https://drive.google.com/open?id=1ir6SW9RA1vdA0oWZ7vlKGf5b5CikFM7s |
| 11/6/2019 | The role of plasma science in enabling new aerospace technologies | John E. Foser | https://drive.google.com/open?id=1y5TX9f-KVNyuSqmOACkozj5VOP7NJcT0 |
| 11/6/2019 | Material-Dependent Charging by Nonequilibrium Plasma | Elijah Thimsen | https://drive.google.com/open?id=1W7_vYWmH2eURULTLrqGRDZhB2HQYuEoF |
| 11/6/2019 | Enhanced Adsorption for Plasma Catalysis | Elijah Thimsen | https://drive.google.com/open?id=1TrlKZ8U-54q7-M_72Jbxp02aeiJZZK1- |
| 11/6/2019 | Self-Organization in Low Temperature Plasmas | Kentaro Hara | https://drive.google.com/open?id=1RxStq3J0Ob_RyjGQTt1Celf--pbr_jlm |
| 11/6/2019 | Advancing Computing Capabilities for Low Temperature Plasma Modeling | Kentaro Hara | https://drive.google.com/open?id=1ioMJpDMQaE9nRpbkXmZ1iS9Gg8EcJpk1 |
| 11/6/2019 | Long Life, High-Current Hollow Cathodes | John D. Williams | https://drive.google.com/open?id=1xZ1MwCwWSxLvQq8rNHQmMzDmxtYsMemU |
| 11/6/2019 | Cross Disciplinary Development of Advanced Plasma Diagnostics | Alexandros Gerakis | https://drive.google.com/open?id=1v8OFcxIrrsDaDD-loXaEuLETD8hUa67y |
| 11/6/2019 | Plasma-Assisted Combustion of Heavy Hydrocarbons at Elevated Pressures and Temperatures | Andrey Starikovskiy | https://drive.google.com/open?id=1ijC3P-wDyr_GHr6npp1DppTijywseoCy |
| 11/6/2019 | Liquid Plasma: Physics of electric breakdown of liquid dielectrics at sub nanosecond time scales | Andrey Starikovskiy | https://drive.google.com/open?id=105C_9H40cGs5kVfCQB_EiBkY3Lzw0hW5 |
| 11/7/2019 | Extreme Plasma Astrophysics | Dmitri Uzdensky | https://drive.google.com/open?id=1sUzpDqEe9jYLtgLavQFpAvbOJfT4BsRN |
| 11/8/2019 | Plasma Propulsion Research in Academia | Benjamin Jorns | https://drive.google.com/open?id=1g7gN9NlfFUDUY056u0bRbZVurLBhM6XM |
| 11/8/2019 | Plasmas for Novel Reconfigurable Electromagnetic Systems | Sergey Macheret | https://drive.google.com/open?id=1PPlt_d3KUwjEB5ucTGOga2RR7UTkh8vq |
| 11/8/2019 | Ultracold Neutral Plasmas: A Controllable Laboratory for Precise Plasma Physics | Jacob Roberts | https://drive.google.com/open?id=1I2fwNl5Spjt_tSVhg_VQy-sFeth5WC-6 |
| 11/8/2019 | The Material Properties of Weakly Collisional, High-Beta Plasmas | Matthew Kunz | https://drive.google.com/open?id=1pRIwl1IOUADcfWkgpXi6zkOabSATbTXL |
| 11/8/2019 | Exposure to Plasma Physics Theory at Undergraduate Institutions | Gregory Howes | https://drive.google.com/open?id=1IXXvlPaW8S8UqpxGEcjYTMNP2fQLciwW |
| 11/9/2019 | "Disruptive Technologies" in Plasma Sciences: Plasmas and the Control of Electromagnetic Waves. | Mark A. Cappelli | https://drive.google.com/open?id=1e2OPGqEpAx7fxMvlyxcCGOM-tqUmppnu |
| 11/11/2019 | Critical Need for a National Initiative in Low Temperature Plasma Research | Philip Efthimion | https://drive.google.com/open?id=1ILNIoEHKtVtTYiDbCQLiZuop5_y-v4-0 |
| 11/11/2019 | Plasma Physics Challenges in Low Temperature Plasma Chemical Conversion for Environment, Biotechnology and Energy | Mark J. Kushner | https://drive.google.com/open?id=1Aq-dMCO23HrlT6knf8xHchoVOVh9tX55 |
| 11/11/2019 | Plasma-NOW | Evdokiya Kostadinova | https://drive.google.com/open?id=1IS_5_WD7PsbWWoA_sfUS5i6okv4MkGGc |
| 11/11/2019 | Controlling charging in dusty plasmas - Revised 11-11-19 | Edward Thomas | https://drive.google.com/open?id=1Qti3hhiy35Y-fnLkRA-yPC2-ou1zOQmD |
| 11/11/2019 | Implementation model for Discovery Plasma Science | Edward Thomas | https://drive.google.com/open?id=1Rdb4BDkQW_ggJc0YL_z3e-sAFcRCMIGp |
| 11/13/2019 | Advancing the Predictive Understanding of Low-Temperature Plasmas | Benjamin T. Yee | https://drive.google.com/open?id=1zzWzNyV2MRDMsggjMFLG4PQF_AohQ036 |
| 11/13/2019 | Inertial Fusion Energy Systems: Repetitive Driver-Target Coupling in Hostile Fusion Chamber Environments | Michael Cuneo | https://drive.google.com/open?id=1s-nQeNJ4cEu780dfLYqmgs-OzlYOvspK |
| 11/17/2019 | Sheath Physics Initiative | Greg Severn | https://drive.google.com/open?id=1TDc4f_ETQ2KFitslLl2aYaIDj_tgou5A |
| 11/17/2019 | The Divertor Test Tokamak (DTT) facility | Piero Martin | https://drive.google.com/open?id=18Qx583ttfDOwag7irACTAH7Tpm1Md_kc |
| 11/18/2019 | USFusionEnergy.org: A public-facing URL for US Fusion | Arturo Dominguez | https://drive.google.com/open?id=1H8NimOM5LoBIMNmRHXhP-jzR3rYzyw8p |
| 11/20/2019 | Assessment and validation of plasma-molecule and atom-photon interaction detached divertor scenarios | Mathias Groth | https://drive.google.com/open?id=1DqQpcJqms3lwBSHO_5T-c5C4nACXtkLa |
| 11/21/2019 | Closing gaps in PMI physics and technology in a mixed-material environment using short-pulse tokamaks v2 | Tyler Abrams | https://drive.google.com/open?id=1AKeJpaOI0P5DHwBLdOl0V6e_fW_d3XuI |
| 11/25/2019 | Advocacy for an inertial fusion energy program within FES | Alex Zylstra | https://drive.google.com/open?id=18bmZ3OWaMP3_VyeYtLIu2ZnqT7Nr8adA |
| 11/26/2019 | Non-Equilibrium Plasma Applied to Biosciences | David Graves | https://drive.google.com/open?id=1O834I1r3Ky1gdhif6d67-7o_uq8E0u3Q |
| 11/27/2019 | Maximizing the Scientific Impact of Laboratory Experiments for Discovery Plasma Science | William Daughton | https://drive.google.com/open?id=1VSIhciD5lMh7FWNxekGTZ9nTJ9A3cuU1 |
| 11/29/2019 | Diagnosing fusion burn using neutron spectrometry v3 | David Schlossberg | https://drive.google.com/open?id=1rmAamZ5k3UC7ytX4xHyc8_lPDDipStNX |
| 11/29/2019 | High Fidelity Simulations of High Gain Cylindrical Pellets for Inertial Fusion Energy using a High-Energy Heavy-Ion Driver (HFSHG) | Alexander T. Burke | https://drive.google.com/open?id=1YEz4kju-RTjoztOxekYlRsPt1d9Zf5RU |
| 11/29/2019 | Cross Cutting Diagnostics: Leveraging expertise across the Entire plasma physics community | David Schlossberg | https://drive.google.com/open?id=1vnBQ9KvnfDotnp82gWKZrcr0249zAlb_ |
| 12/1/2019 | Cross Cutting Diagnostics: Leveraging expertise across the Entire plasma physics community, v2 | Dave Schlossberg | https://drive.google.com/open?id=1kJLevhNZst4FLLWN8xwBGypCK7YD-raC |
| 12/4/2019 | Challenges in the areas of plasma medicine and plasma agriculture | Alexander Fridman and Danil Dobrynin | https://drive.google.com/open?id=1N54EN-rYEGivBu3AxoOvS_RfChme_Ywd |
| 12/4/2019 | Turbulence and order in magnetized flowing plasmas | Fatima Ebrahimi | https://drive.google.com/open?id=1Q3IxpLfmUwNGzdaJq7dqjK6CGNXibWTa |
| 12/12/2019 | Priorities for Fusion Fuel Cycle Technology Development for a Fusion Pilot Plant | Dave Babineau | https://drive.google.com/open?id=15L71j3Ez3Wx-Y_blHOfa48ovNzzSvdrs |
| 1/13/2020 | Letter of Support for the Inclusion of Alternative Magnetic and Magneto-Inertial Fusion Energy R&D in the DOE-FES Portfolio | Derek A. Sutherland | https://drive.google.com/open?id=1C-KYwsDVj7eIJ25dtdzUsVPiF-CyU4Wx |
| 2/27/2020 | Workforce Development - Supplemental Appendix | Workforce Development Cross-Cut Team | https://drive.google.com/open?id=1Kaw2DKTQPFgNxNDjtoMenaGqEFByq5NU |